\documentstyle[psfig,doublespace,vustyle]{report}
\def\gsim{\stackrel{>}{\sim}}
\def\lsim{\stackrel{<}{\sim}}
\def\beq{\begin{equation}}
\def\eeq{\end{equation}}
\def\bea{\begin{eqnarray}}
\def\eea{\end{eqnarray}}
\def\ba{\begin{array}}
\def\ea{\end{array}}
\def\ol{\overline}

\def\djdot{{\mbox{\boldmath $\cdot$}}}

\def\gupo{{\gamma^0}}
\def\gupa{{\gamma^1}}
\def\gupb{{\gamma^2}}
\def\gupc{{\gamma^3}}
\def\gupu{{\gamma^{\mu}}}
\def\gupv{{\gamma^{\nu}}}
\def\gupaa{{\gamma^a}}

\def\gdno{{\gamma_0}}

\def\gdnf{{\gamma_5}}

\def\gdnu{{\gamma_{\mu}}}

\def\lag{{\cal L}}
\def\ab{{\alpha \beta}}
\def\acb{{\alpha, \beta}}

\def\nuL{{\nu_L}}
\def\NR{{N_R}}
\def\nuRc{{\nu^{\!\!\!\!\mbox{\tiny{c}}}_R}}
\def\NLc{{N^{\!\!\!\!\mbox{\tiny{c}}}_L}}
\newcommand{\smc}[1]{{#1}^{\!\!\!\!\mbox{\tiny{c}}}}

\newcommand{\smp}[1]{{#1}^{\!\!\!\!\mbox{\tiny{p}}}}
\newcommand{\smcp}[1]{{#1}^{\!\!\!\!\mbox{\tiny{cp}}}}
\newcommand{\smcpt}[1]{{#1}^{\!\!\!\!\mbox{\tiny{cpt}}}}
\def\nuaRc{{\smc{\nu}_{\alpha R}}}
\def\NaLc{{\smc{N}_{\alpha L}}}

\def\half{{\frac{1}{2}}}
\def\fourth{{\frac{1}{4}}}
\def\openone{\leavevmode\!\!\!\!\hbox{\small1\kern-1.55ex\normalsize1}}

\def\nue{{\nu_{e}}}
\def\numu{{\nu_{\mu}}}
\def\nutau{{\nu_{\tau}}}
\def\nui{{\nu_{i}}}

\def\nua{{\nu_\alpha}}
\def\nub{{\nu_\beta}}
\def\nug{{\nu_\gamma}}

\def\Pab{{P\makebox[9 mm][r]{\raisebox{-1.5ex}{{\scriptsize{$\nua \rightarrow
\nub$}}}}} \!\!\!\!\!\!\!\!\!\!\!(x)\;}
\def\Pmab{{P^{\, M}\!\!\!\makebox[9 mm][r]{\raisebox{-1.5ex}{{\scriptsize{$\nua \rightarrow
\nub$}}}}} \!\!\!\!\!\!\!\!\!\!\!\!(x)\;}
\def\Paa{{P\makebox[9 mm][r]{\raisebox{-1.5ex}{{\scriptsize{$\nua \rightarrow
\nua$}}}}} \!\!\!\!\!\!\!\!\!\!\!(x)\;}
\def\Pba{{P\makebox[9 mm][r]{\raisebox{-1.5ex}{{\scriptsize{$\nub \rightarrow
\nua$}}}}} \!\!\!\!\!\!\!\!\!(x)\;}
\def\Peu{{P\makebox[8 mm][r]{\raisebox{-1.5ex}{{\scriptsize{$\nue \rightarrow
\numu$}}}}} \!\!\!\!\!\!\!\!\!(x)\;}
\def\Put{{P\makebox[8 mm][r]{\raisebox{-1.5ex}{{\scriptsize{$\numu \rightarrow
\nutau$}}}}} \!\!\!\!\!\!\!\!\!(x)\;}
\def\Pet{{P\makebox[8 mm][r]{\raisebox{-1.5ex}{{\scriptsize{$\nue \rightarrow
\nutau$}}}}} \!\!\!\!\!\!\!\!\!(x)\;}
\def\Pucec{{P\!\!\!\makebox[9 mm][r]{\raisebox{-0.8ex}
  {{\tiny{$\smc{\nu}_{\mu} \rightarrow \smc{\nu}_e $}}}}} }
\newcommand{\Pnu}[2]{{P\!\!\!\!\makebox[9 mm][r]{\raisebox{-0.8ex}
  {{\tiny{$\nu_{#1} \!\!\rightarrow \!\!\nu_{#2}$}}}}}}

\newcommand{\B}[2]{\;^{#1}\Box_{#2}\;}
\newcommand{\Bs}[2]{\;^{#1}\Box_{#2}^*\;}
\newcommand{\Baibj}{\B{\alpha i}{\beta j}}
\newcommand{\Bsaibj}{\Bs{\alpha i}{\beta j}}
\newcommand{\R}[2]{\,^{#1}\!\!\!\!\mbox{\footnotesize{R}}_{#2}}
\newcommand{\J}[2]{\,^{#1}\!\!\!\!\mbox{\footnotesize{J}}_{#2}}

\def\Pij{{\Phi_{ij}}}

\def\Pe{{\Phi_{23}}}
\def\Pu{{\Phi_{13}}}
\def\Pt{{\Phi_{12}}}

\newcommand {\V}[1] {V_{#1}}
\newcommand {\Vs}[1] {V^*_{#1}}
\def\Vai{{V_{\alpha i}}}
\def\Vajs{{V_{\alpha j}^*}}
\def\Vbj{{V_{\beta j}}}
\def\Vbis{{V_{\beta i}^*}}

\newcommand {\Vm} {V^M}

\def\Vmai{{\Vm_{\alpha i}}}
\def\Vmajs{{V^{M*}_{\alpha j}}}
\def\Vmbj{{\Vm_{\beta j}}}
\def\Vmbis{{V^{M*}_{\beta i}}}

\newcommand {\co}[1] {c_{#1}}
\newcommand {\s}[1] {s_{#1}}
\newcommand {\ct}[1] {c^2_{#1}}
\newcommand {\st} [1] {s^2_{#1}}
\def\sp{{\hspace{1.in}}}
\def\hsp{{\hspace{1.3in}}}
\def\ssp{\hspace{0.3 in}}
\def\vsp{\vspace{0.5 cm}}

\def\tMnu{\tilde{M}_{\nu}}
\def\tMl{\tilde{M}_l}
\def\tM{\tilde{M}}
\begin{document}
\author{Doris Jeanne Wagner}
\title{Connecting Experiment with Theory:  A Model-Independent \\
Parameterization of Neutrino Oscillations}
\dept{Physics}   
\advisor{Professor Thomas J. Weiler}       
\phdthesis          
\maketitlepage      
\chapter*{Acknowledgements}

This dissertation is not due solely to my efforts.  Its beginnings lie in a
high school physics class taught by Yvonne Waters, one of the most energetic
and inspiring teachers I have encountered.  Were it not for her obvious love
and enthusiasm for the subject, I would never have enrolled in a physics
course in college.  As my interest in the subject grew, it was nurtured by the
wonderful physics faculty at the College of William and Mary, in particular, my
undergraduate advisor, Marc Sher.  At Vanderbilt, 
my research has been shaped and guided by 
my advisor, Tom ``Prof'' Weiler,  whose
insight, advice, and ``Seinfeld'' watching helped bring this thesis
to its present state. 
My graduate career has been greatly
enriched by his support and guidance, and I am deeply appreciative of the
assistance he has provided.  I would also like to thank 
Professors Webster and Kephart for the many 
times they have helped me straighten out a perplexing problem.  The rest of my
committee members, Professors Greiner, Waters, and Ernst, have all supported
me and my work in many different ways.  I would like to thank each one of the
six gentlemen on my committee for their guidance and the faith in me and my
abilities which they have repeatedly displayed.  Many other Vanderbilt faculty,
including Tonnis ter Veldhuis and Professors Pinkston, Brau, Arenstorf, 
and Umar, have
helped me professionally and personally, and I am quite grateful for the role
each has played in bringing me to this point.

My wonderful friends who shared my graduate experience have helped me keep my
sanity and sense of perspective.  Katrina Wagner has stood by my side through
ups and downs too numerous to count.  I would like to thank her for keeping me
honest while solving physics problems, eating chocolate with me when I got
frustrated, giving me advice on oh so many papers and talks, and just being a
fantastic friend.  I would like to thank Alan Calder for his
friendship, computer advice, company studying for the qualifier, and 
perspective on gun control laws.  He too provided moral
support during stressful situations, and I am very grateful.  I 
appreciate greatly the assistance supplied by Russ Kegley. 
Russ was never too busy to help others with homework problems, computer
problems, research problems, or personal problems, and many of us at
Vanderbilt are indebted to him.  I would like to thank all of my other friends
who supported and encouraged me through my graduate career:  Don Lamb, Jerri
Tribble, Vijaya Sankaran, Tom Ginter, Lei Shan, Sung-Hye Park, Julie
Wendt, and the members of the young
adult Sunday School classes at Belmont United Methodist Church. 

I have
also received much encouragement from my home church family, namely the Sharing
Group and the entire congregation of St.\ Luke's United Methodist Church.  I know
I have been in their thoughts and prayers throughout my twenty-two years of
school, and I have frequently been encouraged by that knowledge.  
God has blessed me with so much, and I thank
God for all of the gifts I have received.  In particular, 
I thank God for giving me the abilities and perseverance necessary 
to complete this dissertation, as well as giving me all these friends to support and
encourage my endeavors.

I wish to thank my brothers,
their wives, and my grandparents for their love and encouragement.  And I would
like to give special thanks to
my nephews and nieces:  Frederick, Bethany, Megan, and Richard Wagner. 
Knowing that these children look up to me has encouraged me to reach my goals and
be the best role model I can.  I thank them not only for the sense of purpose
they provide, but for the joy and love they constantly bring to me.  Lastly,
but certainly not leastly, I thank my parents.
I would not be where, or who, I am were it not for my parents, two of the
most intelligent, loving, and Christian 
people I know.  I cannot express how much I appreciate 
their unconditional support for everything I have
done.  They wisely held their tongues, even when I thought I wanted to be an
accountant, and let me choose my own course.  I therefore dedicate my thesis to
them in honor of the support and inspiration they constantly provide.
\begin{abstract}

Many experiments are currently looking for evidence of neutrino mass in the form
of neutrino oscillations.  Oscillation probabilities are non-linear functions
of the neutrino mixing matrix elements, so most comparisons of data to theory
are based on simplifying models of the mixing matrix.  We begin this
dissertation with a review of neutrino interactions 
and a few of the popular models
describing neutrino masses and mixing.  Next we present our model-independent
description of neutrino oscillations and derive the predictions of various
models in terms of our new ``box'' 
parameterization.  Finally, we use our boxes to
find mixing matrices consistent with existing neutrino data.  As more definitive
data becomes available, these solutions will probably need to be adjusted; when
such a need arises, our box notation will provide a convenient method for
finding new solutions.

\end{abstract}
\thispagestyle{empty}
\setstretch{1.7}


\chapter{Introduction}


\begin{verse}
\begin{it}
\sp Neutrinos, they are very small. \\
\hsp They have no charge and have no mass \\
\sp And do not interact at all. \\
\sp The earth is just a silly ball \\
\hsp To them, through which they simply pass, \\
\sp Like dustmaids down a drafty hall \\
\hsp Or photons through a sheet of glass.\\
\hsp They snub the most exquisite gas, \\
\sp Ignore the most substantial wall,\\
\hsp Cold-shoulder steel and sounding brass, \\
\sp Insult the stallion in his stall, \\
\hsp And, scorning barriers of class, \\
\sp Infiltrate you and me!  Like tall \\
\sp And painless guillotines, they fall \\
\hsp Down through our heads into the grass. \\
\sp At night, they enter at Nepal \\
\hsp And pierce the lover and his lass \\
\sp From underneath the bed -- you call \\
\hsp It wonderful; I call it crass. \\
\end{it}

\hspace{2.0in} - COSMIC GALL by John Updike \footnote{From COLLECTED POEMS
(Alfred A. Knopf).  Reprinted by permission; \copyright 1960 John
Updike. Originally in {\it The New Yorker}.  All rights reserved.}
\end{verse}

Updike's description of the neutrino, while quite poetic, could use a bit of
updating to be scientifically accurate.  Neutrinos do travel great distances
through matter before interacting, but to say they ``do not interact at all''
is incorrect.  Bernstein, the author of
reference \cite{bern}, also considers Updike's poem and 
suggests ```And do not
interact a lot . . .' is better science but worse poetry.''  But it is the
claim of the poem's previous line that neutrinos ``have no mass'' which is addressed
by my research.
Bernstein happens to agree here with the poet, claiming
that ``most physicists would be willing to give high betting odds that the mass
is exactly zero.''  Bernstein and Updike, however, were writing in the 1960s,
and a few decades can change the prevailing views in science.  No
concrete experimental evidence of non-zero mass has yet been found, but the
existence of massive neutrinos is currently favored on theoretical and indirect
experimental grounds.  In view of the prevailing scientific opinion, I suggest
the following replacement for the first three lines of Updike's poem:

\begin{verse}
\begin{it}
\sp Neutrinos, they are very small. \\
\hsp They have no charge and not much mass \\
\sp And barely interact at all. \\
\end{it}
\end{verse}

This dissertation will present some of the consequences of
non-zero neutrino mass and will develop a mechanism to connect the theoretical
models with the potential results of experiments.


\section{Some History}


The early 20th century was a time of great change for physics.  Driving much of
that change was the discovery and study of nuclear decays.  Beta decay, in
which an atom changes atomic number and emits an electron, provided some puzzling
results.  The decrease in the atom's energy should be equal to the energy
carried away by the electron.  But not only was the energy of the emitted electron     
in a particular reaction usually less than the expected amount, it
was different for each occurrence of the same reaction.  Energy seemed to be
randomly vanishing in beta decay!  
In 1929, Pauli suggested the existence of an undetected particle which carried
away the missing energy \cite{Pauli}, \cite{Pais}.
In retrospect, Pauli's solution may seem obvious.  But physicists in the
1920s were not yet in the habit of postulating the existence of new particles
to explain surprising results.  Only the proton, the electron, and the 
photon had been
discovered.  The prejudice against new particles was so strong, Bohr preferred
to discard energy conservation.  He suggested that individual reactions need
not conserve energy as long as energy was conserved on the average.  Pauli
hesitantly proposed his ``neutron'' as a possible solution in 1927, and
Fermi renamed it the neutrino when Chadwick discovered what
we now call the neutron \cite{Pais}.

The existence of a chargeless, weakly-interacting
fermion was generally accepted by the mid-1930s, but 
experimental confirmation of the neutrino's existence did not arrive until the
1950s \cite{Reines}, \cite{Pais}.  
Reines eventually shared the 1995 Nobel Prize in physics for this
innovative experiment.  Yet these elusive particles have continued 
continued to raise questions in the decades
following their detection.  Some of the questions which intrigue researchers 
today, such as
whether neutrinos and antineutrinos are distinct particles, have reemerged many
years after they appeared resolved.  Other questions, such as the number of 	
neutrino flavors in nature, and whether the neutrinos are massless, have never
really gone away. 

According to Pauli's proposal, the neutrino and the electron share the energy  
lost by an atom in beta decay.  The minimum energy given to the electron is
proportional to its mass, and the minimum energy carried off by the neutrino
should yield a measurement of the neutrino mass.  But in some of the decays,
the electron seemed to carry off all of the energy, leaving none for the
neutrino.  Thus Perrin commented in 1933 ``that the neutrino has {\it zero
intrinsic mass}, like the photon'' \cite{Perrin}, 
\cite{Pais}.  Yet experiments cannot
demonstrate that the mass of a particle (even the photon) is precisely zero. 
They may merely place upper bounds on the mass.  The bounds on the mass of
Pauli's ``neutron'' (known now as the electron antineutrino) from beta decay
currently require this mass to be more than 50000 times smaller than the mass of
an electron, or less than 10 eV.  Fifty thousand times smaller than the electron
 mass may
sound negligibly small, 
but the existence of even a tiny non-zero neutrino mass has profound
effects. Nevertheless, a prevailing belief in massless
neutrinos continued well into the 1960s, when the Standard Model of particle
physics was developed.  This enormously successful theory describing 
particle interactions requires neutrino masses to be exactly zero.

Starting in 1968 with the Homestake solar neutrino 
experiment \cite{Homestake}, evidence has been
building in favor of non-zero neutrino masses.  The Homestake experiment counts
the neutrinos coming from the sun, and the number it detects is far smaller than
the number predicted by solar theory.  This ``solar neutrino deficit" has since been
confirmed by four other experiments \cite{solardat}.  
In addition, experiments measuring
neutrinos produced in our atmosphere find an ``anomaly" in the ratio of
muon-type to electron-type neutrinos \cite{atmosdat}.  
Both of these results could be explained
by neutrino mass through its quantum-mechanical implications.  
As for the Standard Model, it is not the ``Theory of
Everything" physicists view as their Holy Grail.  Adding masses to
neutrinos is a natural extension to the theory, requiring merely additional
particle fields, without significant formal
changes.  These minimal changes to the theory, however, have profound
physical results.
As will be discussed below, non-zero neutrino mass may lead to mixing between
different flavors of neutrinos and to oscillations of neutrino flavor.  
The mixings and oscillations
resulting from neutrino mass could explain the currently puzzling solar 
neutrino deficit and atmospheric anomaly.  Large-scale neutrino experiments
hope to address these
issues, and my work provides a model-independent path between the data from
such experiments and the answers to the mysteries surrounding neutrinos.


\section{Whence We Start:  A Review of the Standard Model}
\label{SMsec}


The Standard Model describes the interactions of quarks, leptons, gauge fields, 
and a Higgs scalar.  Quarks come in three generations of two quarks each: up and 
down, strange and charm, and top and bottom. Leptons too come in three
generations: electron, muon, and tau.  Each generation contains
a negatively charged lepton and its associated neutral neutrino.  Each quark
and lepton has an associated antiparticle as well.  The gauge
fields include the photon, carrier of the electromagnetic force, the three
vector bosons $(W^+, W^-, Z^0)$, carriers of the weak nuclear force, and eight
colored gluons, carriers of the strong nuclear force.  Of these particles, only
the Higgs scalar and the tau neutrino have yet to be detected by experiments.

Experiments show that only the left-handed particles and right-handed antiparticles 
are affected by the charged weak nuclear force 
(see Appendix \ref{handsec} for a discussion on handedness, or helicity).  
This ``maximal parity violation'' by the charged weak interaction was
discovered in 1957 by Wu, {\it et.\ al.\ }\cite{CSWu}, 
\cite{Grein5}, and it necessitates an 
asymmetry between right-handed and left-handed particles in the Standard Model.
Left-handed particles, which participate in charged weak interactions, 
are placed in ``isospin
doublets.'' Right-handed particles, which do not experience the charged 
weak force, compose ``isospin singlets.''

The Lagrangian of a theory provides information on particle interactions.  Each
term in the Lagrangian describes either a kinetic energy, a mass, or an 
interaction.  Fermion terms contain
at least one field with Dirac conjugation $\ol{\psi}$ (defined in
Appendix~\ref{key}), and one without,
$\psi'$.  The conjugated fields represent outgoing particles or incoming
antiparticles; the unconjugated fields represent incoming particles or
outgoing antiparticles.  Each term in the Lagrangian must follow the
known conservation laws.  For example, electric charge seems to be conserved in
nature, so the total incoming charge of any term in the Lagrangian should equal
the outgoing charge.
Isospin and its component in a chosen direction $(I_3)$ must be conserved, so
the field with the positive $I_3$ in an
incoming isospin doublet (isospin $\half$) must couple
to the field with the positive $I_3$ in 
an outgoing isospin doublet.  Fields in an incoming triplet (isospin $1$) can 
couple to fields of another triplet or to two doublet fields.  
Singlets have isospin $0$ and thus can 
couple to any term without affecting the isospin of that term.

Many of the sources listed in the bibliography of this paper contain 
reviews of the Standard Model.  References \cite{Grein5} and \cite{Nact} treat
the Model in great detail, while references \cite{MP}, \cite{KimPev},
and \cite{BP} consider only the aspects of the Model relevant to neutrinos.  We
will synthesize these treatments in the summary which follows.

Neutrinos do not 
interact by either the strong nuclear force or the
electromagnetic force, so we will ignore the terms in the Lagrangian
corresponding to those forces.  
The terms of the Standard Model  Lagrangian that are of interest to
neutrino physics may be represented as follows: 
\bea
\label{Lnu}
\lag^{\nu}_{SM} & = & 
\sum_{\alpha} \ol{\nua_{L}} i \gupu \partial_{\mu} \nua_L 
+ \sum_{\alpha} \ol{\alpha_L} i \gupu \partial_{\mu} \alpha_L
+ \sum_{\alpha} \ol{\alpha_R} i \gupu \partial_{\mu} \alpha_R 
- \sum_{\alpha} \ol{\alpha_L} m_{\alpha} \alpha_R
- \ \sum_{\alpha} \ol{\alpha_R} m_{\alpha} \alpha_L \nonumber \\
&& - \frac{e}{\co{W}\s{W}}{\bf Z}_{\mu}{\cal J}^{\mu}_{NC} - \frac{e}{\sqrt{2}
\s{W}} \left( {\bf W}^{+}_{\mu} {\cal J}^{\mu}_{CC} + {\bf W}^{-}_{\mu} {\cal
J}^{\mu \dagger}_{CC} \right), \mbox{ where} \nonumber \\
{\cal J}^{\mu}_{NC} & = & \sum_{\alpha} \left[ 
\half \ol{\nua_L} \gupu \nua_L - \half \ol{\alpha_L} \gupu \alpha_L 
+ \sin^2 \theta_W \left( \ol{\alpha_L} \gupu \alpha_L 
   + \ol{\alpha_R} \gupu \alpha_R \right) \right] \\
  & = &
\half \ol{\nuL} \gupu \nuL - \left(\half-\sin^2 \theta_W \right) \ol{l_L} \gupu l_L
  + \sin^2 \theta_W \ol{l_R} \gupu l_R ,
\mbox{ and} \nonumber \\
{\cal J}^{\mu}_{CC} & = &  \sum_{\alpha} \ol{\nua_{L}} \gupu \alpha_{L} =
\ol{\nuL} \gupu l_L.
\nonumber 
\eea
Here $\alpha_{L}$ is the field operator spinor (expressed in full in
Appendix~\ref{wvfcnsec}) for the charged lepton for the 
neutrino spinor $\nua_{L}$, and $\nuL$ and $l_L$ are the vectors containing all
$\nua_L$ and $\alpha_L$, respectively. The subscripts $_R$ and $_L$ indicated
left- and right-handed particles, respectively (defined in
Appendix~\ref{handsec}).  ${\bf Z}_{\mu}, {\bf W}^{+}_{\mu}$, and 
${\bf W}^{-}_{\mu}$ are the fields of the vector bosons.  $e$ is the
dimensionless charge of the electron, which equals the square root of the 
fine structure constant, and $\theta_W$ is the
Weinberg angle:
\beq
\cos^2 \theta_W = \frac{m_W^2}{m_Z^2}.
\label{thetaW}
\eeq
Appendices~\ref{key} and \ref{gammapp}
contain a list of conventions used in this paper
and descriptions of the gamma-matrices.

The Standard Model 
Lagrangian contains mass terms for the charged leptons, but not for the
neutrinos.  Mass terms occur when right-handed particles couple to their
left-handed partners, giving rise to terms of the form
\beq
\ol{\psi_L} m_{\psi} \psi_R,
\eeq
where $m_{\psi}$ is the mass of the fermion field $\psi = \psi_L + \psi_R$. 
In such a term, the outgoing left-handed field has 
isospin $\half$, and the incoming right-handed term has isospin $0$.  
We know most fermions have a mass, so any physical theory must
include mass terms for electrons, quarks, etc., but these terms seem to violate
isospin conservation.  Higgs \cite{Higgs}, drawing from the work of
Goldstone, Guralnik, Hagen, Kibble, Lange, and many others \cite{Higgshist}, 
finally came up
with a solution to this apparent paradox.  Higgs' work was incorporated
into the Standard Model by Weinberg.  

In the Standard 
Model's ``Higgs mechanism,'' particles obtain masses
by interacting with a Higgs field, introduced
to conserve isospin:
\beq
\lag_{\phi l} = \sum_{\alpha, \beta} 
    g^D_{\ab} \ol{\psi_{\alpha L}} \phi_D \beta_R + h.c.
\label{fullHiggs}
\eeq
$\phi_D$ is the Higgs doublet, responsible for giving charged leptons and down-type
quarks their masses:
\beq
\phi_D = \left( \ba{c} \phi_D^- \\ \phi_D^0 \ea \right),
\label{phiD}
\eeq
$\psi_{\alpha L}$ is the left-handed lepton doublet of flavor $\alpha$:
\beq
\psi_{\alpha L} = \left( \ba{c} \nua_L \\ \alpha_L \ea \right),
\label{psiaL}
\eeq
$\beta_R$ is the right-handed charged lepton of flavor $\beta$, 
and the $g^D_{\ab}$ are coupling constants.  Up-type quarks obtain masses in 
the Standard Model by coupling with the conjugate
Higgs doublet \cite{BP}
\beq
\tilde{\phi}_D \equiv i \sigma_2 \phi_D^* = 
  \left( \ba{c} \phi_D^{0*} \\ -\phi_D^{+*} \ea \right),
\label{HiggsDup}
\eeq
where $\sigma_2$ is the second Pauli matrix as given in Appendix~\ref{gammapp}.

The neutral Higgs field has all the quantum numbers of the vacuum, so it may 
obtain a ``vacuum expectation value,'' or VEV:
\beq
\langle \phi_D^0 \rangle = \langle 0 | \phi_D^0 | 0 \rangle.
\eeq
The charged Higgs field cannot obtain a VEV since the vacuum is electrically
neutral, so the ground state of the Higgs
doublet is
\beq
\langle \phi_D \rangle = \left( \ba{c} 0 \\ \langle \phi_D^0 \rangle \ea
\right).
\eeq
The Higgs field is dynamical, so we
include a term representing the deviation from the vacuum value
and arrive at our final form for the Higgs doublet:
\beq
\phi_D^{SM} = \left( \ba{c} 0 \\ 
   \langle \phi_D^0 \rangle + \frac{H^0(x)}{\sqrt{2}} \ea \right).
\eeq
$H^0(x)$ is the massive Higgs scalar, and
the VEV $\langle \phi_D^0 \rangle$ gives mass to particles through interactions
such as
\beq
g^D_{\ab} \ol{\alpha_L} \langle \phi_D^0 \rangle \beta_R \equiv 
\ol{\alpha_L} M_{\ab} \beta_R.
\eeq
Because these terms involving the VEV do not conserve isospin, but the original
interactions in equation (\ref{fullHiggs}) do, the Higgs mechanism {\it
spontaneously breaks} the SU(2) symmetry in the ground state.  The extra fields
present in the original Higgs doublet (\ref{phiD}) do not simply vanish in the
symmetry breaking; they are transformed into extra spin states of the vector
bosons.  Such a discussion is irrelevant to the focus of this paper, but the
interested reader can consult the sources \cite{MS} or \cite{Itzykson} 
for more information.

In the Standard Model, the charged lepton mass matrix is diagonal, 
$M_{\ab} = m_{\alpha} \delta_{\ab}$, and mass terms for neutrinos do not arise
because the Model has no right-handed
singlet neutrino fields to participate in the coupling of equation
(\ref{fullHiggs}).  We will examine the consequences of the addition of
a neutrino mass term in Chapter~\ref{intro2}.

The Lagrangian in equation (\ref{Lnu}) does not explicitly display 
terms involving the right-handed antineutrino
fields $\nuRc \equiv C 
\ol{\nuL}^T$, where $C$ is
the charge-conjugation operator, discussed in
Appendix~\ref{Csec}.  These fields are
CP-conjugates of the neutrino fields $\nuL$.\footnote{
The term antineutrino can be a bit confusing, since the
charge-conjugate neutrino field is different from the physical CPT-conjugate 
antineutrino field.  These subtleties are discussed in Appendix~\ref{Csec}
}  
The Lagrangian does, however, contain antiparticle fields
implicitly.
$\lag^{\nu}_{SM}$ above is CP-invariant, so if you replace
the particles in equation (\ref{Lnu}) with their CP-conjugates, you get the same
terms as you started with.  Take, for example, a kinetic term for
antineutrinos $\ol{\nuRc} \gupu \partial_{\mu} \nuRc$: 
\beq
\ol{\nuRc} \gupu \partial_{\mu} \nuRc = \ol{C \ol{\nuL}^T } 
\gupu \partial_{\mu} C \ol{\nuL}^T = \left( \gupo
\nuL \right)^T C^{\dagger} \gupo \gupu C \partial_{\mu}  \ol{\nuL}^T.
\label{antiL}
\eeq
The term in equation (\ref{antiL}) is a number, so it is equal to its transpose:
\beq
\left( \gupo \nuL \right)^T C^{\dagger} \gupo \gupu \partial_{\mu} C 
\ol{\nuL}^T = -(\partial_{\mu} \ol{\nuL}) C \gupu^T \gupo C^{-1} \gupo
\nuL,
\eeq
where we have added the minus sign because we switched the order of two
fermion fields $\nuL$ and $\nuL^{\dagger}$,\footnote{
When taking the transpose
of a product of fermion fields, it is necessary to account for the
anticommutation of those fields, so $(\psi_1 \psi_2)^T = -\psi_2^T \psi_1^T$ if
$\psi_1$ and $\psi_2$ are fermions.  When taking the hermitian conjugate,
however, no minus sign appears: $(\psi_1 \psi_2)^{\dagger}=\psi_2^{\dagger}
\psi_1^{\dagger}$.  The hermitian conjugate of a product is defined by the
previous expression, and the anticommutator is never taken.  Consider the
number operator $N=a^{\dagger}a$ from quantum mechanics, which must be
hermitian since it is observable:  $N^{\dagger}=N$, so 
$(a^{\dagger}a)^{\dagger}$ must equal $a^{\dagger}a$, and not
$-a^{\dagger}a$.
} 
and we have used the relations (\ref{Cprop}) from Appendix \ref{Csec}.  
The operator $C$ anticommutes with $\gupo$, and $C \gupu^T C^{-1} = 
-\gupu$, so
\beq
\ol{\nuRc} \gupu \partial_{\mu} \nuRc = -(\partial_{\mu} \ol{\nuL}) \gupu \nuL.
\eeq
The last step is to bring the partial derivative $\partial_{\mu}$ to the right
of the spinor $\ol{\nuL}$.  We may do this by remembering that $\lag^{\nu}_{SM}$
is a Lagrangian {\it density} and thus always appears in an integral.  We may
therefore use the technique of integrating by parts to obtain
\beq
-(\partial_{\mu} \ol{\nuL}) \gupu \nuL = -(\partial_{\mu} \ol{\nuL} \gupu \nuL)
+ \ol{\nuL} \gupu \partial_{\mu} \nuL,
\label{intbypart}
\eeq
where we have moved the partial derivative through the coordinate-independent
$\gupu$ in the second term.  The first term on the right-hand side of 
equation (\ref{intbypart}) becomes a surface
term and vanishes for integrals over all space-time.  So we are left with
\beq
\ol{\nuRc} \gupu \partial_{\mu} \nuRc = -(\partial_{\mu} \ol{\nuL}) \gupu \nuL
= \ol{\nuL} \gupu \partial_{\mu} \nuL,
\eeq
which shows the kinetic term is invariant under CP-conjugation.

The CP-invariance of the weak current terms in $\lag^{\nu}_{SM}$  may be 
shown by a similar treatment,
realizing that under a parity transformation, the vector field becomes its
negative, so under CP, $Z \rightarrow -Z$, and $W^{\pm} \rightarrow
-W^{\mp}$.  The neutral current term, like the kinetic term, becomes itself. 
The $W^+$ term becomes
the $W^-$ term, and vice versa:
\bea
\smcp{(W^+_{\mu} \ol{\nuL} \gupu l_L +W^-_{\mu} \ol{l_L} \gupu \nuL )} & = & 
\smcp{(W^+_{\mu})} \ol{\nuRc} \gupu \smc{l}_R + 
\smcp{(W^-_{\mu})} \ol{\smc{l}_R} \gupu \nuRc  \nonumber \\
& = &- W^-_{\mu} \left( - \nuL^T C^{\dagger} \right) \gupu C \ol{l_L}^T - 
W^+ \left( - l_L^T C^{\dagger} \right) \gupu C \ol{\nuL}^T \nonumber \\
& = & -W^-_{\mu} \ol{l_L} C^T \gupu^T C^{\dagger T} \nuL - 
W^+_{\mu} \ol{\nuL} C^T \gupu^T C^{\dagger T} l_L \nonumber \\
& = & 
W^-_{\mu} \ol{l_L} \gupu \nuL + 
W^+_{\mu} \ol{\nuL} \gupu l_L. 
\eea
where we have used equation~(\ref{olnuc}) from Appendix~\ref{Csec} to
substitute for $\ol{\nuRc}$, we have obtained a minus sign by switching
the order of fermion fields in taking the transpose as before, and we have
again used the properties of $C$ given in equation (\ref{Cprop}) in the Appendix.
The charged lepton mass terms also swap under CP,
\bea
\ol{\smc{\alpha}_R} m_{\alpha} \smc{\alpha}_L + 
\ol{\smc{\alpha}_L} m_{\alpha} \smc{\alpha}_R & = & 
-\alpha_L^T C^{\dagger} m_{\alpha} C \ol{\alpha_R}^T - 
\alpha_R^T C^{\dagger} m_{\alpha} C \ol{\alpha_L}^T \nonumber \\
& = & 
\ol{\alpha_R} C^T C^{\dagger T} m_{\alpha} \alpha_L +
\ol{\alpha_L} C^T C^{\dagger T} m_{\alpha} \alpha_R =
\ol{\alpha_R} m_{\alpha} \alpha_L +
\ol{\alpha_L} m_{\alpha} \alpha_R,
\label{massCP}
\eea
so the entire $\lag^{\nu}_{SM}$ is
invariant under a CP transformation; including separate terms for antiparticles
would be redundant.  In fact, the CP-invariance of the
Lagrangian tells us that the antiparticle states are already contained in the
particle spinors.

\begin{figure}[htb]
\vsp
\centerline{\hbox{
\psfig{figure=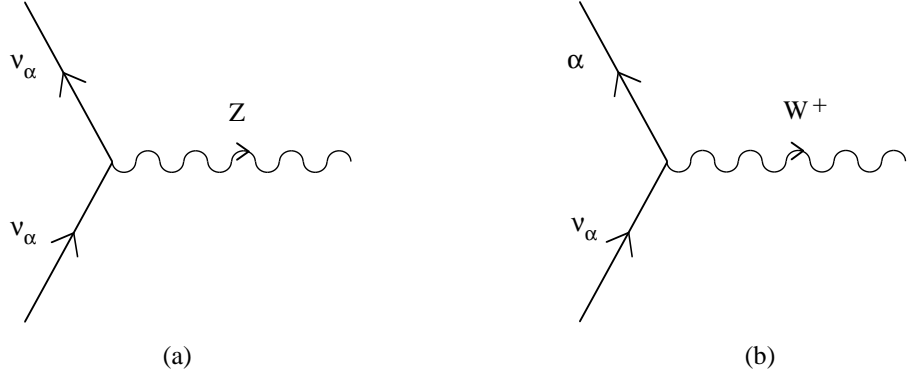,height=2.1 in}
}}
\caption{Feynman diagrams of neutrino vertices for the (a) neutral current, and
(b) charged current.\label{vertfig}}
\vsp
\end{figure}

\begin{figure}[htb]
\vsp
\centerline{\hbox{
\psfig{figure=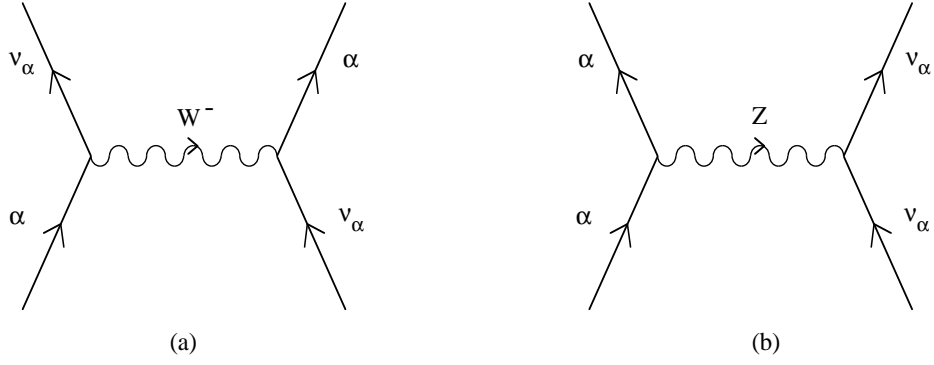,height=2.1 in}
}}
\caption{Feynman diagrams for (a) charged current $\alpha-\nua$ scattering,
and (b) neutral current $\alpha-\nua$ scattering. \label{ccncfig}}
\vsp
\end{figure}

\begin{figure}[bh!]
\vsp
\centerline{\hbox{
\psfig{figure=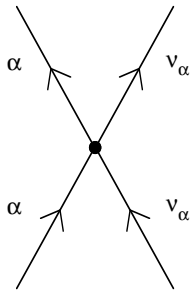,height=1.8 in}
}}
\caption{A Feynman diagram for the Fermi four-fermion 
neutrino scattering \label{fermfig}}
\vsp
\end{figure}

Terms in the Lagrangian describe single vertices,
such as those shown in Figure~\ref{vertfig}.  But many 
physical processes, such as
neutrino-electron scattering, consist of two
vertices, as shown in Figure~\ref{ccncfig}. So the matrix 
element for the charged current
interaction shown in Figure~\ref{ccncfig}a has the form \cite{Nact}
\beq
\label{2Lag} 
\frac{e^2}{2 \sin^2{\theta_{W}}} {\bf W}^{+}_{\mu} {\cal J}^{\mu}_{CC}  
{\bf W}^{- \sigma} {\cal J}_{CC \sigma}^{\dagger}. 
\eeq
As before, $e$ is defined in Appendix~\ref{key} and 
is only equal to the magnitude of the electron charge when $\hbar=c=1$.
A similar term occurs for the neutral current scattering in
Figure~\ref{ccncfig}b. 
If the momentum transferred by the W boson in Figure~\ref{ccncfig}a is small 
compared
to the mass of the boson, then the propagator for the boson collapses into
$\frac{-g_{\mu \nu}}{m_W^2}$ \cite{Nact}, and the interaction looks like Fermi's four-fermion
coupling illustrated in Figure \ref{fermfig}.  The resulting matrix element 
displays the low-energy symmetries 
of a Lagrangian and provides the same analytic results
for the S-matrix as the second order term in equation 
(\ref{2Lag}) when the momentum transferred by the boson is 
much less than the mass of the boson \cite{Nact}.  Together
with the appropriate neutral-current terms, this matrix element is known as the
``effective Lagrangian": 
\beq
\label{Leff}
\lag_{eff} =  -\frac{4 G}{\sqrt{2}} \left( {\cal J}^{\mu}_{NC}{\cal
J}^{\ }_{NC \mu} + {\cal J}^{\mu}_{CC}{\cal J}^{\dagger}_{CC \mu} \right ).
\eeq 
G is the Fermi constant, and may be found by comparing equations (\ref{Leff}) and
(\ref{2Lag}) and substituting $-\frac{1}{m_W^2} g_{\mu \sigma}$ for the product of
${\bf W}$ fields:
\beq
G = \sqrt{2} \frac{e^2}{8 m_W^2 \sin^2 \theta_W}.
\eeq
The effective Lagrangian has the same form as Fermi's four-fermion coupling;  
this particular effective Lagrangian for electron-neutrino scattering  is
responsible for matter-induced
oscillations for massive neutrinos, as will be discussed in Section~\ref{mattersec}.


\section{Whither We Go:  The Changes Wrought By Neutrino Oscillations}


The Standard Model contains only left-handed isospin $\half$ neutrinos and one
Higgs doublet, so it
allows no neutrino mass terms.  As
mentioned above, the neutrino mass was thought to be exactly zero in the 1960s,
when Updike wrote his poem, and the Standard Model was developed during this
same period.  The Model, like the poem, reflects the prevailing views of its
time and therefore yields a neutrino mass of zero.  In the last thirty years,
however, both the solar neutrino deficit and the atmospheric neutrino anomaly
have surfaced.  Experiments counting neutrinos from the sun see far fewer
electron neutrinos than theory predicts, and experiments measuring neutrinos
created in the atmosphere measure a muon-type-to-electron-type-neutrino ratio
much smaller than expected.  These surprising 
experimental results may be explained by making neutrino masses unequal
and therefore not all zero.  Non-zero neutrino mass is now preferred for symmetry
reasons as well.  Every fermion but the neutrino in the Standard Model has a
right-handed singlet state and thus a mass, so the absence of right-handed 
neutrinos in the Standard Model seems
unnatural and rather arbitrary to those who seek consistency in nature.  Most
extensions of the Standard Model, including Grand Unified Theories not based
on an SU(5) symmetry, give the neutrinos mass. 
Some of the mechanisms by which neutrino mass
may appear are discussed in Chapter~\ref{intro2}.  The existence of mass terms
in the neutrino Lagrangian may lead to neutrino oscillations between flavor
states, as presented in Chapter~\ref{osc}.  Chapter~\ref{boxes} develops a
parameterization to describe neutrino oscillations which is independent of the
model used to generate neutrino mass.  This parameterization is then used to
describe the predictions of many popular neutrino mass models 
(Chapter~\ref{massbox})  The predictions are compared to the existing neutrino
oscillation data in Chapter~\ref{expt}.  Chapter~\ref{summary} examines some of the 
implications of neutrino mass and oscillations and addresses the exciting future for 
this field of study.


\chapter{Giving Neutrinos Mass}
\label{intro2}


  The Lagrangian in equation (\ref{Lnu}) does not contain a neutrino mass term.  
Masses of fermions arise in the Standard Model through the coupling of the 
fermion with a
Higgs scalar doublet, as described in Section~\ref{SMsec}.  The Higgs doublet 
interacting
with fermions such as the electron has isospin $\half$, and it couples an
isospin $\half$ state with an isospin singlet.  Such interactions flip the
helicity of the fermion involved, so both a right- and a 
left-handed state of
each massive fermion are necessary in the Standard Model.  This type of
interaction is described by a 
Dirac mass term. Neutrinos may acquire a Dirac mass simply
by adding a right-handed neutrino (and a left-handed antineutrino) to the
Standard Model.  Right-handed neutrinos do not participate in the charged weak
interaction and are therefore dubbed ``sterile''.  In this work, sterile
neutrinos will be denoted by $\NR$ as in \cite{Lang}, while the active neutrinos
will be represented by the traditional $\nuL$.  The associated
helicity states of the CP-conjugate 
antineutrino will be given by $\nuRc$ and 
$\NLc$.\footnote{
I prefer $\NR$ rather than
the more apparent $\nu_R$ to represent the right-handed component 
for two reasons: it
clearly distinguishes active from sterile neutrinos, and it 
avoids confusion between $\ol{\nu_R}$ and $(\ol{\nu})_R = \ol{\nu_L}$. 
In my notation, the former is denoted $\ol{\NR}$, which is clearly different 
from the latter.
}

Because neutrinos are neutral, they may obtain mass by an additional method.  
Unlike charged particles, whose particle and antiparticle
states differ by the sign of the charge, neutrinos may be their own
antiparticle, like the photon is.  This type of fermion is known as a 
Majorana
particle, after the physicist who first examined the
possibility and ramifications of a self-conjugate fermion \cite{Maj}.  If
neutrinos are Majorana particles, they could form a mass term from the 
left-handed
$\nuL$ and the right-handed $\nuRc$.  Such a combination has isospin $1$, so
Majorana mass terms must be generated by a triplet 
Higgs interaction, which would be an addition to the Standard Model.


\section{The Differences Between Majorana and Dirac Neutrinos}
\label{MajvsDiracsec}

\begin{figure}[h]
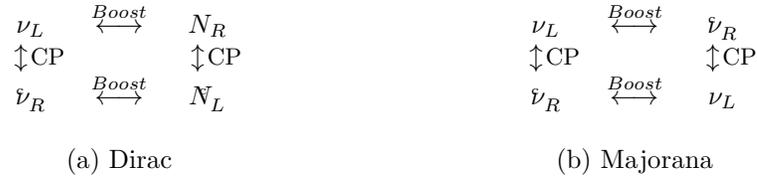

\begin{centering}
$$
\ba{ccc}
\ba{ccc} \nuL & \stackrel{Boost}{\longleftrightarrow} & \NR \\ 
\large{\ } \updownarrow \! \mbox{\small{CP}} & & 
\large{\ } \updownarrow \! \mbox{\small{CP}}  \\ 
\nuRc & \stackrel{Boost}{\longleftrightarrow} & \NLc \ea
& \hspace{1 in} &
\ba{ccc} \nuL & \stackrel{Boost}{\longleftrightarrow} & \nuRc \\ 
\large{\ } \updownarrow \! \mbox{\small{CP}} & & 
\large{\ } \updownarrow \! \mbox{\small{CP}}  \\ 
\nuRc & \stackrel{Boost}{\longleftrightarrow} & \nuL \ea
\\
\\
\mbox{(a) Dirac} & & \mbox{(b) Majorana}
\ea
$$
\caption{A pictorial representation of the relationships between neutrino 
states
for a) Dirac mass terms, and b) Majorana mass terms.\label{Kayserfig}}
\end{centering}
\end{figure}

Two different types of neutrino states have been experimentally observed:  
$\nuL$
and its CP-conjugate $\nuRc$.  The Dirac mass term introduces the
right-handed neutrino $\NR$ and its CP-conjugate $\NLc$, and has the form
\beq
{\cal L}_{Dirac} = -\left( \overline{\nuL} M_D \NR + \overline{\NR}
M_D^{\dagger} \nuL \right).
\label{Ldirac}
\eeq
As discussed in Section \ref{SMsec}, terms with the antiparticles $\nuRc$ and
$\NLc$ are redundant and therefore not explicitly written.  
A Lorentz boost or spin flip will transform $\nuL$ to $\NR$ and $\nuRc$ to $\NLc$,
as shown in Figure~\ref{Kayserfig}a, which is adapted from reference \cite{Kayser}. 
CP conjugation, however, transforms $\nuL$ to $\nuRc$ and $\NR$ to $\NLc$, as
discussed in Appendix~{\ref{Csec}}, so four distinct neutrino states are
necessary for Dirac mass terms.
\vsp
\vsp

Majorana neutrinos, however, are self-conjugate, so the sterile states are not
necessary \cite{Kayser}, \cite{BP}:
\beq
\label{Maj}
\smc{\psi}_{\left(maj\right)}(x) = C \gdno \psi_{\left(maj\right)}^*(x) = 
e^{i \theta} \psi_{\left(maj\right)}(x),
\eeq
where C is the charge-conjugation operator.  
Acting on a Majorana field with 
CP only changes the phase of the field and takes one 
helicity state
into the other, so it has the same effect as boosting the state, 
as illustrated in Figure~\ref{Kayserfig}b.  Thus only two distinct neutrino
states are necessary for Majorana mass terms.

Majorana mass terms have the general form
\beq
\ol{\smc{\psi}_R} M_M \psi_L + \ol{\psi_L} M_M^{\dagger} \smc{\psi}_R.
\label{genMmass}
\eeq
Because the Majorana field is self-conjugate, its mass matrix $M_M$ is 
symmetric
and so may be diagonalized by a transformation involving only one
unitary matrix, as we will show. 
Following the treatment found
in reference \cite{BP}, we set the Lagrangian in equation (\ref{genMmass}) 
equal to its
transpose, which is allowed since the Lagrangian is a scalar:
\beq
\ol{\smc{\psi}_R} M_M \psi_L + h.c. =
\left(\ol{\smc{\psi}_R} M_M \psi_L + h.c. \right)^T = 
-\left( \psi_L^T M_M^T \ol{\smc{\psi}_R}^T  \right) + h.c.,
\eeq
where the minus sign comes from swapping the order of fermion fields.  
Using the
definitions of Dirac conjugates found in Appendix~\ref{key}, 
the Majorana condition
(\ref{Maj}), and the expression (\ref{olpsic}) for $\ol{\psi_{L,R}}$ derived 
in
Appendix~\ref{Csec}, we find
\bea
\ol{\smc{\psi}_R} M_M \psi_L + h.c. & = & 
- \left( \psi_L^T M_M^T \left( - \psi_L^T C^{\dagger} \right)^T 
\right) + h.c. = 
\psi_L^T M_M^T C^{\dagger T} \psi_L + h.c. \\
& = & \psi_L^T  \left(- C^{\dagger} \right) M_M^T\psi_L + h.c.
= \ol{\smc{\psi}_R} M_M^T \psi_L + h.c.
\eea
For the first expression to equal the last, $M_M$ must be symmetric.

Still following the treatment of source \cite{BP}, we remember that an 
arbitrary
square matrix $M$ is diagonalized by a biunitary rotation:
\bea
D & = & V^{\dagger} M U, \mbox{\ \ or } \\
M & = & V D U^{\dagger},
\label{DtoM}
\eea
with $D$ diagonal and $V$ and $U$ unitary.
For a symmetric matrix, $M_M = M_M^T$, so 
\beq
M_M M_M^{\dagger} = M_M^T M_M^{\dagger T}.
\eeq
Using equation (\ref{DtoM}), we find
\bea
V D U^{\dagger} \left(V D U^{\dagger} \right)^{\dagger} & = &
    \left(V D U^{\dagger} \right)^T \left(V D U^{\dagger} \right)^{\dagger T},
\mbox{\ \ so} \\ 
V D U^{\dagger} U D^{\dagger} V^{\dagger} = V D^2 V^{\dagger} & = &
 U^{\dagger T} D^T V^T V^{\dagger T} D^{\dagger T} U^T = U^{\dagger T} D^2 U^T.
\eea
Thus, multiplying by $U^T$ on the left and $V$ on the right,
\beq
U^T V D^2 = D^2 U^T V.
\eeq
Since the matrix $U^T V$ commutes with the matrix $D^2$, they must be
simultaneously diagonalizable.  $D^2$ is already diagonalized, so $U^T V$ must
be diagonal too.  A diagonal unitary matrix has the form of a 
diagonal matrix of phases, $X$, so
\beq
U^T V = X, \mbox{\ \ with \ \ } X_{ij} = \delta_{ij} e^{i \xi_i}.
\eeq
We may define a new matrix $V'$ which incorporates the phases of $X$:
\beq
V'=VX^{\dagger},
\eeq
giving $U^T V' = \openone$.
We may thus diagonalize $M_M$ with the single
matrix $U$ and its transpose $V'^{\dagger}=U^T$:
\beq
D = V^{\dagger} M_M U = X^{\dagger} V'^{\dagger} M_M U, \mbox{\ \ so \ \ }
\eeq
\beq
U^{T} M_M U = X D,
\eeq
which is diagonal, as claimed above.


\section{The Dirac Mass Term}
\label{Diracsec}


	The Dirac mass term couples left-handed active neutrinos with their
right-handed sterile partners through an interaction with the same 
Higgs
doublet which gives mass to up-type quarks in the Standard Model.
\beq
{\cal L}_{Dirac} = -\sum_{\acb} g^{D'}_{\ab}\ol{\psi_{\alpha L} }
\tilde{\phi}_D N_{\beta R} + h.c.,
\label{LDnu}
\eeq
with $\psi_{\alpha L}$ and 
$\tilde{\phi}_D$ defined in equations (\ref{psiaL}) and (\ref{HiggsDup}),
respectively,
and $g^{D'}_{\ab}$ the Higgs-doublet coupling constants for the neutrino sector.
Once the symmetry is broken, the
Lagrangian term in equation (\ref{LDnu}) contains a mass term 
of the form
\beq
{\cal L}_{Dirac} = -\sum_{\acb}  \overline{\nua_L} M_{\ab} N_{\beta R} + h.c.
\eeq
The mass matrix $M_D$ contains all of the coupling constants:
\beq
M_{D_{\ab}} = g^{D'}_{\ab} \langle {\phi^0_D}^* \rangle.
\eeq
This type of mass term is analogous to the quark mass terms and requires only 
the
addition of sterile neutrinos to the Standard Model.  Generating small
neutrino masses by such a term would require the Yukawa coupling between
the Higgs and the neutrinos to be many orders of magnitude smaller
than the similar coupling between the Higgs and the quarks.  While this
difference of coupling constants may indeed exist, no ready explanation for the
difference of scale has been suggested.  This unexplained lack of symmetry is
unattractive, so other mechanisms for generating neutrino masses have been
proposed.


\section{The Majorana Triplet Mass Term}


	If neutrinos possess a Majorana nature, no sterile neutrinos are 
necessary to
produce a non-zero neutrino mass.  The Majorana triplet mass term couples
active left-handed neutrinos to their CP-conjugate right-handed partners as
follows:
\beq
{\cal L}_{Triplet} = -\half \left( \overline{\nuL} M_T \nuRc + \overline{\nuRc}
M_T^{\dagger} \nuL \right).
\label{Ltrip}
\eeq
Because both $\nuL$ and $\nuRc$ have $I_3$ of $+\half$, the third
component of isospin changes
by one unit in such a term. An SU(2) triplet operator is therefore needed to
conserve isospin.  Such an operator may be a single object such as a 
Higgs
triplet or a combination of objects yielding a single, isospin
$1$ ``effective'' operator.


\subsection{The Triplet Higgs}


Gelmini and Roncadelli \cite{GR} proposed a model
using a triplet Higgs $\phi_T$ with a Lagrangian given by
\beq
{\cal L}_{Triplet} = - \sum_{\acb} g^{T'}_{\ab} 
\left( \ba{cc}
\ol{\smc{\alpha}_R} & \ol{\nuaRc}
\ea \right)
\left(
\ba{cc}
\phi_T^+    &   \sqrt{2} \phi_T^{++}   \\
 -\sqrt{2} \phi_T^0   &   \phi_T^+
\ea
\right)
\left(
\ba{c}
\nub_L \\  \beta_L 
\ea
\right)
+ h.c.
\eeq
The neutrino mass matrix in this model is
\beq
M_{T\ab}^{\dagger} = - 2 \sqrt{2} g^{T'}_{\ab} \langle \phi_T^0 \rangle,
\eeq
and the charged leptons do not obtain any mass contribution from the triplet
since they do not couple to the neutral Higgs triplet field.
To conserve total lepton number before symmetry breaking, the 
Gelmini-Roncadelli
model assigns a lepton charge of 2 to the Higgs triplet.  When SU(2) symmetry
is broken, lepton-number symmetry is broken as well, resulting in the formation
of an additional Goldstone boson, the Majoron.  This Majoron obtains a mass
far smaller than the mass of the Z boson, so it provides new final states
affecting the measured value of
the Z-decay width.  Such an effect contradicts experimental measurements of the
width, so the Gelmini-Roncadelli triplet model with a lepton-number-conserving
Lagrangian has been ruled out \cite{BP}, \cite{KimPev}.

All is not lost, however, for the Higgs triplet model.  If lepton number is
not conserved by the mass terms, the masses of the physical Higgs particles
produced have no upper bound \cite{BP}.  Thus the Higgs triplet remains a
viable possibility for generation of Majorana neutrino masses without the
addition of sterile neutrinos, but at the cost of introducing new Higgs
particles and abandoning lepton number conservation entirely.
We shall not concern ourself further with this model, but the
interested reader
may consult sources \cite{BP}, \cite{GR}, \cite{Georgi}, and \cite{KimPev}.


\subsection{Radiative Mass Terms and the Zee Model}
\label{Zeesec}


	Additional mechanisms for generating neutrino masses exist at the one-loop
level.  Masses generated by such a mechanism are called ``radiative masses.''  
One of the best-known
radiative models using additional Higgs fields was proposed by Zee \cite{Zee}
(for discussions of the Zee model, 
see also \cite{WolfZee}, \cite{SmirZee}, \cite{BP}, \cite{PetZee}, and \cite{MP}).  
Zee's model in its simplest form uses only two Higgs doublets,
$\phi_a$ and $\phi_b$, and a charged singlet Higgs field $h$.  Such a singlet arises
naturally from the simplest SU(5) theories, and it interacts with lepton
doublets and Higgs doublets, but not quark fields \cite{Zee}.  The scalar $h$
may be given lepton number $-2$, so its interactions with the lepton fields
\beq
{\cal L}_{h \psi_L} = f^{\ab} \left( \psi^i_{\alpha L} 
C \psi^j_{\beta L} \right) 
\epsilon_{ij} h + h.c.
\label{hpsiL}
\eeq
conserves lepton number.\footnote{Reference \cite{Zee} 
incorrectly states that individual lepton
numbers $L_e$, $L_{\mu}$, and $L_{\tau}$ are conserved by this term.  Only 
the
total lepton number $L=L_e + L_{\mu} + L_{\tau}$ is conserved by the Zee
interaction.  Zee corrects this misstatement in his later work 
\cite{Zee2}.}
Here $\psi_{\alpha L}$ is the left-handed lepton 
doublet of flavor $\alpha$,
containing both neutrinos and charged leptons. $i$ and $j$ represent the SU(2)
index:  $\psi^1_{\alpha L} = \nua_L$, and $\psi^2_{\alpha L} = \alpha_L$
\cite{Zee}.  $\epsilon_{ij}$ is of course the antisymmetric tensor, and $C$ is
the charge-conjugation operator.  The coupling constant $f^{\ab}$
must be antisymmetric because of Fermi statistics; interchanging two fermion
spinors must change the overall sign of any fermion coupling, so
\beq
f^{\ab} \left( \psi^i_{\alpha L} C \psi^j_{\beta L} \right) 
\epsilon_{ij} h =
- f^{\beta \alpha} \left( \psi^j_{\beta L} C \psi^i_{\alpha L} \right) 
   \epsilon_{ji} h.
\eeq
Obtaining the left-hand side from the right-hand side involves swapping the
order of two fermion fields, which provides one sign change; swapping the
indices of the antisymmetric tensor, thereby providing a second sign change;
and swapping the order of the indices of the coupling $f^{\ab}$.  The coupling
must therefore be antisymmetric in its indices to arrive at the required 
overall sign change.

The coupling between the charged scalar $h$ and the doublets $\phi_n$
takes the form \cite{Zee}
\beq
{\cal L}_{h \phi}=M_{ab}\epsilon_{ij} \phi_a^i \phi_b^j h + h.c.
\label{hphiL}
\eeq
The $\epsilon_{ij}$ couples the doublets to conserve $I_3$.  
This interaction violates lepton number by $-2$.
Most treatments of the Zee model \cite{MP}, 
\cite{BP}, \cite{WolfZee}, \cite{PetZee},  
treat only the simplest case, in which only one of the Higgs
doublets $\phi_a$ couples to leptons, and then only diagonally in 
isospin space:
\beq
{\cal L}_{\phi \beta} = \sum_{\beta} 
g^{D'}_{\ab} \ol{\psi^i_{\beta L}} \beta_R \phi^i_a + h.c.
\label{phipsiL}
\eeq

\begin{figure}[htb]
\centerline{\hbox{
\psfig{figure=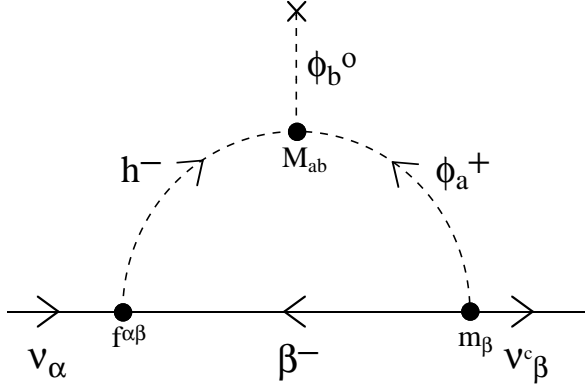,height=2.3 in}
}}
\caption[A Feynman diagram for the Zee model.]
{A Feynman diagram for the Zee model.  The arrows denote the direction of
the flow of the quantum numbers associated with the given particle.  Charge is
conserved at each vertex, but lepton number is not conserved at the top
Higgs-Higgs interaction vertex. 
\label{Zeefig}}
\end{figure}
\vsp

None of the interactions (\ref{hpsiL}), (\ref{hphiL}), or (\ref{phipsiL}) alone
will give the neutrino a mass, but the combination of these three terms at the
one-loop level as shown in Figure~\ref{Zeefig} produces a Majorana mass term
for the neutrino.  The neutrino mass term under this Zee model ultimately has
the form of equation (\ref{Ltrip}) \cite{PetZee}:
\beq
-{\cal L}_{\nu}^{eff}= \half \sum_{\acb} \ol{\smc{\nu}_{R\alpha}} 
M_{\ab}^{Zee} 
\nuL_{\beta} + h.c., \mbox{\ \ with \ \ }
\label{LZee}
\eeq
\beq
M_{\ab}^{Zee} = 2 g^{Zee}_{\ab} \left( m^2_{\alpha} - m^2_{\beta} \right).
\label{Zeemass}
\eeq
$g^{Zee}_{\ab}$ is a complicated function of Higgs masses and Standard Model
parameters.  An expression for it may be found in reference 
\cite{PetZee}, but it is not
necessary for our purposes.  Because the mass term in equation (\ref{LZee}) is
a Majorana term, the matrix $M_{\ab}$ in equation (\ref{Zeemass}) must 
be
symmetric, so $g^{Zee}_{\ab} = -g^{Zee}_{\beta \alpha}$.  
The observant reader may  note that
the diagonal entries $M_{\alpha \alpha}$ are therefore equal to zero.

As mentioned above, the assumption that the lepton-Higgs doublet coupling has
the diagonal form (\ref{phipsiL}) is generally, but not always, made.  Zee
himself did not make this simplification in reference \cite{Zee}, and the
authors of \cite{FY} have chosen a different Zee model to present.  The mass
matrix in reference \cite{FY} is ``motivated by the empirical hierarchy in the
[quark mixing] matrix,'' and has the approximate form
\beq
M_{\ab} \approx \frac{m_{\tau}}{m_h} A \left( \ba{ccc}
\epsilon^3 & \epsilon^2 & \epsilon \\
\epsilon^2 & \epsilon & 1 \\
\epsilon & 1 & 0 \ea \right).
\eeq
This alternative form has its virtues, but for the rest of this paper we will
consider only the form (\ref{Zeemass}), remembering that it is merely the
simplest, rather than the only, possibility for a mass term generated 
according
to the Zee model.


\section{The Majorana Singlet Mass Term}


The sterile neutrinos $\NR$, being isospin singlets, may obtain a 
Majorana mass
without Higgs fields.  Such a ``bare mass'' term has the form
\beq
{\cal L}_{Singlet} = -\half \left( \overline{\NLc} M_S \NR + \overline{\NR}
M_S^{\dagger} \NLc \right).
\label{Lsing}
\eeq
This term is explicitly SU(2)-invariant, so it may 
be included in the overall Lagrangian
before symmetry breaking.  Such a term may also be generated by interactions
with a Higgs singlet, but the use of such an extra Higgs in not necessary. 
The singlet mass term does not generate any type of mass for the active
neutrinos and
thus is somewhat uninteresting on its own, but it is possible in models with
right-handed neutrinos and so it must be included in any full treatment of
neutrino mass possibilities.


\section{Putting Them All Together}
\label{allflavsec}


	Dirac terms and Majorana terms are not mutually exclusive, and
mass terms of all three types (\ref{Ldirac}), (\ref{Ltrip}), and (\ref{Lsing}) 
may occur simultaneously.  We
may also add additional sterile neutrinos $\chi_{\eta R}$ which do not have active
counterparts.  These additional neutrinos may be Majorana or Dirac particles. 
In the most general case, we can have $n_L$ flavors of active neutrinos
$\nua_L$, $n_R$ flavors of their right-handed sterile partners 
$N_{\beta R}$, 
and $n_{\chi}$ flavors of additional sterile
neutrinos $\chi_{\eta}$.  Most reasonable scenarios set $n_L = n_R$, but one
can imagine other scenarios, so we will not {\it a priori} set them equal.  
The
neutrino mass terms of (\ref{Ldirac}), (\ref{Ltrip}), and (\ref{Lsing}) may be
combined with the mass terms involving the additional sterile neutrinos 
in the matrix equation
\beq
{\cal L}_{mass}=-\half \left( \ol{\nuL} \; \ol{\NLc} \; 
\ol{\smc{\chi}_L} \right)
M_{\nu} \left( \ba{c} \nuRc  \\ \NR \\ \chi_R
\ea
\right) + h.c.
\label{Ltot}
\eeq
$M_{\nu}$ is the $n \times n$ neutrino mass matrix, where $n=n_L+n_R+n_{\chi}$, 
$\nu$ contains all $n_L$ active neutrino spinors, $\NR$ contains the $n_R$
Dirac counterparts, and $\chi$ contains all $n_{\chi}$ sterile
neutrino spinors.  $M_{\nu}$ may be broken into block form:
\beq
M_{\nu} = \left( \ba{ccc} M_T & M_D & M_{\nu \chi} \\ M_D^T & M_S & M_{N \chi}
\\ M_{\nu \chi}^T & M_{N \chi}^T & M_{\chi} \ea \right),
\label{blockmass}
\eeq
where $M_T$ is $n_L \times n_L$, $M_D$ is $n_L \times n_R$, {\it et cetera}. 

The singlet and triplet Majorana terms fall directly out of equations
(\ref{Ltot}) and(\ref{blockmass}) 
when the other couplings are set to zero, but the Dirac term
requires a bit of algebra.  For pure Dirac neutrinos, the matrices $M_T$,
$M_S$, and all $M_{\chi}$ are zero.  We are left with the Lagrangian
\bea
{\cal L}_{Dirac}&=&-\half \left( \ol{\nuL} \; \ol{\NLc} \; \ol{\smc{\chi}_L} 
\right)
\left( \ba{ccc} 0 & M_D & 0 \\ M_D^T & 0 & 0
\\ 0 & 0 & 0 \ea \right)
\left( \ba{c} \nuRc  \\ \NR \\ \chi_R
\ea
\right) + h.c. \nonumber \\
&=& 
-\half \left[ \ol{\nuL} M_D N_R + \ol{\NLc} M_D^T \nuRc \right] + h.c.
\eea
Using the relations (\ref{olpsic}) and (\ref{nuRcdef}) from
Appendix~\ref{Csec}, we see that the two terms in the brackets are identical:
\beq
\ol{\NLc} M_D^T \nuRc = -N_R^T C^{\dagger} M_D^T C \ol{\nuL}^T =
\ol{\nuL} C^T M_D C^{\dagger T} N_R = \ol{\nuL} M_D N_R.
\eeq
The Dirac mass Lagrangian becomes
\beq
{\cal L}_{Dirac}= - \ol{\nuL} M_D N_R + h.c.,
\eeq
in agreement with the definition of a Dirac term (\ref{Ldirac}).
The extra $\half$ in the definition of the Majorana terms (\ref{Ltrip}) and
(\ref{Lsing}) does not appear in Dirac terms because the
particle and antiparticle Dirac mass terms sum when we use the general notation
(\ref{Ltot}).  This factor is also consistent with what one expects for the
equation of motion resulting from variation of ${\cal L}$ with respect to the
neutrino fields.  For Dirac fields, $\psi$ and $\psi^{\dagger}$ are varied
independently, whereas for Majorana fields the two are not independent and their
variation consequently returns a factor of 2.


\section{The See-Saw Model}
\label{seesawsection}


A popular class of neutrino mass models which includes both Majorana and Dirac 
mass terms is that of seesaw models.  Seesaw models have achieved their
popularity because they naturally give rise to small
neutrino masses with minimum additions to the Standard Model.  In such models,
active neutrinos compose Dirac mass terms with their sterile
counterparts, and the sterile neutrinos also form a bare Majorana mass term. 
Equation (\ref{Ltot}) becomes
\beq
{\cal L}_{mass}=-\half \left( \ol{\nuL} \; \ol{\NLc} \right) \left( \ba{cc}
0 & M_D \\ M_D^T & M_S \ea \right) \left( \ba{c} \nuRc \\ \NR \ea \right).
\eeq
In see-saw models, the elements of $M_D$, which arise from coupling
to the Higgs doublet and the standard SU(2)$\times$U(1) symmetry
breaking, 
are on the order of lepton masses, while the SU(2)$\times$U(1)-invariant 
elements of
$M_S$ are on the order of a larger scale, perhaps the scale of grand unified 
theories.  Thus we may expand to leading order in $M_S^{-1}$.  Roughly
following the treatment in \cite{Lang}, we find that $U_{\nu}$ has the form
\beq
U_{\nu} = \left( \ba{cc} A & M_D M_S^{-1} B \\ 
-M_S^{-1 \dagger} M_D^{\dagger} A & B \ea \right),
\eeq
where $A$ and $B$ diagonalize the ``light'' and ``heavy'' mass matrices:
\beq
-A^{\dagger} M_D M_S^{-1} M_D^T A^{T \dagger} = D_{light}, \mbox{\ \  and \ \ }
B^{\dagger} M_S B^{T \dagger} = D_{heavy}.
\eeq
The mass eigenstates are
\beq
\left( \ba{c} \nu^m \\ N^m \ea \right)_L = \left( \ba{c}
A^{\dagger} \left( \nuL - M_D M_S^{-1} \NLc \right) \\
B^{\dagger} \left( M_S^{-1 \dagger} M_D^{\dagger} \nuL + \NLc \right) \ea
\right).
\eeq
These are Majorana states, so the right-handed component is also the
charge-conjugate of the left-handed state:
\beq
\left( \ba{c} \nu^m \\ N^m \ea \right)_R = 
\left[\left( \ba{c} \nu^m \\ N^m \ea \right)_L \right]^c = \left( \ba{c}
A^T \left( \nuRc - M_D^{T \dagger} M_S^{-1 \dagger} \NR \right) \\
B^T \left( M_S^{-1} M_D^T \nuRc + \NR \right) \ea \right).
\eeq
The seesaw model is so-named because it naturally gives small masses to the
states $\nu^m$, which are composed mainly of active neutrinos, while giving
large masses to the states $N^M$, which are largely made up of sterile
neutrinos.

In the above treatment, $M_T$ has been set to zero.  The seesaw model still
works
for non-zero $M_T$, provided the elements of $M_T$ are still much smaller than
the elements of $M_S$.  Such an inclusion of $M_T$ has negligible effects on
the heavy mass states, and changes the light mass matrix from 
$-M_D M_S^{-1} M_D^T$ to $M_T - M_D M_S^{-1} M_D^T$ \cite{Lang}.


\section{Mass States vs.\ Flavor States}
\label{CKMsec}


Any mass term, regardless of the origin of the term, has the form
$\ol{\psi_L} M_{\nu}
\psi_R$, where $\psi$ is a $n$-dimensional vector and $M$ is an $n \times n$
matrix.  $\psi$ can contain the active $\nua$, the sterile $N_{\alpha}$, the
additional $\chi_{\eta}$, or all three.
As is the case for quarks, the mass matrix $M$ for neutrinos is not 
necessarily diagonal.  Neutrino states of definite mass correspond to
eigenvectors of the mass matrix, which are generally distinct from the states of
definite flavor.  Adding neutrino masses, the lepton Lagrangian of equation
(\ref{Lnu}) in matrix form is
\bea
\label{wholeL}
\lag^{\nu} & = & {\ol \psi }i\partial_{\mu} \gupu \psi 
+ {\ol l} i \partial_{\mu} \gupu l 
- c_1 Z_{\mu} \left( {\ol \nuL} \gupu \nuL 
  - (1-2\sin^2 \theta_W){\ol l_L} \gupu l_L 
  + 2 \sin^2 \theta_W {\ol l_R} \gupu l_R \right) \\
&&- \ c_2 \left( W^+_{\mu} \ol{\nuL} \gupu l_L 
       + W^-_{\mu} \ol{l} \gupu \nuL \right) 
  +\ol{l_L} M_l l_R + \ol{l_R} M^{\dagger}_l l_L + \ol{\psi_L} M_{\nu} \psi_R +
\ol{\psi_R} M^{\dagger}_{\nu} \psi_L. \nonumber
\eea
The subscripts on the mass matrices $\nu$ and $l$ label a matrix by the type of
particle for which it provides mass; they do not denote flavor indices.  
The lepton spinors in equation~(\ref{wholeL}) correspond to states of definite
{\it flavor}.  States of definite mass (to be denoted by the superscript $^m$) 
are achieved by diagonalizing the mass
matrix, thereby rotating from the flavor basis to the mass basis.  
The diagonal
mass matrices, $D_l$ and $D_{\nu}$, are given by
\beq
D_l = U_{l_L}^{-1} M_l U_{l_R}, \mbox{\ \ and \ \ }
D_{\nu} = U_{\nuL}^{-1} M_{\nu} U_{\NR}.
\eeq
As discussed in Section~\ref{MajvsDiracsec}, when $\psi$
represents a Majorana spinor, $M_{\nu}$ is symmetric and we may take 
$U_{\NR} = U_{\nuL}^{\dagger T}$.
At this point, however, we will consider a three-generation case without
Majorana terms.  The following
treatment is extended to allow higher generations in Section~\ref{pretracesec}.

The rotation from flavor states to mass states leaves the kinetic
and neutral current terms 
unchanged, but the charged current is not invariant under the
transformation:
\bea
\label{CKMderiv} 
\lag^{\nu} & = &
\ol{\nuL} U_{\nuL} U_{\nuL}^{-1} i\partial_{\mu} 
  \gupu U_{\nuL} U_{\nuL}^{-1} \nuL
+ \ol{\NR} U_{\NR} U_{\NR}^{-1} i\partial_{\mu} 
  \gupu U_{\NR} U_{\NR}^{-1} \NR \nonumber \\
&& + \ \ol {l_L}  U_{l_L} U_{l_L}^{-1} i \partial_{\mu} 
  \gupu  U_{l_L} U_{l_L}^{-1} l_L 
 + \ol {l_R} U_{l_R} U_{l_R}^{-1} i \partial_{\mu} 
  \gupu U_{l_R} U_{l_R}^{-1} l_R \nonumber \\
&& - \ c_1 Z_{\mu} \left( {\ol \nuL} U_{\nuL} U_{\nuL}^{-1} 
  \gupu U_{\nuL}^{-1} U_{\nuL} \nuL 
- \left(1-2\sin^2{\theta_W}\right) 
    {\ol l_L} U_{l_L} U_{l_L}^{-1} \gupu U_{l_L} U_{l_L}^{-1} l_L 
    \right.\nonumber \\
&& \left. + \ 2 \sin^2 \theta_W \ol{l_R} U_{l_R} U_{l_R}^{-1} \gupu U_{l_R}
 U_{l_R}^{-1} l_R
\right) \nonumber \\
&& - \ c_2 \left( W^+_{\mu} \ol{\nuL} U_{\nuL} U_{\nuL}^{-1} 
  \gupu U_{l_L} U_{l_L}^{-1} l_L 
+ W^-_{\mu} \ol{l_L} U_{l_L} U_{l_L}^{-1} 
  \gupu U_{\nuL} U_{\nuL}^{-1} \nuL \right)
+ \ol{l_L} U_{l_L} U_{l_L}^{-1} M_l U_{l_R} U_{l_R}^{-1} l_R \nonumber \\
&& + \ \ol{l_R} U_{l_R} U_{l_R}^{-1} M^{\dagger}_l U_{l_L} U_{l_L}^{-1} l_L 
+ \ol{\nuL} U_{\nuL} U_{\nuL}^{-1}  M_{\nu} U_{\NR} U_{\NR}^{-1} \NR 
+ \ol{\NR} U_{\NR} U_{\NR}^{-1} M^{\dagger}_{\nu} U_{\nuL} U_{\nuL}^{-1} \nuL
\nonumber \\
& = &
\ol{\nu_L^m} U_{\nuL}^{-1} U_{\nuL} i\partial_{\mu} \gupu \nu_L^m
+ \ol{N_R^m} U_{\NR}^{-1} U_{\NR} i\partial_{\mu} \gupu N_R^m \nonumber \\
&& + \ \ol {l_L^m}  U_{l_L}^{-1} U_{l_L} i \partial_{\mu} \gupu l_L^m 
+ \ol {l_R^m} U_{l_R}^{-1} U_{l_R} i \partial_{\mu} \gupu l_R^m \nonumber \\
&& - \ c_1 Z_{\mu} \left( \ol{\nu_L^m} U_{\nuL}^{-1} U_{\nuL} \gupu \nu_L^m 
-\left(1-2\sin^2\theta_W\right){\ol l_L^m} U_{l_L}^{-1} U_{l_L} \gupu l_L^m 
+ 2 \sin^2 \theta_W {\ol l_R^m} U_{l_R}^{-1} U_{l_R} \gupu l_R^m\right)
\nonumber \\
&& - \ c_2 \left( W^+_{\mu} \ol{\nu_L^m} U_{\nuL}^{-1} U_{\l_L} \gupu l_L^m 
+ W^-_{\mu} \ol{l_L^m} U_{l_L}^{-1} U_{\nuL} \gupu \nu_L^m \right)
+ \ol{l_L^m} D_l l_R^m 
+ \ol{l_R^m} D^{\dagger}_l l_L^m \nonumber \\
&&+ \ \ol{\nu_L^m} D_{\nu} N_R^m  
+ \ol{N_R^m} D^{\dagger}_{\nu} \nu_L^m \nonumber \\
& = & 
\ol{\nu_L^m} i\partial_{\mu} \gupu \nu_L^m
+ \ol{N_R^m}i\partial_{\mu} \gupu N_R^m
+ \ol {l_L^m} i \partial_{\mu} \gupu l_L^m
+ \ol {l_R^m} i \partial_{\mu} \gupu l_R^m \nonumber \\
&& - \ c_1 Z_{\mu} \left( {\ol \nu_L^m} \gupu \nu_L^m 
- \left(1-2\sin^2\theta_W\right) {\ol l_L^m} \gupu l_L^m 
+ 2 \sin^2 \theta_W {\ol l_R^m} \gupu l_R^m \right)
\nonumber \\ 
&& - \ c_2 \left( W^+_{\mu} \ol{\nu_L^m} U_{\nuL}^{-1} U_{\l_L} \gupu l_L^m 
+ W^-_{\mu} \ol{l_L^m} U_{l_L}^{-1} U_{\nuL} \gupu \nu_L^m \right) \nonumber \\
&& + \ \ol{l_L^m} D_l l_R^m 
+ \ol{l_R^m} D^{\dagger}_l l_L^m 
+ \ol{\nu_L^m} D_{\nu} N_R^m  
+ \ol{N_R^m} D^{\dagger}_{\nu} \nu_L^m. 
\eea
The charged current term in equation (\ref{CKMderiv}) is not simultaneously
diagonalizable in both the flavor basis and mass basis; upon rotation to the mass
basis, it acquires a mixing matrix, $V=U_{l_L}^{-1} U_{\nuL}$.  For Dirac
neutrinos, the lepton mixing matrix $V$ has the same properties as the CKM matrix
in the quark sector.  If the Lagrangian does not contain a neutrino mass term,
as in the Standard Model, or if the neutrino mass eigenvalues are degenerate, 
the neutrino states may undergo an arbitrary
rotation, so we may set $U_{\nuL} = U_{l_L}$, which yields $V= \openone$. 
In such models, flavor states and mass states are indistinguishable, so we may
assume a diagonal charged lepton 
mass matrix for the Standard Model, as we did in Section
\ref{SMsec}, with no loss of generality.

Even when neutrino mass terms are present, the matrices $U_{\nuL}$ and
$U_{l_L}$ are not unique.  Multiplying each by any unitary matrix $R$ leaves
\beq
V=\left(RU_{lL}\right)^{\dagger} \left(RU_{\nuL}\right) = U_{lL}^{\dagger}
U_{\nuL}
\eeq
invariant.  In the quark sector, an implicit choice $R=U_{upL}^{\dagger}$ is
made so that quark mixing is constrained to the down-type quarks, with the
up-type flavor states assumed identical to mass states.  Mixing in the lepton 
sector is typically similarly restricted to
neutrinos.  Under this implicit assumption, the neutrino flavor states are
connected to the neutrino mass states by the mixing matrix $V$ which appears in
the weak charged current term:
\beq
\nua_L = \sum_{i=1}^n \Vai \nui_L, \mbox{ with }
\eeq
\beq
{\cal J}_{CC} = \sum_{\alpha=flavors} \ol{\nua_L} \gupu l_{\alpha L} + h.c. =
\sum_{i,j=1}^n \ol{\nui_L} V^{-1}_{ij} \gupu l_{jL} +h.c.
\eeq
The mixing matrix occurring in the charged current term has two mass
indices, since it couples the neutrino mass states to the lepton mass states. 
Under the assumption that charged leptons have a diagonal mass matrix, the
lepton mass state $l_{iL}$ is equivalent to the lepton flavor state $l_{\alpha
L}$, and the mixing matrix may be given a mass index and a flavor index and
identified with the neutrino rotation matrix $U_{\nuL}$.  In the
following Chapters, we will assume the lepton matrix is diagonal and give the
mixing matrix one mass index and one flavor index.


\chapter{Oscillations}
\label{osc}


If the eigenvalues of the neutrino mass matrix are non-degenerate, then
neutrinos may change flavors as they propagate.  We will show that the flavor
of a neutrino changes in a cyclic fashion, so the neutrino flavor {\it 
oscillates} with distance traveled.  To see how these oscillations arise, let
us consider the neutrino Lagrangian and derive the appropriate equations of 
motion.


\section{The Equations of Motion}
\label{eomsec}


For simplicity, we will consider a Dirac
mass term as in equation (\ref{Ldirac}).  The neutrino Lagrangian in
vacuum is then 
given by
\beq
\lag^{\nu}  =  \sum_{\alpha = flavors} \left[ \ol{\nua_{L}} i \partial_{\mu} 
  \gupu \nua_{L} 
  + \ol{N_{\alpha R}} i \partial_{\mu} \gupu N_{\alpha R}
  -\sum_{\beta=flavors} \left( \ol{\nua_L} {M_D}_{\alpha \beta} N_{\beta R} 
  + \ol{N_{\alpha R}} {M_D^{\dagger}}_{\alpha \beta} \nub_L \right) \right ]
\label{nulag}
\eeq
Varying (\ref{nulag}) with respect to $\ol{\nua_L}$ and then with respect to
$\ol{N_{\alpha R}}$ yields the following equations of motion:
\bea
\frac{\delta \lag^{\nu}}{\delta \ol{\nua_L}} & = & i \partial_{\mu} \gupu 
  \nua_{L}
  - \sum_{\beta} {M_D}_{\alpha \beta} N_{\beta R} = 0, \mbox{ and} 
\label{varynu}\\
\frac{\delta \lag^{\nu}}{\delta \ol{N_{\alpha R}}} & = & 
  i \partial_{\mu} \gupu N_{\alpha R} 
  -\sum_{\beta} {M_D^{\dagger}}_{\alpha \beta} \nub_L = 0.
\label{varyN}
\eea
Solving for $N_{\alpha R}$ in the first equation and then substituting into
the second, we obtain an equation involving only the active neutrino $\nug_L$:
\beq
i \partial_{\mu} \gupu N_{\alpha R}  = i \partial_{\mu} \gupu
\sum_{\gamma}  M^{-1}_{D \alpha \gamma}  
 \left( i \partial_{\nu} \gupv \nug_L \right) =
 \sum_{\gamma} {M_D^{\dagger}}_{\alpha \gamma} \nug_L.
\label{eos}
\eeq
Following the treatment of Kuo and Pantaleone in \cite{KP}, we make the
following simplification using the properties of the gamma matrices:
\beq
\partial_{\mu} \gupu \partial_{\nu} \gupv = \half \left[ \partial_{\mu}
\partial_{\nu} \gupu \gupv + \partial_{\nu} \partial_{\mu} \gupv \gupu  \right]
= \partial_{\mu} \partial_{\nu} g^{\mu \nu} = \partial_{\mu}
\partial^{\mu}.
\label{gammasimp}
\eeq
With this substitution, equation (\ref{eos}) may be written as
\beq
\partial_{\mu} \partial^{\mu} \nub_L + \sum_{\gamma}
(M_DM_D^{\dagger})_{\beta \gamma} \nug_L = 0.
\label{vaceos}
\eeq
This has the same form as a Klein-Gordon equation.
Following the traditional treatment of a Klein-Gordon equation, we assume a
plane wave solution of the form
\beq
\nua_L (x_{\mu}) = \sum_{\beta} e^{-i p_{\ab}^{\mu} x_{\mu}} \nub.
\label{planewave}
\eeq
$\nub$ is a $\beta$-flavor neutrino spinor, with no time or spatial 
dependence.  The 4-momentum is represented by a matrix because it is not
diagonal in flavor space.  Eventually, we will diagonalize the 4-momentum and
move into the mass basis, but first a word about the plane wave assumption.

As mentioned above, a given flavor of neutrino created at a source
does not necessarily have a definite mass, and therefore the 4-momentum is not
diagonal in flavor space.  Most discussions of neutrino oscillations, such as 
those found in \cite{MP} and \cite{BP}, treat the neutrino of definite flavor 
as a particle with definite momentum, composed of a superposition of energy 
states
corresponding to the different mass states.  Reference \cite{Kays2} shows that
this treatment yields the same results as a treatment using a definite energy
and a superposition of momentum states, and that both yield results consistent
with a more rigorous treatment of the problem using wave packets, provided the
neutrinos are relativistic.  Appendix~\ref{wvpktsec} summarizes the
wave-packet approach and the validity of its approximation by the plane wave
solution (\ref{planewave}).  The energy-momentum assumptions of the standard
oscillation treatment are examined in reference \cite{Goldman}, and we present
those arguments in Appendix~\ref{relativsec}.  In what follows, we will use the
plane wave solution (\ref{planewave}) rather than wave packets for simplicity,
but we will use the relativistic results of Appendix~\ref{relativsec} to conserve
both energy and momentum.

Examining the plane wave $\nua_L$ in equation (\ref{planewave}), we find
\beq
\partial_0 \partial^0 \nua_L  =  
\ - \!\!\!\! \sum_{\beta=flavors} \left(E_{\ab}\right)^2 \nub_L, \mbox{ and}
\label{partialt}
\eeq
\beq
\partial_a \partial^a \nua_L  = 
 - \sum_{a=1}^3 \sum_{\beta=flavors} p_{\ab a} p_{\ab}^a \nub_L
\equiv - \sum_{\beta} (|\vec{p}|^2)_{\ab} \nub_L.
\label{partials}
\eeq
Note that  the index $\alpha$ is not summed over in equations (\ref{partials})
and (\ref{partialt}), but $a$ is.  As usual, $\alpha$ and $\beta$ are flavor 
indices, and
$a$ is a spatial Lorentz index.  The conventions we use for summing and indices 
are described in Appendix~\ref{key}.  Using the 
plane wave solution in (\ref{vaceos}), we obtain an equation for the matrix
$p^{\mu}_{\ab}p_{\ab \mu}$.
\beq
\partial_{\mu} \partial^{\mu} \nua_L 
= \sum_{\beta} \left( E_{\ab}^2 - |\vec{p}|^2_{\ab} \right) \nub_L = 
\sum_{\beta} (M_D M_D^{\dagger})_{\ab} \nub_L.
\label{vacham}
\eeq
Neutrinos propagate in {\it mass} states, not flavor states, so we must
diagonalize the mass matrix in (\ref{vacham}) to find an equation of motion for
the mass states.  The same mixing matrix $V$ diagonalizes both
the mass matrix and the four-momentum matrix, leaving us with an expression 
for the four-momentum of the $i$th mass state $\nui_L$:
\bea
\sum_{\alpha,\beta,\gamma} V_{j \alpha}^{\dagger}
\left(E^2 - |\vec{p}|^2\right)_{\alpha \gamma} 
\left( V V^{\dagger} \right)_{\gamma \beta}\nub_L & = &
\sum_{\alpha,\beta, \gamma} V_{j \alpha}^{\dagger} 
\left(M_D M_D^{\dagger}\right)_{\alpha \gamma} 
\left( V V^{\dagger} \right)_{\gamma \beta} \nub_L, 
\mbox{\ \ so \ \ } \nonumber \\ 
\sum_{\alpha, \gamma} \sum_i \left[ V^{\dagger}_{j \alpha}
\left(E^2 - |\vec{p}|^2\right)_{\alpha \gamma} V_{\gamma i} \right]
\sum_{\beta} V^{\dagger}_{i \beta} \nub_L & = & 
\sum_{\alpha, \gamma} \sum_i \left[ V^{\dagger}_{j \alpha}
\left(M_D M_D^{\dagger} \right)_{\alpha \gamma} V_{\gamma i} \right]
\sum_{\beta} V^{\dagger}_{i \beta} \nub_L \nonumber \\
& = &
\sum_i m_i \delta_{ij} \nui_L,
\label{rotateM}
\eea
with
\beq
\nui_L \equiv \sum_{\beta} V^{\dagger}_{i \beta} \nub_L
\eeq
and
\beq
\sum_{\alpha \gamma}V^{\dagger}_{i \alpha} (M_D M_D^{\dagger})_{\alpha \gamma} 
V_{\gamma j} = m_i^2 
\delta_{ij}.
\eeq
The matrix on the right-hand side of equation (\ref{rotateM}) is diagonal, so 
the matrix on the left-hand side must be diagonal too.  Consider first the 
rest frame of the neutrino.  The three
momentum is zero, so the left-hand side of equation (\ref{rotateM}) is
\beq
\sum_{\alpha,\gamma} \left[V_{j \alpha}^{\dagger}
\left(E^2_{\alpha \gamma} \right)
V_{\gamma i} \right] \nui_L \equiv E_i \delta_{ij}.
\label{rotateE}
\eeq
Making a Lorentz boost back into the moving frame only changes Lorentz indices,
not flavor indices, so equation (\ref{rotateE}) remains valid in the frames
with non-zero momentum.  Since the left-hand side of equation (\ref{rotateM})
must be diagonal and the first term is diagonal by equation (\ref{rotateE}),
the second term must be diagonal too:
\beq
\sum_{\alpha, \gamma} \left[V_{j \alpha}^{\dagger}
\left(|\vec{p}|^2_{\alpha \gamma}\right) V_{\gamma i} \right] \equiv
|\vec{p}_i|^2 \delta_{ij}.
\label{rotatep}
\eeq
Comparison of equations (\ref{rotateM}) through (\ref{rotatep}) yield the
familiar energy-momentum relationship
\beq 
(E_i^2 - |\vec{p}_i|^2) \nui_L = m_i^2 \nui_L.
\eeq

Having rotated into mass states of definite energy and momentum,
we may now describe the propagation of the neutrino 
mass eigenstates through space:
\beq
\nui_L (t_i, {\bf x}_i) = e^{i \left(E_i t_i - {\bf p}_i \djdot {\bf
x}_i\right)} 
\nui.
\label{propDirac}
\eeq
We have placed mass indices on the time and position, since the propagation of
the state in space and time depends on the mass.  (Different mass states will
travel the same distance in different times or different distances in the same
time).  An equation similar to (\ref{propDirac}) may be derived for the sterile 
neutrino $N_R$, and the two
may be combined in vacuum to describe the behavior of a Dirac neutrino $\nu$.


\section{Vacuum Oscillations}
\label{vacoscsec}


Weak interactions at a neutrino source produce neutrinos of a definite flavor,
$\nua$, so the wavefunction of a neutrino when it is produced is
\beq
\nu_L (t=0,x=0) = \nua_L = \sum_i V_{\alpha i} \nui_L,
\label{nuinit}
\eeq
The wavefunction a distance $x$ from the source, and a time $t$ after emission,
(taking the x-axis along the
momentum direction) is
\beq
\nu_L (t, x) = \sum_{i=1}^{n} \Vai \nui e^{-i\left(E_i t_i - p_i x_i\right)} 
\equiv \sum_{i=1}^{n} \Vai \nui e^{-i \phi_i},
\label{xfinal}
\eeq
where $\phi_i \equiv E_i t_i-p_i x_i$ is the 
{\it kinematic phase} of the propagating mass state $\nu_i$. 
Although neutrinos travel in mass states, they are detected in flavor states,
so we must rotate back to the flavor basis to compare with experiment: 
\beq
\nu_L (t, x) = \sum_{\beta=flavors} \left[ \sum_{i=1}^{n} \Vai {\Vbis}
e^{-i \phi_i} \right] \nub 
\equiv \sum_{\beta} a_{\beta \alpha} \nub.  
\label{relativistic}
\eeq
$a_{\beta \alpha}$ is the amplitude for the transformation $\nua \rightarrow 
\nub$, and
the probability of the transformation is
\beq
\Pab = |a_{\beta \alpha}|^{2} = \sum_{i=1}^{n} \sum_{j=1}^{n} (\Vai \;
\Vbis \; \Vajs \; \Vbj ) e^{-i \left( \phi_i - \phi_j \right)}. 
\label{Probab}
\eeq
We call the phase difference the {\it relative phase}
\beq
\Phi_{ij} \equiv \half \left( \phi_i-\phi_j \right),
\eeq
where the factor of $\half$ is included for future convenience.

Because the mixing matrix is unitary, 
\beq
\sum_{i=1}^{n} \sum_{j=1}^{n} \left(\Vai {V_{\beta i}}^{\!\!\!\!\!\!*}
\;{V_{\alpha j}}^{\!\!\!\!\!\!*} \;V_{\beta j}\right) = \delta_{\alpha \beta},
\eeq
and (\ref{Probab}) may be written as
\beq
\Pab = \sum_{i=1}^{n} \sum_{j=1}^{n} \left(\Vai
{V_{\beta i}}^{\!\!\!\!\!\!*} \;{V_{\alpha j}}^{\!\!\!\!\!\!*}
\;V_{\beta j}\right) \left( e^{-i 2 \Phi_{ij}} - 1 \right) + 
 \delta_{\alpha \beta}.  
\label{delta}
\eeq
Terms antisymmetric under the interchange $i \leftrightarrow j$ will
cancel in the double sum, leaving only symmetric terms.  
The product of matrix elements in equation (\ref{delta})
becomes its complex conjugate under this interchange, so the real part is
symmetric and the imaginary part antisymmetric.  The exponential may also be
broken into its symmetric cosine and antisymmetric sine contributions.  Keeping
only the symmetric products, equation (\ref{delta}) becomes
\bea
\Pab & = & \sum_{i=1}^{n} \sum_{j=1}^{n} \left[ \mbox{Re}\left(\Vai
{V_{\beta i}}^{\!\!\!\!\!\!*} \;{V_{\alpha j}}^{\!\!\!\!\!\!*}
\;V_{\beta j}\right)  \left( \cos \left( 2 \Phi_{ij} \right) - 1 \right) 
\right. \nonumber \\ 
&&\left.  + \ \mbox{Im} \left(\Vai
{V_{\beta i}}^{\!\!\!\!\!\!*} \;{V_{\alpha j}}^{\!\!\!\!\!\!*}
\;V_{\beta j}\right) \sin \left( 2 \Phi_{ij} \right) \right]
+ \delta_{\alpha \beta} \nonumber \\ 
& = & - \sum_{i=1}^{n} \sum_{j=1}^{n} \left[
\mbox{Re} \left( \Vai
{V_{\beta i}}^{\!\!\!\!\!\!*} \;{V_{\alpha j}}^{\!\!\!\!\!\!*}
\;V_{\beta j}\right) 2 \sin^2 \left( \Phi_{ij} \right)
\right. 
\label{symm} \\
&&\left. - \ \mbox{Im} \left( \Vai
{V_{\beta i}}^{\!\!\!\!\!\!*} \;{V_{\alpha j}}^{\!\!\!\!\!\!*}
\;V_{\beta j}\right) \sin \left( 2 \Phi_{ij} \right) \right]
+ \delta_{\alpha \beta}. \nonumber 
\eea
The terms for which $i=j$ do not contribute since $\Phi_{ii}=0$, so we remove 
them 
to arrive at our final form for the neutrino oscillation probability:
\bea
\Pab & = & -2 \sum_i \sum_{j \neq i} \mbox{Re}(\Vai \Vbis \Vajs \;\Vbj) 
\sin^2 \left( \Phi_{ij} \right) 
\label{oscillation}  \\
&&+ \ \sum_i \sum_{j \neq i} \mbox{Im}(\Vai \Vbis \Vajs \;\Vbj)  
\sin \left( 2 \Phi_{ij} \right) + \delta_{\ab} 
\nonumber 
\eea

For relativistic neutrinos, $\Phi_{ij}$ is derived in Appendix~\ref{relativsec}
to be
\beq
\Phi_{ij} \approx \frac{m_i^2-m_j^2}{4p} x,
\label{Phiij}
\eeq
which is linear in $x$, the distance traveled.
As a neutrino propagates, its transition probability thus oscillates in 
distance
as the square of a sine function with an oscillation length
\beq
L_{ij} \equiv \frac{\pi x}{\Phi_{ij}} = 
\frac{4 \pi p}{m_{i}^{2} - m_{j}^{2}} \equiv \frac{4 \pi p}{\Delta
m_{ij}^{2}}.
\label{length}
\eeq
$\Delta m_{ij}^2 \equiv m_i^2 - m_j^2$ is called the 
{\it mass-squared difference} for mass states $i$ and $j$.

We see upon examination of equations (\ref{oscillation}) and (\ref{Phiij}) that
the oscillation probability when all $m_i=m_j$ is $\delta_{\ab}$.  Thus if the
neutrino masses are all zero, the probability for a neutrino to change flavors
is $0$ and the probability for the neutrino to stay the same flavor is $1$. 
Detection of neutrino oscillations would therefore be proof of non-zero
neutrino masses and potentially the first contradiction of the Standard Model.


\subsection{Two-Way Oscillations}


Consider the simplified case of two neutrino flavors, $n=2$.  An
arbitrary $2\times2$ unitary matrix will have one angle parameterizing
the real degree of freedom, and three
phases corresponding to the three imaginary degrees of freedom.
But all three phases may be absorbed into the definitions of
Dirac fermion fields (or cancel in oscillation probabilities for Majorana
neutrinos as described later in Section~\ref{dogsec}), 
so a $2\times2$ mixing matrix has only one (real) degree of freedom, which we
will describe by the angle $\theta$.  The mixing
matrix V then has the form
\beq
V = \left( 
\begin{array}{cc}
       \mbox{cos} \,\theta    &   -\mbox{sin} \,\theta  \\
       \mbox{sin} \,\theta    &    \mbox{cos} \,\theta
\end{array}
\right).
\label{twoflavorV}
\eeq
Since the matrix is explicitly real, the term in equation (\ref{oscillation})
involving imaginary matrix elements disappears, and the oscillation probability
in the two flavor case is simply
\bea
\Pab & = & + 4\, \mbox{cos}^2
\,\theta\; \mbox{sin}^2 \,\theta \;\sin^2 \left( \frac{\Delta
m^{2}_{12}}{4 p} x \right) + \delta_{\ab} \nonumber \\ &&  
     = \mbox{sin}^2 \,2 \theta \; \sin^2 \left( \frac{\Delta m^{2}_{12}}{4 p}
x \right) + \delta_{\ab}
\label{twoflavorP}
\eea


\subsection{Three-Way Oscillations}
\label{3waysec}


The three-generation case provides a bit of freedom in parameterizing the
mixing matrix.  An arbitrary $3 \times 3$ unitary matrix has three real
degrees of freedom, and six phases.  As will be shown in Section~\ref{dogsec},
$2n-1=5$ relative phases are not observable, so the three-generation mixing
matrix needs three mixing angles and one phase to describe it.
The quark sector contains such a matrix, called the CKM matrix after Cabbibo,
Kobayashi, and Maskawa, who developed the theory of quark mixing.
The original choice of these four parameters, by Kobayashi and Maskawa, 
is perhaps the best known parameterization.  Their choice of $V$ is
\cite{Nact}
\beq
\left( \ba{ccc} 
\co{1} & \;\s{1} \co{3}\; & \s{1} \s{3} \\
-\s{1}\co{2} & \; \co{1}\co{2}\co{3}-\s{2}\s{3}e^{i\delta}\; &
\co{1}\co{2}\s{3}+\s{2}\co{3}e^{i\delta} \\
-\s{1}\s{2} & \co{1}\s{2}\co{3}+\co{2}\s{3}e^{i\delta} &
\co{1}\s{2}\s{3}-\co{2}\co{3}e^{i\delta}  \ea \right),
\label{NactCKM}
\eeq
where $\co{a}\equiv \cos \theta_a$, and $\s{a} \equiv \sin \theta_a$.  We will
refer to this particular choice of angles as the ``standard'' or ``KM''
parameterization, while the name ``CKM matrix'' refers to any three-generation
mixing matrix parameterization.

In the standard KM parameterization, the phase only appears in the lower
right-hand ``corner'' of the matrix.  The placement of the phase is somewhat
arbitrary, since we absorb five relative phases into the field definitions. 
Choosing to absorb a different set of relative phases repositions the remaining
phase in the mixing matrix.  Clearly the location of the phase cannot be
measurable!  Indeed, as discussed in the next chapter, individual mixing matrix 
elements cannot be measured independently, and the measurable probabilities 
include a single function of the phase, called ${\cal J}$, not the phase 
itself.  In the standard KM
parameterization, the invariant ${\cal J}$ has the form
\beq
{\cal J} = \co{1} \st{1} \co{2} \s{2} \co{3} \s{3} \sin {\delta}.
\label{Jdef}
\eeq
This function has a maximum value of $\frac{1}{6\sqrt{3}}$ \cite{Nir}:
\beq
{\cal J}_{max} = \left(\cos{\theta_1} - \cos{\theta_1}^3\right)_{max} 
\left(\half \sin 2\theta_2\right)_{max} \left(\half \sin 2\theta_3\right)_{max}
\left(\sin \delta \right)_{max}
= \frac{2}{3\sqrt{3}} \half \half \left(1\right) = \frac{1}{6\sqrt{3}}.
\label{Jmax}
\eeq

Another popular parameterization chooses slightly different mixing angles and
places the phases elsewhere in the mixing matrix \cite{BargPhil}, \cite{Nir}:
\beq
\left( \ba{ccc}
\co{12}\co{13} & \s{12}\co{13} & \s{13} e^{-i \delta'} \\
-\co{23}\s{12} - \co{12}\s{23}\s{13} e^{i\delta'} & 
  \; \co{12}\co{23}-\s{12}\s{23}\s{13} e^{i\delta'} \; & \co{13}\s{23} \\
\s{12}\s{23} - \co{12}\co{23}\s{13} e^{i\delta'} & 
  \; - \co{12}\s{23}-\co{23}\s{12}\s{13} e^{i\delta'} \; & \co{13}\co{23}
\ea \right).
\label{altCKM}
\eeq
Both of these parameterizations (\ref{NactCKM}) and (\ref{altCKM}) are equally
valid, but they are not equivalent.  In addition, we will demonstrate in
Chapter~\ref{massbox} 
how complicated the observables are in terms of the standard
KM parameterization.  Our development of a model-independent parameterization
has been motivated by the arbitrariness and complexity of the traditional
approach.


\section{Matter Effects}
\label{mattersec}


When neutrinos travel through matter, they may interact with the
electrons, protons, and neutrons contained in the matter.  
Once these interactions
are averaged, the resulting terms in the effective Lagrangian resemble mass
terms, as shown below.  All flavors of neutrinos will scatter off
of electrons, protons, and neutrons by the neutral current interaction, so the 
mass matrix receives a contribution proportional to the identity matrix which 
does not affect the 
mixing matrix.  (Rotating this identity term by the mixing matrix leaves it
diagonal while diagonalizing the mass terms.  Thus the same mixing matrix that
diagonalizes the vacuum Lagrangian will diagonalize the Lagrangian with
neutral-current terms added.)  However, only      
electron-type neutrinos and electrons may scatter via the charged current. 
This type of
interaction produces a non-trivial contribution to the mass matrix, affecting
the mixing matrix and oscillation parameters.

As discussed in Section~\ref{SMsec}, neutrino-electron charged current
scattering may be
represented by the effective Lagrangian of equation (\ref{Leff}).  Substituting
for ${\cal J}^{CC}$, we obtain
\beq
\lag_{eff}^{CC} = \frac{-4 G}{\sqrt{2}} \left[ \ol{\nu_{eL}} \gupu e_L
\right] \left[ \ol{e_L} \gdnu \nu_{eL} \right].
\eeq
The order of the spinors may be changed by a {\it Fierz transformation} (see,
for example, Appendix F of reference \cite{Nact} or Exercise 2.12 of reference
\cite{Grein5} for details on these 
transformations), resulting in
\beq
\lag_{eff}^{CC} = -\frac{4G}{\sqrt{2}} \left[ \ol{\nue_L} \gupu \nue_L
\right] \left[ \ol{e_L} \gdnu e_L \right].
\label{mLeff}
\eeq
The electron factor may be reduced by averaging over the electron field and
assuming non-relativistic electrons \cite{MP} to give
\beq
\langle \ol{e_L} \gupu e_L \rangle \approx \langle \ol{e_L} \gupo e_L \rangle
= \langle e_L^\dagger e_L \rangle = n_{e_L} \approx \half n_e,
\eeq
where $n_e$ is the electron number density in the medium interacting with the
neutrinos, and we assume that the medium contains equal numbers of left- and
right-handed electrons.  Thus the contribution to the Lagrangian from the 
charged current neutrino-electron scattering is
\beq
\lag_{eff}^{CC} = -\sqrt{2} G n_e \ol{\nu_{eL}} \gdno \nu_{eL}.
\eeq
A similar treatment may be applied to the neutral current contribution to the
effective Lagrangian, resulting in
\beq
\lag_{eff}^{NC} = \sum_{\alpha} \frac{\sqrt{2}G}{2} \left(n_n + n_p - n_e
\right) \ol{\nu_{\alpha L}} \gdno \nu_{\alpha L} = 
\sum_{\alpha} \frac{1}{\sqrt{2}} G n_n \ol{\nu_{\alpha L}} 
\gdno \nu_{\alpha L},
\eeq
where the factor of $\half$ comes from the difference between charged and
neutral currents, and we have set $n_e = n_p$ for electrically neutral matter.
As we shall see, this neutral current term does not affect neutrino 
oscillations since it is uniform in flavor.

Because these effective Lagrangian terms contain only the spinor products
$\ol{\nu} \nu$, they are similar to mass terms and affect the diagonalization
of the mass matrix.  When these interaction terms are added to the vacuum
neutrino Lagrangian (\ref{nulag}), equation (\ref{varynu})
becomes
\beq
\frac{\delta \lag}{\delta \ol{\nua_L}} = i \partial_{\mu} \gupu \nua_{L}
  - \sum_{\beta} {M_D}_{\alpha \beta} N_{\beta R}
  + \frac{1}{\sqrt{2}} G n_n \gdno \nu_{\alpha L}
  - \sqrt{2} G n_e \delta_{\alpha e} \gdno \nua_L = 0,
\eeq
while equation (\ref{varyN}) remains unchanged:
$$ 
\frac{\delta \lag^{\nu}}{\delta \ol{N_{\alpha R}}} = 
  i \partial_{\mu} \gupu N_{\alpha R} 
  -\sum_{\beta} {M_D^{\dagger}}_{\alpha \beta} \nub_L = 0. \eqno{(\ref{varyN})}
$$ 
  The equation of motion for the
active neutrino (\ref{eos}) becomes
\bea
i \partial_{\mu} \gupu N_{\alpha R} & = &
\sum_{\gamma} M^{-1}_{D \alpha \gamma} i \partial_{\mu} \gupu 
\left[ \left( i \partial_{\nu} \gupv + \frac{1}{\sqrt{2}} G n_n \gdno 
- \sqrt{2} G n_e \delta_{\gamma e} \gdno \right) \nug_L \right] \nonumber \\
& = & \sum_{\gamma} {M_D^{\dagger}}_{\alpha \gamma} \nug_L.
\label{meos}
\eea
If the densities $n_e$ and $n_n$ do not fluctuate quickly in space or time, we may
approximate
\beq
i \partial_{\mu} n_{e,n}  \approx  n_{e,n} i \partial_{\mu}.
\eeq
This approximation is valid if $\frac{1}{n E}\frac{dn}{dt} << 1$ \cite{KP}. 

From the Dirac equation we know
\beq
\partial_{\mu} \gupu \nua_L \approx m_{\nu} \nua_L \approx 0.
\eeq
The flavor states don't have definite mass, but we can assume $m_{\nu}$, the
effective mass of a flavor state, to be on the order of the individual masses
$m_i$, and therefore small.  From this approximation, we find
\beq
\partial_0\gupo \nua_L \approx \partial_a \gupaa \nua_L, \mbox{\ \ so \ \ }
\eeq
\beq
\partial_{\mu} \gupu \gdno \nua_L = \gdno ( \partial_0 \gupo + \partial_a
\gupaa) \nua_L  \approx  \gdno (2 \partial_0 \gupo) \nua_L = 2 \partial_0
\nua_L. 
\label{consapprox}
\eeq
With these substitutions and the simplification of equation (\ref{gammasimp}),
equation (\ref{meos}) becomes
\beq
\partial_{\mu} \partial^{\mu} \nua_L 
+ \sum_{\beta} (M_DM_D^{\dagger})_{\alpha \beta} \nub_L
+ iG \sqrt{2} \left(2 n_e \delta_{\alpha e} - n_n \right) \partial_0 \nua_L = 0.
\eeq
This looks like the Klein-Gordon equation (\ref{vaceos}) with an amount 
$G \sqrt{2} \partial_0 (2 n_e \delta_{\alpha e} - n_n )$ 
added to the diagonal elements of the mass matrix. 
Again we try a plane-wave solution of the form (\ref{planewave}).  The
energy-momentum relationship, however, is affected by the matter, becoming
\beq
\sum_{\beta} \left( E_{\ab}^2 - |\vec{p}|_{\ab}^2 \right) \nub_L = 
\sum_{\beta} \left[ \left( M_D M_D^{\dagger} \right)_{\ab} +
iG \sqrt{2} \left(2 n_e \delta_{\alpha e} - n_n \right) \left(-iE_{\ab}\right)
\right] \nub_L.
\eeq
The neutral current term will automatically be diagonalized by the same mixing
matrix which diagonalizes $E_{\ab}$, but
the charged current term will affect the diagonalization.  We may define a new
mass matrix $M^M$ which includes this charged current term and is diagonalized
by the matter mixing matrix, $\Vm$:
\beq
V^{M\dagger}_{i \alpha} \left[ (M_D M_D^{\dagger})_{\ab} 
+ 2 E_{\ab} \sqrt{2}Gn_e\delta_{\alpha e} \right] \Vmbj 
\equiv V^{M\dagger}_{i \alpha} \left(M^M M^{M \dagger}\right)_{\ab} \Vmbj
= \left({m^M_i}\right)^2 \delta_{ij},
\label{massCKM}
\eeq
where ${m^M_i}$ are the effective neutrino masses in matter.  
The mixing matrix is independent of the neutral current 
term, as predicted before.  As for the vacuum case, the
momentum and energy are also diagonalized by the mixing matrix:
\beq
V^{M\dagger}_{i \alpha} |\vec{p}|^2_{\ab} \Vmbj = |\vec{p}_i|^2 \delta_{ij},
\mbox{\ \ and \ \ }
\eeq
\beq
V^{M \dagger}_{i \alpha} E^2_{\ab} \Vmbj = E_i^2 \delta_{ij}.
\eeq

The oscillation equations in matter are the same as they were for neutrinos
traveling through a vacuum, provided the mixing matrix $V$ is replaced with 
$\Vm$, and the masses $m_i^2$ are also replaced with their corresponding
$\left(m^M_i\right)^2$.  For example, equation (\ref{oscillation}) would become
\bea
\Pmab & = & -2 \sum_i \sum_{j \neq i} \mbox{Re}(\Vmai \Vmbis \Vmajs \;\Vmbj) 
\sin^2 \left( \frac{\left(m_i^M\right)^2-\left(m_j^M\right)^2}{4p}\right) \\
&&+ \ \sum_i \sum_{j \neq i} \mbox{Im}(\Vmai \Vmbis \Vmajs \;\Vmbj)  
\sin \left( \frac{\left(m_i^M\right)^2-\left(m_j^M\right)^2}{2p}\right) 
+ \delta_{\ab}.
\nonumber 
\eea


\chapter{A New Parameterization}
\label{boxes}


\section{The Unmeasurables}
\label{relphasesec}


The individual elements of the mixing matrix $V$ which appear in the Lagrangian
are unmeasurable.  Consider
the two different parameterizations of the three-generation mixing matrix
presented in equations (\ref{NactCKM}) and (\ref{altCKM}) of 
Section~\ref{3waysec}.  The freedom to absorb phases into the
definitions of fermion fields produces a freedom in the placement of the
remaining phase.  For a Dirac mass term, we may absorb
$2n$ phases into the definitions of the $n$ left-handed neutrino 
fields and $n$ left-handed fields of the charged leptons.  Such an absorption
removes $2n-1$ relative phases from 
the mixing matrix without changing any other
terms in the Lagrangian, provided we absorb conjugate phases in the
right-handed fields:
\bea
{\cal L} 
& = &\ol{\nuL} M_{\nu} N_R + \ol{l_L} M_l l_R + c_2 W^+_{\mu} \ol{\nuL} \gupu 
  l_L + h.c. \nonumber \\
& = &\ol{\nu_L^m} D_{\nu} N_R^m + \ol{l_L^m} M_l l_R^m + 
c_2 W^+_{\mu} \ol{\nu_L^m} V \gupu l_L^m + h.c. \\
& = & \ol{\nu_L^m} X_{\nu} X_{\nu}^{-1} D_{\nu} X_{N} X_{N}^{-1} N_R^m + 
  \ol{l_L^m} X_{l_L} X_{l_L}^{-1} M_l X_{l_R} X_{l_R}^{-1} l_R^m \nonumber \\
&& + \ c_2 W^+_{\mu} \ol{\nu_L^m} X_{\nu} X_{\nu}^{-1} V X_{l_L} X_{l_L}^{-1} 
    \gupu l_L^m + h.c.,\nonumber 
\eea
with
\beq
\left(X_{\psi}\right)_{ij} = \delta_{ij} e^{-i \left(\xi_{\psi}\right)_i}    
\eeq
Setting $X_N = X_{\nu}$ and $X_{l_R} = X_{l_L}$, which leaves the diagonalized
matrices $D_{\nu}$ and $M_l$ unchanged, and redefining
$\psi'=X_{\psi}^{-1} \psi^m$, we get
\beq
\ol{\nuL'} D_{\nu} \NR' + \ol{l_L'} M_l l_R' + 
  c_2 W^+_{\mu} \ol{\nuL'} \left[X_{\nu}^{-1} V X_{l_L} \right] \gupu l_L' + h.c.
\eeq
So we may absorb $n$ phases through $X_{\nu}$ and $n$ phases through
$X_{l_L}$ in the field definitions, thereby removing $2n-1$ relative phases from
the mixing matrix, without changing the rest of the Dirac neutrino Lagrangian. 
Note that it is the phase {\it differences} between $X_{\nu}^{-1}$ and $X_l$
that enter into the new mixing matrix, and so $2n-1$, rather than $2n$, phases
are removed from $V$.
  
If the neutrinos are Majorana particles, however, we do not have the freedom to
set the phase matrix $X_N$ independently of $X_{\nu}$ \cite{BP}.  The Majorana mass term 
has the form
\beq
\ol{\nuL} M_{\nu} \nuRc, \mbox{\ \ \ with \ \ \ }
\nuRc = C \ol{\nuL}^T.
\eeq
Making the transformation $\nuL \rightarrow \nuL' = X_{\nu}^{-1} \nuL^m$ results in
the transformation $\nuRc \rightarrow \nuRc' = X_{\nu}^{-1 *} \nuRc^m = X_{\nu}
\nuRc^m$; the
phase matrices in the mass term do not cancel, but combine, and the Lagrangian
is not invariant under such a transformation:
\beq
\ol{\nuL} M_{\nu} \nuRc \neq \ol{\nuL'} D_{\nu} \nuRc'.
\eeq 
Therefore phases may be absorbed only in the charged lepton fields, and just
$n_l$ (which may be different from $n=n_{\nu}$ in models with sterile
neutrinos) phases may be removed from a Majorana mixing matrix.

These $2n-1-n_l$ extra phases in a mixing matrix for Majorana neutrinos are not,
however, observable in oscillation experiments \cite{BP}.  Each row and each
column of $V$ may be multiplied by an arbitrary phase; {\it id est},  
any mixing matrix element $V_{\alpha i}$ may be replaced with
$e^{i \xi_{\alpha}} V_{\alpha i} e^{-i \xi_i}$ without changing the product of
four matrix elements found in the oscillation equations:
\beq
\Vai \Vajs \Vbis \Vbj = e^{i \xi_{\alpha}} \Vai e^{-i \xi_i}
e^{-i \xi_{\alpha}} \Vajs e^{i \xi_j} e^{-i \xi_{\beta}} \Vbis e^{i \xi_i}
e^{i \xi_{\beta}} \Vbj e^{-i \xi_j}.
\eeq
Thus whether the mass matrix is Majorana or Dirac, $2n-1$ relative phases will
not appear in the oscillation probabilities even if the number of
neutrinos is greater than the number of charged leptons.  
Processes sensitive to the ``extra'' Majorana phases include double-beta decay
and the radiative decay of unstable neutrinos contained 
in some models; in these processes
the {\it same} vertex occurs twice (as opposed to a vertex and its hermitian
conjugate) in the relevant Feynman diagrams.  A discussion of neutrinoless
double-beta decay and its relation to Majorana masses is contained in
Appendix~\ref{appd}.


\section{The Measurables}
\label{boxdefsec}


The immeasurability of the mixing matrix elements in the quark sector has been
addressed by numerous authors, such as those of references \cite{Jarls1},
\cite{Wu}, \cite{BD}, and \cite{DDW}.  Measurable quantities include only the
magnitudes of mixing matrix elements, the products of four mixing matrix
elements appearing in the oscillation equation (\ref{oscillation}), and
particular higher-order functions of mixing matrix elements \cite{Wu}.  
Neutrino oscillation probabilities depend linearly on the fourth-order 
objects,
\beq
\Baibj \equiv \Vai V^{\dagger}_{i \beta} \Vbj V^{\dagger}_{j \alpha} 
= \Vai \V{\alpha j}^* \V{\beta i}^* \Vbj,
\label{boxdef}
\eeq
which we will call ``boxes'' since each contains as factors the 
corners of a submatrix, or ``box," of the mixing matrix.
For example,
\beq
\B{11}{22} = \V{11} \V{12}^* \V{21}^* \V{22}.
\eeq
These boxes are the neutrino equivalent of the ``plaques'' used by 
Bjorken and Dunietz for another purpose in the quark sector in reference
\cite{BD}.  In this work, we will examine the applications of using boxes
to describe neutrino oscillations.

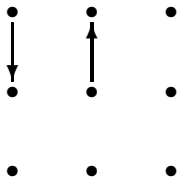
\begin{figure}[htb]
\vsp
\begin{centering}
\begin{picture}(70,70)(0,0)
\thicklines
\multiput(5,65)(30,0){3}{\circle*{4}}
\multiput(5,35)(30,0){3}{\circle*{4}}
\multiput(5,5)(30,0){3}{\circle*{4}}
\put(5,61){\vector(0,-1){22}}
\put(35,39){\vector(0,1){22}}
\end{picture}
\caption[The graphical representation for 
the box $\B{11}{22}$ in three generations.]{The graphical representation for 
the box $\B{11}{22}$ in three generations.  Vertical arrows point from the
matrix elements which are not complex conjugated in the box to the 
complex-conjugated elements.
\label{b1122fig}}
\end{centering}
\end{figure}

We have developed a graphical representation (discussed in detail in
Appendix~\ref{graphapp}) which proves useful when considering relationships 
between boxes.  Figure~\ref{b1122fig} illustrates the representation of 
$\B{11}{22}$.  This representation will be used throughout this Chapter to
illustrate various relationships.

In general, each of the box indices $i$, $j$, $\alpha$, and $\beta$ may be any
number between $1$ and $n$, the number of neutrino flavors.  We therefore
initially have $n^4$ possible boxes.
Examination of equation~(\ref{boxdef}), however, 
reveals a few symmetries in the indexing:
\beq
\Baibj = \B{\beta j}{\alpha i} = \Bs{\beta i}{\alpha j} = 
\Bs{\alpha j}{\beta i}.
\label{symmetries}
\eeq
If the order of either set of indices is reversed ({\it id est}, 
$j \leftrightarrow i$
or $\beta \leftrightarrow \alpha$), the box turns into its complex conjugate; 
if both sets of indices are reversed, the box returns to its original value.  
Equation~(\ref{symmetries}) demonstrates that boxes with $\alpha = \beta$ or 
$i = j$, are real.  Indeed, these are given from equation~(\ref{boxdef}) as
\bea
\B{\alpha i}{\alpha j} & = & |\V{\alpha i}|^{2} |\V{\alpha j}|^{2}, 
\mbox{\ \ and}
\label{samea} \\ 
\B{\alpha i}{\beta i} & = & |\V{\alpha i}|^{2} |\V{\beta i}|^{2}.
\label{samei} 
\eea
Those boxes with both sets of indices equal are
\beq
\B{\alpha i}{\alpha i} = |\Vai|^4.
\eeq

We call boxes with one or two repeated indices ``singly-degenerate'' and
``doubly-degenerate,'' respectively.
As can be seen from equation (\ref{oscillation}), singly-degenerate boxes 
with repeated flavor indices enter into the formulae for flavor-conserving
survival probabilities, but not for flavor-changing transition
probabilities.  Degenerate boxes with repeated mass indices (including 
the doubly-degenerate boxes) do not appear in
any oscillation formula.  Degenerate boxes may be expressed in terms of 
the nondegenerate boxes with
$\alpha \neq \beta$ and $i \neq j$, as will be shown shortly. This
possibility and the symmetries expressed in 
equation~(\ref{symmetries}) allow us to express combinations of boxes in terms
of only the    
nondegenerate ``ordered'' boxes for which $\alpha < \beta$ and    
$i < j$. 

The number of flavor-index pairs satisfying $\alpha < \beta$ ordering is 
$N \equiv\left( \ba{c} n \\ 2 \ea \right) = 
\frac{n\left(n-1\right)}{2}$.  $N$ 
mass-index
pairs similarly satisfy $i<j$ ordering.  Thus, the number of ordered
nondegenerate boxes is $N^2$.  The number of flavor-degenerate boxes is $n^3$,
and that of mass-degenerate boxes is also $n^3$.  The $n^2$ doubly-degenerate
boxes  appear in both these counts, so the total number of degenerate boxes is
$2 n^3 - n^2 = n^2(2n-1)$, implying $n^2(n-1)^2$ nondegenerate boxes.  From the
nondegenerate boxes, we want to select only the ordered ones for which $j>i$ and
$\beta>\alpha$, which comprise one fourth of the nondegenerate boxes.  The
number of nondegenerate ordered boxes is therefore $\left( \frac{n(n-1)}{2}
\right)^2$, as argued above.  Applying ordering to the singly-degenerate boxes
gives a count of $\half n^2 \left(n-1\right) = nN$ for flavor-degenerate boxes,
and the same for mass-degenerate boxes, yielding $2nN$ total singly-degenerate
ordered boxes.  For $n=3$, $N=3$ too, and
we have nine ordered nondegenerate boxes, nine ordered singly-degenerate boxes,
and nine doubly-degenerate boxes.  These counts are recapped
later in Table~\ref{counttab}, along with other box counts.

Using the symmetries expressed in equation (\ref{symmetries}), equation 
(\ref{oscillation}) becomes
\bea
\Pab & = & -\sum_{i=1}^{n} \sum_{j>i} 
\left[ 2  \mbox{Re}\left( \Baibj \right) \sin^2 \Pij - 
\mbox{Im} \left( \Baibj \right) \sin 2 \Pij \right] \nonumber \\ 
&&- \ \sum_{i=1}^{n} \sum_{j<i}
\left[ 2  \mbox{Re}\left( \Baibj \right) \sin^2 \Pij - 
\mbox{Im} \left( \Baibj \right) \sin 2 \Pij \right] \label{boxorig} \\
&=& -\sum_{i=1}^{n} \sum_{j>i} \left[ 2 \mbox{Re}\left( \Baibj + \Bsaibj \right)
\sin^2 \Pij - \mbox{Im} \left( \Baibj - \Bsaibj \right) \sin 2 \Pij \right] +
\delta_{\alpha \beta} \nonumber \\
 & = & -2 \sum_{i=1}^{n} \sum_{j>i} \left[ 2 \mbox{Re}\left( \Baibj \right)
\sin^2 \Pij - \mbox{Im} \left( \Baibj \right) \sin 2 \Pij \right] +
\delta_{\alpha \beta},
\nonumber 
\eea
with $\Pij$ defined in equation (\ref{Phiij}). 
The survival probabilities $\Paa$ may be found from the transition
probabilities $\Pab$ by
\beq
\Paa = 1 - \sum_{\beta \neq \alpha} \Pab.
\eeq
Interchanging $\alpha \leftrightarrow \beta$ in equation (\ref{boxorig}) gives the
time-reversed reactions $\Pba$:
\beq
\Pba =  -2 \sum_{i=1}^{n} \sum_{j>i} \left[ 2 \mbox{Re}\left( \Baibj \right)
\sin^2 \Pij + \mbox{Im} \left( \Baibj \right) \sin 2 \Pij \right] +
\delta_{\alpha \beta},
\eeq
so a measure of T-violation, or equivalently CP-violation, in the neutrino
sector is 
\beq
\Pab - \Pba = 
4 \sum_{i=1}^{n} \sum_{j>i}\mbox{Im} \left( \Baibj \right) \sin 2 \Pij.
\label{CPprob}
\eeq
 
Equation (\ref{boxorig}) may be expressed in matrix form.  For three flavors,
we find
\beq
{\cal P}(n=3) \equiv
\left( \ba{c}
\Peu \\ \Put \\ \Pet
\ea \right) =  
-4 \mbox{ Re}({\cal B})\; S^2(\Phi) + 2 \mbox{ Im} ({\cal B})\; S(2\Phi),
\label{boxprob}
\eeq
where
\beq
{\cal B} \equiv 
\left( \ba{ccc}
\B{11}{22} & \B{12}{23} & \B{11}{23} \\
\B{21}{32} & \B{22}{33} & \B{21}{33} \\
\B{11}{32} & \B{12}{33} & \B{11}{33}
\ea \right), \mbox{\ \ and \ \ } 
S^k(\Phi) \equiv  
\left( \ba{c}
\sin ^k\Pt \\  \sin^k \Pe \\ \sin^k \Pu
\ea \right).
\label{boxbox}
\eeq
Here the operation Re(${\cal B}$) is defined so that $[\mbox{Re}({\cal
B})]_{kl} \equiv \mbox{Re}({\cal B}_{kl})$; Im(${\cal B}$) is defined in a
similar manner.  $\cal{B}$ and $S^k(\Phi)$ are defined for three flavors in 
equation (\ref{boxbox}), and
these definitions may be extended to any number of flavors.  This extension,
however, rapidly becomes difficult to manage.  Adding a fourth, perhaps
sterile, neutrino flavor increases the number of boxes from nine
to thirty-six, since there are now six independent transition probabilities and
six mass differences.  
For the six flavors of neutrinos in the see-saw mechanism, ${\cal B}$ is a
fifteen-by-fifteen matrix, and we have two hundred twenty-five boxes! 

If CP is conserved, the boxes are real.  CP conservation also implies 
time-reversal symmetry if CPT is a good symmetry, so $\Pab = \Pba$. These
symmetries leave us with just three independent probabilities (assuming
three flavors): \beq
\left( \ba{c} \Peu \\ \Put \\ \Pet \ea \right) = 
-4 \left( \ba{ccc} \B{11}{22} & \B{12}{23} & \B{11}{23} \\
		  \B{21}{32} & \B{22}{33} & \B{21}{33} \\
		  \B{11}{32} & \B{12}{33} & \B{11}{33} \ea \right)
\left( \ba{c} \sin^2 \Pt \\ \sin^2 \Pe \\ \sin^2 \Pu \ea \right).
\eeq
Our three probabilities are, however, still functions of $E$ and $x$, so
different experiments may not measure the same values for a given oscillation
probability, regardless of whether CP is conserved.


\subsection{Mixing Matrix Elements and Boxes}
\label{boxVsec}


Neutrino oscillation experiments will measure the boxes in equation 
(\ref{boxorig}),  
not the individual mixing matrix elements, $\Vai$.  But one would like to obtain
the fundamental $\Vai$ from the measured boxes.  Some tautologous relationships
between the degenerate and nondegenerate boxes are easily confirmed using
equation  (\ref{boxdef}); they hold for any number of generations:
\bea
|V_{\alpha i}|^2 |V_{\alpha j}|^2 & = & \B{\alpha i}{\alpha j} =
   \frac{\Bs{\alpha i}{\eta j} \B{\alpha i}{\lambda j}}{\B{\eta i}{\lambda j}},
 \;\;\;\;\;\;\;\;\;\;\;\;
 (\eta \neq \lambda \neq \alpha), 
 \label{vproda} \\
\nonumber \\
|V_{\alpha i}|^2 |V_{\beta i}|^2 & = & \B{\alpha i}{\beta i} = 
  \frac{\Bs{\alpha i}{\beta x} \B{\alpha i}{\beta y}}{\B{\alpha x}{\beta y}}, 
 \;\;\;\;\;\;\;\;\;\;\;
 (x \neq y \neq i),  \mbox{\ \ and \ \ } 
 \label{vprodi} \\
\nonumber \\
\frac{|\V{\alpha i}|^2}{|\Vbj|^2} & = & 
   \frac{\Bs{\alpha i}{\eta j}\B{\alpha i}{\beta x}}
   {\B{\alpha j}{\beta x} \Bs{\beta i}{\eta j}}, 
 \;\;\;\;\;\;\;\;\;\;(\eta \neq \alpha \neq \beta,
 \mbox{ and } x \neq i \neq j).
 \label{vratio}
\eea
In these equations and what follows, $\alpha$, $\beta$, $\gamma$, $i$, $j$, and
$k$ will usually 
be reserved for indices that are chosen at the start of a calculation, 
while other indices such as $x$, $y$, $\eta$, and $\lambda$ primarily 
represent ``dummy''
indices which are chosen arbitrarily except to respect the inequality 
constraints following equations such as equation (\ref{vratio}).  

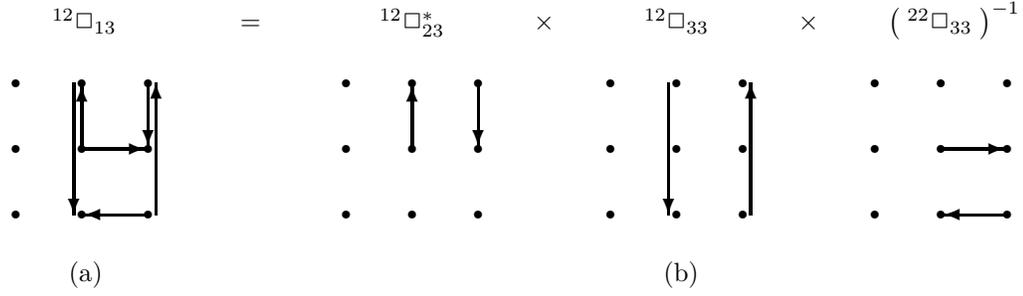
\begin{figure}[thb]
\vsp
\begin{picture}(425,105)(0,95)
\thicklines
\multiput(25,175)(25,0){3}{\circle*{3}}
\multiput(25,150)(25,0){3}{\circle*{3}}
\multiput(25,125)(25,0){3}{\circle*{3}}
\put(47,175){\vector(0,-1){50}}
\put(50,151){\vector(0,1){23}}
\put(78,125){\vector(0,1){50}}
\put(75,174){\vector(0,-1){23}}
\put(51,150){\vector(1,0){23}}
\put(74,125){\vector(-1,0){23}}
\put(36,195){$\B{12}{13}$}
\put(45,100){(a)}
\multiput(150,175)(25,0){3}{\circle*{3}}
\multiput(150,150)(25,0){3}{\circle*{3}}
\multiput(150,125)(25,0){3}{\circle*{3}}
\put(175,151){\vector(0,1){23}}
\put(200,174){\vector(0,-1){23}}
\multiput(250,175)(25,0){3}{\circle*{3}}
\multiput(250,150)(25,0){3}{\circle*{3}}
\multiput(250,125)(25,0){3}{\circle*{3}}
\put(272,175){\vector(0,-1){50}}
\put(303,125){\vector(0,1){50}}
\multiput(350,175)(25,0){3}{\circle*{3}}
\multiput(350,150)(25,0){3}{\circle*{3}}
\multiput(350,125)(25,0){3}{\circle*{3}}
\put(376,150){\vector(1,0){23}}
\put(399,125){\vector(-1,0){23}}
\put(110,195){$=$}
\put(160,195){$\Bs{12}{23}$}
\put(221,195){$\times$}
\put(260,195){$\B{12}{33}$}
\put(321,195){$\times$}
\put(355,195){$\left(\B{22}{33}\right)^{-1}$}
\put(270,100){(b)}
\end{picture}
\caption{The graphical representation in three generations for 
(a) the degenerate box $\B{12}{13}$, \ and 
(b) the combination of nondegenerate boxes which ``create'' it. 
\label{vprodafig}}
\end{figure}

The
tautologies (\ref{vproda}) to (\ref{vratio}) become evident not only by
algebraically using the definitions of the boxes, but also by considering
the graphical representation described in Appendix~\ref{graphapp}.  For
example, the degenerate 
box $\B{12}{13}$ is ``created'' when one uncanceled vertical
arrow enters and another leaves each of the matrix elements
$V_{12}$ and $V_{13}$, while arrows at all other points cancel.  This is
illustrated in Figure~\ref{vprodafig}a.  Those
arrows are the result of the combination of ordered boxes given in
(\ref{vproda}), as shown in Figure~\ref{vprodafig}b.

Equations (\ref{vproda}) and (\ref{vprodi}) 
are themselves special cases of the more general
\bea
\label{2boxij}
\Baibj \B{\gamma i}{\delta j} &=& \left[ \Vai \Vajs \Vbj \Vbis \right]
\left[ V_{\gamma i} \Vs{\gamma j} V_{\delta j} \Vs{\delta i} \right] \\
& = &\left[ \Vai \Vajs V_{\delta j} \Vs{\delta i}\right]
\left[ V_{\gamma i} \Vs{\gamma j} \Vbj \Vbis \right] = 
\B{\alpha i}{\delta j} \B{\gamma i}{\beta j}, \nonumber
\eea
and the analogous relation
\beq
\Baibj \B{\alpha k}{\beta l} = \B{\alpha i}{\beta l} \B{\alpha k}{\beta j}.
\label{2boxab}
\eeq
These relations hold for both degenerate boxes and nondegenerate boxes. 
Including disordered boxes adds no new information, so we will as usual consider
only ordered boxes on one side of the equation. (Swapping indices to produce the
other side of the equation may lead to disordered boxes on that side, but that
can quickly be remedied by using equations~(\ref{symmetries})).
If $\alpha = \gamma$ or $\beta = \delta$ for equation~(\ref{2boxij}), 
or if $i = k$ or $j = l$ for equation~(\ref{2boxab}), the relationships are
trivial.  Ordering of these pairs, however, 
should not be imposed.  

Consider the second relation, (\ref{2boxab}).  Each of 
the $\half n(n+1)$ ordered plus degenerate 
pairings of $(\alpha, \beta)$ contains 
$\frac{1}{12} n (n+1)(3n-2)(n-1)$ nontrivial relations with $j \neq l$, 
$i \neq k$, $i \le j$, and 
$k \le l$.\footnote{
We arrive at this number by considering all $\left(\half n(n+1)\right)^2$ 
$\left(i \le j, \ k \le l\right)$ pairings, then subtracting the 
$\half n(n+1)$ $\left(l=j, \  k=i, \ i \le j\right)$ possibilities, 
followed by the $\left(\ba{c}n\\3\ea\right) = \frac{1}{3}n(n-1)(n-2)$ 
$\left(l \ne j, \ k=i, \ i < \mbox{min}(j,l)\right)$ 
combinations and the same number of 
$\left(j=l, \ k \ne i, \ j>\mbox{max}(i,k)\right)$ combinations. 
 We must also subtract the $2\left(\ba{c}n\\2\ea\right)=n(n-1)$ 
$\left(l \ne j, \ k=i=\mbox{min}(j,l)\right)$ 
possibilities and the same number of $\left(k \ne i, \ j=l=\mbox{max}(i,k)
\right)$ possibilities.  So the number of terms in the sum
$\sum_{j \neq l, i \neq k, i \le j, k \le l}$ is $\left(\half n(n+1)\right)^2-
\half n(n+1)-2 \frac{1}{3}n(n-1)(n-2) - 2 n(n-1)= \frac{1}{12} n
(n+1)(3n-2)(n-1).$
}
In all,
equation~(\ref{2boxab}) implies $\frac{1}{24} n^2 (n+1)^2(n-1)(3n-2)$ 
nontrivial relations, as does equation~(\ref{2boxij}).  On the other hand, 
if we restrict ourselves to considering only the $N$ boxes nondegenerate in
$\left(\acb\right)$, $\left(i,j\right)$, and $\left(k,l\right)$, the number of
constraints
drops to $\frac{1}{12} n(n-1)(n-2)(3n-5)$ $j \neq l$, $i \neq k$, $i < j$, $k<l$
combinations for each of the $N$ $\alpha < \beta$ 
pairings,\footnote{
To obtain this number, we remove the $i=j$ and $k=l$ possibilities.  So we start
with only $\left(\half n(n-1)\right)^2$ $\left(i<j, \ k<l\right)$ 
pairings and subtract $\half n(n-1)$ $\left(l=j, \  k=i, \ i < j\right)$
combinations followed by $\frac{1}{3}n(n-1)(n-2)$ twice as above.  The $2n(n-1)$
pairings for  $\left(l \ne j, \ k=i=\mbox{min}(j,l)\right)$ and 
$\left(k \ne i, \ j=l=\mbox{max}(i,k) \right)$ do not occur in our original sum
since $i\ne j$ and $k\ne l$ in that sum. The number of terms in the sum
$\sum_{j \neq l, i \neq k, i < j, k < l}$ is therefore 
$\left(\half n(n-1)\right)^2-\half n(n-1)-2 \frac{1}{3}n(n-1)(n-2)= 
\frac{1}{12} n(n-1)(n-2)(3n-5).$
}
leading to a total number of nondegenerate constraints given by
$\frac{1}{24} n^2(n-1)^2(n-2)(3n-5)$.
Degenerate boxes 
participate in the majority of the relations implied by equations~(\ref{2boxij})
and(\ref{2boxab}).
For $n=3$, the number of $i$, $j$, $k$, $l$ combinations in
equation~(\ref{2boxab}) including degenerate
boxes is $14$; without degenerate boxes this number decreases to $2$.  These
numbers are doubled once the relationships implied by equation~(\ref{2boxij})
are considered as well.

This index rearrangement of equations~(\ref{2boxij}) and (\ref{2boxab}) may
straightforwardly be extended to obtain higher-order relationships (trilinear,
quadrilinear, {\it et cetera}) in the boxes.  These higher-order relationships
do not enter in our treatment of oscillation probabilities, so we will not
examine them here.

We may find $|V_{\alpha i}| = \left(\B{\alpha i}{\alpha i}\right)^{\frac{1}{4}}$ 
in terms of three singly-degenerate boxes by
setting $\alpha=\beta$ in equation~(\ref{vprodi}).  Using equation (\ref{vproda})
to substitute for the singly-degenerate boxes yields an expression for the
doubly-degenerate box in terms of nine nondegenerate boxes: 
\bea 
|V_{\alpha i}|^4 = \B{\alpha i}{\alpha i} =  
\frac{\B{\alpha i}{\alpha x} \B{\alpha
i}{\alpha y}}{\B{\alpha x}{\alpha y}} = 
\frac{\B{\alpha x}{\tau i} \B{\alpha i}{\sigma x}  \B{\alpha y}{\rho i}
\B{\alpha i}{\zeta y} \B{\omega x}{\mu y}} {\B{\tau i}{\sigma x} \B{\rho
i}{\zeta y} \B{\alpha y}{\omega x}  \B{\alpha x}{\mu y}}, \;\;\;& 
\left\{
\ba{c}
\tau \neq \sigma \neq \alpha \\
\zeta \neq \rho \neq \alpha \\
\mu \neq \omega \neq \alpha \\
x \neq y \neq i 
\ea .
\right.
\label{vfour}
\eea
As noted earlier, flavor-degenerate boxes give the probabilities for
flavor-conserving oscillations, while nondegenerate boxes give the
probabilities for flavor-changing oscillations.  Thus individual 
$|V_{\alpha i}|$ may be deduced from a set of measurements of either kind.
One can obtain a relationship similar to equation~(\ref{vfour}) 
by setting $i=j$ in
equation~(\ref{vproda}) and then using equation (\ref{vprodi}) to substitute
for the singly-degenerate boxes: \bea
|V_{\alpha i}|^4 = \B{\alpha i}{\alpha i} =
\frac{\B{\alpha i}{\lambda i} \B{\alpha i}{\eta i}}{\B{\lambda i}{\eta i}} = 
\frac{\B{\alpha i}{\lambda n} \B{\lambda i}{\alpha p} 
\B{\alpha i}{\eta r} \B{\eta i}{\alpha s} \B{\eta t}{\lambda u}}
{\B{\lambda n}{\alpha p} \B{\eta r}{\alpha s} \B{\lambda i}{\eta t} 
\B{\eta i}{\lambda u}}, \;\;\;& 
\left\{
\ba{c}
n \neq p \neq i \\
r \neq s \neq i \\
t \neq u \neq i \\ 
\lambda \neq \eta \neq \alpha 
\ea .
\right.
\label{vfour2}
\eea

\begin{figure}[htb]
\vsp
\begin{centering}
\begin{picture}(420,115)(-10,10)
\thicklines
\multiput(0,40)(40,0){3}{\circle*{4}}
\multiput(0,80)(40,0){3}{\circle*{4}}
\multiput(0,120)(40,0){3}{\circle*{4}}
\put(-10,118){\vector(0,-1){76}}
\put(-6,42){\vector(0,1){76}}
\put(-2,82){\vector(0,1){36}}
\put(2,118){\vector(0,-1){36}}
\put(38,82){\vector(0,1){36}}
\put(42,118){\vector(0,-1){76}}
\put(82,118){\vector(0,-1){36}}
\put(86,42){\vector(0,1){76}}
\put(38,78){\vector(0,-1){36}}
\put(78,42){\vector(0,1){36}}
\put(78,122){\vector(-1,0){36}}
\put(42,118){\vector(1,0){36}}
\put(2,82){\vector(1,0){76}}
\put(38,78){\vector(-1,0){36}}
\put(42,78){\vector(1,0){36}}
\put(2,42){\vector(1,0){36}}
\put(78,42){\vector(-1,0){36}}
\put(78,38){\vector(-1,0){76}}
\multiput(160,40)(40,0){3}{\circle*{4}}
\multiput(160,80)(40,0){3}{\circle*{4}}
\multiput(160,120)(40,0){3}{\circle*{4}}
\put(158,118){\vector(0,-1){36}}
\put(162,118){\vector(0,-1){36}}
\put(158,78){\vector(0,-1){36}}
\put(162,78){\vector(0,-1){36}}
\put(198,118){\vector(0,-1){76}}
\put(202,82){\vector(0,1){36}}
\put(202,42){\vector(0,1){36}}
\put(238,82){\vector(0,1){36}}
\put(238,42){\vector(0,1){36}}
\put(242,42){\vector(0,1){76}}
\put(162,122){\vector(1,0){76}}
\put(162,118){\vector(1,0){36}}
\put(202,118){\vector(1,0){36}}
\put(238,82){\vector(-1,0){36}}
\put(202,78){\vector(1,0){36}}
\put(198,42){\vector(-1,0){36}}
\put(238,42){\vector(-1,0){36}}
\put(238,38){\vector(-1,0){76}}
\multiput(320,40)(40,0){3}{\circle*{4}}
\multiput(320,80)(40,0){3}{\circle*{4}}
\multiput(320,120)(40,0){3}{\circle*{4}}
\put(314,42){\vector(0,1){76}}
\put(318,42){\vector(0,1){36}}
\put(322,78){\vector(0,-1){36}}
\put(326,118){\vector(0,-1){76}}
\put(358,42){\vector(0,1){76}}
\put(362,118){\vector(0,-1){36}}
\put(362,78){\vector(0,-1){36}}
\put(398,82){\vector(0,1){36}}
\put(398,42){\vector(0,1){36}}
\put(402,118){\vector(0,-1){76}}
\put(398,122){\vector(-1,0){76}}
\put(322,118){\vector(1,0){36}}
\put(362,118){\vector(1,0){36}}
\put(358,82){\vector(-1,0){36}}
\put(398,82){\vector(-1,0){36}}
\put(322,78){\vector(1,0){76}}
\put(398,42){\vector(-1,0){36}}
\put(362,38){\vector(1,0){36}}
\put(35,15){(a)}
\put(195,15){(b)}
\put(355,15){(c)}
\end{picture}
\caption[The graphical representation for 
(a) $|\V{11}|^4$, \ (b) $|\V{21}|^4$, \ and (c) $|\V{31}|^4$.]{The graphical
representation for 
(a) $|\V{11}|^4$, \ (b) $|\V{21}|^4$, \ and (c) $|\V{31}|^4$.  Notice that each matrix
point except the one being represented has an equal number of vertical arrows as
horizontal arrows entering and leaving.  The element being represented has two
vertical arrows leaving (representing $V_{\alpha i}^2$) and two entering
(representing $V_{\alpha i}^{*2}$) to produce $|\Vai|^2$.  
Appendix~\ref{graphapp} describes how these were produced, using $|V_{21}|^4$ as
an example.
\label{V4fig}}
\end{centering}
\end{figure}
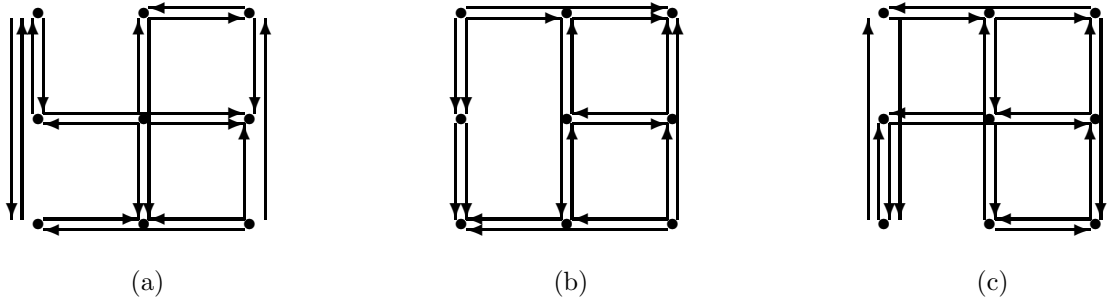

In the three-generation case, these equations are unique and equivalent to each
other, since all nine boxes are used in both equations.
This equivalence may be seen by choosing $s=n=u=x$, $r=p=t=y$, 
$\sigma=\rho=\omega=\lambda$, and $\tau=\zeta=\mu=\eta$, which yields for
either expression (\ref{vfour}) or (\ref{vfour2})
\beq
|\V{11}|^4 = \frac{\B{11}{22} \B{21}{13} \B{11}{33} \B{31}{12} \B{33}{22}}
{\B{22}{13} \B{33}{12} \B{21}{33} \B{31}{22}} = 
\frac{\B{11}{22} \Bs{11}{23} \B{11}{33} \Bs{11}{32} \B{22}{33}}
{\Bs{12}{23} \B{12}{33} \B{21}{33} \Bs{21}{32}}
\label{V11four}
\eeq
for the case $\alpha=i=1$.  The second equality merely represents a
substitution of ordered boxes for disordered boxes.
Two other examples are
\bea
|\V{21}|^4 &=& \frac{\B{21}{32} \B{11}{22} \B{21}{33} \B{11}{23} \B{12}{33}}
{\B{11}{32} \B{11}{33} \B{22}{33} \B{12}{23}}, \mbox{\ \ and \ \ } 
\label{V21four} \\
\nonumber \\
|\V{31}|^4 &=& \frac{\B{11}{32} \Bs{21}{32} \B{21}{33} \Bs{11}{33} \B{12}{23}}
{\B{11}{22} \Bs{11}{23} \B{12}{33} \Bs{22}{33}}.
\label{V31four}
\eea 
The equalities~(\ref{V11four}), (\ref{V21four}), and (\ref{V31four}) are
illustrated in Figure~\ref{V4fig}.

Before continuing, we should note that all of the relationships in this Section 
follow from the definitions of the
boxes in equation (\ref{boxdef}) and so are valid for any matrix, unitary or
otherwise.  The unitarity of $V$ will impose further relations among the boxes
which may not hold for models which only approximate unitarity of the mixing
matrix.
Equations (\ref{vproda}) to (\ref{2boxab}), however, will always hold.


\subsection{Using Unitarity to Reduce the Number of Independent Boxes}
\label{dogsec}


An $n \times n$ arbitrary matrix has $n^2$ elements, which may be complex,
leading to $2 n^2$ parameters.  Our mixing matrix, however, is unitary, and has
$n^2$ unitarity conditions as constraints.  Of the $n^2$ remaining parameters,
$\half n(n-1)$ (the number of parameters in a real unitary, or orthogonal,
matrix) are real, and the remaining $\half n(n+1)$ are imaginary phases. 
As discussed in Section~\ref{relphasesec}, we may 
remove $2n-1$ relative phases from 
the mixing matrix, leaving 
$\half n(n+1) - (2n-1) = \half (n-1)(n-2)$ phases measurable in oscillations.
If the phases are all zero, all of the boxes are real, so CP violation, as
expressed in equation~(\ref{CPprob}) disappears.  Similarly, if any of the
phases are not zero, CP is violated in the neutrino sector.

The total number of CKM mixing
matrix parameters is $n^2-(2n-1)=(n-1)^2$.  The number of real and imaginary
independent box parameters should be the same as the number of real and
imaginary CKM parameters \cite{KS}, but the reduction to a basis is
complicated for boxes.  The
number of ordered boxes $N^2$ grows much faster than the 
number of mixing matrix elements $n^2$ and exceeds that number once
$n>3$, so we need more box constraints than mixing matrix constraints
for the higher numbers of
generations.  The parameter counts for the mixing matrix and the boxes are
summarized in Table~\ref{counttab}.

\begin{table}[htb]
\caption[Parameter counting for the mixing matrix and boxes.]
{Parameter counting for the mixing matrix and boxes. $N^2$ is the
number of ordered, nondegenerate boxes, with $N \equiv \half n(n-1)$. 
\label{counttab}}
\centering
\begin{tabular}{|c||c|c|c|c|c|} \hline
Number of generations        & $n$                & 2  & 3  & 4   & 6 
  \\ \hline \hline
Params. for arbitrary matrix & $2 n^2$            & 8  & 18 & 32  & 72 
  \\ \hline
Unitarity constraints        & $n^2$              & 4  & 9  & 16  & 36 
  \\ \hline
Relative phases              & $2n-1 $            & 3  & 5  & 7   & 11 
  \\ \hline
Real params for unitary V    & $N$                & 1  & 3  & 6   & 15 
  \\ \hline
Remaining phases in V        & $\half(n-1)(n-2)$  & 0  & 1  & 3   & 10
  \\ \hline \hline
Initial boxes                & $n^4$              & 16 & 81 & 256 & 1296
  \\ \hline
Doubly-degenerate boxes      & $n^2$              & 4  &  9 &  16 & 36
  \\ \hline
Ordered singly-degenerate boxes & $2nN$           & 4  & 18 &  48 & 180
  \\ \hline
Ordered nondegenerate boxes  & $N^2$              & 1  & 9  & 36  & 225
  \\ \hline
Independent Im$(\Baibj)$s    & $\half(n-1)(n-2)$  & 0  & 1  & 3   & 10
  \\ \hline
Independent Re$(\Baibj)$s    & $N$                & 1  & 3  & 6   & 15
\\ \hline
\end{tabular}
\end{table}

Unitarity requires  that
\bea
\sum_{\eta=1}^n \V{\eta i} \V{\eta j}^* & = & \delta_{ij}, 
\label{unitrow} \mbox{ \ \ and \ \ } \\
\sum_{y=1}^n \V{\alpha y} \V{\beta y}^* & = & \delta_{\alpha \beta}.
\label{unitcol}
\eea
These equations are not independent sets of constraints at this 
point.\footnote{
The number of unitarity constraints $n^2$ becomes clear from examination of
either equation~(\ref{unitrow}) or (\ref{unitcol}).  Equation~(\ref{unitrow})
gives a real constraint for each of the $n$ $i=j$ choices, $N$ constraints from the 
real parts of the sum when $i \neq j$, and $N$ constraints from the imaginary 
parts of the $i \neq j$ sum.  Thus we have $n+2N = n^2$ unitarity constraints
from either one of the equations~(\ref{unitrow}) and (\ref{unitcol}).  As argued
next, the two equations are redundant, so the total number of constraints on
the mixing matrix is
just $n^2$.
}
Equation~(\ref{unitrow}) states \mbox{$V^{\dagger}V=\openone$,} and
equation~(\ref{unitcol})
states \mbox{$VV^{\dagger}=\openone$,} 
which is implied by the previous expression and
the associativity of matrix multiplication:
\beq
VV^{\dagger}=\openone \Rightarrow (VV^{\dagger})V = V(V^{\dagger}V) = V
\Rightarrow (V^{\dagger}V) = \openone.
\eeq
We can, however, use these equivalent equations to obtain two separate sets of
constraints on boxes by multiplying equation (\ref{unitrow}) by 
$V_{\lambda i}^* V_{\lambda j}$ and equation (\ref{unitcol}) by 
$V_{\alpha x}^* V_{\beta x}$:
\beq
\sum_{\eta=1}^n V_{\lambda i}^* V_{\lambda j} \V{\eta i} \V{\eta j}^* = 
 V_{\lambda i}^* V_{\lambda j} \delta_{ij}, \mbox{\ \ so \ \ }
\eeq
\beq
\hspace{1.0 cm} \sum_{\eta=1}^n \B{\eta i}{\lambda j} 
  =\sqrt{\B{\lambda i}{\lambda i}} \delta_{ij}, \mbox{\ \ and\ \ }
\label{urow} 
\eeq
\beq
\sum_{y=1}^n V_{\alpha x}^* V_{\beta x} \V{\alpha y} \V{\beta y}^* =
 V_{\alpha x}^* V_{\beta x} \delta_{\ab}, \mbox{\ \ or \ \ }
\eeq
\beq
\sum_{y =1}^n \B{\alpha y}{\beta x} 
  = \sqrt{\B{\alpha x}{\alpha x}}\delta_{\ab}.
\label{ucol}
\eeq
Separating the manifestly degenerate boxes and the nondegenerate boxes,
equation~(\ref{urow}) becomes
\beq
\sum_{\eta \neq \lambda} \B{\eta i}{\lambda j} = 
\sqrt{\B{\lambda i}{\lambda i}} \delta_{ij} - \B{\lambda i}{\lambda j},
\eeq
and equation~(\ref{ucol}) becomes
\beq
\sum_{y\neq x} \B{\alpha y}{\beta x} = 
\sqrt{\B{\alpha x}{\alpha x}}\delta_{\ab} - \B{\alpha x}{\beta x}.
\eeq

Summing equation~(\ref{urow}) over $\lambda$ in the $i\neq j$ case, we find
\beq
0=\sum_{\lambda=1}^n \sum_{\eta=1}^n \B{\eta i}{\lambda j} =
\sum_{\lambda=1}^n \left[\B{\lambda i}{\lambda j} + \sum_{\eta \neq \lambda}
\B{\eta i}{\lambda j}\right] = \sum_{\lambda=1}^n \B{\lambda i}{\lambda j}
+ 2 \sum_{\lambda=1}^n \sum_{\eta < \lambda} \R{\eta i}{\lambda j}.
\label{sumrowB}
\eeq
Comparison with equations~(\ref{boxbox}) and (\ref{vproda}) shows that the sum
of a column of $\mbox{Re}\left({\cal B}\right)$ 
which has a fixed $i$ and $j$ should equal 
$-\half \sum_{\lambda=1}^n |V_{\lambda i}|^2 |V_{\lambda j}|^2$.  We may in a
similar manner show that the sum of a row of $\mbox{Re}\left({\cal B}\right)$ 
with fixed
$\alpha$ and $\beta$ equals 
$-\half \sum_{x=1}^n |V_{\alpha x}|^2|V_{\beta x}|^2$.

An alternative way to obtain constraints from unitarity is to start with the
definition of the boxes (\ref{boxdef}) and use the unitarity of the mixing
matrix:
\beq
\Baibj = \left(V_{\alpha i} \Vs{\alpha j} \right)
\left(V_{\beta j} \Vs{\beta i} \right) = 
\left(\delta_{ij} - \sum_{\eta \neq \alpha} V_{\eta i} \Vs{\eta j}\right)
\left(\delta_{ij} - \sum_{\lambda \neq \beta} V_{\lambda j}\Vs{\lambda
i}\right).
\label{profunit}
\eeq
After a bit of algebra, this becomes
\beq
\sum_{\eta \neq \alpha} \sum_{\lambda \neq \beta} \B{\eta i}{\lambda j}
=\Baibj - \delta_{ij} \left(-1+|V_{\alpha i}|^2 + |V_{\beta i}|^2\right).
\eeq
This relation is equivalent to using our result (\ref{urow}), as shown below:
\bea
\sum_{\lambda \neq \beta} \sum_{\eta \neq \alpha} \B{\eta i}{\lambda j} &=& 
\sum_{\lambda \neq \beta} \left(\sum_{\eta=1}^n \B{\eta i}{\lambda j} - 
\B{\alpha i}{\lambda j}\right) = \sum_{\lambda \neq \beta} 
\left(|V_{\lambda i}|^2 \delta_{ij} - \B{\alpha i}{\lambda j}\right) \nonumber \\
&=& \delta_{ij} \sum_{\lambda \neq \beta} |V_{\lambda i}|^2 - 
\left(\sum_{\lambda=1}^n \B{\alpha i}{\lambda j} - \Baibj\right) \\
&=& \delta_{ij} \left(1-|V_{\beta i}|^2\right) -|V_{\alpha i}|^2 \delta_{ij} 
+ \Baibj
\nonumber 
\eea 

Equations~(\ref{urow}) and (\ref{ucol}) each present $\half n^2(n+1)$ relations
among the $N^2$ complex ordered nondegenerate boxes, the $2nN$ 
real ordered singly-degenerate boxes, and the $n^2$ doubly-degenerate boxes.  
Not all of these constraints are independent.  
We can identify one source of redundancy by summing equation~(\ref{urow}) over
$\lambda$:
\beq
\sum_{\lambda=1}^n \sum_{\eta =1}^n \B{\eta i}{\lambda j}  = 
\sum_{\lambda=1}^n \sqrt{\B{\lambda i}{\lambda i}} \delta_{ij}=
\sum_{\lambda=1}^n |V_{\lambda i}|^2 \delta_{ij}=\delta_{ij},
\eeq
which is pure real, so
\beq
\sum_{\lambda=1}^n \sum_{\eta=1}^n \J{\eta i}{\lambda j} = 0,
\label{xtrastraint}
\eeq
where we have introduced the notation $\J{\alpha i}{\beta j} \equiv 
\mbox{Im}\left(\Baibj\right)$.  We will similarly define \linebreak
\mbox{$\R{\alpha i}{\beta j} \equiv \mbox{Re}\left(\Baibj\right)$.}  

Since the imaginary sum in equation~(\ref{xtrastraint}) vanishes, 
at most $n-1$ of the values for $\lambda$ give independent imaginary
constraints.  The real constraints exhibit no such obvious dependency, so all
$n$ values of $\lambda$ appear at this point to offer independent constraints.
Equations~(\ref{urow}) and (\ref{ucol}) therefore each offer at most 
$\half n(n-1)^2$ imaginary constraints and $\half n^2(n-1)$ real constraints for
$i\neq j$ and $\alpha \neq \beta$, respectively.

The number of unitarity constraints from equations~(\ref{urow}) and 
(\ref{ucol}), both real and imaginary, grows as $n^3$, while 
the number of ordered boxes grows as $n^4$, so additional relationships
between ordered boxes must be used to identify the $(n-1)^2$ independent $J$s
and $R$s.  
These additional identities will come from the definitions of the boxes, and the
resulting tautologies (\ref{2boxij}) and (\ref{2boxab}), each of which grows as
$n^6$, implying a high degree of redundancy.  Some of these
relationships are explored below in equations~(\ref{taut1}) to
(\ref{realconstraint}).

The constraints (\ref{urow}) and (\ref{ucol}) 
hold independently for the real and imaginary parts of each sum.  
Because the right-hand sides of the equations
are manifestly real, as are terms on the left-hand side involving 
degenerate boxes, the sums of nondegenerate boxes in 
equations (\ref{ucol}) and (\ref{urow}) 
must be real.  This leads to imaginary constraints of the form\footnote{
The constraints (\ref{imagfixrow}) and (\ref{imagfixcol}) 
may be written exclusively in terms of ordered boxes as
\bea
\sum_{\eta < \lambda} \J{\eta i}{\lambda j} - 
\sum_{\eta > \lambda} \J{\lambda i}{\eta j} & = & 0, 
\mbox{\ \ \ and \ \ \ } \label{imagsumrow}\\
\sum_{y < x} \J{\alpha y}{\beta x} - 
\sum_{y > x} \J{\alpha x}{\beta y} & = & 0.
\label{imagsumcol}
\eea
While expressing everything in terms of ordered boxes is our goal, we will
find it more convenient to use the complete sums of equations~(\ref{imagfixrow})
and (\ref{imagfixcol}) in mathematical manipulations and switch to ordered boxes
at the end rather than to deal with the two separate terms in
equations~(\ref{imagsumrow}) and (\ref{imagsumcol}).
}
\bea
\sum_{\eta \neq \lambda} \J{\eta i}{\lambda j} & =& 0, 
\mbox{\ \ \ and \ \ \ } \label{imagfixrow}\\
\sum_{y \neq x} \J{\alpha y}{\beta x} &=& 0.
\label{imagfixcol}
\eea

In three generations, each sum in equation~(\ref{imagfixrow}) or
(\ref{imagfixcol}) contains only two terms, leading to the equality (up to a
sign) of two $J$s.  For example,
choosing $\lambda=1$ in three generations yields
\beq
\J{1j}{2i} = -\J{1j}{3i}.
\eeq
Choosing $\lambda=2$ relates the imaginary parts of $\B{2j}{3i}$ and
$\Bs{1j}{2i}$.  As discussed above, at least one value of $\lambda$ leads to
redundant constraints from equation~(\ref{imagfixrow}); choosing 
$\lambda=3$, which relates the imaginary parts of 
$\Bs{2j}{3i}$ and $\Bs{1j}{3i}$, is not independent of the first two
constraints.  
For each of the $n-1$ choices of $\lambda$ not manifestly redundant, we may choose
$\half n(n-1)$ combinations for $i<j$.  Equation (\ref{imagfixcol}) has
$\half n(n-1)^2$ possible index combinations as well, giving $n(n-1)^2$ total
imaginary constraints.  In three generations, this results in $12$ constraints
on $9$ ordered $J$s; clearly more redundancies among
relations~(\ref{imagfixrow}) and (\ref{imagfixcol}) must exist!

The first set of imaginary constraints (\ref{imagfixrow})
relates boxes that are part of a column of ${\cal B}$,
which was defined in equation (\ref{boxprob});
the second set (\ref{imagfixcol}) 
relates boxes that are part of a row of ${\cal B}$.  Given this
separation, it becomes clear that the constraints derived from
equation~(\ref{urow}) and those derived from 
equation~(\ref{ucol}) are both necessary.  Once
one combines the two sets of constraints, however, more redundancies appear.
 
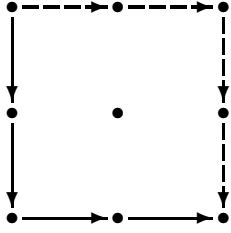
\begin{figure}[htb]
\begin{centering}
\vsp
\begin{picture}(90,90)(-5,-5)
\thicklines
\multiput(0,80)(40,0){3}{\circle*{4}}
\multiput(0,40)(40,0){3}{\circle*{4}}
\multiput(0,0)(40,0){3}{\circle*{4}}
\put(0,76){\vector(0,-1){32}}
\put(0,36){\vector(0,-1){32}}
\put(4,0){\vector(1,0){32}}
\put(44,0){\vector(1,0){32}}
\multiput(4,80)(8,0){3}{\line(1,0){6}}
\put(28,80){\vector(1,0){8}}
\multiput(44,80)(8,0){3}{\line(1,0){6}}
\put(68,80){\vector(1,0){8}}
\multiput(80,76)(0,-8){3}{\line(0,-1){6}}
\put(80,52){\vector(0,-1){8}}
\multiput(80,36)(0,-8){3}{\line(0,-1){6}}
\put(80,12){\vector(0,-1){8}}
\end{picture}
\caption[An illustration of two different paths relating $^{11}J_{22}$ with
$^{11}J_{33}$.]{An illustration of two different paths relating $\J{11}{22}$ 
with $\J{11}{33}$.  An equality (up to a minus sign) between two $J$s is
represented by an arrow connecting those two $J$s.  
The solid arrows represent the path of
equation~(\ref{path1}), and the dashed arrows represent the path of
equation~(\ref{path2}).  The matrix represented here is 
Im$\left({\cal B}\right)$, not $V$;
this illustration is not an application of the graphical representation
developed in Appendix~\ref{graphapp}.
\label{Jfig}}
\end{centering}
\end{figure}

For example, we may relate $\J{11}{22}$ to $\J{11}{33}$ by
using equation (\ref{imagfixrow}) for $i=1,j=2$ 
with $\lambda=2$ and then $\lambda=3$,
followed by equation (\ref{imagfixcol}) for $\alpha=1,\beta=3$ 
with $x=2$ and then $x=3$:
\beq
\J{11}{22} = \J{21}{32} = -\J{11}{32}  =
-\J{12}{33} = \J{11}{33}.
\label{path1}
\eeq 
This path is indicated in Figure~\ref{Jfig} by the solid arrows.  Note that the
matrix points in this Figure represent boxes in the matrix 
Im$\left({\cal B}\right)$, not mixing matrix elements.
We may alternatively
use equation~(\ref{imagfixcol}) for $\alpha=1,\beta=2$ with 
$x=2$ and then $x=3$, followed by equation~(\ref{imagfixrow}) for $i=1,j=3$
with $\lambda=2$ and then $\lambda=3$:
\beq
\J{11}{22} = \J{12}{23} = -\J{11}{23} = 
-\J{21}{33} = \J{11}{33}.
\label{path2}
\eeq
This path is represented by the dashed arrows in Figure~\ref{Jfig}.

\begin{figure}[htb]
\begin{centering}
\vsp
\begin{picture}(250,110)(-5,-25)
\thicklines
\multiput(0,80)(40,0){3}{\circle*{4}}
\multiput(0,40)(40,0){3}{\circle*{4}}
\multiput(0,0)(40,0){3}{\circle*{4}}
\put(3,83){\line(1,0){74}}
\put(3,80){\line(1,0){34}}
\put(43,80){\line(1,0){34}}
\put(3,43){\line(1,0){74}}
\put(3,40){\line(1,0){34}}
\put(43,40){\line(1,0){34}}
\put(3,-3){\line(1,0){74}}
\put(3,0){\line(1,0){34}}
\put(43,0){\line(1,0){34}}
\put(-3,78){\line(0,-1){74}}
\put(0,77){\line(0,-1){34}}
\put(0,37){\line(0,-1){34}}
\put(43,77){\line(0,-1){74}}
\put(40,77){\line(0,-1){34}}
\put(40,37){\line(0,-1){34}}
\put(83,77){\line(0,-1){74}}
\put(80,77){\line(0,-1){34}}
\put(80,37){\line(0,-1){34}}
\multiput(160,80)(40,0){3}{\circle*{4}}
\multiput(160,40)(40,0){3}{\circle*{4}}
\multiput(160,0)(40,0){3}{\circle*{4}}
\put(163,80){\line(1,0){34}}
\put(203,80){\line(1,0){34}}
\put(160,77){\line(0,-1){34}}
\put(160,37){\line(0,-1){34}}
\put(200,77){\line(0,-1){34}}
\put(200,37){\line(0,-1){34}}
\put(240,77){\line(0,-1){34}}
\put(240,37){\line(0,-1){34}}
\put(34,-20){(a)}
\put(194,-20){(b)}
\end{picture}
\caption[An illustration of a) all imaginary constraints, and b) the eight
independent imaginary constraints.]{An illustration of a) all imaginary
constraints, and b) eight
independent imaginary constraints.
\label{Jfig2}}
\end{centering}
\end{figure}
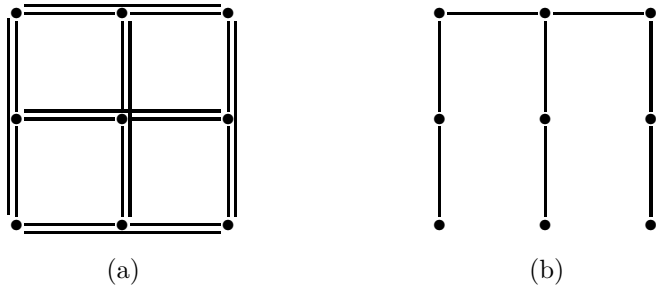

The number of independent constraints contained in equations~(\ref{imagfixrow})
and (\ref{imagfixcol}) may be determined in three generations 
by counting the number of non-redundant
connecting lines in a graph like Figure~\ref{Jfig}.  Figure~\ref{Jfig2}a contains
all $18$ connecting lines implied by equations~(\ref{imagfixrow}) and
(\ref{imagfixcol}) for $n=3$.  Figure~\ref{Jfig2}b illustrates the minimum 
number (8) of
independent lines required to connect all nine imaginary parts.  Thus eight of
our original eighteen constraints are independent, leaving us 
with only one imaginary part as expected, which we'll call ${\cal J} \equiv
\J{11}{22}$.  
Using these remaining constraints, we find
\beq
\mbox{Im} ({\cal B}) = \left( \ba{ccc} 
{\cal J} &  {\cal J} & -{\cal J} \\
{\cal J} &  {\cal J} & -{\cal J} \\
-{\cal J} &  -{\cal J} & {\cal J}
\ea \right) 
\eeq
for three generations.  

For $n>3$ generations, the sums in
equations~(\ref{imagfixrow}) and (\ref{imagfixcol}) contain more than two terms,
so individual $J$s become equal to sums of other $J$s rather than
just equal to another individual $J$.  Our illustrative method
is therefore not applicable to the $n>3$ case.  In addition, the reduction of
imaginary parameters using equations~(\ref{imagfixrow}) and (\ref{imagfixcol})
is no longer sufficient to eliminate all the extra degrees of freedom.  In four
generations, the total number of sums relating the $36$ 
$J$s is $48$, but we can demonstrate the redundancy of $21$ constraints,
leaving $9$ apparently independent imaginary parameters.  But we know from
consideration of the mixing matrix and from the work of \cite{KS} that only $3$
imaginary parameters are truly independent.  The additional $6$ constraints must
come from the expressions relating real and imaginary parameters discussed
below.  Such a reduction to a basis for the four-generation case, while possible, is
fodder for another's dissertation and therefore will not be presented here.

The equality of imaginary boxes for $n=3$ 
has an interesting implication for CP
invariance.  The CP-violating contribution to oscillation probabilities 
(\ref{CPprob}) becomes
\beq
\pm 4 {\cal J} \left( \sin 2 \Pt + \sin 2 \Pe - \sin 2 \Pu \right),
\label{CPsmall}
\eeq
where the plus sign applies for $\Peu$ and $\Put$, and the minus sign for
$\Pet$.  CP is violated if two conditions are met \cite{Nir}:  
i) ${\cal J} \neq 0$, and
ii) $m_i^2 \neq m_j^2$, for all $i,j$.  The first condition should be obvious,
since a vanishing ${\cal J}$ clearly makes the expression (\ref{CPsmall})
vanish.  The second condition follows rather simply also.  If any two masses
are degenerate, $m_i^2=m_j^2$, the corresponding $\Pij$ vanishes, and the
remaining two terms in equation~(\ref{CPsmall}) 
will be equal but opposite.  For example, if $m_1^2 = m_2^2$, then $\Pt=0$,
$\Pe=\frac{x}{4p} \left(m_2^2-m_3^2\right)$, 
$\Pu=\frac{x}{4p} \left(m_1^2-m_3^2\right)=\frac{x}{4p} \left(m_2^2-m_3^2\right)
=\Pe$, and the CP-violating term disappears as predicted.

Even if the conditions i) and ii) are met, CP-violation may be difficult to
directly observe in experiments.  The CP-violating term contains the sum
\beq
\sin 2 \Pt + \sin 2\Pe - \sin 2\Pu = 
\sin \frac{x \Delta m^2_{12}}{2p} + \sin \frac{x \Delta m^2_{23}}{2p} 
- \sin \frac{x \left(\Delta m^2_{12}+\Delta m^2_{23}\right)}{2p}.
\eeq
Using trigonometric identities, we may write
\beq
\sin \frac{x \left(\Delta m^2_{12}+\Delta m^2_{23}\right)}{2p} = 
\sin \frac{x \Delta m^2_{12}}{2p} \cos \frac{x \Delta m^2_{23}}{2p}+ 
\cos \frac{x \Delta m^2_{12}}{2p} \sin \frac{x \Delta m^2_{23}}{2p},
\eeq
so the CP-violating term is
\bea
&&\hspace{-0.8 in}
\sin \frac{x \Delta m^2_{12}}{2p} + \sin \frac{x \Delta m^2_{23}}{2p} - 
\sin \frac{x \Delta m^2_{12}}{2p} \cos \frac{x \Delta m^2_{23}}{2p}- 
\cos \frac{x \Delta m^2_{12}}{2p} \sin \frac{x \Delta m^2_{23}}{2p} \nonumber \\
&&\hspace{0.5 in} = \sin\frac{x\Delta m^2_{12}}{2p} 
\left(1-\cos\frac{x\Delta m^2_{23}}{2p}\right) +
\sin\frac{x\Delta m^2_{23}}{2p} \left(1-\cos\frac{x\Delta m^2_{12}}{2p}\right) 
\\ 
&&\hspace{0.5 in} =
2 \sin 2 \Pt \sin^2 \Pe + 2 \sin 2 \Pe \sin^2 \Pt.  \nonumber
\eea
If $\Pt$ is small, CP violation is of the order
\beq
4 \Pt \sin^2 \Pe.
\eeq

If all phases are small, $\sin 2 \Pij \sim 2 \Pij - \frac{4}{3} \Pij^3$, 
and the CP-violating term becomes
\bea
&\pm 4 {\cal J} \left( \frac{x}{2p} \left[
\left(m_1^2-m_2^2\right)+\left(m_2^2-m_3^2\right)-\left(m_1^2-m_3^2\right) 
\right] \right. & \nonumber \\
& \left. - \ \frac{x^3}{48 p^3} \left[
\left(m_1^2-m_2^2\right)^3+\left(m_2^2-m_3^2\right)^3-\left(m_1^2-m_3^2\right)^3 
\right] + \cdots \right) & \nonumber \\
& = \pm {\cal J} \frac{3x^3}{4p^3} \left( 
(m_1^2-m_2^2) (m_2^2-m_3^2) (m_3^2-m_1^2)
\right) + \cdots . &
\eea
In particular, the first order term in $\Pij$ has vanished; even though the
kinematic phases scale with the square of neutrino mass, the CP-violating linear
combination of phases scales with the sixth power of neutrino mass.  

This phenomenon holds even for higher numbers of generations:
the CP-violating term (\ref{CPprob}) is
\bea
\label{midstepCP}
4 \sum_i \sum_{j > i} \J{\alpha i}{\beta j} \sin 2 \Pij &  = &
4 \sum_i \sum_{j > i} \J{\alpha i}{\beta j} \left( 2 \Pij + {\cal O}\left( m^6
\right) \right)  \nonumber \\
&\approx & 4 \frac{x}{p} \sum_i \sum_{j > i} \J{\alpha i}{\beta j}
\left(m_i^2-m_j^2\right)  \\
& = & 4 \frac{x}{p}\left[\sum_i m_i^2 \sum_{j > i} \J{\alpha i}{\beta j}
- \sum_i \sum_{j > i} m_j^2 \J{\alpha i}{\beta j} \right].
\nonumber 
\eea
The double sum in equation~(\ref{midstepCP}) 
$\sum_i \sum_{j > i}$ contains all $i < j$
combinations, as does the double sum $\sum_j \sum_{i < j}$, so we may
substitute the latter for the former in the second term of
equation~(\ref{midstepCP}).  We may then exchange the dummy indices $i,j$ in the
second term to
obtain an overall sum which is equivalent to the sum in equation~(\ref{imagfixrow}):
\bea
\sum_i m_i^2 \sum_{j > i} \J{\alpha i}{\beta j}
- \sum_i \sum_{j > i} m_j^2 \J{\alpha i}{\beta j} &=& 
\sum_i m_i^2 \sum_{j > i} \J{\alpha i}{\beta j}
- \sum_j \sum_{i<j} m_j^2 \J{\alpha i}{\beta j} \nonumber \\
&=& 
\sum_i m_i^2 \sum_{j > i} \J{\alpha i}{\beta j}
- \sum_i \sum_{j<i} m_i^2 \J{\alpha j}{\beta i} \\
&=& 
\sum_i m_i^2 \sum_{j > i} \J{\alpha i}{\beta j}
+ \sum_i \sum_{j<i} m_i^2 \J{\alpha i}{\beta j} \nonumber \\
&=& 
\sum_i m_i^2 \sum_{j \neq i} \J{\alpha i}{\beta j} =0. \nonumber
\eea
Thus, for
any number of generations, the CP-violating contribution to neutrino
oscillations when the phases are small becomes
on the order of $\left(\Pij\right)^3$, which is the order of $\left(\Delta
m_{ij}\right)^6$ and therefore negligible for small $m$.\footnote{
The manipulation of the double sum performed here can also be used on the exact
CP-violating term (\ref{CPprob}).  The result is
\beq
4 \sum_{i=1}^n \sum_{j>i} \J{\alpha i}{\beta j} \sin 2\Pij = 
8 \sum_{i=1}^n \sin \frac{m_i^2 x}{2 p} \sum_{j\neq i} \cos \frac{m_j^2 x}{2p}
\J{\alpha i}{\beta j}.
\eeq
}

In Chapter~\ref{expt} we show that reasonable values for the
mass-squared differences yield phases too small to produce a CP-violating
effect observable at current accelerator experiments.  Longer baseline
experiments would, of course, allow larger phases, so CP violation too small to
be detected at accelerator sites could possibly be detected by
long-baseline experiments unless matter effects conspire to reduce its effect 
\cite{Sato}.  We will discuss below the possibility of detecting
the effects of CP violations, even when CP violation is not directly
observable.

The real parts of the constraints (\ref{urow}) and (\ref{ucol}) also lead to
interesting conclusions.  Consider first the homogeneous constraints 
for which the Kronecker delta
is zero.  These relationships give the singly-degenerate boxes as sums of
ordered boxes:
\bea
|V_{\lambda i}|^2|V_{\lambda j}|^2 = 
\B{\lambda i}{\lambda j} &=& - \sum_{\eta \neq \lambda} \B{\eta i}{\lambda j}
=- \sum_{\eta \neq \lambda} \R{\eta i}{\lambda j},
\hspace{0.4 in} i \neq j, \mbox{\ \ and \ \ } 
\label{1degrow}\\
|V_{\alpha x}|^2|V_{\beta x}|^2 = 
\B{\alpha x}{\beta x} &=& - \sum_{y \neq x} \B{\alpha y}{\beta x} = 
- \sum_{y \neq x} \R{\alpha y}{\beta x} ,
\ssp \alpha \neq \beta.
\label{1degcol}
\eea
For three generations, each of the sums contains two terms, so we now have
expressions for our singly-degenerate boxes in terms of two nondegenerate boxes
measurable in neutrino oscillations.

For fixed $(i,j)$ in equation~(\ref{1degrow}), $\lambda$ can take $n$ 
possible values, implying $n$ constraint equations.  $N$
ordered nondegenerate boxes appear in these $n$ equations.  Thus, for $N\le n$,
which is true for $n\le 3$, 
the unitarity constraint (\ref{1degrow}) and (\ref{1degcol}) may be inverted to
find a nondegenerate box in terms of singly-degenerate boxes.  
Adding the sums in equation~(\ref{1degcol}) 
for $x=i$ and $x=j$, then subtracting the $x=k$ sum 
yields the expression:
\beq
\R{\alpha i}{\beta j} = -\half\left(
|V_{\alpha i}|^2|V_{\beta i}|^2 + |V_{\alpha j}|^2|V_{\beta j}|^2 - 
|V_{\alpha k}|^2|V_{\beta k}|^2\right),
\eeq
where we have used the property that $R$s are unchanged under mass (or flavor) 
index interchanges.

Many sources, such as reference~\cite{KimPev}, use these sums as the
coefficients of the oscillatory terms in oscillation probabilities.  In terms of
the $|V_{\alpha i}|^2|V_{\beta i}|^2$ for three generations,
\bea
\label{KimPevsum}
\Pab &=& 2 \left( \sin^2 \Pt + \sin^2 \Pu - \sin^2 \Pe \right) 
  |V_{\alpha 1}|^2 |V_{\beta 1}|^2 \nonumber \\
&&+ \ 2 \left( \sin^2 \Pt - \sin^2 \Pu + \sin^2 \Pe \right)
  |V_{\alpha 2}|^2 |V_{\beta 2}|^2 \\
&&+ \ 2 \left( -\sin^2 \Pt + \sin^2 \Pu + \sin^2 \Pe \right)
  |V_{\alpha 3}|^2 |V_{\beta 3}|^2. \nonumber
\eea
Such an expression does not include the possibility of CP-violation.  Our boxes
are clearly preferable for their compactness and versatility.

Similar manipulation of equation~(\ref{1degcol}) gives an
alternate expression in term of the cyclic triad $(\alpha, \beta, \gamma)$:
\beq
\R{\alpha i}{\beta j} = -\half \left(
|V_{\alpha i}|^2|V_{\alpha j}|^2 + |V_{\beta i}|^2|V_{\beta j}|^2 - 
|V_{\gamma i}|^2|V_{\gamma j}|^2\right).
\eeq

The unitarity constraints (\ref{1degrow}) and (\ref{1degcol}) 
greatly simplify our expressions for a
doubly-degenerate box:
\bea
\B{\alpha i}{\alpha i} & = & 
\frac{\B{\alpha i}{\alpha x} \B{\alpha i}{\alpha y}} {\B{\alpha x}{\alpha y}} = 
\frac{\left(- \sum_{\eta \neq \alpha} \R{\alpha i}{\eta x} \right)
\left(- \sum_{\lambda \neq \alpha} \R{\alpha i}{\lambda y} \right)}
{\left(- \sum_{\tau \neq \alpha} \R{\alpha x}{\tau y} \right)},
\ssp x \neq y \neq i, \mbox{\ \ and \ \ } 
\label{2degxy} \\
{\B{\alpha i}{\alpha i}} & = &
\frac{\B{\alpha i}{\lambda i} \B{\alpha i}{\eta i}}{\B{\lambda i}{\eta i}} =
\frac{\left(- \sum_{x \neq i} \R{\alpha x}{\lambda i}  \right)
\left(- \sum_{y \neq i} \R{\alpha y}{\eta i}  \right)}
{\left( - \sum_{z \neq i} \R{\lambda z}{\eta i}  \right)},
\hspace{0.45 in} \lambda \neq \eta \neq \alpha.
\label{2degetlam}
\eea
For three generations, doubly-degenerate boxes are expressible 
in terms of the real parts of 
six ordered boxes, rather than the nine complex boxes 
used in equations (\ref{vfour}) and (\ref{vfour2}).  For example, in three
generations
\beq
|V_{11}|^4 = \B{11}{11} = 
-\frac{\left(\R{11}{22}+\R{11}{32}\right) \left(\R{11}{23}+\R{11}{33}\right)}
{\R{12}{23}+\R{12}{33}}.
\label{B1111}
\eeq

Summing equation~(\ref{1degrow}) over $j\neq i$ yields another 
expression for $|V_{\lambda i}|^2$ in terms of nondegenerate boxes:
\beq
|V_{\lambda i}|^2 \sum_{j\neq i} |V_{\lambda j}|^2 = 
|V_{\lambda i}|^2 \left(1-|V_{\lambda i}|^2\right) = 
-\sum_{j\neq i} \sum_{\eta \neq \lambda} \R{\eta i}{\lambda j}.
\eeq
The same result is obtained by summing equation~(\ref{urow}) over all $j$.  
The explicit solution, valid for any number of generations, of the above
equation has a two-fold ambiguity:
\beq
|V_{\lambda i}|^2 = \half\left[
1 \pm \sqrt{1+4\sum_{j\neq i} \sum_{\eta \neq \lambda} \R{\eta i}{\lambda
j}}\right].
\label{profeq}
\eeq
This solution gives the doubly-degenerate box in terms of the $(n-1)^2$
nondegenerate boxes of the double sum.
When the double sum is small, an approximation to the exact equation
(\ref{profeq}) is
\beq
|V_{\lambda i}|^2 \approx \left(
- \sum_{j\neq i} \sum_{\eta \neq \lambda} \R{\eta i}{\lambda j}, 
1+\sum_{j\neq i} \sum_{\eta \neq \lambda} \R{\eta i}{\lambda j}\right).
\eeq
For three generations, this yields $|V_{\lambda i}|^2$ as a linear equation of
four nondegenerate boxes.  For example,
\bea
|V_{12}|^2 &=& \half \left[ 1 \pm
\sqrt{1+4\left(\R{11}{22}+\R{11}{32}+\R{12}{23}+\R{12}{33}\right)}\right]
\nonumber \\
& \approx & \left(-\left(\R{11}{22}+\R{11}{32}+\R{12}{23}+\R{12}{33}\right), 
1+\left(\R{11}{22}+\R{11}{32}+\R{12}{23}+\R{12}{33}\right)\right). \nonumber 
\eea

We may use the homogeneous unitarity conditions 
(\ref{1degrow}) and (\ref{1degcol}), along with the
tautologies (\ref{vproda}) and (\ref{vprodi}) to obtain constraints between
ordered boxes, thereby reducing the number of real degrees of freedom. 
Recognizing that the tautologies (\ref{vproda}) and (\ref{vprodi}) give
\bea
\B{\alpha i}{\alpha j} & = & 
\mbox{Re} \left(\frac{\Bs{\alpha i}{\eta j} \B{\alpha i}{\lambda j}}
{\B{\eta i}{\lambda j}} \right) \nonumber \\
& = & 
\mbox{\small{$\frac{\R{\alpha i}{\eta j} \R{\alpha i}{\lambda j} \R{\eta i}{\lambda j} +
\J{\alpha i}{\eta j} \J{\alpha i}{\lambda j} \R{\eta i}{\lambda j} - 
\J{\alpha i}{\eta j} \R{\alpha i}{\lambda j} \J{\eta i}{\lambda j} +
\R{\alpha i}{\eta j} \J{\alpha i}{\lambda j} \J{\eta i}{\lambda j}}
{\left(\R{\eta i}{\lambda j}\right)^2 + \left(\J{\eta i}{\lambda j}\right)^2}$}},
\mbox{\ \ and \ \ } 
\label{raa} \\
\B{\alpha i}{\beta i} & = & 
\mbox{Re} \left(\frac{\Bs{\alpha i}{\beta x} \B{\alpha i}{\beta y}}
{\B{\alpha x}{\beta y}} \right) \nonumber \\
& = & 
\mbox{\small{$\frac{\R{\alpha i}{\beta x} \R{\alpha i}{\beta y} \R{\alpha x}{\beta y} +
\J{\alpha i}{\beta x} \J{\alpha i}{\beta y} \R{\alpha x}{\beta y} - 
\J{\alpha i}{\beta x} \R{\alpha i}{\beta y} \J{\alpha x}{\beta y} +
\R{\alpha i}{\beta x} \J{\alpha i}{\beta y} \J{\alpha x}{\beta y}}
{\left(\R{\alpha x}{\beta y}\right)^2 + \left(\J{\alpha x}{\beta y}\right)^2}$}},
\label{rii}
\eea
our unitarity constraints (\ref{1degrow}) and (\ref{1degcol}) become
{\small{
\bea
\hspace{-0.5 in}
\R{\alpha i}{\eta j} \R{\alpha i}{\lambda j} \R{\eta i}{\lambda j} +
\J{\alpha i}{\eta j} \J{\alpha i}{\lambda j} \R{\eta i}{\lambda j} - 
\J{\alpha i}{\eta j} \R{\alpha i}{\lambda j} \J{\eta i}{\lambda j} +
\R{\alpha i}{\eta j} \J{\alpha i}{\lambda j} \J{\eta i}{\lambda j}
 = && \nonumber \\
\hspace{1 in} - \left(\left(\R{\eta i}{\lambda j}\right)^2 + 
  \left(\J{\eta i}{\lambda j}\right)^2 \right)
  \sum_{\tau \neq \alpha} \R{\alpha i}{\tau j},  &&
\label{realdegaa}
\eea
}}
and
{\small{
\bea
\hspace{-0.5 in}
\R{\alpha i}{\beta x} \R{\alpha i}{\beta y} \R{\alpha x}{\beta y} +
\J{\alpha i}{\beta x} \J{\alpha i}{\beta y} \R{\alpha x}{\beta y} - 
\J{\alpha i}{\beta x} \R{\alpha i}{\beta y} \J{\alpha x}{\beta y} +
\R{\alpha i}{\beta x} \J{\alpha i}{\beta y} \J{\alpha x}{\beta y} 
  = && \nonumber \\
\hspace{1 in} 
- \left(\left(\R{\alpha x}{\beta y}\right)^2 + 
  \left(\J{\alpha x}{\beta y}\right)^2 \right)
  \sum_{z \neq i} \R{\alpha i}{\beta z}, &&
\label{realdegii}
\eea
}}
with the usual inequalities $\eta \neq \lambda \neq \alpha$, $i \neq j$, in
equations~(\ref{raa}) and (\ref{realdegaa}), and 
$\alpha \neq \beta$, and $x \neq y \neq i$ in equations~(\ref{rii}) and
(\ref{raa}) satisfied.  

We may rename some
indices and solve both equations
(\ref{realdegaa}) and (\ref{realdegii}) for one $\R{\alpha i}{\beta j}$:
\bea
\R{\alpha i}{\beta j} & = &
-\frac{\J{\alpha i}{\beta j} 
\left(\J{\alpha i}{\lambda j} \R{\beta i}{\lambda j} - 
\R{\alpha i}{\lambda j} \J{\beta i}{\lambda j}\right)
+ \left(
\left(\R{\beta i}{\lambda j}\right)^2 + \left(\J{\beta i}{\lambda j}\right)^2
\right) \sum_{\tau \neq \acb} \R{\alpha i}{\tau j}}
{\J{\beta i}{\lambda j} 
\left(\J{\alpha i}{\lambda j} + \J{\beta i}{\lambda j} \right) + 
\R{\beta i}{\lambda j} 
\left(\R{\alpha i}{\lambda j} + \R{\beta i}{\lambda j} \right)} 
\label{Raibjaa}\\
& = &
-\frac{\J{\alpha i}{\beta j} 
\left(\J{\alpha i}{\beta y} \R{\alpha j}{\beta y} - 
\R{\alpha i}{\beta y} \J{\alpha j}{\beta y}\right)
+ \left(
\left(\R{\alpha j}{\beta y}\right)^2 + \left(\J{\alpha j}{\beta y}\right)^2
\right) \sum_{z \neq i,j} \R{\alpha i}{\beta z}}
{\J{\alpha j}{\beta y} 
\left(\J{\alpha i}{\beta y} + \J{\alpha j}{\beta y} \right) + 
\R{\alpha j}{\beta y} 
\left(\R{\alpha i}{\beta y} + \R{\alpha j}{\beta y} \right)}.
\label{Raibjii}
\eea
Similar expressions could be found for $\J{\alpha i}{\beta j}$ if so desired.

These constraints interrelate imaginary and real parts of boxes through
unitarity for any number
of generations.  If CP is a good symmetry, all of the $\J{\alpha i}{\beta j}$
vanish, and the above expressions for $\R{\alpha i}{\beta j}$ will be
different than if CP were not conserved.  
For three generations, $\tau$ must equal $\lambda$, and $\J{\alpha i}{\beta j}
= -\J{\alpha i}{\lambda j} = \J{\beta i}{\lambda j}$ by
equation~(\ref{imagfixrow}).  Similarly, $z=y$, and $\J{\alpha i}{\beta j} = 
- \J{\alpha i}{\beta y} = 
\J{\alpha j}{\beta y}$ by equation (\ref{imagfixcol}).  Equations
(\ref{Raibjaa}) and (\ref{Raibjii}) for $n=3$ become
\bea
\R{\alpha i}{\beta j} & = & \frac{\left(\J{\alpha i}{\beta j}\right)^2 - 
\R{\alpha i}{\lambda j}\R{\beta i}{\lambda j}}
{\R{\beta i}{\lambda j}+\R{\alpha i}{\lambda j}} 
\label{R3genaa} \\
& = & \frac{\left(\J{\alpha i}{\beta j}\right)^2 - 
\R{\alpha j}{\beta y}\R{\alpha i}{\beta y}}
{\R{\alpha j}{\beta y}+\R{\alpha i}{\beta y}}
\label{R3genii} 
\eea
Rearranging the equations gives a measure of CP-violation in three generations:
\bea
\left(\J{\alpha i}{\beta j}\right)^2 & = &
\R{\alpha i}{\beta j} \R{\beta i}{\lambda j} + 
\R{\alpha i}{\beta j} \R{\alpha i}{\lambda j} + 
\R{\alpha i}{\lambda j} \R{\beta i}{\lambda j} 
\label{CPone} \\
& = & \R{\alpha i}{\beta j} \R{\alpha j}{\beta y} + 
\R{\alpha i}{\beta j} \R{\alpha i}{\beta y} + 
\R{\alpha j}{\beta y} \R{\alpha i}{\beta y}.
\label{CPtwo}
\eea
These relationships between $R$s and $J$s in three generations
exhibit a simple parameter symmetry.  $(\J{\alpha i}{\beta j})^2$ in
equation~(\ref{CPone}) equals $\R{\alpha i}{\beta j} \R{\lambda i}{\beta j}+ $ 
terms cyclic in $(\alpha, \beta, \lambda)$; in equation~(\ref{CPtwo}), 
it equals $\R{\alpha i}{\beta j} \R{\alpha y}{\beta j}+$ terms cyclic in
$(i, j, y)$.

If CP is violated, $\J{\alpha i}{\beta j} \neq 0$, so the combination of real
parts on the right-hand sides of equations~(\ref{CPone}) and (\ref{CPtwo}), measurable with CP-conserving averaged neutrino
oscillations (discussed in detail in Chapter~\ref{expt}), cannot be zero.
Thus, even if CP violation is not
directly observable in an experiment because of a short baseline or small mass
differences, the effects of CP violation may be seen
through the relationships among the real parts of different boxes!

We may infer similar constraints  
by setting the imaginary part of a singly-degenerate box to zero:
{\small{
\bea
\mbox{Im} \left(\B{\alpha i}{\alpha j}\right) = 
\mbox{Im} \left( \frac{\Bs{\alpha i}{\eta j} \B{\alpha i}{\lambda j}}
{\B{\eta i}{\lambda j}} \right)=  \hspace{3 in}&& \label{taut1} \\
\frac{\R{\alpha i}{\eta j} \J{\alpha i}{\lambda j} \R{\eta i}{\lambda j}
     - \J{\alpha i}{\eta j} \R{\alpha i}{\lambda j} \R{\eta i}{\lambda j}
     - \R{\alpha i}{\eta j} \R{\alpha i}{\lambda j} \J{\eta i}{\lambda j}
     - \J{\alpha i}{\eta j} \J{\alpha i}{\lambda j} \J{\eta i}{\lambda j}}
 {\R{\eta i}{\lambda j}^2 + \J{\eta i}{\lambda j}^2} = 0, && \nonumber
\eea
}}
and
{\small{
\bea 
\mbox{Im} \left(\B{\alpha i}{\beta i}\right) = 
\mbox{Im} \left( \frac{\Bs{\alpha i}{\beta x} \B{\alpha i}{\beta y}}
{\B{\alpha x}{\beta y}} \right) = \hspace{3 in} && \\
\frac{\R{\alpha i}{\beta x} \J{\alpha i}{\beta y} \R{\alpha x}{\beta y}
     - \J{\alpha i}{\beta x} \R{\alpha i}{\beta y} \R{\alpha x}{\beta y}
     - \R{\alpha i}{\beta x} \R{\alpha i}{\beta y} \J{\alpha x}{\beta y}
     - \J{\alpha i}{\beta x} \J{\alpha i}{\beta y} \J{\alpha x}{\beta y}}
 {\R{\alpha x}{\beta y}^2 + \J{\alpha x}{\beta y}^2} = 0, && \nonumber
\eea
}}
so
\beq
\R{\alpha i}{\eta j} \J{\alpha i}{\lambda j} \R{\eta i}{\lambda j}
  - \J{\alpha i}{\eta j} \R{\alpha i}{\lambda j} \R{\eta i}{\lambda j}
  - \R{\alpha i}{\eta j} \R{\alpha i}{\lambda j} \J{\eta i}{\lambda j}
  - \J{\alpha i}{\eta j} \J{\alpha i}{\lambda j} \J{\eta i}{\lambda j} =0,
\eeq
and
\beq
\R{\alpha i}{\beta x} \J{\alpha i}{\beta y} \R{\alpha x}{\beta y}
  - \J{\alpha i}{\beta x} \R{\alpha i}{\beta y} \R{\alpha x}{\beta y}
  - \R{\alpha i}{\beta x} \R{\alpha i}{\beta y} \J{\alpha x}{\beta y}
  - \J{\alpha i}{\beta x} \J{\alpha i}{\beta y} \J{\alpha x}{\beta y} = 0.
\eeq
With a bit of index switching, these lead to constraints of the form
\beq
\R{\alpha i}{\beta j} = 
  \frac{\J{\alpha i}{\beta j} \R{\alpha i}{\lambda j} \R{\beta i}{\lambda j}
    + \J{\alpha i}{\beta j} \J{\alpha i}{\lambda j} \J{\beta i}{\lambda j}}
  {\J{\alpha i}{\lambda j} \R{\beta i}{\lambda j}  
    - \R{\alpha i}{\lambda j} \J{\beta i}{\lambda j}} = 
   \frac{\J{\alpha i}{\beta j} \R{\alpha i}{\beta y} \R{\alpha j}{\beta y}
    + \J{\alpha i}{\beta j} \J{\alpha i}{\beta y} \J{\alpha j}{\beta y}}
  {\J{\alpha i}{\beta y} \R{\alpha j}{\beta y}
    - \R{\alpha i}{\beta y} \J{\alpha j}{\beta y}}.
\label{realconstraint}
\eeq

For three generations, unitarity enforces the simple relations~(\ref{imagfixrow})
and (\ref{imagfixcol}) among the $J$s, and we find that 
equation~(\ref{realconstraint}) reproduces 
the same expressions (\ref{R3genaa}) and (\ref{R3genii}) which were derived from
the unitarity constraints~(\ref{Raibjaa}) and (\ref{Raibjii}) above.
For $n>3$, however,
equation (\ref{realconstraint}) is different from the prior equations,
since the sums $\sum_{\tau \neq \acb}$ and $\sum_{z \neq i,j}$ in
equations~(\ref{Raibjaa}) and (\ref{Raibjii}) ensure that
boxes with all $n$ flavor or mass indices enter into equations (\ref{R3genaa})
or (\ref{R3genii}), respectively.  Equation (\ref{realconstraint}) involves
only three such indices, so it will be different from the unitarity constraints 
in higher
generations, providing additional identities among the real parts of boxes.

We have not yet used the inhomogeneous 
unitarity constraints (\ref{urow}) and (\ref{ucol}) that occur
when the Kronecker delta is
not zero.  The homogeneous unitarity constraints cannot provide the absolute
normalization of the $V_{\alpha i}$ or the boxes, so we must need the
inhomogeneous unitarity constraints.  These inhomogeneous 
constraints are functions of degenerate boxes:\footnote{
In terms of mixing-matrix elements, equations~(\ref{2degsuma}) and
(\ref{2degsumi}) are trivial.  For example, equation~(\ref{2degsuma}) is
\beq
|\Vai|^4 + \sum_{\eta \neq \alpha} |\Vai|^2|V_{\eta i}|^2 = |\Vai|^2,
\eeq
or
\beq
\sum_{\eta=1}^n |V_{\eta i}|^2 = 1.
\eeq
}
\bea
\B{\alpha i}{\alpha i} + \sum_{\eta \neq \alpha} \B{\alpha i}{\eta i} &= &
\sqrt{\B{\alpha i}{\alpha i}}, \mbox{\ \ and \ \ }
\label{2degsuma} \\
\B{\alpha i}{\alpha i} + \sum_{z \neq i} \B{\alpha i}{\alpha z} &=& 
\sqrt{\B{\alpha i}{\alpha i}}.
\label{2degsumi}
\eea
Parenthetically, we note by 
comparing equations~(\ref{2degsuma}) and (\ref{2degsumi}) that a sum
over mass-degenerate boxes equals a sum over flavor-degenerate boxes:
\beq
\sum_{\eta \neq \alpha} \B{\alpha i}{\eta i} =
 \sum_{z \neq i} \B{\alpha i}{\alpha z}.
\eeq

Equations~(\ref{2degsuma}) and (\ref{2degsumi}) 
can be rewritten strictly in terms of nondegenerate boxes 
by using the homogeneous unitarity constraints
(\ref{1degrow}) and (\ref{1degcol}). (If these expressions were not available,
we would need to use the tautologies~(\ref{vfour}) and (\ref{vfour2}) for the
doubly-degenerate boxes, and equations~(\ref{vproda}) and (\ref{vprodi}) for the
singly-degenerate boxes, resulting in an incredibly ugly mess (as compared to
the merely unincredibly ugly mess derived below)).  We find
\bea
&& \hspace{-2.5 cm} 
\frac{\left(- \sum_{\lambda \neq \alpha} \R{\alpha i}{\lambda x} \right)
\left(- \sum_{\sigma \neq \alpha} \R{\alpha i}{\sigma y} \right)}
{\left(- \sum_{\tau \neq \alpha} \R{\alpha x}{\tau y} \right)} \nonumber \\
&& - \ \sqrt{\frac{\left(- \sum_{\lambda \neq \alpha} 
\R{\alpha i}{\lambda x} \right)
\left(- \sum_{\sigma \neq \alpha} \R{\alpha i}{\sigma y} \right)}
{\left(- \sum_{\tau \neq \alpha} \R{\alpha x}{\tau y} \right)}} 
+ \sum_{\eta \neq \alpha} \left(- \sum_{z \neq i} \R{\alpha z}{\eta i} \right) 
= 0,
\eea
with $x \neq y \neq i$, and
\bea
&& \hspace{-2.5 cm} 
\frac{\left(- \sum_{x \neq i} \R{\alpha x}{\lambda i}  \right)
\left(- \sum_{y \neq i} \R{\alpha y}{\eta i}  \right)}
{\left( - \sum_{t \neq i} \R{\lambda t}{\eta i}  \right)} \nonumber \\
&& - \ \sqrt{\frac{\left(- \sum_{x \neq i} \R{\alpha x}{\lambda i}  \right)
\left(- \sum_{y \neq i} \R{\alpha y}{\eta i}  \right)}
{\left( - \sum_{t \neq i} \R{\lambda t}{\eta i}  \right)}}
+ \sum_{z \neq i} \left(- \sum_{\eta \neq \alpha} \R{\alpha z}{\eta i} \right)
= 0,
\eea
with $\lambda \neq \eta \neq \alpha$.

In three generations, each of the sums has only two terms.  Eliminating the
square root, and multiplying through by the denominator,
we find constraints of the form
{\small{ 
\beq
\ba{l}
\left(\R{\alpha i}{\lambda x} + \R{\alpha i}{\eta x}\right)
\left(\R{\alpha i}{\lambda y}+\R{\alpha i}{\eta y}\right)
\left(\R{\alpha x}{\lambda y}+ \R{\alpha x}{\eta y}\right)
\left[1+ 2 \left(\R{\alpha i}{\lambda x} + \R{\alpha i}{\eta x}
+ \R{\alpha i}{\lambda y}+\R{\alpha i}{\eta y}\right) \right] +
 \\
\hspace{0.2 cm} 
\left( \R{\alpha i}{\lambda x} + \R{\alpha i}{\eta x} \right)^2
\left(\R{\alpha i}{\lambda y}+\R{\alpha i}{\eta y}\right)^2 + 
\left(\R{\alpha i}{\lambda x} + \R{\alpha i}{\eta x}
+ \R{\alpha i}{\lambda y}+\R{\alpha i}{\eta y}\right)^2
\left(\R{\alpha x}{\lambda y}+ \R{\alpha x}{\eta y}\right)^2=0,
\ea
\label{Vfouraa}
\eeq
}}
and
{\small{
\beq
\ba{l}
\left(\R{\alpha i}{\lambda x} + \R{\alpha i}{\lambda y}\right)
\left(\R{\alpha i}{\eta x}+\R{\alpha i}{\eta y}\right)
\left(\R{\lambda i}{\eta x} +\R{\lambda i}{\eta y}\right)
\left[1+ 2 \left(\R{\alpha i}{\lambda x} + \R{\alpha i}{\lambda y}
+ \R{\alpha i}{\eta x}+\R{\alpha i}{\eta y}\right)\right] + 
\\
\hspace{0.2 cm} 
\left(\R{\alpha i}{\lambda x} + \R{\alpha i}{\lambda y}\right)^2
\left(\R{\alpha i}{\eta x}+\R{\alpha i}{\eta y}\right)^2 + 
\left(\R{\alpha i}{\lambda x} + \R{\alpha i}{\lambda y}
+ \R{\alpha i}{\eta x}+\R{\alpha i}{\eta y}\right)^2
\left(\R{\lambda i}{\eta x} +\R{\lambda i}{\eta y}\right)^2=0,
\ea
\label{Vfourii}
\eeq
}}
where $x \neq y \neq i$, and $\lambda \neq \eta \neq \alpha$.  So, for
$\alpha=2$ and $i=1$, we find
\beq
\ba{l}
\left(\R{11}{22}+\R{21}{32}\right)\left(\R{11}{23}+\R{21}{33}\right)
\left(\R{12}{23}+\R{22}{33}\right) 
\left[1+2\left(\R{11}{22}+\R{21}{32}+\R{11}{23}+\R{21}{33}\right)\right] +
\\
\hspace{0.2 cm} 
\left(\R{11}{22}+\R{21}{32}\right)^2 \left(\R{11}{23}+\R{21}{33}\right)^2 + 
\left(\R{11}{22}+\R{21}{32}+\R{11}{23}+\R{21}{33}\right)^2
\left(\R{12}{23}+\R{22}{33}\right)^2 = 0,
\ea
\label{V11fouraa}
\eeq
and
\beq
\ba{l}
\left(\R{11}{22}+\R{11}{23}\right)\left(\R{21}{32}+\R{21}{33}\right)
\left(\R{11}{32}+\R{11}{33}\right) 
\left[1+2\left(\R{11}{22}+\R{11}{23}+\R{21}{32}+\R{21}{33}\right)\right] +
\\
\hspace{0.2 cm} 
\left(\R{11}{22}+\R{11}{23}\right)^2 \left(\R{21}{32}+\R{21}{33}\right)^2 + 
\left(\R{11}{22}+\R{11}{23}+\R{21}{32}+\R{21}{33}\right)^2
\left(\R{11}{32}+\R{11}{33}\right)^2 = 0.
\ea
\eeq

As with the constraints on the imaginary parts of boxes, many of the
constraints (\ref{Raibjaa}) to (\ref{Vfourii})
are redundant, but an independent set may be used to reduce the number of
parameters significantly.  The homogeneous constraints are much simpler than the
inhomogeneous constraints, so we want to use as many of those as are possible.
For the three-generation case, we may
eliminate $\R{12}{23}$ by either equation~(\ref{R3genaa}) or
equation~(\ref{R3genii})
\beq
\R{12}{23} = \frac{\R{12}{33}\R{22}{33} - {\cal J}^2}{-\R{22}{33} - \R{12}{33}} 
           = \frac{\R{11}{22}\R{11}{23} - {\cal J}^2}{-\R{11}{23} - \R{11}{22}}.
\label{1223out}
\eeq
We may similarly eliminate $\R{11}{32}$ and $\R{21}{33}$
\beq
\R{11}{32} = \frac{\R{11}{22}\R{21}{32} - {\cal J}^2}{-\R{21}{32} - \R{11}{22}}
           = \frac{\R{11}{33}\R{12}{33} - {\cal J}^2}{-\R{12}{33} - \R{11}{33}},
\label{1132out}
\eeq
and
\beq
\R{21}{33} = \frac{\R{11}{23}\R{11}{33} - {\cal J}^2}{-\R{11}{33} - \R{11}{23}}
           = \frac{\R{21}{32}\R{22}{33} - {\cal J}^2}{-\R{22}{33} -\R{21}{32}}.
\label{2133out}
\eeq
$\R{12}{33}$ may be eliminated from equation (\ref{1223out}), and $\R{11}{33}$
may be eliminated from equation (\ref{2133out}):
\bea
\R{12}{33} &=& \frac{\R{22}{33} \R{11}{22} \R{11}{23} 
  + {\cal J}^2 \left( \R{11}{23} + \R{11}{22} - \R{22}{33} \right)}
  {\R{22}{33} \R{11}{23} + \R{22}{33} \R{11}{22} - \R{11}{22} \R{11}{23} +
  {\cal J}^2}, \mbox{ \ \ and \ \ } \label{1233out} \\
\nonumber \\
\R{11}{33} &=& \frac{\R{22}{33} \R{11}{23} \R{21}{32} 
  + {\cal J}^2 \left( \R{21}{32} + \R{22}{33} - \R{11}{23} \right)}
  {\R{11}{23} \R{21}{32} + \R{22}{33} \R{11}{23} - \R{21}{32} \R{22}{33} +
  {\cal J}^2}. \label{1133out}
\eea
Equation (\ref{1132out}) will not provide an additional constraint; it is
redundant to the other two.  We must turn to equations~(\ref{Vfouraa}) and
(\ref{Vfourii}) to eliminate the last degree of freedom.  This necessity is
expected, since  
without this set of inhomogeneous constraints, the rows/columns of $V$ will not
be normalized and our set of boxes will not yield a unitary $V$.  
We will here choose the constraint (\ref{V11fouraa}),
with $\alpha=1$ and $i=1$ since its expression in terms of our four remaining
boxes is the least complicated.  We may substitute the second equality
from equation~(\ref{2133out}) for $\R{21}{33}$ and the second equality of
equation~(\ref{1223out}) for $\R{12}{23}$, leaving a constraint which contains
the four parameters $\R{11}{22}$, $\R{11}{23}$, $\R{21}{32}$, and $\R{22}{33}$:
\bea
\left(\R{11}{22}+\R{21}{32}\right)
\left(\R{11}{23}+
  \frac{\R{21}{32}\R{22}{33}-{\cal J}^2}{-\R{22}{33}-\R{21}{32}}\right)
\left(\frac{\R{11}{22}\R{11}{23}-{\cal J}^2}{-\R{11}{23}-\R{11}{22}}
  + \R{22}{33}\right)
\left[1+ 2\left(\R{11}{22} \right. \right. && \nonumber \\
\left. \left. +\R{21}{32}+\R{11}{23}+
  \frac{\R{21}{32}\R{22}{33}-{\cal J}^2}{-\R{22}{33}-\R{21}{32}}\right)\right]
+\left(\R{11}{22}+\R{21}{32}\right)^2\left(\R{11}{23}+
  \frac{\R{21}{32}\R{22}{33}-{\cal J}^2}{-\R{22}{33}-\R{21}{32}}\right)^2 && \\
+\left(\R{11}{22}+\R{21}{32}+\R{11}{23}+
  \frac{\R{21}{32}\R{22}{33}-{\cal J}^2}{-\R{22}{33}-\R{21}{32}}\right)^2
\left(\frac{\R{11}{22}\R{11}{23}-{\cal J}^2}{-\R{11}{23}-\R{11}{22}}
  + \R{22}{33}\right)^2 & = & 0. \nonumber
\eea

Multiplying through to place all of the terms in the numerator, we are left with
\bea
&&\hspace{-0.5 cm} 
0=\left(\R{11}{22}+\R{11}{23}\right)^2 \left(\R{11}{22}+\R{21}{32}\right)^2 
\left(-\R{21}{32} \R{22}{33} + 
\R{11}{23}\left(\R{21}{32} + \R{22}{33}\right) + {\cal J}^2\right)^2 \nonumber \\
&& \left( \left(\R{21}{32}\right)^2 + \left(\R{11}{23} + \R{11}{22} \right) 
\left( \R{21}{32} + \R{22}{33} \right) + {\cal J}^2 \right)^2
\nonumber \\
&& \ssp \times \left(\R{11}{23} \R{22}{33} + \R{11}{22} \left(-\R{11}{23}+\R{22}{33} \right)
{\cal J}^2 \right)^2 
\label{quartic} \\
&&- \ \left(\R{11}{22}+\R{21}{32}\right) \left(\R{11}{22}+\R{11}{23}\right)
\left(\R{11}{22} \left(\R{11}{23}-\R{22}{33}\right) - \R{11}{23} \R{22}{33}
- {\cal J}^2\right) \nonumber \\
&& \ssp \times \left(\R{21}{32} + 2 \R{11}{22} \R{21}{32} + 
2 \left(\R{21}{32}\right)^2 + 2 \R{21}{32} \R{11}{23} \right. \nonumber \\
&& \left. \ssp \ssp + \ \R{22}{33} +
2 \R{11}{22} \R{22}{33} + 2 \R{11}{23} \R{22}{33} + 2 {\cal J}^2\right).
\nonumber
\eea
This equation is quartic in $\R{22}{33}$; we may eliminate this box by
solving the equation
\beq
A + B \left(\R{22}{33}\right) 
+ C \left( \R{22}{33} \right)^2 + D \left( \R{22}{33} \right)^3
+ E \left( \R{22}{33} \right)^4 = 0,
\eeq
with
\bea
A & = & \left(\R{11}{22}\right)^6 \left(\R{21}{32}\right)^2 - 
2 \left(\R{11}{22}\right)^5 \R{21}{32} \left(
  2 \R{21}{32} \R{11}{23} + {\cal J}^2\right)\nonumber \\
&& - \ \left(\R{11}{22}\right)^4 \left[
  9 \left(\R{21}{32}\right)^2 \left(\R{11}{23}\right)^2 + 
  10 \R{21}{32} \R{11}{23} {\cal J}^2 + {\cal J}^4 \right] \nonumber \\ 
&& - \ \left(\R{11}{22}\right)^3 \left[ 
  6 \left(\R{21}{32}\right)^3 \left(\R{11}{23}\right)^2 + 
  \left(\R{21}{32}\right)^2\left(\R{11}{23}\right)^2\left(1+10\R{11}{23}\right)
  \right. \nonumber \\
&& \left. + \ \R{21}{32} \left(
    \R{11}{23} + 20 \left(\R{11}{23}\right)^2 - 2{\cal J}^2\right){\cal J}^2 + 
  6 \R{11}{23} {\cal J}^4 \right] \nonumber \\
&& + \ \R{11}{22}\left\{ 
  \left(\R{21}{32}\right)^2 \left(1+10 \R{11}{23} \right) {\cal J}^4 + 
  4 \R{11}{23} {\cal J}^4 \left(-\left(\R{11}{23}\right)^2 + {\cal J}^2\right)
  \right. \nonumber \\
&& \left. - \ 2 \left(\R{21}{32}\right)^4 \left[
      \left(\R{11}{23}\right)^3 - 2 \R{11}{23} {\cal J}^2 \right. \right.
      \nonumber \\
&& \left.\left. + \ \R{21}{32} \R{11}{23} {\cal J}^2 \left(
        -2 \left(\R{11}{23}\right)^3 + {\cal J}^2+6 \R{11}{23} 
        {\cal J}^2 \right) \right.\right. \\
&& \left.\left. + \ \left(\R{21}{32}\right)^3 \R{11}{23}\left(
    -\left(\R{11}{23}\right)^2-2\left(\R{11}{23}\right)^3+{\cal J}^2 + 
    8 \R{11}{23} {\cal J}^2 \right) \right] \right. \nonumber \\
&& \left. - \ \left(\R{11}{22}\right)^2 \left[
  3 \left(\R{21}{32}\right)^4 \left(\R{11}{23}\right)^2 + 
  \left(\R{21}{32}\right)^3 \R{11}{23} \left(
    \R{11}{23}+8\left(\R{11}{23}\right)^2-6 {\cal J}^2\right)\right]
    \right.\nonumber \\
&& \left.+ \ \left(9 \left(\R{11}{23}\right)^2 - 2 {\cal J}^2\right) 
     {\cal J}^4 + \R{21}{32} {\cal J}^2 \left[
    \left(\R{11}{23}\right)^2 + 14 \left(\R{11}{23}\right)^3 - {\cal J}^2 - 
    8 \R{11}{23} {\cal J}^2\right]  \right.
  \nonumber \\
&& \left.  + \ \left(\R{21}{32}\right)^2 \left[
    \left(\R{11}{23}\right)^3 + 4 \left(\R{11}{23}\right)^4 + 
    2 \left(\R{11}{23}\right)^2 {\cal J}^2 - 3 {\cal J}^4 \right] \right\}
    \nonumber \\
&& + \ {\cal J}^2 \left[
  \left(\R{21}{32}\right)^3 \left(\R{11}{23}\right)^2 
  \left(1 + 2 \R{11}{23}\right) + \left(\R{21}{32}\right)^4 \left(
    2 \left(\R{11}{23}\right)^2 - {\cal J}^2 \right) \right.\nonumber \\
&& \left. + \ \left(\R{21}{32}\right)^2 \left(
    \R{11}{23} + 3 \left(\R{11}{23}\right)^2 - 2 {\cal J}^2\right) {\cal J}^2-
  {\cal J}^2\left(\left(\R{11}{23}\right)^4 + {\cal J}^4 \right) \right],
  \nonumber \\
\nonumber \\
B & = & -\left(\R{11}{22}+\R{11}{23} \right)\left\{
  2 \left(\R{11}{22}\right)^5 \R{21}{32} - \left(\R{21}{32}\right)^4 \left(
    \left(\R{11}{23}\right)^2-2 {\cal J}^2 \right) 
  \right. \nonumber \\
&&\left. - \ \left(\R{21}{32}\right)^3 \left(1+2 \R{11}{23}\right)\left(
    \left(\R{11}{23}\right)^2-{\cal J}^2\right) + 
  4 {\cal J}^6 \right. \nonumber \\
&& \left. - \ \R{21}{32} {\cal J}^2 \left(
    2 \left(\R{11}{23}\right)^3 + {\cal J}^2\right) + 
    \left(\R{11}{22}\right)^4 \left(
      -2 \left(\R{21}{32}\right)^2+6 \R{21}{32} \R{11}{23} + 2{\cal J}^2\right)
    \right. \nonumber \\
&& \left. + \ \left(\R{21}{32}\right)^2 {\cal J}^2 \left(
    -3 \R{11}{23} - 6 \left(\R{11}{23}\right)^2 + 8 {\cal J}^2 \right)
     \right. \nonumber \\
&& \left. - \ 4 \left(\R{11}{22}\right)^3 \left[
    \left(\R{21}{32}\right)^2 \R{11}{23}-2 \R{1}{23}{\cal J}^2+\R{21}{32}\left(
      -3\R{11}{23} + {\cal J}^2 \right) \right] \right. \nonumber \\
&& \left. - \ \R{11}{22} \left[
    6 \left(\R{21}{32}\right)^4 \R{11}{23} + 2 \left(\R{21}{32}\right)^3 \left(
      \R{11}{23} + 5 \left(\R{11}{23}\right)^2 - 3 {\cal J}^2 \right)
     \right.\right. \\
&& \left. \left. + \ 2\R{21}{32} \left(\R{11}{23}+9\left(\R{11}{23}\right)^2
     -2{\cal J}^2\right) {\cal J}^2 + {\cal J}^2 \left(
       -2 \left(\R{11}{23}\right)^3+{\cal J}^2 
       +12 \R{11}{23} {\cal J}^2 \right)
      \right. \right. \nonumber \\
&& \left. \left. + \ \left(\R{21}{32}\right)^2 \R{11}{23} \left(
      -\R{11}{23}+2\left(\R{11}{23}\right)^2+18 {\cal J}^2\right)
    \right] \right. \nonumber \\
&& \left. - \ \left(\R{11}{22} \right)^2 \left[
     12\left(\R{21}{32}\right)^3\R{11}{23}+2\left(\R{21}{32}\right)^2
     \R{11}{23} \left(1+4 \R{11}{23}\right) \right]
     \right. \nonumber \\
&& \left. + \ {\cal J}^2 \left[
       -\R{11}{23}-12 \left(\R{11}{23}\right)^2 + 4{\cal J}^2 
       \right.\right. \nonumber \\
&&\left.\left. + \ \R{21}{32} \left(
  -2 \left(\R{11}{23}\right)^2-8\left(\R{11}{23}\right)^3+{\cal J}^2
    + 24 \R{11}{23} {\cal J}^2 \right)\right]\right\}, \nonumber \\
\nonumber \\    
C & = & -\left(\R{11}{22}+\R{11}{23}\right) \left\{
  \left(\R{11}{22}\right)^5  + 3 \left(\R{21}{32}\right)^4 \R{11}{23}
  + \left(\R{21}{32}\right)^3 \R{11}{23}\left(1+2\R{11}{23}\right)  \right.
  \nonumber \\
&& \left. 
  + \ \left(\R{11}{22}\right)^4 \left(-2\R{21}{32}+\R{11}{23}\right) 
  - 2 \R{21}{32}\R{11}{23}{\cal J}^2 + 6 \R{11}{23} {\cal J}^4
  \right. \nonumber \\
&& \left. + \ \left(\R{11}{22}\right)^3 \left[
    4 \left(\R{21}{32}\right)^2-16 \R{21}{32}\R{11}{23}
    + 6 \left(\R{21}{32}\right)^2-2 {\cal J}^2\right]
    \right.
  \nonumber \\
&& \left. + \ \left(\R{21}{32}\right)^2 \left[
    -2 \left(\R{11}{23}\right)^2-2\left(\R{11}{23}\right)^3+{\cal J}^2
    +10 \R{11}{23} {\cal J}^2\right] \right. \nonumber \\
&& \left. + \ \R{11}{22} \left[
     3 \left(\R{21}{32}\right)^4 + \left(\R{21}{32}\right)^3 
     \left( 1+ 8\R{11}{23}\right) \right.\right. \\
&&\left.\left. - \ \R{21}{32} \R{11}{23} \left(
       \R{11}{23} + 8 \left(\R{11}{23}\right)^2-8 {\cal J}^2 \right)
     \right. \right. \nonumber \\ 
&&\left.\left. - \ 2 \left(\R{21}{32}\right)^2 \left(
       \R{11}{23}+4\left(\R{11}{32}\right)^2-5{\cal J}^2\right)
     -2\left(\R{11}{23}+6\left(\R{11}{23}\right)^2-3{\cal J}^2\right){\cal J}^2
     \right] \right. \nonumber \\
&&\left. + \ \left(\R{11}{22}\right)^2 \left[
  6 \left(\R{21}{32}\right)^3
  + \left(\R{11}{22}\right)^2\left(1-2\R{11}{23}\right)
  +\left(\R{11}{23}\right)^2+4\left(\R{11}{23}\right)^3-{\cal J}^2
  \right.\right. \nonumber \\
&&\left.\left. - \ 14 \R{11}{23} {\cal J}^2 + \R{21}{32} \left(
    -3\R{11}{23}-22\left(\R{11}{23}\right)^2+8{\cal J}^2\right)\right]\right\},
    \nonumber \\
\nonumber \\
D & = & -\left(\R{11}{22}+\R{11}{23}\right)^2 \left\{
  4\left(\R{11}{22}\right)^2\left(\R{21}{32}-\R{11}{23}\right) 
  \right.\nonumber \\
&& \left. - \ \R{21}{32} \R{11}{23}+\left(\R{21}{32}\right)^2
  \left(1+4\R{11}{23}\right)+ 4\R{11}{23}{\cal J}^2 
  \right. \\
&& \left. + \ \R{11}{22}\left[
    \R{21}{32}+4\left(\R{21}{32}\right)^2 -\R{11}{23}
    +4\R{21}{32}\R{11}{23}-4\left(\R{11}{23}\right)^2+4{\cal J}^2\right]\right\},
    \mbox{\ \ and \ \ } \nonumber \\
\nonumber \\
E & = & -\left(\R{11}{22}+\R{11}{23}\right)^4.
\eea
Eliminating $\R{22}{33}$ with this constraint leaves us with three
real parameters $\R{11}{22}$, $\R{11}{23}$, and $\R{21}{32}$, 
and the one imaginary component ${\cal J}$ as our basis.  This final
constraint relating the four boxes, while not as ugly as it would be without the
use of equations~(\ref{1degrow}) and (\ref{1degcol}), is quite complicated.
The algebraic solutions (which are obtainable) are too messy to be particularly
illuminating, so we will choose to eliminate our final spurious parameter by
solving equation~(\ref{V11fouraa})
numerically rather than algebraically when we compare to experiment in
Chapter~\ref{expt}. 


\section{More Unmeasurables}
\label{pretracesec}


The previous Sections have developed relationships between the lepton 
mixing matrix and
the boxes.  Now we turn our attention to relationships between the neutrino mass
matrix and the boxes.

We showed in
Section~\ref{CKMsec} that we cannot distinguish experimentally between theories
containing a diagonal $M_l$ with $U_{\nuL} = V$ and theories containing mixing
in both the neutrino sector and the charged lepton sector.  We thus lose no
physical information by restricting the mixing to the neutrino sector.
Most models for fundamental 
lepton mass matrices, however, do not make this {\it ansatz} of
a diagonal mass matrix for the charged leptons.  On the contrary, many models
postulate  as a symmetry principle 
the same initial form for the charged lepton mass matrix as for the
neutrino mass matrix.  
Such models will yield predictions equivalent to those of many other
models, including a model that contains all of the lepton mixing in the
neutrino sector.  

Consider a model for lepton mass matrices that contains a
neutrino mass matrix $M_{\nu}$ and a lepton mass matrix $M_l$.  We will first 
assume a Dirac mass term
for the neutrinos for simplicity, then extend our treatment to 
the general case.
The Lagrangian in the mass basis (\ref{CKMderiv}) which we derived in
Section~\ref{CKMsec} contains the terms
\beq \ol{\nuL^m} D_{\nu} \NR^m + \ol{l_L^m} D_l l_R^m + 
c_2 W^+_{\mu} \ol{\nuL^m} V^{-1} \gupu l_L^m + h.c..
\label{massLag}
\eeq
We illustrated in that Section how the Lagrangian is invariant under a rotation
of the mixing matrices $U_{\nuL, l_L} \rightarrow RU_{\nuL, l_L}$ by any
unitary matrix $R$.  Such a transformation is equivalent to rotating the mass
matrices $M_{\nu}$ and $M_l$ by the same
arbitrary unitary matrix.  In particular, we may
choose this arbitrary rotation such that one of the mass matrices is diagonal
from the beginning.  This implies the mixing matrices
$U_{\nuL}= V$ and $U_{l_L}= \openone$, which we assumed when we derived our box
relationships.  The charged lepton mass matrix is 
diagonal in this basis, which we will call the {\it standard basis}.
The neutrino mass matrix $\hat{M}_{\nu}$ in this basis
is related to the original mass matrix $M_{\nu}$ by 
$\hat{M}_{\nu} = U^{-1}_{l_L} M_{\nu}$, which can be seen by equating the
diagonal mass matrices in the two bases:
\beq
D_{\nu} = U^{-1}_{\nuL} M_{\nu} U_{\NR} = V^{-1} \hat{M}_{\nu} U_{\NR}
= U^{-1}_{\nuL} U_{l_L} \hat{M} U_{\NR}, \mbox{\ \ so}
\label{DMM'}
\eeq
\beq
\hat{M}_{\nu} = V U^{-1}_{\nuL} M_{\nu} = U_{l_L}^{-1} U_{\nuL} U^{-1}_{\nuL}
M_{\nu}
= U^{-1}_{l_L} M_{\nu}.
\label{MM'}
\eeq

When more generations of neutrinos are introduced, the left-handed 
neutrino mixing matrix $U_{\nu_L}$ no longer has the same dimensions 
as the left-handed
charged-lepton mixing matrix.  We will use the general block form of the neutrino
mass matrix from Section~\ref{allflavsec},
$$
M_{\nu} = \left( \ba{ccc} M_T & M_D & M_{\nu \chi} \\ M_D^T & M_S & M_{N \chi}
\\ M_{\nu \chi}^T & M_{N \chi}^T & M_{\chi} \ea \right).
\eqno{(\ref{blockmass})}
$$
The mass term (\ref{Ltot}) containing this matrix 
is of the Majorana form, so the matrix
$M_{\nu}$ must be symmetric and may be diagonalized by 
$U_{\nu}^{-1} M_{\nu} \left(U_{\nu}^{T}\right)^{-1}$, 
as described in Section~\ref{MajvsDiracsec}.  
We may combine the spinors of all three types of neutrinos
into one $n$-dimensional vector $\psi_{\nu}$:
\beq
\psi_{\nu} = \left( \ba{c} \nu \\ N \\ \chi \ea \right).
\eeq
In addition, we will sometimes 
break the neutrino rotation matrix $U_{\nu}$ into three
submatrices, $U_1$, $U_2$, and $U_3$ such that
\beq
U_{\nu}= \left( \ba{c} U_1 \\ U_2 \\ U_3 
\ea \right).
\eeq
$U_1$ is $n_L \times n$, $U_2$ is $n_R \times n$, and $U_3$ is $n_{\chi} \times
n$, with $n=n_L+n_R+n_{\chi}$.  
Such a decomposition is necessary because the charged current term contains
only $\nu_L$, which is related to the $n$ mass states contained in 
$\psi_{\nuL}^m$ by $\nu_L = U_1^{-1} \psi_{\nuL}^m$.  
These $U_i$ are not square and
so are not unitarity, but $U_i U_i^{\dagger} = \openone$ ($\openone$ here has
dimensions $n_L \times n_L$), so $U_i^{\dagger}$ is
the right inverse of $U_i$.

Using these definitions, the mass-basis Lagrangian (\ref{massLag}) is
\beq \ba{c}
\ol{\psi_{\nu L}} U_{\nu} U_{\nu}^{-1} M_{\nu} U_{\nu}^{T-1} U_{\nu}^T 
\psi_{\nu R} + \ol{l_L} U_{l_L} U_{l_L}^{-1} M_l U_{l_R} U_{l_R}^{-1} l_R 
+ \ol{\nuL} U_1 U_1^{-1} \gupu U_{l_L} U_{l_L}^{-1} l_L
+ h.c. \\
=\ol{\psi_{\nu L}^m} D_{\nu} \psi_{\nu R}^m + \ol{l_L^m} D_l l_R^m +
\ol{\psi_{\nu L}^m} V^{-1} \gupu \l_L^m + h.c.
\ea \eeq
The matrix in the charged current term $V = U_{l_L}^{-1} U_1$ is not square,
but has dimension $n_L \times n$.

We still have the freedom to incorporate all of the lepton 
mixing in the neutrino sector, effectively replacing $U_1$ with
$V$, so the neutrino mixing matrix in this standard basis is
\beq
\hat{U}_{\nu} = \left( \ba{c} V \\ U_2 \\ U_3 \ea \right) = 
\left( \ba{ccc} U_{l_L}^{-1} & 0 & 0 \\
0 & 1 & 0 \\ 0 & 0 & 1 \ea \right) 
U_{\nu}.
\eeq
This replacement implies a different non-diagonalized 
neutrino mass matrix $\hat{M_{\nu}}$:
\bea
\hat{M}_{\nu} = \hat{U}_{\nu} D \hat{U}^T_{\nu} & = &
\left( \ba{ccc} 
U_{l_L}^{-1} & 0 & 0 \\ 0 & 1 & 0 \\ 0 & 0 & 1 \ea \right) U_{\nu} D U_{\nu}^T
 \left( \ba{ccc} U_{l_L}^{T -1} & 0 & 0 \\ 0 & 1 & 0 \\ 0 & 0 & 1 \ea \right)
\nonumber \\
& = &
\left( \ba{ccc} 
U_{l_L}^{-1} & 0 & 0 \\ 0 & 1 & 0 \\ 0 & 0 & 1 \ea \right) M_{\nu}
 \left( \ba{ccc} U_{l_L}^{T -1} & 0 & 0 \\ 0 & 1 & 0 \\ 0 & 0 & 1 \ea \right).
\label{MM'gt3}
\eea
This transformation includes both a right and left rotation because it involves
Majorana neutrinos.  The transformation for $n_L$ generations developed in
equations~(\ref{DMM'}) and (\ref{MM'}) assumes the right-handed fields $N_R$ are
independent of the left-handed fields $\nuL$, so the right-handed rotation
matrix $U_{\NR}$ appearing in equation~(\ref{DMM'}) 
is unaffected by our new choice of $U_{\nuL}$.  For the Majorana mass term
indicated in equation~(\ref{blockmass}), this is no longer true.  The
right-handed fields are related to the left-handed fields, and the right-handed
rotation matrix must equal $\left(U_{\nu}^T\right)^{-1}$.  So changing to the
standard basis rotates the mass matrix on the right as well on the left.  Such
an additional rotation could be performed in the three-generation Dirac case for
consistency's sake (since we cannot measure $U_{\NR}$ in the absence of a
right-handed current), but it is not necessary on mathematical grounds.  We note
that under the transformation in equation~(\ref{MM'gt3}) the Dirac subblocks
$M_D$ and $M_D^T$ in the matrix~(\ref{blockmass}) do transform to $U_{l_L} M_D$
and $M_D^T U_{l_L}^T$, in agreement with equations~(\ref{DMM'}) and (\ref{MM'}). 
The issue here
will become moot in the next Section when we restrict ourselves to the hermitian
mass-squared matrix $MM^{\dagger}$.  The extra rotation cancels in this product.

Even the diagonal mass matrix 
$D$ is not completely measurable.  The mass matrix $M$ does
not have to be hermitian, so the $m_i$ in $D$ 
are not necessarily real.  We have the
freedom to perform a chiral rotation $e^{i \theta \gdnf}$ on the fermion field
definitions; under $\psi \rightarrow e^{i \theta \gdnf} \psi$, mass terms
become
\bea
&&\ol{\psi_L} M \psi_R + h.c. \rightarrow
\ol{\psi_L} M e^{2 i \theta \gdnf} \psi_R + h.c. = \\
&&\hspace{0.5 in} 
\ol{\psi_L} M \left(\cos 2 \theta + i \gdnf \sin 2 \theta\right) \psi_R + 
\ol{\psi_R} M^{\dagger} \left(\cos 2\theta+i \gdnf \sin 2\theta \right)\psi_L,
\eea 
where we have used the properties $\gdnf^{2k} = \openone$ and
$\gdnf^{2k+1}=\gdnf$ for integer k in the Taylor expansions of 
$e^{2i\theta\gdnf}$, $\sin \theta$, and $\cos \theta$.  
Since $\gdnf P_L = -P_L$, and $\gdnf P_R=P_R$, we have
\beq
\ol{\psi_L} M i \gdnf \psi_R +
\ol{\psi_R} M^{\dagger} i \gdnf \psi_L =
\ol{\psi_L} M i \psi_R - 
\ol{\psi_R} M^{\dagger} i \psi_L,
\eeq
so our whole mass term has become
\beq
\ol{\psi_L} M e^{2i\theta} \psi_R + 
\ol{\psi_R} M^{\dagger} e^{-2i\theta}\psi_L = 
\ol{\psi_L} e^{2i\theta} M \psi_R + h.c. 
\eeq
A chiral rotation by $\theta = \frac{\pi}{2}$ has the same effect as the
transformation $M \rightarrow -M$; a rotation by $\theta=\frac{\pi}{4}$ effects
$M \rightarrow i M$.  We can separately perform chiral rotations on particular
flavor fields, so the phases of individual mass eigenvalues are arbitrary.  
These rotations lead to corresponding effects
on the diagonal $D$; the phase of the mass matrix, and the
phases of the fermion masses contained in the diagonal mass matrix $D$, 
cannot be measurable since we may absorb these
phases into the fermion field definitions.

In the next Section, 
we will derive equations expressing the elements of the matrix
$\hat{M}_{\nu}\hat{M}^{\dagger}$ in terms of the observable boxes.  To use these relationships to
place constraints on the elements of the mass matrices given by
particular models, the transformation (\ref{MM'}) 
must be invoked.  Once extra neutrinos are included, the CKM matrix
$V=\hat{U_1}$ appearing in the charged-current term 
is no longer equivalent to the full neutrino mixing
matrix $U_{\nu}$.  The matrix $V$ knows about all $n$ of the neutrino {\it mass}
states,
since it connects $\nuL$ with $\psi_{\nu}^m$, but it knows nothing of the
sterile {\it flavor} neutrinos and therefore cannot be used to describe mixing
between active and sterile flavors.  The entire matrix $U_{\nu}$ is necessary
to completely describe neutrino mixing.  In what follows, however, we will use
the symbol $V$ to denote the left-handed neutrino mixing matrix, as we did in
the previous Sections of this Chapter.  
For three generations, this assignment presents no
ambiguity; for $n>3$, the neutrino mixing matrix $V$
will have different dimensions than the CKM matrix $V$, so again there should be
no confusion.


\section{More Measurables}


We showed in Section~\ref{relphasesec} that $2n-1$ relative 
phases can
not be observed.  For Dirac neutrinos, 
we also have the freedom to rotate $U_{\NR}$ {\it ad nauseum}
since it is not observable in the absence of right-handed currents.  The
effects of this last degree of freedom may be ignored if we consider only the
hermitian product $M M^{\dagger}$ rather than the generally non-hermitian 
mass matrix $M$.  Should right-handed currents be found, one may explore
$U_{\NR}$ via the hermitian matrix $M^{\dagger}M$.  Note that in the absence of
the symmetry imposed by Majorana neutrinos, $M^{\dagger}M = 
\hat{M}^{\dagger}\hat{M}$, since the rotation by $U_{\nuL}$ in
$\hat{M}=U_{\nuL}^{-1}M U_{\NR}$ is canceled;
this matrix is diagonalized by $U_{\NR}$ alone into $D^2$:
\beq
U_{\NR}^{\dagger} M^{\dagger} M U_{\NR} = D^2.
\eeq

The phase ambiguity  of the mixing matrix requires that any observables which
are functions of the left-handed neutrino mixing matrix $V$ and/or 
the product $M
M^{\dagger}$  must be invariant under the simultaneous
transformations
\beq
V \rightarrow Y V X, \mbox{ \ \ and \ \ } 
\hat{M} \hat{M}^{\dagger} \rightarrow X^{-1} \hat{M} \hat{M}^{\dagger}X,
\label{invar}
\eeq
where $X$ and $Y$ are diagonal
matrices of phases.
These invariances are identical to those possesed by traces of products of
$MM^{\dagger}$ and diagonal basis matrices, or of traces of products of $V$ and
diagonal basis matrices.
Such traces offer an elegant method of obtaining the measurable functions of the
mixing matrix and the mass matrix.

We introduce the notation
\beq
H \equiv \hat{M}\hat{M}^{\dagger},
\eeq
and the diagonal basis matrices
\beq
\left(E_i\right)_{mn} \equiv \delta_{mi} \delta_{mn},
\eeq
We may express mass-matrix invariants in terms of these newly-defined quantities
as
\beq
\mbox{Tr}\left[\prod_{\alpha} \left(HE_{\alpha}\right)\right].
\eeq
Thus,
\beq
\mbox{Tr} \left[HE_{\alpha}\right] = H_{\alpha \alpha}
\eeq
is an invariant, as is
\beq
\mbox{Tr} \left[HE_{\alpha}HE_{\beta}\right] = |H_{\alpha \beta}|^2.
\eeq
Higher-order traces such as $\mbox{Tr}
\left[HE_{\alpha}HE_{\beta}HE_{\gamma}\right]$
are also invariants.

Turning to the mixing matrix, we see that equal numbers of $V$s and
$V^{\dagger}$s provide invariants.  $V$ placed into a trace by itself is not
invariant, since the phase matrices $X$ and $Y$ occuring on either side of the
rotated mixing matrix do not cancel in the trace:
\beq
\mbox{Tr}\left[VE_i\right] \neq \mbox{Tr} \left[YVE_iX\right].
\eeq  
Here we have used the commutativity of the diagonal matrices $X$ and $E_i$.
The product $V^{\dagger}V$,
however, is rotated by the single transformation $X$, which will cancel in the
trace, so
\beq
\mbox{Tr}\left[V^{\dagger}E_{\alpha}VE_i\right] =
\mbox{Tr}\left[X^{-1}V^{\dagger}E_{\alpha}Y^{-1}YVE_iX\right]= |\Vai|^2
\eeq
is invariant \cite{Jarls1},  
as is the fourth-order trace
\beq
\mbox{Tr} \left[VE_{\alpha}V^{\dagger}E_iVE_{\beta}V^{\dagger}E_j\right]
=\Vai V^{\dagger}_{i \beta}\Vbj V^{\dagger}_{j \alpha},
\eeq
which is a box.  Again, traces of products 
higher-order in even powers of $V$ are also
measurable.

The mass-matrix invariant traces are related to the mixing-matrix invariant
traces by
\beq
H=VD^2V^{\dagger} =V\left[\sum_{j=1}^n E_j m_j^2\right] V^{\dagger} = 
\sum_{j=1}^n m_j^2 VE_jV^{\dagger}.
\eeq
So we find
\beq
H_{\alpha \alpha} = \mbox{Tr}\left[HE_{\alpha}\right] = 
\sum_{j=1}^n m_j^2 \mbox{Tr} \left[VE_jV^{\dagger}E_{\alpha}\right] = 
\sum_{j=1}^n m_j^2 |V_{\alpha j}|^2 = 
\sum_{j=1}^n m_j^2 \sqrt{\B{\alpha j}{\alpha j}},
\label{Haa}
\eeq
and
\beq
|H_{\ab}|^2 = \mbox{Tr}\left[HE_{\alpha}HE_{\beta}\right] = 
\sum_{i=1}^n \sum_{j=1}^n m_i^2 m_j^2
\mbox{Tr}\left[VE_{\alpha}V^{\dagger}E_iVE_{\beta}V^{\dagger}E_j\right] = 
\sum_{i=1}^n \sum_{j=1}^n m_i^2 m_j^2 \Baibj.
\label{Hab2}
\eeq
The next-order expression becomes
\beq
H_{\ab} H_{\beta \gamma} H_{\gamma \alpha} = 
\mbox{Tr}\left[HE_{\beta}HE_{\gamma}HE_{\alpha}\right] = 
\sum_{i=1}^n \sum_{j=1}^n \sum_{k=1}^n m_i^2 m_j^2 m_k^2
V_{\alpha i} V^{\dagger}_{i\beta} V_{\beta j} V^{\dagger}_{j \gamma}
V_{\gamma k} V^{\dagger}_{k\alpha}.
\label{hiorder}
\eeq
These products of six mixing matrix elements and even the product of eight
mixing matrix elements could be measured in processes such
as the decay $\mu^{\pm} \rightarrow e^{\pm}e^+e^-$ which proceeds through
penguin and box graphs with internal neutrino lines.

The right-hand side of equation~(\ref{Hab2}) may be written as
\beq
\sum_{i=1}^n m_i^4 \Baibj + \sum_{i=1}^n \sum_{j\neq i} m_i^2 m_j^2 \Baibj = 
\sum_{i=1}^n m_i^4 |V_{\alpha i}|^2 |V_{\beta i}|^2 + 
2 \sum_{i=1}^n \sum_{j>i} m_i^2 m_j^2 \R{\alpha i}{\beta j}.
\eeq
From this we see that not only is  information about relative phases lost, but
that the measurable $J$s do not enter either.  The missing information is
carried in the higher-order observables, such as those found in
equation~(\ref{hiorder}).  An alternative approach is suggested by the unitarity
constraints~(\ref{Raibjaa}) and (\ref{Raibjii}) which may be rearranged to express
$\J{\alpha i}{\beta j}$ in terms of $R$s.  Unitarity therefore provides a
mechanism for implicitly obtaining the $J$s from real trace relations.

Equations~(\ref{Haa}) and (\ref{Hab2}) involve the neutrino mass matrix in the
standard basis, with all of the mixing in the neutrino sector.
If we want to compare the predictions of mass models specified in a
non-standard basis to the measurable
quantities in the right-hand side of equations~(\ref{Haa}) and (\ref{Hab2}),
we need to rotate $\hat{M}$ back
to the $M_{\nu}$ of the specific model as described in equation~(\ref{MM'}) 
Performing this rotation on $H_{\ab}$,
we find
\beq
|H_{\ab}|^2 = 
\left|\left(\hat{M}\hat{M}^{\dagger}\right)_{\ab}\right|^2 =\sum_{i,j=1}^n 
| \left(U_{lL}\right)_{\alpha i}^{-1}\left(M_{\nu}M_{\nu}^{\dagger}\right)_{ij}
 \left(U_{lL}\right)_{j\beta}|^2.
\label{Mrot}
\eeq
$U_{lL}$ and $M_{\nu}$ are given by the model, while the
right-hand side of equation (\ref{Hab2}) (which equals (\ref{Mrot})) contains
measurable quantities.  Finding individual elements of
$M_{\nu}M^{\dagger}_{\nu}$ in terms of
the observables might be messy, but it is possible using the above equations.

Consider next the converse issue of constructing the boxes from a given $M$
(and therefore a given $H$).
Equations~(\ref{Haa}) and (\ref{Hab2}) present equations for each of the $n^2$
possible $\alpha$, $\beta$.  These equations involve $N^2 \sim n^4$ ordered
boxes.  Equations~(\ref{Haa}) and (\ref{Hab2}) therefore cannot
be simply inverted to obtain the boxes from the mass matrix in the general case.
\cite{DocA}.  The unitarity constraints presented in the earlier Sections of
this Chapter reduce the number of independent boxes so the inversion could in
principle be done.  But the relations would be highly non-linear and
intractable.  To obtain the boxes from a given mass matrix, we will diagonalize
the specified matrix and solve for the eigenvectors.  These eigenvectors may
then be used to construct the mixing matrix $V$, which gives the boxes.  
This procedure is illustrated in Chapter~\ref{massbox} 
for a few selected mass matrix textures.  


\chapter{Applications of the Boxes to Specific Phenomenological Models}
\label{massbox}


Having developed a parameterization for neutrino oscillation which is
independent of neutrino mass and mixing models, we will now examine how
different models may be expressed in terms of the boxes.


\section{Models of the Mixing Matrix}


\subsection{The Standard KM Parameterization}


	For Dirac neutrinos, the mixing matrix $V$ becomes the familiar CKM
matrix from quark mixing (although the mixing angles need not have the same
values as for the quark case), and $V$ may be characterized by three
mixing angles $\theta_a$ and a phase $\delta$.  A popular choice of $V$ is the
Kobayashi-Maskawa matrix of equation~(\ref{NactCKM}):
$$
\left( \ba{ccc} 
\co{1} & \;\s{1} \co{3}\; & \s{1} \s{3} \\
-\s{1}\co{2} & \; \co{1}\co{2}\co{3}-\s{2}\s{3}e^{i\delta}\; &
\co{1}\co{2}\s{3}+\s{2}\co{3}e^{i\delta} \\
-\s{1}\s{2} & \co{1}\s{2}\co{3}+\co{2}\s{3}e^{i\delta} &
\co{1}\s{2}\s{3}-\co{2}\co{3}e^{i\delta}  \ea \right).
\eqno{(\ref{NactCKM})}
$$
In terms of these angles, we find that ${\cal J}$, the imaginary part of the
boxes is
\bea
{\cal J} & = & \co{1} \st{1} \co{2} \s{2} \co{3} \s{3} \sin \delta \nonumber \\
&=& \frac{1}{8} 
\sin \theta_1 \sin 2 \theta_1 \sin 2 \theta_2 \sin 2 \theta_3 \sin \delta
\equiv J \sin \delta,
\eea
as predicted in Chapter~\ref{osc}'s equation~(\ref{Jdef}).
Using this definition of $J$, we may express our boxes as
\bea
\B{11}{22} & = & -\ct{1} \st{1} \ct{2} \ct{3} + J e^{i \delta} \nonumber \\
\B{12}{23} & = & \st{1} \ct{3} \st{3} (\ct{1} \ct{2} - \st{2}) + J \cos \delta
	(\ct{3} - \st{3}) + i J \sin \delta \nonumber \\
\B{11}{23} & = & -\ct{1} \st{1} \ct{2} \st{3} - J e^{i \delta} \nonumber \\
\B{21}{32} & = & \st{1} \ct{2} \st{2}(\ct{1} \ct{3} - \st{3}) + J \cos \delta
	(\ct{2} - \st{2}) + i J \sin \delta \nonumber \\
\B{22}{33} & = & \ct{3} \st{3} \left[ \st{2} \ct{2} (s_{1}^4 + 6
\ct{1} + 2 \ct{1} \cos 2 \delta) - \ct{1} \right]
\nonumber \\
 & &  + \frac{J}{\st{1}} (1+\ct{1})(\ct{2}-\st{2})(\st{3} - 
\ct{3}) \cos \delta + i J \sin \delta 
\label{stanbox} \\
\B{21}{33} & = & \st{1} \ct{2} \st{2} (\ct{1} \st{3} - \ct{3}) + J \cos \delta
	(\st{2} - \ct{2}) - i J \sin \delta  \nonumber \\
\B{11}{32} & = & -\ct{1} \st{1} \st{2} \ct{3} - J e^{i \delta} \nonumber \\
\B{12}{33} & = & \st{1} \ct{3} \st{3} (\ct{1} \st{2} - \ct{2}) + J \cos \delta 
	(\st{3} - \ct{3}) - i J \sin \delta \nonumber \\
\B{11}{33} & = & -\ct{1} \st{1} \st{2} \st{3} + J e^{i \delta} \nonumber 
\eea

In terms of the standard parameterization the observable
probabilities, presented here as boxes, are big ugly messes.  It is no wonder
that approximations have been commonly used when comparing experimental results 
to the standard mixing angles and phase!

In principle, one can solve equations~(\ref{stanbox}) with $\delta=0$ for
$\theta_{1}$, $\theta_{2}$, and $\theta_{3}$ in terms of a basis of 
the real parts of any three independent nondegenerate boxes.  Then
substitutions into equations~(\ref{stanbox}) give the remaining six
nondegenerate real boxes in terms of these chosen three.  $\delta$ is then
given in terms of these three boxes and the imaginary part $\pm {\cal J}$ of any
box from
\beq
\delta = \sin^{-1} \left( \frac{{\cal J}}
{ J \left( \theta_1,\theta_2,\theta_3 \right)} \right).
\eeq
In practice, we have been unable to analytically invert equation~(\ref{stanbox})
with only three basis 
boxes.\footnote{
Allowing four nondegenerate boxes as a dependent basis does provide an analytic
inversion of equation~(\ref{stanbox}).  We find
\bea
\sin^2 2\theta_1 &=& -4
\left(\B{11}{22}+\B{11}{23}+\B{11}{32}+\B{11}{33}\right),
\\
\tan^2 \theta_2 &=& \frac{\B{11}{32}+\B{11}{33}}{\B{11}{22}+\B{11}{23}},
\mbox{\ \ and \ \ } \\
\tan^2 \theta_3 &=& \frac{\B{11}{23}+\B{11}{33}}{\B{11}{22}+\B{11}{32}}.
\eea
We see, however, no way to reduce this basis to three independent boxes.  We
choose the four basis boxes by examination of the relationships~(\ref{stanbox});
the boxes which are most simply expressed in terms of the mixing angles are the
boxes we choose for our basis.  Thus the choice of 
which four boxes to isolate as the basis depends on the
unitary parameterization used for $V$.  $V$ is constructed by three successive
rotations and a phase matrix;
one may choose the rotation axes $(\hat{i}, \hat{j}, \hat{k})$ in 
$V(\delta=0) =
R_{\hat{i}}(\theta_1) R_{\hat{j}}(\theta_2) R_{\hat{k}}(\theta_3)$ 
such that each of the desired four 
basis boxes has only two terms, being
composed from three of the simplest $V$s and one binomial $V$.
}
Thus we have another reason to eschew the arbitrary unitary parameterization of
$V$ and work directly with the box algebra.


\subsection{The Wolfenstein Parameterization}
	
	
	Wolfenstein suggested \cite{Wolf} a now-popular parameterization of  
the CKM matrix based on data from the quark sector.  The parameter $\lambda$
is defined by $\lambda = V_{12}$, and the rest of the matrix is determined by
comparing matrix elements with powers of this small angle \cite{Wolf}. (For
example, the element $V_{23}$ is on the order of $V_{12}^2$ in the quark
sector, so it is expressed in terms of $\lambda^2$.  Keeping the matrix unitary
to the desired order of $\lambda$ yields the higher-order terms in $V_{11}$ and
$V_{22}$. 
The Wolfenstein mixing matrix to ${\cal{O}}(\lambda^3)$ is
\beq
V =  \left( \ba{ccc}
1 - \half \lambda^2       & \lambda             & \lambda^3 A (\rho - i\eta) \\
-\lambda                  & 1 - \half \lambda^2 & \lambda^2 A \\
\lambda^3 A(1-\rho-i\eta) & -\lambda^2 A        & 1 \\
\ea  \right).
\label{wolfV}
\eeq
This matrix is only unitary to order $\lambda^3$, so any relationships between
our boxes based on the unitarity of the mixing matrix, such as the
relationships proving the equality of the imaginary parts of
boxes, will hold only to order $\lambda^3$. 
Wolfenstein describes the mixing matrix in terms of 4 parameters:  $\lambda,
A, \rho$, and $\eta$.  We choose the three boxes $\B{11}{22}$, $\B{21}{32}$, and
$\B{11}{23}$ as our basis.  
Combining the definitions of our boxes (\ref{boxdef})
with Wolfenstein's mixing matrix (\ref{wolfV}), 
we find
\beq
\B{11}{22} =  -\lambda^2(1-\half\lambda^2)^2  \simeq  \;\; -\lambda^2, 
\eeq
\beq
\B{21}{32} = A^2\lambda^6(1-\half\lambda^2)(1-\rho-i\eta)  \simeq 
	A^2\lambda^6(1-\rho-i\eta), \mbox{ and}
\eeq
\beq
\B{11}{23} = -A^2\lambda^6(1-\half\lambda^2)(1-\rho-i\eta) \simeq 
-A^2\lambda^6(\rho-i\eta). 
\eeq
$\B{11}{22}$ is real, and the imaginary component of $\B{21}{32}$ has equal
magnitude but opposite sign as the imaginary component of
$\B{11}{23}$, so the three boxes act as four independent parameters.  (Again,
the boxes do not appear to have the same imaginary parts because Wolfenstein
has thrown out terms higher order in $\lambda$.  If we were to keep the
leading-order imaginary term for all of the boxes,
 all $\J{\alpha i}{\beta j}$ would be $\pm A^2 \lambda^6
\eta$.)  We can  
now find the rest of the boxes and express them in terms of our chosen
three boxes.
\beq
\B{11}{33} \simeq A^2 \lambda^6 (\rho-i\eta)(1-\rho-i\eta) \simeq 
	\frac{\B{11}{23} \B{21}{32}}{\Bs{11}{23} - \B{21}{32}}
\eeq
\beq
\label{Wolfeq}
\B{12}{23} \simeq -\B{12}{33} \simeq A^2\lambda^6(\rho-i\eta) 
\simeq -\B{11}{23}
\eeq
\beq
\B{11}{32} \simeq +\B{21}{33} \simeq -A^2\lambda^6(1-\rho-i\eta) \simeq
-\B{21}{32}
\eeq
$\cal{B}$, our matrix of boxes in (\ref{boxprob}) becomes the following for
the Wolfenstein parameterization:
\beq
\cal{B} = \left( \ba{ccc}
\B{11}{22} & -\B{11}{23} & \B{11}{23} \\
\B{21}{32} & \frac{\B{21}{32} - \Bs{11}{23}}{\B{11}{22}} & -\B{21}{32} \\
-\B{21}{32} & \B{11}{23} & \frac{\B{11}{23} \B{21}{32}}{\Bs{11}{23}-\B{21}{32}}
\ea
\right).
\eeq

The dominant box in this approximation is $\B{11}{22}$.  Thus the $\nue
\leftrightarrow \numu$ transition would be most probable, with its primary
contribution coming from the $\Delta m_{12}^2$ term.  The transition $\numu
\leftrightarrow \nutau$ is next-probable, down by the order of $\lambda^2$ from
the $\nue \leftrightarrow \numu$ transistion, and it receives its dominant
contribution from the $\Delta m_{23}^2$ term.  The least likely transition
according to the Wolfenstein model, down the order of $\lambda^4$ from the 
$\nue \leftrightarrow \numu$ transition, is the $\nue \leftrightarrow \nutau$
transition, and it receives equal contributions from all three mass
differences.


\subsection{The One-Angle Approximation}


A related near-unitary 
parameterization fitting the quark data uses only one real parameter, 
$\theta$, and a phase $\delta$.  Experimental determinations of 
the quark mixing matrix  
elements are consistent with the following form for $V$ \cite{BargPhil}:
\beq
V = \left( \ba{ccc}
1 & \theta & \theta^3 e^{-i\delta} \\
-\theta & 1 & \theta^2 \\
-\theta^3 e^{i\delta} & -\theta^2 & 1
\ea \right).
\label{smalltheta}
\eeq
This approximation
is a special case of the Wolfenstein parameterization, setting $A = |\rho -
i\eta|=1$ and
results in the following matrix of boxes:
\beq
{\cal B} = \left( \ba{ccc}
\B{11}{22} & -\Bs{21}{32} & \Bs{21}{32} \\
\B{21}{32} & -\left(\B{11}{22}\right)^2 & -\B{21}{32} \\
-\B{21}{32} & \Bs{21}{32} & \left(\B{11}{22}\right)^3
\ea \right),
\eeq
with
\bea
\B{11}{22} = - \theta^2, \mbox{ and}
\B{21}{32} = -\theta^6 e^{i \delta}.
\eea

The mixing matrix in (\ref{smalltheta}) contains only two independent
parameters, but our matrix ${\cal B}$ seems to contain three: $\B{11}{22}$,
which is real,
$(\R{11}{32})$, and $(\J{11}{32})$.  But the three are related by
\beq
|\B{21}{32}|^2 = \left(\R{21}{32}\right)^2 + \left(\J{21}{32}\right)^2
 = \left(\B{11}{22}\right)^6
\eeq
So we are indeed left with two independent parameters.

Because this one-angle approximation is a special case of the Wolfenstein
parameterization, it produces the same predictions for the relative
significance of transition probabilities:  $\nue \leftrightarrow \numu$ is the
most probable, and $\nue \leftrightarrow \nutau$ is the least probable. 
Discerning between the general Wolfenstein parameterization 
and this extension would require measurements of probabilities at different
lengths so boxes within a row may be measured individually.


\subsection{The Dominant Mass Scale Approximation}
\label{onemasssec}


	A popular approximation for neutrino masses and mixing assumes 
$m_3 >> m_1, m_2$ \cite{Fogli}, \cite{Card}, so 
$\Pe \approx \Pu \approx \frac{m_3^2}{4p}$.  Under this approximation,
\beq
\mbox{Re}\left({\cal B}\right) \ S^2(\Phi)  =  
\left( \ba{cc}
\R{11}{22} & \R{12}{23} + \R{11}{23} \\
\R{21}{32} & \R{22}{33} + \R{21}{33} \\
\R{11}{32} & \R{12}{33} + \R{11}{33}
\ea \right) \left( \ba{c}
\sin^2 \Pt \\ \sin^2 \Pu
\ea \right), 
\label{smallangleR}
\eeq
and
\bea
\mbox{Im}\left({\cal B}\right) \ S(2\Phi)& = &\left( \ba{ccc}
\J{11}{22} & \J{12}{23} + \J{11}{23} \\
\J{21}{32} & \J{22}{33} + \J{21}{33} \\
\J{11}{32} & \J{12}{33} + \J{11}{33}
\ea \right) \left( \ba{c}
\sin (2\Pt) \\ \sin (2\Pu) \ea \right) \nonumber \\
&=& \left( \ba{c}
\J{11}{22}  \\ \J{21}{32}  \\ \J{11}{32} 
\ea \right) \sin (2\Pt) 
=  {\cal J} \sin (2\Pt) \left( \ba{c}
1 \\ 1 \\ -1 \ea \right). 
\label{1massImag}
\eea
The second column of Im$\left({\cal B}\right)$ vanishes because the
sums of boxes are real by equations (\ref{imagfixrow}) and (\ref{imagfixcol}).  

For small $x$,  
\beq
\sin^2 \Pt << \sin^2 \Pu
\eeq
due to the mass hierarchy.
Thus experiments with a small $x$ will measure the factors multiplying the
bigger
phase $\Pu$, such as  the sum
\beq
\R{12}{23} + \R{11}{23}.
\label{smalsum}
\eeq
In Chapter~\ref{expt}, we apply a dominant-mass-scale approach to the
oscillation data currently available; in that treatment, the LSND experiment
measures the sum (\ref{smalsum}).
This sum is equal to the degenerate box $\R{13}{23}$ by
equation~(\ref{1degcol}).

Sources producing neutrinos with uncertainties in $p$ or $x$ 
large enough that all the oscillatory terms 
average will provide measurements of a sum of all three boxes in a row
of ${\cal B}$:
\beq
\mbox{Re}{\cal B} \langle S^2\left(\Phi\right)\rangle = 
\half \left( \ba{c} \R{11}{22} + \R{12}{23} + \R{11}{23} \\
\R{21}{32} + \R{22}{33} + \R{21}{33} \\
\R{11}{32} + \R{12}{33} + \R{11}{33}
\ea \right) = -\fourth \left( \ba{c} 
\sum_{x=1}^n |V_{1x}|^2|V_{2x}|^2 \\
\sum_{x=1}^n |V_{2x}|^2|V_{3x}|^2 \\
\sum_{x=1}^n |V_{1x}|^2|V_{3x}|^2 \ea \right),
\eeq
where the last equality comes from equation~(\ref{sumrowB}).
This type of asymptotic measurement could not detect 
CP violation directly since
the average of
the $\sin 2\Pij$ function is zero.  Our analysis in Chapter~\ref{expt} applies
this averaging to both solar neutrinos and atmospheric neutrinos.

CP violation is a very small effect in this dominant mass-scale 
approximation, even when the
oscillatory terms do not vanish. As seen in equation~({\ref{1massImag}), the
CP-violating term depends on $\sin (2\Pt)$, so the CP-violating
effects for local neutrino sources are of order $\frac{m_1^2 - m_2^2} 
{2p} x$.  Oscillatory behavior would eventually develop over longer baselines.


\section{Models of the Mass Matrix}
\label{massboxsec}


\subsection{Using a Fritzsch Mass Matrix}
\label{Fritzschsec}


In the late 1970s, Fritzsch \cite{Fritzsch} proposed a form of the quark mass
matrices which has proven to be consistent with experiment.  Extending this
model to the lepton sector means the mass matrices $M_{\nu}$ and $M_l$ for
neutrinos and charged leptons will both be Dirac matrices and have the 
Fritzsch form \cite{Fritzsch}:
\bea
M_{\nu} & = & \left( \ba{ccc}
0 & a e^{i \xi_1} & 0 \\
a e^{i \xi_2} & 0 & b e^{i \xi_3} \\
0 & b e^{i \xi_4} & c e^{i \xi_5}
\ea \right), \mbox{ and} \label{Mnu}\\
M_l & = & \left( \ba{ccc}
0 & A e^{i \Xi_1} & 0 \\
A e^{i \Xi_2} & 0 & B e^{i \Xi_3} \\
0 & B e^{i \Xi_4} & C e^{i \Xi_5}
\ea \right).
\eea
The parameters $a$, $b$, $c$, $A$, $B$, and $C$ may be taken to be real 
positive and will be 
determined by the lepton masses.  The postulated zero entries are today known
as ``texture zeros.''  They can be motivated by invoking symmetries in the
Higgs sector; such symmetries may be natural in grand unified theories.  
Each of the mass matrices originally has five 
independent phases, but we may rotate the mass matrix
and absorb the phases in the lepton fields:
\bea
\ol{\nuL} M_{\nu} \NR + \ol{l_L} M_l l_R + h.c.& =& \ol{\nuL}  
X_{\nuL} X_{\nuL}^{-1} M_{\nu} X_{\NR} X_{\NR}^{-1} \NR + \ol{l_L} 
X_{l_L} X_{l_L}^{-1} M_{l} X_{l_R} X_{l_R}^{-1} l_R + h.c \nonumber \\
&=&\ol{\tilde{\nuL}} \tMnu \tilde{\NR} + \ol{\tilde{l_L}} \tMl \tilde{l_L} +
h.c., 
\mbox{ with}
\eea
\beq
\tilde{\nuL} = X_{\nuL}^{-1} \nuL, \  \tilde{\NR} = X_{\NR}^{-1} \NR, 
\  \tilde{l_L}= X_{l_L}^{-1} l_l, \mbox{ and } \tilde{l_R} = X_{l_R}^{-1} l_R.
\eeq
As before, $\nuL, l_L, \NR,$ and $l_R$ are the $1 \times 3$ vectors 
containing all three generations of spinors.  Following Fritzsch's treatment 
in \cite{Fritzsch}, we choose
\bea
X_{\nuL}& =& \left( \ba{ccc} 1&0&0 \\ 0&e^{-i(\xi_1-\xi_3-\xi_4+\xi_5)}&0 \\
0&0&e^{-i(\xi_1-\xi_4)} \ea \right), \mbox{ and }   \\
X_{\NR}&=&\left( \ba{ccc} e^{-i(\xi_1+\xi_2-\xi_3-\xi_4+\xi_5)}&0&0
\\ 0&e^{-i\xi_1}&0 \\ 0&0&e^{-i(\xi_1-\xi_4+\xi_5)} \ea \right),
\eea
with $X_{l_L}$ and $X_{l_R}$ found by substituting $\Xi$ for $\xi$.  Under
this transformation,
\beq
M_{\nu} \rightarrow \tMnu = X_{\nuL}^{-1}M_{\nu}X_{\NR} = \left( \ba{ccc} 0 & 
a & 0 \\ a & 0 & b \\ 0 & b & c \ea \right), \mbox{ and}
\eeq
\beq
M_l \rightarrow \tMl = X_{l_L}^{-1}M_{l}X_{\l_R}= \left( \ba{ccc} 0 & A & 0 
\\ A & 0 & B \\ 0 & B & C \ea \right).
\eeq
These real
symmetric matrices
 may be diagonalized by the orthogonal matrices $R_{\nu}$
and $R_l$:
\beq
R_{\nu}^{-1} \tMnu R_{\nu} = 
	\left( \ba{ccc} m_1 & 0 & 0 \\ 0 & -m_2 & 0 \\ 0 & 0 & m_3 \ea \right),
	\mbox{  and } 
R_l^{-1} \tMl R_l  =  
	\left( \ba{ccc} m_e & 0 & 0 \\ 0 & -m_{\mu} & 0 \\ 0 & 0 & m_{\tau}
	\ea \right).
\label{Rdef}
\eeq
The $m_i$ are positive, and the minus signs in front of the second mass
eigenstates occur because the determinants of the $\tM$ are negative.  Such a 
sign factor may be absorbed by a chiral 
redefinition of the appropriate lepton fields.  Under $\psi
\rightarrow e^{i \frac{\pi}{2} \gdnf} \psi = i\gdnf\psi$,  $m\ol{\psi}\psi
\rightarrow -m\ol{\psi}\psi$, as discussed in Section~\ref{pretracesec}.  This
chiral rotation, however, would put the mass matrix into a non-Fritzsch form. 
The issue of the sign of the fermion mass is obviated if we work with
$MM^{\dagger}$ as in the previous Chapter, but we have here chosen instead 
to parallel Fritzsch's development.

These transformations $R$ and $X$ take us to the mass basis:  
$\nuL^m = R_{\nu}^{-1} X_{\nuL}^{-1} \nuL, \ \ \NR^m = 
R_{\nu}^{-1} X_{\NR}^{-1} N_R,$ {\it et cetera}.
As discussed in 
Section~\ref{CKMsec}, the mixing matrix arises in the charged-current
interaction:
\beq
{\cal J}_{CC} = \ol{\nuL} \gupu l_L = \ol{\nuL} X_{\nuL} R_{\nu} 
\left( R_{\nu}^{-1} X_{\nuL}^{-1} X_{l_L} R_l \right) R_l^{-1} X_{l_L}^{-1} 
\gupu l_L = \ol{\nuL^m} V \gupu l_L^m,
\eeq
Once $R_{\nu}$ and $R_l$ are found from (\ref{Rdef}),the mixing matrix may be
calculated from
\beq
V=R_{\nu}^{-1} X_{\nuL}^{-1} X_{l_L} R_l = R_{\nu}^{-1} \left( \ba{ccc} 
1 & 0 & 0 \\ 0 & e^{i \sigma} & 0 \\ 0 & 0 & e^{i\eta} \ea \right) R_l. 
\label{VFr}
\eeq
The two phases $\sigma$ and $\eta$ contain all of the phases in $X_{\nuL}$ and
$X_{l_L}$:  $\sigma = (\xi_1 - \Xi_1) - 
(\xi_3 - \Xi_3) - (\xi_4 - \Xi_4) + (\xi_5 - \Xi_5),$ and $\eta =
(\xi_1 - \Xi_1) - (\xi_4 - \Xi_4)$. One of these phases may be 
determined by
measurement of the standard CKM phase $\delta$ in (\ref{NactCKM}). 
The other only appears in real terms (see below) and therefore may
be absorbed in the definition of the mixing angles in the standard
parameterization, as shown in \cite{Fritzsch}.  

This extra phase occurs because Fritzsch postulates a set form for the mass
matrices, thereby depriving us of some extra degrees of freedom.  As discussed
above, any measurable quantities may be described by the masses and the mixing
matrix, $V$.  This matrix is unitary and therefore may have at most six phases.
Five relative phases may be absorbed in the definitions of the left-handed
leptons and right-handed neutrinos, leaving one phase in this observable
matrix.  Fritzsch, however, assumes a form for the mass matrices which yields
ten phases.  Only eight relative phases may be absorbed in the lepton fields
(the left-handed neutrinos and charged leptons must be rotated together, so
nine rotations may be performed on the left-handed leptons, the right-handed
neutrinos, and the right-handed charged leptons), leaving two phases in the
mass matrix.  Under this method, the rotation matrices have the form
\bea
X'_{\nuL} = X'_{l_L} & = & \left( \ba{ccc} 
e^{i \left( \xi_1+\xi_2-\xi_3+\xi_4-\xi_5\right)} & 0 & 0 \\
0 & e^{i \left( \xi_2 + \sigma \right) }& 0 \\
0 & 0 & e^{i \left( \xi_2-\xi_3+2\xi_4-\xi_5+\eta \right)} \ea \right), \\
X'_{\NR} & = & \left( \ba{ccc}
1 & 0 & 0 \\
0 & e^{i \left( \xi_2 - \xi_3 + \xi_4 - \xi_5 \right)} & 0 \\
0 & 0 & e^{i \left( \xi_2 - \xi_3 \right)} \ea \right), 
\mbox{\  and \ }  \\
X'_{l_R} & = & \left( \ba{ccc}
e^{i \left( \xi_2 - \Xi_2 \right)} & 0 & 0 \\
0 & e^{i \left( \xi_1 - Xi_1 + \xi_2 - \xi_3 + \xi_4 - \xi_5 \right)} & 0 \\
0 & 0 & e^{i \left( \xi_2 - \Xi_3 \right)} \ea \right).
\eea
These rotations yield the matrices $\tMl$ and $X^{-1} \tMnu$, with $X$ equal to
the net phase matrix above:
\beq
X = \left( \ba{ccc} 1 & 0 & 0 \\ 0 & e^{i\sigma} & 0 \\ 0 & 0 & e^{i\eta} 
\ea \right).
\eeq

Once we have simplified the mass matrices to $\tMnu$ and $\tMl$ which have the 
same form, we need only solve
for $R_{\nu}$ by diagonalizing $\tMnu$ in (\ref{Rdef}) and then obtain $R_l$ by
substitution.
The first step in diagonalizing a matrix is to solve the eigenvalue equation, 
a task which is simplified because we know the eigenvalues, $m_1$, $-m_2$, and
$m_3$, and want only the unknown $a$, $b$, and $c$ in terms of the masses:
\beq
\mbox{Det}(\tMnu - m I) = -m^3 + c m^2 + (a^2+b^2) m - a^2 c = 
(m_1-m)(-m_2-m)(m_3-m).
\label{cubic}
\eeq
Solving for the unknown $a$, $b$, and $c$ by matching coefficients of $m$, we
find
\bea
c & = & m_1 - m_2 + m_3 \nonumber\\
b^2 + a^2 & = & m_1 m_2 + m_2 m_3 - m_3 m_1 \\
a^2 c & = & m_1 m_2 m_3. \nonumber 
\eea
The first of these relations gives $c$ explicitly, and we can solve for $a$ and
$b$:
\bea
\label{ab}
a^2 & = & \frac{m_1 m_2 m_3}{m_1-m_2+m_3} \\
b^2 & = & m_1 m_2 + m_2 m_3 - m_3 m_1 - a^2. \nonumber
\eea
Because $a$ and $b$ were defined to be positive real numbers, we must always
take the positive square root of the right sides of (\ref{ab}). 

We may construct the diagonalizing matrix $R_{\nu}$ out of the eigenvectors
$\psi$ of $\tMnu$.  These eigenvectors satisfy the relationship
\beq
\tMnu \psi = m \psi,
\eeq
which provides three constraints on the elements $\psi_1$, $\psi_2$, and 
$\psi_3$ of $\psi$:
\bea
\label{Frvector} 
a \psi_2 & = & m \psi_1, \nonumber \\
a \psi_1 + b \psi_3 & = & m \psi_2, \mbox{ and} \\
b \psi_2 + c \psi_3 & = & m \psi_3. \nonumber
\eea
The first and the second of the equations in (\ref{Frvector}) describe the form
of $\psi$.  The third, combined with the conclusions of the first two,
reiterates the eigenvalue equation (\ref{cubic}).  We thus find that $\psi$
has the form
\beq
\psi = n \left( \ba{c} a b \\ m b \\ m^2 - a^2 \ea \right),
\eeq
where $n$ is a normalization constant, chosen to make the length of $\psi$
equal to $1$:
\beq
\frac{1}{n^2} = a^2 b^2 + m^2 b^2 + (m^2 -a^2)^2.
\eeq  
These relations hold for $m$ equal to each of the eigenvalues $m_1$, $-m_2$, 
or $m_3$, giving the apparently simple form for $R_{\nu}$ 
\beq
R_{\nu} = \left( \ba{ccc} n_1ab & n_2ab & n_3ab \\
n_1 m_1 b & -n_2 m_2 b & n_3 m_3 b \\
n_1(m_1^2 - a^2) & n_2(m_2^2 - a^2) & n_3(m_3^2 - a^2) \ea \right).
\label{Rnu}
\eeq
By analogy, $R_l$ is given by
\beq
R_l = \left( \ba{ccc} N_eAB & N_{\mu}AB & N_{\tau}AB \\
N_em_e B & -N_{\mu}m_{\mu} B & N_{\tau}m_{\tau} B \\
N_e(m_e^2 - A^2) &N_{\mu}(m_{\mu}^2 - A^2) & N_{\tau}(m_{\tau}^2 - A^2) \ea 
\right), \mbox{ with}
\label{Rl}
\eeq
\bea
C & = & m_e - m_{\mu} + m_{\tau}, \nonumber\\
A^2 & = & \frac{m_e m_{\mu} m_{\tau}}{m_e - m_{\mu} + m_{\tau}}, \label{ABC} \\
B^2 & = & m_e m_{\mu} + m_{\mu} m_{\tau} - m_{\tau} m_e - A^2, \mbox{ and} 
\nonumber \\
\frac{1}{N_l^2} & = & A^2 B^2 + m_l^2 B^2 + (m_l^2 - A^2)^2, \ 
(l=e,{\mu},{\tau}).
\eea
Using $R_{\nu}$ and $R_l$ from (\ref{Rl}) and (\ref{Rnu})in (\ref{VFr}), we 
find the
following form for the elements of the mixing matrix:
\beq
V_{\alpha l} = n_{\alpha} N_l \left( abAB + 
(-1)^{\left(\delta_{\alpha 2}+\delta_{l\mu}\right)} 
  m_{\alpha} m_l bB e^{i\sigma} +
  (m_{\alpha}^2-a^2)(m_l^2-A^2)e^{i\eta} \right),
\label{ValFr}
\eeq
for $\alpha=1,2,3$ and $l=e,\mu,\tau$.\footnote{
Our assignment of numbers to the neutrino index and lepton flavors to the
charged lepton index is slightly different from our normal convention of $\alpha
= e, \mu, \tau$ and $l=1,2,3$.  Because the charged lepton flavor states are not
equivalent to mass states in the Fritzsch model, we have chosen to use flavor
indices to represent the charged lepton flavor.  To avoid complicated indexing
of the neutrino states, we have labeled them by numbers.  In retrospect, this
appears a bit inconsistent, but it seemed like a good idea at the time.
}
  
Obviously the boxes from the Fritzsch mixing matrix will be rather complicated
functions of the masses. 
In a parameterization based on the mixing matrix elements, we would have to
invert equations quartic in the $V$s given by equation~(\ref{ValFr}) 
to find the Fritzsch
parameters $a, b,$ {\it et cetera} from the observable probabilities.  The
boxes, however, are the probabilities (more or less), and the Fritzsch
parameters may be found by using equation~(\ref{Hab2}).  Unfortunately, we must
first rotate the Fritzsch $M_{\nu}$ to $\hat{M}$ using the charged lepton
rotation matrix (\ref{Rl}) and equation~(\ref{Mrot}).  This again gives daunting
equations, so we will content ourselves with considering only two limiting
cases.

We know that the
charged lepton masses obey a strong hierarchy: 
$m_{\tau}~\gg~m_{\mu}~\gg~m_e$.  With these constraints, $R_l$ becomes
\beq
R_l = \left( \ba{ccc} 1 & \Lambda_1 & \Lambda_1 \Lambda_2^3 \\
\Lambda_1 & -1 & \Lambda_2 \\
-\Lambda_1 \Lambda_2 & \Lambda_2 & 1 
\ea \right), \mbox{\ \  where}
\label{Rlhier}
\eeq
\beq
\Lambda_1 = \sqrt{\frac{m_e}{m_{\mu}}}, \mbox {\ \ and \ \ }
\Lambda_2 = \sqrt{\frac{m_{\mu}}{m_{\tau}}}.
\eeq
If neutrinos also obey a strong mass hierarchy, $m_3~\gg m_2~\gg~m_1$,
$R_{\nu}$ will have the same form (\ref{Rlhier}), with $\lambda$ substituted for
$\Lambda$ and
\beq
\lambda_1 = \sqrt{\frac{m_1}{m_2}}, \mbox {\ \ and \ \ }
\lambda_2 = \sqrt{\frac{m_2}{m_3}}.
\eeq
The diligent reader may notice that the matrix in equation
(\ref{Rlhier}) does not exactly diagonalize $\tMl$.  But the mass eigenvalues
do appear on the diagonal, and the off-diagonal elements are negligible under
the mass hierarchy approximation. The mixing matrix for a Fritzsch model with
dual mass hierarchies takes the form
\beq
V = \left( \ba{ccc}
1+\lambda_1 \Lambda_1 e^{i\sigma} & \Lambda_1 - \lambda_1 e^{i\sigma} &
  \lambda_1 \Lambda_2 e^{i\sigma} - \lambda_1 \lambda_2 e^{i\eta} \\
\lambda_1 - \Lambda_1 e^{i\sigma} & \lambda_1 \Lambda_1 + e^{i\sigma} +
\lambda_2 \Lambda_2 e^{i\eta} & -\Lambda_2 e^{i\sigma} + \lambda_2 e^{i\eta} \\
\lambda_2 \Lambda_1 e^{i\sigma} - \Lambda_1 \Lambda_2 e^{i\eta} & -\lambda_2
e^{i\sigma} + \Lambda_2 e^{i\eta} & \lambda_2 \Lambda_2 e^{i\sigma} + e^{i\eta}
\ea \right)
\label{Vhier}
\eeq
This result disagrees with the CKM matrix given in \cite{Fritzsch} and
\cite{FY} by the sign of $\V{12}$, $\V{21}$, $\V{23}$, and $\V{32}$. 
One possible explanation of this discrepancy is that the authors of
\cite{Fritzsch} and \cite{FY} have absorbed an extra phase $e^{i\pi}$ into the
fields of the muon and muon-neutrino.  Such an absorption will not affect the
measurable boxes.   Indeed, every box has
an even number of the terms with the disputed sign, so the
extra minus signs cancel.  

The boxes formed from the CKM matrix in (\ref{Vhier}) are  rather
complicated functions of mass ratios.  For example,
\bea
\B{11}{22} & = & e^{i\sigma} \left( \Lambda_1\lambda_1 - \Lambda_1^2 e^{i\sigma}
- \lambda_1^2 e^{i\sigma} + \lambda_1\Lambda_1 e^{2i\sigma} \right),\
\mbox{ and } \\
\B{12}{33} & = & e^{i\eta} \left(
- \lambda_1\lambda_2\Lambda_1\Lambda_2e^{2i\sigma}
- \left( \lambda_1\lambda_2^2\Lambda_1 + \lambda_1\Lambda_1\Lambda_2^2 \right) 
    e^{i\left(\sigma+\eta\right)}
- \left( \lambda_2^2\Lambda_1^2 + \Lambda_1^2\Lambda_2^2 \right)
    e^{i\left(2\sigma+\eta\right)} \right. \\
&& \left. + \ \lambda_2\Lambda_1^2\Lambda_2e^{i\left(\sigma+2\eta\right)}
- \lambda_1\lambda_2\Lambda_1\Lambda_2e^{2i\eta}
- \lambda_2\Lambda_1^2\Lambda_2e^{3i\sigma}   \right).
\eea
Under this double-mass hierarchy, the dominant boxes are $\B{11}{22}$ 
$\B{22}{33}$, which are on
the order of a mass ratio (two powers of $\Lambda$ or $\lambda$).  The other boxes are on the 
order of a mass ratio squared.
$\B{11}{22}$ appears in the expression (\ref{boxprob}) for the probability of a 
$\nue \leftrightarrow \numu$ transition, multiplied by $\sin^2 \Pt$, where 
$\Pt$ is given by (\ref{Phiij}) to be
\beq
\Pt = \frac{m_1^2 - m_2^2}{4 p} x,
\eeq
which is on the order of $m_2^2 \frac{x}{p}$ for small $x$, so its
contribution is lessened by $\lambda_2^4$ compared to boxes multiplied by $\Pe$
or $\Pu$.  This reduction more than cancels out the effect from the box's
dominance. $\B{22}{33}$, on the other hand, appears in the probability of a
$\numu \leftrightarrow \nutau$ transition, multiplied by $\sin^2 \Pe$, making
the $\numu \leftrightarrow \nutau$ transition the most probable transition in
the double-hierarchy Fritzsch model.

Another popular scenario for neutrino mass is the reverse hierarchy: 
$m_1~\gg~m_2~\gg~m_3$.  Using this approximation in the Fritaxch model 
yields a rotation matrix given by
\beq
R_{\nu} = \left( \ba{ccc}
\lambda'_2 \lambda_1^{'3} & \lambda'_2 & 1 \\
\lambda_1' & -1 & \lambda'_2 \\
1 & \lambda'_1 & -\lambda'_2 \lambda'_1
\ea \right), \mbox{ with }
\eeq
\beq
\lambda'_1 = \sqrt{\frac{m_2}{m_1}}, \mbox{\ \ and \ \ }
\lambda'_2 = \sqrt{\frac{m_3}{m_1}}.
\eeq
Using a straight hierarchy for the charged leptons and this reverse hierarchy
for neutrinos, we arrive at the following form for the mixing matrix:
\beq
V = \left( \ba{ccc}
\lambda'_1 \Lambda_1 e^{i\sigma} - \Lambda_1 \Lambda_2 e^{i\eta} &
-\lambda'_1 e^{i\sigma} + \Lambda_2 e^{i\eta} & 
\lambda'_1 \Lambda_2 e^{i\sigma} + e^{i\eta} \\
\lambda'_2 - \Lambda_1 e^{i\sigma} & 
\lambda'_2 \Lambda_1 + e^{i\sigma} + \lambda'_1 \Lambda_2 e^{i\eta} & 
-\Lambda_2 e^{i\sigma} + \lambda'_1 e^{i\eta} \\
1 + \lambda'_2 \Lambda_1 e^{i\sigma} & \Lambda_1 - \lambda'_2 e^{i\sigma} &
\lambda'_2 \Lambda_2 e^{i\sigma} - \lambda'_1 \lambda'_2 e^{i\eta}
\ea \right).
\eeq

Under this approximation, the dominant boxes are $\B{21}{32}$ and $\B{12}{23}$, 
which contribute
to $\numu \leftrightarrow \nutau$ mixing scaled by $\sin^2 \Pt$ and to $\nue
\leftrightarrow \numu$ mixing scaled by $\sin^2 \Pe$, respectively.  In this
case, $\sin^2 \Pt$ is large at small distances, so once again
the $\numu \leftrightarrow \nutau$ mixing predominates for neutrinos from local 
sources.  These predictions of the Fritzsch model will be compared against
experimental data in Chapter~\ref{expt}.

If neutrinos are not pure
Dirac particles, then the matrix in (\ref{VFr}) does not necessarily apply. 
But if a light-neutrino $3 \times 3$ matrix $M_{\nu}$ is obtained by the see-saw
mechanism $M_{\nu}=M_D^T M_S^{-1} M_D$, as in Section \ref{seesawsection}, then 
the mixing
matrix may have the form in (\ref{VFr}), as long as $M_S$, the mass matrix for
the right-handed neutrino singlet term, is proportional to the unit matrix in
the basis where $M_D$ is diagonal \cite{FY}.


\subsection{Using a Democratic Mass Matrix}


Another texture of mass matrix popularized by the disparities in quark masses
is the ``democratic'' mass matrix \cite{Kaus}:
\beq
M = \left( \ba{ccc} 1&1&1 \\ 1&1&1 \\ 1&1&1 \ea \right).
\eeq
This matrix has eigenvalues $0$, $0$, and $3$.  The mass matrix must be 
normalized to set the non-zero eigenvalue equal to the heaviest mass.  We know
from experiments the mass of the top quark is much heavier 
than those of the up and charm quarks, and the bottom quark 
mass is much heavier than the other down-type quark masses, so the democratic
mass matrix is a reasonable approximation for the quark sector.  The charged
leptons also have one dominant mass scale, and it is not unreasonable to assume
the neutrino masses might follow the same pattern.  So we have
\beq
M_{\nu} = \frac{m_3}{3} \left( \ba{ccc} 1&1&1 \\ 1&1&1 \\ 1&1&1 \ea 
\right), \mbox{\ \ \ and \ \ \ } 
M_l = \frac{m_{\tau}}{3} \left( \ba{ccc} 1&1&1 \\ 1&1&1 \\ 1&1&1 \ea \right).
\eeq
Because of the degeneracy of two of the masses in each matrix, the eigenvectors
depend on an undetermined parameter which we call $a$.  They have the forms
\beq
\psi^1 = n_1 \left( \ba{c} a \\ 1 \\ -(a+1) \ea \right), \;\;\;\;\;
\psi^2 = n_2 \left( \ba{c} a+2 \\ -(2a +1) \\ a-1 \ea \right), 
\mbox{\ \ \  and\ \ \ }
\psi^3 = 3^{-\frac{1}{2}} \left( \ba{c} 1 \\ 1 \\ 1 \ea \right).
\eeq
$n_1$ and $n_2$ are fixed by normalizing the eigenvectors:
\bea
n_1 & = & 2 (a^2 + a + 1); \\
n_2 & = & 2 (3a^2 + 4a + 3).
\eea
As above, we will choose lowercase $a$ and $n_i$ for the neutrino matrices, and
uppercase $A$ and $N_l$ for the charged lepton matrices.  These eigenvectors
may be used to form the $R_{\nu}$ and $R_l$ that diagonalize $M_{\nu}$ and
$M_l$.  Constructing the CKM matrix from the $R$ matrices, we find a rather
simple result:
\beq
V = \left( \ba{ccc} n_1 N_e (2 + 2aA + a + A) & n_1 N_{\mu} (3a - 3A - 2) & 0 \\
n_2 N_e (3A - 3a - 2) & 3 n_2 N_{\mu} (2 + 2aA + a + A) & 0 \\
0 & 0 & 1 \ea \right).
\label{demckm}
\eeq
$a$ and $A$ are completely arbitrary; the eigenvectors corresponding to the
lighter particles are degenerate, so one can rotate freely between the two 
states with no observable effect.  We can represent this freedom by
parameterizing the mixing matrix with rotation angles $\theta$ for the
neutrino sector and $\Theta$ for the charged-lepton sector.  Orthonormality of
the eigenvectors is met by setting
\bea
n_1 a & = & \frac{\cos \theta}{\sqrt{2}} + \frac{\sin \theta}{\sqrt{6}}, 
  \mbox{ and} \\
N_1 A & = & \frac{\cos \Theta}{\sqrt{2}} + \frac{\sin \Theta}{\sqrt{6}}.
\eea
The mixing matrix has a quite obvious form with this choice of parameterization:
\beq
V = \left( \ba{ccc} \cos(\theta-\Theta) & \sin(\theta-\Theta) & 0 \\
-\sin(\theta-\Theta) & \cos(\theta-\Theta) & 0 \\
0 & 0 & 1 \ea \right).
\eeq
The lighter states are rotated by a combination $\theta - \Theta$.    If the
charged lepton sector is described by the same angle as the neutrino sector, no
mixing occurs at all.  Also, if the
lighter-mass degeneracy is exact, there is complete freedom to choose
$\theta-\Theta=0$.  
One expects, however, that ``accidental'' symmetries will be broken
by higher-order effects.  Parameterizing this symmetry breaking of the
$2\times2$ light mass matrix as
\beq
\left(\ba{cc}
\sigma_1 & \epsilon \\
\epsilon^* & \sigma_2 \ea \right),
\eeq
one finds 
\beq
\tan 2\theta = \frac{2|\epsilon|}{\sigma_1-\sigma_2},
\eeq
 and
\beq
\Delta m_{12}^2 = \left(\sigma_1+\sigma_2\right)
\sqrt{\left(\sigma_1-\sigma_2\right)^2+4|\epsilon|^2}.
\eeq

Whether one chooses $a$ and $A$ or $\theta$ and $\Theta$, 
the mixing matrix as given has only one non-zero box:
\beq
\B{11}{22} = n_1^2 n_2^2 N_1^2 N_2^2 3(2+2aA+a+A)^2 (3a-3A-2)(3A-3a-2) =
-\frac{1}{4} \sin^2(2(\theta-\Theta)).
\eeq
This box appears in the expression (\ref{boxprob}) for the probability of a $\nue
\rightarrow \numu$ transition, multiplied by $\sin^2 \Pt$, where $\Pt$ is given
by (\ref{Phiij}) to be
\beq
\Pt = \frac{m_1^2 - m_2^2}{4 p} x,
\eeq
which is very small for a nearly exact 
democratic mass matrix except in the case of extremely
large $x$.  Symmetry breaking may also induce small nonzero boxes mixing the
third mass with the first two.  Here $\Pu$ and $\Pe$ would be large, but the
amplitude of oscillation would be small.  
Oscillations may therefore not be measurable in
this scenario for terrestrial experiments.  Experiments looking for
oscillations from cosmologically distant objects might, however, be sensitive
to the $\nu_1\leftrightarrow\nu_2$ transition \cite{us}.

The democratic mass matrix is only an approximation.  The masses of
the electron and muon are not exactly zero, and we reasonably expect
non-zero masses for the two lighter neutrinos too.  The masses of the muon and
electron are, however, much smaller than the tau mass, so the democratic mass
matrix may be a good first approximation, with the smaller masses arising from
corrections to it.


\subsection{Using a Zee Mass Matrix}
\label{Zeeboxsec}


As discussed in Section \ref{Zeesec}, the Zee mechanism with diagonal coupling
between a Higgs doublet and charged leptons gives rise to a
neutrino mass matrix of the form (\ref{Zeemass}).  Such a matrix may be
parameterized in terms of three parameters $\alpha, \sigma$, and $m_0$
\cite{WolfZee}, \cite{SmirZee}, \cite{MP}:
\beq
M_{\nu} = m_0 \left( \ba{ccc} 0 & \sigma & \cos{\alpha} \\
                          \sigma & 0 & \sin{\alpha} \\
                          \cos{\alpha} & \sin{\alpha} & 0 \ea \right).
\label{Zeematrix}
\eeq
These new parameters are related to the charged lepton masses and the coupling
constants of (\ref{Zeemass}) by 
\bea
\cos{\alpha} & = & \frac{f_{e \tau}}{m_0} \left( m_{\tau}^2 - m_e^2 \right)
  \approx \frac{f_{e \tau}}{\sqrt{f_{e \tau}^2 + f_{\mu \tau}^2}}, \nonumber \\
\sin{\alpha} & = & \frac{f_{\mu \tau}}{m_0} \left( m_{\tau}^2 - m_{\mu}^2 \right) 
  \approx \frac{f_{\mu \tau}}{\sqrt{f_{e \tau}^2 + f_{\mu \tau}^2}}, \\
\sigma & \approx & \frac{f_{e \mu} m_{\mu}^2}{f_{e \tau} m_{\tau}^2} \cos{\alpha},
   \mbox{\ \ and \ \ } \nonumber \\
m_0 & \approx & m_{\tau}^2 \sqrt{f_{e \tau}^2 + f_{\mu \tau}^2}. \nonumber  
\eea
These relationships ignore $m_e^2$ and $m_{\mu}^2$ with respect to
$m_{\tau}^2$, which is a fairly good approximation given the mass hierarchy of
the charged leptons.  We will also assume $\sigma \ll 1$, which holds unless
$f_{e \mu} \gsim 10^4 f_{e \tau}$ \cite{MP}.  

Upon diagonalizing the matrix in equation (\ref{Zeematrix}), we find the
neutrino masses \cite{MP}
\bea
m_1 & = & - m_0 \sigma \sin{2 \alpha} \nonumber \\
m_2 & = & m_0 \left(1-\half \sigma \sin{2 \alpha} \right) \\
m_3 & = & m_0 \left(1+\half \sigma \sin{2 \alpha} \right), \nonumber
\eea
where an overall minus sign on $m_3$ has been absorbed into the definition of
the neutrino field through the chiral rotation $\psi \rightarrow i\gdnf \psi$.  
The mass differences for this Zee model are
\bea
m_1^2-m_2^2 & \approx & -m_0^2 \left(1-2\sigma\cos{\alpha}\sin{\alpha} \right),
   \nonumber \\
m_1^2-m_2^3 & \approx & -m_0^2 \left(1+2\sigma\cos{\alpha}\sin{\alpha} \right),
  \mbox{\ \ and \ \ } \\
m_2^2-m_2^3 & \approx & -4 m_0^2 \sigma\cos{\alpha}\sin{\alpha}. \nonumber
\eea
$m_3$ and $m_2$ are nearly degenerate and much larger than $m_1$.

Because the lepton mass matrix has been assumed
diagonal for this model, the weak-interaction
mixing matrix $V$ is equivalent to the neutrino
mixing matrix $U_{\nuL}$, and is found by diagonalizing the mass matrix in
equation (\ref{Zeematrix}) \cite{SmirZee}:
\beq
V = \frac{1}{\sqrt{2}} \left( \ba{ccc} 
\sqrt{2} \cos{\alpha} & \sin{\alpha} + \sigma \cos{\alpha} 
      & \sin{\alpha} + \sigma \cos{\alpha} \\
-\sqrt{2} \sin{\alpha} & \cos{\alpha} & \cos{\alpha} \\
-\sqrt{2} \sigma & 1 & -1 \ea \right).
\label{ZeeV}
\eeq
The boxes from this mixing matrix have the forms
\bea
\B{11}{22} & = & -\half \sin{\alpha} \cos^2{\alpha} \left( \sin{\alpha} + 
   \sigma \cos{\alpha} \right) \nonumber \\
\B{11}{23} & = & -\half \sin{\alpha} \cos^2{\alpha} \left( \sin{\alpha} - 
   \sigma \cos{\alpha} \right) \nonumber \\
\B{11}{32} & = & -\half \sigma \cos{\alpha} \left( \sin{\alpha} + 
   \sigma \cos{\alpha} \right) \nonumber \\
\B{11}{33} & = & -\half  \sigma\cos{\alpha} \left( \sin{\alpha} - 
   \sigma \cos{\alpha} \right) \nonumber \\
\B{12}{23} & = & \frac{1}{4} \cos^2{\alpha} \left( \sin^2{\alpha} - 
   \sigma^2 \cos^2{\alpha} \right) \label{Zeebox} \\
\B{12}{33} & = & -\frac{1}{4} \left( \sin^2{\alpha} - 
   \sigma^2 \cos^2{\alpha} \right) \nonumber \\
\B{21}{32} & = & \half \sigma \cos{\alpha} \sin{\alpha} \nonumber \\
\B{21}{33} & = & -\half \sigma \cos{\alpha} \sin{\alpha} \nonumber \\ 
\B{22}{33} & = & -\frac{1}{4} \cos^2{\alpha} \nonumber
\eea

If $f_{e \tau} \sim f_{\mu \tau}$, then $\sin \alpha \gg \sigma \cos \alpha$ and
the boxes have the following relative magnitudes:
\bea
\B{11}{22}  \sim  \B{11}{23}  \sim  -\B{12}{23} & \sim & 
-\sin^2\alpha \cos^2 \alpha, 
\nonumber \\
\B{11}{32}  \sim  \B{11}{33} \sim  -\B{21}{32}  \sim  \B{21}{33} & \sim &
-\sigma \sin \alpha \cos \alpha,
\label{Zee1} \\
\B{12}{33} & \sim & -\sin^2 \alpha, \mbox{\ \ and \ \ }
\nonumber \\
\B{22}{33} & \sim & -\cos^2 \alpha. 
\eea

If, on the other hand, $f_{e \tau} \gg f_{\mu \tau}$ we may 
neglect the first terms of the sums in the box
relationships (\ref{Zeebox}).  In this scenario we find
\bea
\B{11}{22}  \sim  - \B{11}{23} & \sim & -\sigma \sin \alpha \cos^3 \alpha,
\nonumber \\
\B{11}{32}  \sim  - \B{11}{33}  \sim  \B{12}{33} & \sim & 
-\sigma^2 \cos^2 \alpha, 
\label{Zee2} \\
\B{12}{23} & \sim & - \sigma^2 \cos^4 \alpha \\
\B{21}{32} \sim  -\B{21}{33} & \sim &  \sigma \sin \alpha \cos \alpha, 
\mbox{\ \ and \ \ } \nonumber \\
\B{22}{33} & \sim & -\cos^2 \alpha. 
\eea
\newpage 
The predictions of both of these scenarios are compared with experimental
results in Chapter~\ref{expt}.


\chapter{The Connection to Experiment}
\label{expt}


\renewcommand{\textfraction}{0}
\renewcommand{\topfraction}{1}
\renewcommand{\bottomfraction}{1}

The advantages of the box notation become clear when drawing conclusions from
experimental results.  To that purpose, we will examine the currently existing 
data.
No direct evidence of neutrino oscillation has yet been confirmed.  The LSND
experiment at Los Alamos has preliminary data which may be explained by
oscillations of muon antineutrinos into electron antineutrinos \cite{LSND}.  
This oscillation interpretation has not
yet been confirmed by another experiment, but we will consider LSND's result as
a possible measurement of an oscillation probability.  Other accelerator
experiments put bounds on different oscillation probabilities, and we will use
these bounds as constraints in what follows.  Indirect evidence for neutrino
oscillations comes from the solar neutrino deficit 
\cite{Homestake},\cite{solardat}
and the atmospheric neutrino anomaly \cite{atmosdat}, 
and we will use the oscillation interpretation for these phenomena
also.  The probabilities derived from such indefinite results are not meant to
be definitive, but they will allow us to demonstrate the usefulness of the box
notation.  Once more conclusive data becomes available, the numbers used below
may
be adjusted, but the procedure will remain the same.


\section{The Data}
\label{datasec}

	
Solar neutrino experiments measure the flux of neutrinos from the sun
and compare it to the flux predicted by solar models such as Pinisonneault and 
Bahcall's \cite{Bahcall}.  With
the exception of Kamiokande, the older solar neutrino detectors are sensitive
only to electron type neutrinos, so the ratio of observed flux to predicted
flux gives the survival probability for electron neutrinos, $\Pnu{e}{e}$.\ssp  
The Kamiokande experiment measures neutral current events too, so it may detect
muon and tau neutrinos.   The neutral-current cross section in water is 0.17
(about one-sixth) times
the charged-current cross section
for the neutrino
energies measured by Kamiokande \cite{Bahbook}.
Thus the probability measured
by Kamiokande (and now by Super K) 
is 
\beq
\Pnu{e}{e}+0.17 (\Pnu{e}{\mu}+\Pnu{e}{\tau}+\Pnu{e}{e}),
\label{NCacct}
\eeq 
where $\Pnu{e}{e}$ is
the probability as measured by charged-current-only detectors.  The
analysis of solar neutrinos used below \cite{AP} takes into account this additional
contribution to the Kamiokande data.  The quoted solar deficits contain hidden
uncertainties, because the expected neutrino spectrum given by solar models is
itself somewhat uncertain.

Atmospheric neutrino experiments measure the flux of muon and electron
neutrinos produced in the atmosphere.  The chain of events producing these
neutrinos should create roughly two muon-type ($\numu$ or $\smc{\nu}_{\mu}$) 
neutrinos for every electron-type ($\nue$ or $\smc{\nu}_e$)
neutrino.  Experiments, however, have measured nearly equal amounts of
electron-type and muon-type neutrinos.  These results are expressed as a ratio
of ratios, $R_{atm}=r_{expected}/r_{measured}$, where the ratios 
$r$ are of electron-type to 
muon-type.  For three generations, this anomaly could be due 
to $\numu
\leftrightarrow \nue$ mixing, $\numu \leftrightarrow \nutau$ mixing, or
three-way mixing $\numu \leftrightarrow \nue \leftrightarrow \nutau$, and these
three possibilities are described in detail in the paper which first suggested
oscillations as a solution to the atmospheric neutrino anomaly, \cite{LPW}. 
We will abbreviate
$r_{expected}$ as simply $r$, and take its value to be $0.48$ \cite{Gaisser}, 
the result of sophisticated Monte Carlo simulations assuming no oscillations.  
This number is somewhat dependent on detector efficiencies, but the
differences between detectors will not affect our conclusions significantly.

If the atmospheric anomaly is caused solely by $\numu
\leftrightarrow \nue$ mixing, the ratio depends only on $\Pnu{e}{\mu}$ and its
time-reversal conjugate $\Pnu{\mu}{e}$:
\beq
R_{atm} = \frac{1-\Pnu{\mu}{e} + r\Pnu{e}{\mu}}
{1-\Pnu{e}{\mu} + r^{-1}\Pnu{\mu}{e}}.
\eeq
When CP is effectively conserved, such as
when $\sin 2\Pij$ averages to zero,
the two probabilities are equal, and the measured ratio is simply
\beq
R_{atm} = \frac{1-\Pnu{e}{\mu} \left(1-r\right)}
{1-\Pnu{e}{\mu} \left(1-r^{-1}\right)}.
\eeq
The transition probability may be found from data by inverting this expression:
\beq
\Pnu{e}{\mu}  = \frac{y}{x},
\eeq
where $x$ and $y$ are defined for our use as
\bea
x & = & 1-r-R_{atm}+r^{-1}R_{atm}, \mbox{\ \ and \ \ } \nonumber \\
y & = & 1-R_{atm}.
\label{xydef}
\eea
For the value of $r=0.48$ used here, 
\beq
x=0.52 + 1.1 R_{atm}.
\eeq

If the atmospheric anomaly is caused by $\numu \leftrightarrow \nutau$ mixing,
the ratio is simply expressed:
\beq
R_{atm} = 1 - \Pnu{\mu}{\tau}.
\eeq
This possibility is similar to one with $\numu \leftrightarrow
\nu_{\chi}$ mixing, or oscillations to both $\nutau$ and $\nu_{\chi}$, since
$\nutau$ is not measured by atmospheric neutrino experiments and therefore
acts like a sterile neutrino.  The
transition probabilities in these scenarios are related to the measured ratio
by
\bea
R_{atm} & = & 1 - \Pnu{\mu}{\chi}, \mbox{\ \ and \ \ } \\
R_{atm} & = & 1 -\Pnu{\mu}{\tau} - \Pnu{\mu}{\chi},
\eea
respectively.
Probabilities may be found from
\beq
\Pnu{\mu}{\tau} \ + \Pnu{\mu}{\chi} \ = 1-R_{atm} = y,
\eeq
using the definition of $y$ given in equation~(\ref{xydef})

The final possibility for the atmospheric anomaly involves all neutrino
flavors.  The measured ratio in this case, including a possible sterile
neutrino, is
\beq
R_{atm} = \frac{1-\Pnu{\mu}{e}-\Pnu{\mu}{\tau}-\Pnu{\mu}{\chi}+r\Pnu{e}{\mu}}
{1-\Pnu{e}{\mu}-\Pnu{e}{\tau}-\Pnu{e}{\chi}+r^{-1}\Pnu{\mu}{e}}.
\label{fourwayR}
\eeq
Once some of the
probabilities in equation (\ref{fourwayR}) are measured directly by experiment,
this four-way possibility could become a preferred solution.  In what follows,
we will confine our consideration of this possibility 
to cases for which $\Pnu{e}{\chi}$ and 
$\Pnu{\mu}{\chi}$ are negligible, and for which CP is conserved by atmospheric
neutrinos.  
The expression for the measured ratio in this three-way mixing scheme is
\beq
R_{atm} = \frac{1-\Pnu{e}{\mu}-\Pnu{\mu}{\tau}+r\Pnu{e}{\mu}}
{1-\Pnu{e}{\mu}+r^{-1}\Pnu{e}{\mu}-\Pnu{e}{\tau}},
\eeq
which may be rearranged to give
\beq
\Pnu{\mu}{\tau}+ x \Pnu{e}{\mu} - R_{atm} \Pnu{e}{\tau} = y.
\label{Patav}
\eeq

The accelerator experiments considered here, LSND, CHORUS, KARMEN, and
CHARM~II, are, with the exception of KARMEN's $\nue\rightarrow\nu_x$
``disappearance'' (of an $\nue$ to any other flavor $x=\mu,\tau,\chi$)
measurement, ``appearance'' experiments, so they measure 
(or constrain) individual
transition probabilities.  Analysis of these experiments' results is thus 
simpler than the analysis of the ``disappearance'' solar neutrino and
atmospheric neutrino experiments.  Many other experiments (Bugey in France,
Gosgen in Switzerland, NOMAD and BEBC at Cern, CCFR and E531 at Fermilab, E776
at Brookhaven, and Krasnoyarsk in Russia) have searched (and/or are searching) 
for neutrino oscillations as well, but none of these have yet 
found any evidence of oscillations.

The data from the experiments which we will use is summarized in
Table \ref{datatab}.
Only the data from Kamiokande, SuperK, 
and IMB have been listed under atmospheric
neutrinos; other experiments (Fr\'{e}jus, Soudan2, Baksan, and Nusex) have
also measured the fluxes of atmospheric neutrinos, but their uncertainties are
considerably larger than those of the listed experiments, which makes them less
useful.

\begin{table}[htb!]
\caption[Experimental data pertaining to neutrino oscillations.]
{Experimental data pertaining to neutrino oscillations.  Sources for the data
are listed under reference~\cite{tabledat}.  When both systematic and
statistical uncertainties were given for an experiment, they were added in
quadrature to obtain the total uncertainty listed below.  The measured
ratio for solar experiments is the measured flux divided by the flux predicted
by Bahcall's solar model
\protect\cite{Bahcall}\protect.  Kamiokande and SuperK do not measure
$\Pnu{e}{e}$ but the combination given in equation~(\ref{NCacct}).  The addition
to their measured ratios is noted by ``$+NC$'' in the Table.
For atmospheric experiments, the measured ratio is the ratio of
ratios, $R_{atm}$.  The probabilities for the atmospheric experiments are found
from $R_{atm}$ by assuming CP invariance.  CP invariance is also assumed for
the combined probability from the CHARM II experiment.  The limits on measured
probabilities from the accelerator experiments are reported for the $90$ \%
confidence level.  Energies marked with a $\sim$ represent peak energies in the
energy spectrum for the reaction specified.  Accelerator neutrino energies with
no such demarcation correspond to monoenergetic neutrinos.  LSND-1 refers to the
decay-at-rest results; LSND-2 refers to the newly-presented decay-in-flight
results, for which values of $L$ and $E$ are not presently available.}
\label{datatab}
\begin{centering}
\begin{tabular}{|c|c||c|c|c||c|c|c|} \hline
Type   & Experiment & Length               & Energy       & Measured
    &    $\nua$        &  $\nub$      & $\Pab$  \\
       &            &   (m)                & (MeV)        & Ratio         
    &                   &              &         \\ 
\hline\hline
       & Homestake  & $1.5 \times 10^{11}$ & $>0.814$     & $0.28 \pm 0.04$ 
    & $\nue$           & $\nue$                    & $0.3$   \\ \cline{2-8}
       & Kamiok. & $1.5 \times 10^{11}$ & $>7.5$       & $0.50 \pm 0.06$ 
    & $\nue$           & $\!\!\nue+NC \!\!\!$      & $0.4$   \\ \cline{2-8}
Solar  & SAGE       & $1.5 \times 10^{11}$ & $>0.233$     & $0.53 \pm 0.10$
    & $\nue$           & $\nue$                           & $0.5$   \\
    \cline{2-8}
       & Gallex     & $1.5 \times 10^{11}$ & $>0.233$     & $0.51 \pm 0.06$ 
    & $\nue$           & $\nue$                    & $0.5$   \\  \cline{2-8}
       & SuperK     & $1.5 \times 10^{11}$ & $>7.5$       & $0.44 \pm 0.04$
    & $\nue$           & $\!\!\nue+NC\!\!\!$       & $0.4$   \\  \cline{2-8}
\hline\hline
&&&&& $\numu$          & $\nutau$, 
                         $\nu_{\chi}$  & $0.46$ \\ \cline{6-8}
       &IMB         &  $10^4$ to $10^7$    & $\gsim 10^3$ & $0.54 \pm 0.09$
    & $\numu$          & $\nue$        & $0.37$ \\  \cline{6-8}
&&&&& \multicolumn{2}{c|}{3-way}       & $x=1.1$, \\
&&&&&  \multicolumn{2}{c|}{}    & $y=0.46$  \\ \cline{2-8}
&&&&& $\numu$          & $\nutau$, 
                         $\nu_{\chi}$  & $0.40$ \\  \cline{6-8}
& Kamiok. &  $10^4$ to $10^7$    & $\gsim 10^3$ & $0.60 \pm 0.06$
    & $\numu$          & $\nue$        & $0.30$ \\  \cline{6-8}
Atmos. &&&&& \multicolumn{2}{c|}{3-way}       & $x=1.2$, \\ 
&&&&& \multicolumn{2}{c|}{}            & $y=0.40$  \\  \cline{2-8}
&&&&& $\numu$          & $\nutau$, 
                         $\nu_{\chi}$  & $0.33$ \\  \cline{6-8}
& SuperK     &  $10^4$ to $10^7$    & $\gsim 10^3$ & $0.67 \pm 0.08$
    & $\numu$          & $\nue$        & $0.27$ \\  \cline{6-8}
&&&&& \multicolumn{2}{c|}{3-way}       & $x=1.2$, \\ 
&&&&& \multicolumn{2}{c|}{}            & $y=0.33$  \\  
\hline\hline
    & LSND-1     & 30                    & $\sim 50$     &                 
  & $\smc{\nu}_{\mu}$ & $\smc{\nu}_e$ & $3.1\pm 1.3\times 10^{-3}$\\ \cline{2-8}
    & LSND-2     &                       & $   $     &                 
  & $\smc{\nu}_{\mu}$ & $\smc{\nu}_e$ & $2.7\pm 1.4\times 10^{-3}$\\ \cline{2-8}
    &&                                    &$30$       &
  & $\numu$           & $\nue$        & $<0.026$ \\ \cline{4-8}
     & KARMEN     & 17.6                  &$\sim 35$      &
  & $\nue$            &$\nutau$, $\numu$,
                        $\nu_{\chi}$   & $<0.197$ \\ \cline{4-8}
Accel.&&                                   &$\sim 50$      &
  & $\smc{\nu}_{\mu}$ & $\smc{\nu}_e$ & $<.0038$ \\ \cline{2-8}
     & CHORUS     & 590                   & $\sim27000$ &
    & $\numu$           & $\nutau$      & $<0.004$ \\ \cline{2-8}
&&&&& $\numu$           & $\nue$        & $<0.0047$ \\ \cline{6-8}
     & CHARM II   & 650                   & $\sim20000$ &
    & $\smc{\nu}_{\mu}$ & $\smc{\nu}_e$ & $<0.0026$ \\ \cline{6-8}
&&&&& \multicolumn{2}{c|}{both combined}& $<0.0028$ \\ \hline 
\end{tabular}
\end{centering}
\end{table}

If the uncertainty in a phase $\Pij$ of a neutrino 
is $\gsim$ one-fourth an oscillation length, the oscillatory terms in the
probability equation average, with $\sin^2 \Pij$ averaging to one-half, and
$\sin 2 \Pij$ averaging to zero.   This assumption that the 
oscillations average thus becomes an
assumption about the 
magnitude of the mass-squared differences:
\beq
\Delta\Pij \gsim \frac{\pi}{2} \Rightarrow m_i^2-m_j^2 \gsim \frac{2 \pi} 
{\Delta \left(\frac{x}{E}\right)},
\eeq
where the symbol $\Delta$ represents the uncertainty of a quantity.  In our data
analysis below, we will
consider only mass-squared differences for which this type of averaging occurs 
for both solar neutrinos and
atmospheric neutrinos.  

According to stellar evolution theory, 
solar neutrinos are produced in the inner part of the
sun below $0.2 R_{\odot}$ \cite{Bahbook}, so the distance traveled varies between 
$1~a.u.-(0.2)R_{\odot}$ and $1~a.u.+(0.2)R_{\odot}$.  ($1~a.u.=~1$~astronomical
unit~$=~1.5~\times~10^8$~km, and $R_{\odot}=$~solar 
radius~$=7.0~\times~10^5$~km.)  This leads to
an uncertainty in distance of less than one percent.  The energy of
solar neutrinos, however, ranges from $0.2$~MeV to $25$~MeV
\cite{Bahbook}.  Thus $\frac{x}{E}$ varies from $3.7~\times~10^{12}$~eV$^{-2}$ 
to $3~\times~10^{10}$~eV$^{-2}$, and we find 
\beq
m_i^2-m_j^2 \gsim 2 \pi \left( 2.7 \times 10^{-13} \right) \mbox{ eV}^2 \approx 1.7 
\times 10^{-12} \mbox{ eV}^2
\label{solave}
\eeq
as a rough criterion for the solar neutrino oscillations to average.  A more
refined calculation would include an energy integration over the neutrino
spectrum and the detection cross section starting from the detector energy
threshold, but the above estimate is sufficient for
our purposes.

Making this same assumption for atmospheric neutrinos results in a stronger
constraint on mass-squared differences.  The
detected neutrinos separate by energy into three distinct groups:  contained,
stopped, and through-going
.  Neutrinos in 
contained events produce charged
leptons in the detector that are absorbed by, and thereby contained in, the
detector.  These events are caused by lower-energy neutrinos, with energies
ranging from $0.2$~GeV to $1.2$~GeV \cite{Gaisser}, 
\cite{atmos}, \cite{LPW}.    Because the lepton is formed in the detector,
down-going background events from cosmic rays are easily identified.  Down-going
signal events dominate the contained data, 
due to the closeness to the Earth's surface. 
 
Stopped events
are caused by upward-going muon neutrinos which create a muon outside the
detector which travels into the detector and is stopped.  Electron neutrinos do
not contribute to this type of event since electrons will not travel far
through the material surrounding the detector, and down-going events are vetoed
because the down-going background is significantly larger than the signal.  
The muon neutrinos producing
the stopped muons are more energetic than those causing contained events 
(since the muons from stopped events may travel further than
the width of the detector), having energies between $1$ and $100$~GeV
\cite{atmos}.  Still higher-energy up-going 
muon neutrinos may produce muons outside the detector which travel
through the detector and out the other side.  These through-going events result
from neutrinos with energies between $1.1$~GeV and $10^4$~GeV \cite{atmos}.

In our data analysis, we follow the lead of reference~\cite{AP} and consider
only the contained events, which are identified roughly ten times as often as 
events of the other
two types \cite{atmos}.  These events have an energy uncertainty of about a GeV.
Because down-going events dominate, the majority of contained events' neutrinos
originate in a cone of about 90$^{\circ}$ about the vertical \cite{AP}.  Thus
the distance traveled by a contained-event neutrino varies from $30$ km, the
depth of the atmosphere, to $6.5 \times 10^3$ km, the radius of the Earth. 
Contained events therefore have an uncertainty in $\frac{x}{E}$ of 
$\frac{6.5 \times 10^3 \mbox{km}}{0.2 \mbox{ GeV}}-\frac{30 \mbox{km}}{1.2
\mbox{ GeV}}
\approx 1.7~\times 10^{6} \mbox{ eV}^{-2}$.  This uncertainty causes the
oscillations to smear out if
\beq
m_i^2-m_j^2 > 2 \pi \left(5.9 \times 10^{-7} \right) 
\mbox{ eV}^2 \approx 3.7 \times 10^{-6} \mbox{ eV}^2.
\label{atmosave}
\eeq

When this average is made, the probabilities become simple sums of boxes:
\beq
\langle \Pnu{\alpha}{\beta} \rangle =  
-2 \left(\R{\alpha 1}{\beta 2} + \R{\alpha 2}{\beta 3} + 
\R{\alpha 1}{\beta 3} \right).
\label{Pav}
\eeq
As discussed following equation~(\ref{sumrowB}) 
in Chapter~\ref{boxes}, this row sum equals the sum
$\sum_{x=1}^3 |V_{\alpha x}|^2 |V_{\beta x}|^2$.

We cannot average oscillations for accelerator experiments.  Upon observation of
Table~\ref{datatab},
we see that the oscillatory behavior must occur if the LSND results
are to be consistent with the limit placed by CHARM~II.


\section{Interpreting the Data}


The atmospheric and solar data may be analyzed independent of neutrino masses,
provided the mass-squared differences are large enough to yield the averaging
described above.  The probabilities measured at accelerators, however, are
functions of these mass-squared differences.  We
will follow the authors of reference
\cite{AP} and choose the mass-squared difference $\Delta m_{23}^2$ which puts the
LSND experiment at an oscillation maximum, and the difference $\Delta m_{12}^2$ to
be much smaller:
\beq
m_3^2-m_2^2 \sim m_3^2-m_1^2 \sim 2 \mbox{ eV}^2, \mbox{\ \ and \ \ }
m_2^2-m_1^2 \sim 10^{-2} \mbox{ eV}^2.
\label{APmasses}
\eeq
The smaller difference $\Delta m_{12}^2$ is small enough to be negligible in
accelerator experiments but large enough to be above the limits set in
equations~(\ref{solave}) and (\ref{atmosave}).  With this choice, 
oscillations average and CP is effectively 
conserved for both solar and atmospheric neutrinos.

The oscillatory factors for accelerator experiments are listed in 
Table~\ref{phitable} for our choice of
mass-squared differences. We have used the typical $E$ quoted in
Table~\ref{datatab} to obtain the first column of numbers.  
For each of the $\frac{x}{4E}$ given in
the Table, $\sin^2 \Pt$ and $\sin
2 \Pt$ are much smaller than the sin functions involving the other
mass-squared differences, which are roughly equal.  Under these conditions,
the sum of sin functions in equation (\ref{CPsmall}) is nearly zero.  (The
biggest value for this sum comes from the LSND experiment and is $\pm 0.015
{\cal J}$.)
CP is therefore roughly conserved, and
$\Pnu{\alpha}{\beta}$ measured by an accelerator experiment will be
approximately
\bea
\Pnu{\alpha}{\beta}^{\!\!\!\!\!\!\!accel} 
& = & -2\left( \R{\alpha 1}{\beta 2} \sin^2\Pt
  + \R{\alpha 2}{\beta 3} \sin^2\Pe + \R{\alpha 1}{\beta 3}\sin^2\Pu\right)
\nonumber \\
& \approx & -2 \left( \sim 10^{-4} \R{\alpha 1}{\beta 2} 
  + \R{\alpha 2}{\beta 3} + \R{\alpha 1}{\beta 3} \right) \sin^2 \Pe.
\label{accelstraint}
\eea
Unless the real part of $\B{\alpha 1}{\beta 2}$ is much larger than the other
real boxes, its 
term will not contribute significantly to the transition probability.

\begin{table}[htb]
\caption[Values of oscillatory terms for accelerator experiments, using a
specific choice of mass-squared differences.]
{Values of oscillatory terms for accelerator experiments.  The numbers were
calculated using the mass-squared differences given in equation
(\protect\ref{APmasses}\protect).  The typical values of energy indicated in
Table~\ref{datatab} were used to calculate the second column. 
A more complete treatment would integrate over
the entire energy range, but the approximation we have chosen is sufficient for
our purposes.  \label{phitable}}
\begin{tabular}{|c||c|c|c|c|c|c|c|} \hline
Experiment &$\!\frac{x}{4E}($eV$^2)\!$& $\sin^2 \Pt$         & $\sin^2 \Pe$ 
     & $\sin^2 \Pu$         & $\sin 2 \Pt$       & $\sin 2 \Pe$ & $\sin 2 \Pu$ 
\\ \hline \hline
LSND       & $0.76$               & $5.8 \times 10^{-5}$ & $1.0$ 
     & $1.0$                & $0.015$            & $0.10$       & $0.10$ 
\\ \hline
KARMEN     & $0.64$               & $4.1 \times 10^{-5}$ & $0.92$
     & $0.92$               & $ 0.013$             & $0.55$       & $0.55$  
\\ \hline
CHORUS     & $0.028$              & $7.8 \times 10^{-8}$ & $3.1 \times 10^{-3}$
     & $3.1 \times 10^{-3}$ & $5.6 \times 10^{-4}$ & $0.11$       & $0.11$ 
\\ \hline
CHARM II   & $0.041$              & $3.7 \times 10^{-2}$ & $6.7 \times 10^{-3}$ 
     & $6.7 \times 10^{-3}$ & $8.2 \times 10^{-4}$ & $0.16$       & $0.16$ 
\\ \hline
\end{tabular}
\end{table}

Acker and Pakvasa \cite{AP} have analyzed the solar neutrino data, including
the possible 
contributions from muon and tau neutrino neutral current interactions
to the Kamiokande data.  They find that all the solar neutrino data is
consistent with $\langle \Pnu{e}{e}\rangle$ 
between $0.4$ and $0.55$ for reasonable
values of the initial $^8B$ neutrino flux from the sun.  These values were
attained by assuming the oscillatory part of the probability $\sin^2(\Pij)$
averages to one-half for all three mass-squared differences, so
$\langle \Pnu{e}{e} \rangle = 1-\langle \Pnu{e}{\mu} \rangle-\langle \Pnu{e}{\tau} \rangle$.  The range of
probabilities allowed by the solar neutrino deficit is therefore
\beq
0.45 \lsim \langle \Pnu{e}{\mu} \rangle + \langle \Pnu{e}{\tau} \rangle \lsim 0.60.
\label{solarstraint}
\eeq 

Using Kamiokande's results (which are comparable to IMB's and SuperK's 
numbers but with smaller error bars) with a range
of $1.5\sigma$, the atmospheric neutrinos in this CP-conserving scenario should
obey
\beq
0.51 \lsim \frac{1-\langle\Pnu{e}{\mu}\rangle-\langle\Pnu{\mu}{\tau}\rangle+
r\langle\Pnu{e}{\mu}\rangle}
{1-\langle\Pnu{e}{\mu}\rangle+r^{-1}\langle\Pnu{e}{\mu}\rangle-
\langle\Pnu{e}{\tau}\rangle} \lsim 0.69.
\eeq
Up to this point we have not used our choices for neutrino masses, except to
argue that oscillations average for atmospheric and
solar neutrinos.

The LSND experiment measures $\Pucec$, which equals $\Pnu{e}{\mu}$ by CPT
conservation.  Since only the decay-at-rest (LSND-1) result was available when
this analysis was carried out, we restrict ourselves to consideration of that
data.  Using that measured probability and our  choices of
mass-squared differences, we find the
constraint (again using 1.5 standard deviations as our limit)
\beq
1.2 \times 10^{-3} \lsim -2 \left( 10^{-4} \ \R{11}{22} 
  + \R{12}{23} + \R{11}{23} \right) \sin^2  \Pe \lsim 5.1 \times 10^{-3}
\eeq
from equation (\ref{accelstraint}).  Since 
$10^{-4}~\R{11}{22}~\ll~10^{-3}$ for reasonable values of $\R{11}{22}$ and 
$\sin^2 \Pe=1$ from Table~\ref{phitable}, this constraint requires
\beq
2.5\times 10^{-4} < -\left(\R{12}{23} + \R{11}{23} \right) < 
2.8\times 10^{-3}.
\label{LSNDstraint}
\eeq
KARMEN's 90\% confidence level 
places a tighter upper bound than the $1.5\sigma$ bound
from LSND:
\beq
-\left(\R{12}{23} + \R{11}{23}\right)  <  1.0\times 10^{-3}
\eeq

By a similar argument, the constraints from CHORUS and KARMEN may each be turned 
into a constraint on the sum of two boxes for the other oscillation
possibilities.  From CHORUS, we find a limit on
$\numu$-$\nutau$ mixing:
\beq
- \left( \R{22}{33} + \R{21}{33} \right) < 0.32.
\eeq
KARMEN provides a limit on $\nue$-$\nutau$ mixing:
\beq
- \left( \R{12}{33} + \R{11}{33} \right) < 5.4 \times 10^{-2}.
\eeq

We may combine the constraints on $\nue$-$\numu$ mixing with 
equation (\ref{1223out}) to eliminate $\R{12}{23}$, and solve for
$\R{11}{23}$.  If we call the sum 
$-\left(\R{12}{23} + \R{11}{23}\right)$ as measured
by LSND or other accelerator experiments $s$,
\beq
\R{12}{23} = \frac{\R{11}{22}\R{11}{23} - {\cal J}^2}{-\R{11}{23} - \R{11}{22}}
= - s - \R{11}{23},
\label{LSND1}
\eeq
so
\beq
\R{11}{23} = \frac{ - s \pm \sqrt{s^2 
  - 4 \R{11}{22} s - 4 {\cal J}^2}}{2} \nonumber .
\label{LSND2}
\eeq
If $\R{11}{23}$ is to be real, which it
is by definition, 
\beq
4 {\cal J}^2 < s^2-4 \R{11}{22} s.
\eeq
Assuming $-\R{11}{22}$ has its maximum value of half $\langle \Peu \rangle_{max}$,
or $0.3$, and $s$ equals $10^{-3}$, 
the maximum value allowed by KARMEN, we find the
following minimal constraint on ${\cal J}^2$:
\beq
{\cal J}^2 <3 \times 10^{-4}.
\eeq

Even though CP is essentially undetectable 
due to our choice of masses, the imaginary part of the boxes, ${\cal
J}$, may still play a role in determining the oscillation probabilities.  Using the
above maximum values of $-\R{11}{22}$ and $s$ along with 
this maximum value of ${\cal J}^2$ in equations~(\ref{LSND1}) and
(\ref{LSND2}) yields
\beq
\R{12}{23} = \R{11}{23} = -\frac{s}{2} = -5 \times 10^{-4}.
\eeq
If, on the other hand, ${\cal J}^2=0$,
\beq
\left(\R{11}{23}, \R{12}{23}\right) = \left( 1.7 \times 10^{-2}, 1.8 \times
10^{-2}\right)  \mbox{\ \ or \ \ } \left(1.8 \times 10^{-2}, 
1.7 \times 10^{-2} \right)
\eeq
A set of measurements of probabilities at different distances which pinpointed
the values of individual boxes 
would thus probe the value of ${\cal J}^2$, even if CP were hidden
from observation in the oscillations being observed!
The above ambiguity between the values of $\R{11}{23}$ and $\R{12}{23}$ arises
because $\sin^2\Pe \approx \sin^2\Pu$.  We have chosen one mass scale to
dominate the other, a situation examined in Section~\ref{onemasssec}. 
Consideration of equation~(\ref{smallangleR}) shows that only the sums
$\R{\alpha 2}{\beta 3} + \R{\alpha 1}{\beta 3}$ are measurable.  Unitarity
constraints of the form~(\ref{1223out}) for a known ${\cal J}^2$ and $\R{\alpha
1}{\beta 2}$ allow us to
determine the choices for $\R{\alpha 2}{\beta 3}$ and $\R{\alpha 1}{\beta 3}$
but not to say which has which value.


\section{Finding Solutions}


As promised in Section~\ref{dogsec}, we numerically find boxes which match the data.
Different values of the five parameters $\R{11}{22}$, $\R{11}{23}$, $\R{21}{32}$,
$\R{22}{33}$, and ${\cal J}^2$ are stepped through, the other five $R$s found by
equations~(\ref{1223out}) to (\ref{1133out}), and the resulting oscillation
probabilities compared to the solar neutrino, atmospheric neutrino, and LSND results
discussed above.  The boxes must also produce a mixing matrix which is unitary. 

We find many solution sets which meet all of these conditions.  
For ${\cal J}^2=0$, the two best fits are found with
\beq
{\cal B}_1 = \left( \ba{ccc}
-18 & -1.0 & 0.95 \\
-1.0 & -0.67 & -2.0 \\
0.95 & -2.0 & -1.8 
\ea \right) \times 10^{-2},
\label{B1}
\eeq
which is symmetric, and
\beq
{\cal B}_2 = \left( \ba{ccc}
-10 & 0.91 & -1.0 \\
-13 & -1.0 & -0.93 \\
5.6 & -10 & -13 
\ea \right) \times 10^{-2}.
\label{B2}
\eeq
The first solution gives a solar ratio $\langle\Pnu{e}{e}\rangle$ 
of $0.42$, an atmospheric ratio
$R_{atm}$ of $0.55$, and an LSND probability of $2.1 \times 10^{-3}$.  The second
gives a solar ratio of
$0.55$, an atmospheric ratio of $0.69$, and an LSND probability of $3.6 \times
10^{-3}$.

We may obtain the magnitudes $|V_{\alpha i}|$ from the boxes from
equation~(\ref{vfour}).  Using the notation
$|V|$ for the matrix of these magnitudes, the two solutions yield
\beq
|V|_1 = \left(\ba{ccc}
0.88 & 0.47 & 0.20 \\
0.50 & 0.88 & 0.12 \\
0.12 & 0.20 & 0.99
\ea \right),
\eeq
and 
\beq
|V|_2 = \left(\ba{ccc}
0.50 & 0.41 & 0.74 \\
0.65 & 0.73 & 0.04 \\
0.53 & 0.51 & 0.65 
\ea \right).
\eeq
Inspection reveals appropriate places to insert minus signs to make these
matrices unitary to within 
the tolerance required by our numerical technique.  Such an assignment, however,
is not unique due to the phase ambiguities discussed in
Section~\ref{relphasesec}.

Comparing the solutions~(\ref{B1}) and (\ref{B2}), we see that by the first
solution, $\nue$-$\numu$ mixing comprises most of the solar neutrino deficit, while
the second solution uses $\nue$-$\nutau$
mixing to account for most of the solar deficit.

Acker and Pakvasa in \cite{AP} expressed their solution in terms of the mixing
matrix:
\beq
V_{AP} = \left( \ba{ccc}
0.70 & 0.70 & 0.14 \\
-0.71 & 0.69 & 0.12 \\
-0.010 & -0.19 & 0.98
\ea\right).
\eeq
We find the corresponding box matrix to be
\beq
{\cal B}_{AP} = \left( \ba{ccc}
-24 & 0.84 & -0.87 \\
-0.92 & -1.6 & 0.87 \\
-0.92 & -1.8 & -0.96
\ea \right) \times 10^{-2}.
\label{BAP}
\eeq
This solution was not selected by our technique because the resulting
atmospheric ratio, $R_{atm}=0.50$, is not within the $1.5 \sigma$ range of
Kamiokande's data. 
Acker and Pakvasa's solution yields
$\langle\Pnu{e}{e}\rangle=0.48$ 
for the solar neutrino deficit, and an LSND probability of $1.2\times
10^{-3}$.

The above solutions all contain no CP violation.  We found solution sets for
CP-violating cases too, with
${\cal J}^2$ going as high as $2.7 \times 10^{-4}$.  The only solution for that
maximum ${\cal J}^2$ is
\beq
{\cal B}_{3} = \left( \ba{ccc}
-22 & -0.12 & 5 \times 10^{-4} \\
-1.0 & -1.0 & 0.85 \\
0.84 & -2.3 & -3.2
\ea \right) \times 10^{-2},
\label{BhiJ}
\eeq
yielding
\beq
|V|_3 = \left( \ba{ccc}
0.73 & 0.63 & 0.25 \\
0.65 & 0.74 & 0.14 \\
0.21 & 0.19 & 0.95
\ea\right).
\eeq
This CP-non-conserving solution gives a solar ratio of $0.53$, an atmospheric
ratio of $0.51$, and an LSND probability of $4.9 \times 10^{-3}$.  None of the
solutions which conserved CP found this combination of high solar ratio and
low atmospheric ratio.  The CP-conserving
solution set which most closely reproduces these CP-violating probabilities is
\beq
{\cal B}_{4} = \left( \ba{ccc}
-10 & -1.1 & 1.0 \\
5.0 & -16 & 7.3 \\
-10 & 1.0 & -1.1
\ea \right) \times 10^{-2},
\label{B4}
\eeq
which produces a solar ratio of $0.40$, an atmospheric ratio of $0.52$, and an
LSND probability of $4.5 \times 10^{-3}$.  One can see that CP violation may
have a measurable effect, even if the mass-squared differences conspire to hide
direct evidence of it.


\section{How Do Specific Models Stack Up?}


None of the models presented in Chapter~\ref{massbox} are
compatible with the solutions we have presented here.  Wolfenstein's
parameterization requires 
\beq
\R{11}{23} \approx -\R{12}{23} \approx \R{12}{33},
\eeq
as seen by examination of equation~(\ref{Wolfeq}).  The first approximate
equality is met by all three solutions, (\ref{B1}), (\ref{B2}), and (\ref{BAP}),
but the second is not met by any of them.  The small-angle approximation is merely a
special case of the Wolfenstein parameterization, so it is not consistent with the
current data either.

Equations~(\ref{Zee1}) and (\ref{Zee2}) of Chapter~\ref{massbox} present 
two possibilities for Zee-model-induced
oscillations.  Both possibilities predict $\B{11}{22} \approx \B{11}{23}$, a
condition which is not met by any of the above solutions.  This equality is very
hard to reconcile with both the LSND results and the solar neutrino
measurements, but it could become relevant if matter effects enhance the effect
of an otherwise small $\B{11}{22}$.  And as discussed in Section~\ref{Zeesec},
we have only examined the Zee mechanism for two limits of coupling constants;
our work does not address the applicability of other coupling-constant choices.

The Fritzsch mass matrix
model of Section~\ref{Fritzschsec} predicted that for a mass hierarchy, the
diagonal elements $\B{11}{22}$ and $\B{22}{33}$ should be much larger than the
other
elements.  These two boxes are of the same order in solution~2~(\ref{B2}), but
they
are not significantly larger than the off-diagonal $\B{21}{32}$ and
$\B{12}{33}$. 
The reverse mass hierarchy considered in that Section predicts the same two
dominant boxes, once we swap the $3$
and $1$ mass indices to account for the new labeling of mass states, so it is
inconsistent with our solutions too.  Again, our work does not comment on the
more general Fritzsch model, but only on its limits when one mass scale
dominates.

Before concluding this Chapter we must remind the Gentle Reader that the 
solutions derived above assume an energy-independence of the solar neutrino
data and the atmospheric neutrino data.  Choosing this interpretation allows the
fitting of all current data with only three neutrino flavors, but we do not mean
to present this choice as the only one possible.  Many other excellent
analyses of the data, such as those by Fogli and Lisi \cite{Lisi} and those by
Cardall and Fuller \cite{CF}, take different approaches which are equally valid.

The neutrino-oscillation
interpretations of all three data sets considered here (solar, atmospheric, and
LSND) are still unconfirmed and controversial.  If an energy
dependence of solar neutrino oscillations is unequivocally demonstrated by
future
experiments, or if any of the probabilities used here change dramatically (or
keep
the same mean value with smaller error bars), the solutions we have found 
will no longer be
valid, and another neutrino generation and/or matter effects on oscillations 
may be implicated.  At the present, however, our solution sets fit all of the
existing data and provide a demonstration of the use of the boxes.


\chapter{Outlook}
\label{summary}


Neutrino oscillations address many current physics questions.  Observation of
them would present the first contradiction of the Standard Model and provide a
clue as to the direction that extensions of the Model should take. 
Understanding what role, if any, oscillations play in the solar neutrino deficit
will help refine theories of stellar evolution.  And indications of the
magnitude of neutrino masses from oscillation experiments would indicate how
much neutrinos contribute to dark matter.  With approximately 
115 low-energy ``relic''
neutrinos of each flavor occupying every cubic centimeter of space \cite{KT}, a
single neutrino mass on the order of 7~eV (or all neutrino masses summing to
7~eV) provides enough mass to close the universe \cite{MP}.

Once neutrino oscillations are detected and the boxes and neutrino masses
determined, neutrinos may be used to probe the universe.  Neutrino oscillations
could help determine the density of different regions of the earth
\cite{Glashow} and
bring information about cosmologically distant objects \cite{us}. 
Neutrinos can act as a long-wavelength
telescope, peering further out in space (and therefore further 
back in time) than conventional telescopes.  The distance for which neutrino
telescopes will be sensitive depends on the values of mass-squared differences. 
This dependence is illustrated in Figure~\ref{Learnfig}, a plot popularized by
J.\ G.\ Learned.  The neutrino telescope AMANDA is already taking data, hoping to use
gamma-ray bursters and supernovae to measure neutrino masses \cite{Halzen}. 
Other neutrino telescopes are under construction and will soon join the hunt.

\begin{figure}[htb!]
\centerline{\hbox{
\psfig{figure=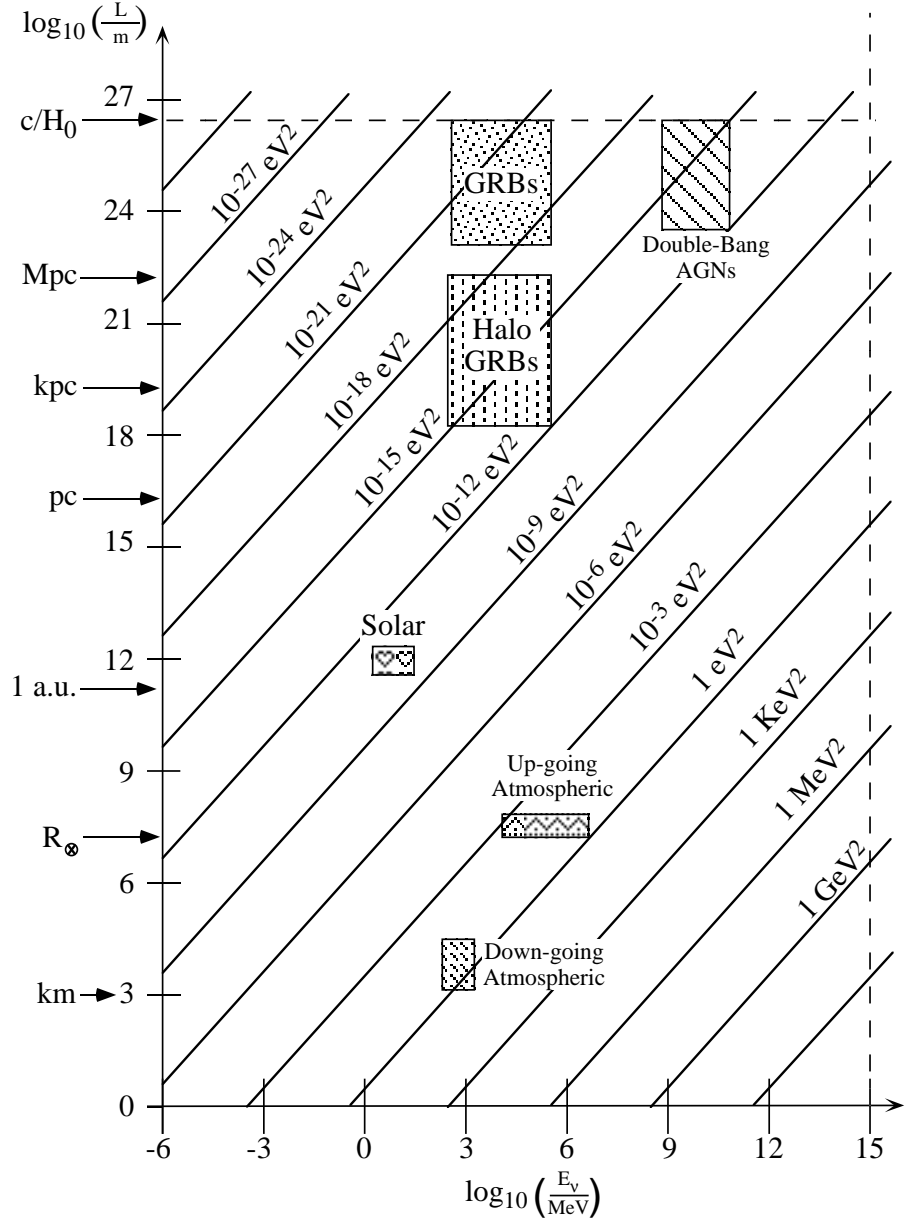,height=6.5 in}
\vspace{-0.5 in}
}}
\caption[Lengths and energies probed by neutrino oscillations for different
mass-squared differences.]{Lengths and energies probed by neutrino oscillations
for different mass-squared differences, obtained by setting 
$\Pij=\frac{\pi}{2}$
so that the Earth sits on a maximum of an oscillation length.  
Lengths of interest are labeled on the
vertical axis, with $R_{\otimes}$ representing the radius of Earth and $H_0$
representing the present value of the Hubble constant.  
Shaded regions represent energies and
distances typical for the indicated cosmological neutrino source.  Included are
solar neutrinos, both up- and down-going atmospheric neutrinos, gamma-ray
bursters both within the galactic halo and outside our Galaxy, and active
galactic nuclei.  Double-bang AGNs refer to those AGNs producing tau neutrinos
energetic enough to produce a ``double-bang'' signature in detectors. \label{Learnfig}}
\end{figure}

Neutrino physics has entered a golden age of research.  New experiments all
over the globe promise an unequaled amount of data, from the sun, the
atmosphere, accelerators, supernovae, and other cosmic sources. 
SuperK in Japan amassed as much data in its first 102 days of running
as all of the previous solar neutrino experiments did over their combined
lifetimes \cite{SuperK}.
Analyzing such refined data requires a consistent, model-independent
approach.  In this dissertation, we have developed such an approach and then
illustrated its use by comparison with the low-statistics data currently
available.
 

\appendix

\chapter{Notation Key}
\label{key}

Greek Lorentz indices $\mu, \nu, \sigma,$ {\it et cetera} may take the values
$0$ to $3$.  Latin Lorentz indices $a, b,$ {\it et cetera} refer to the spatial
components $1$ to $3$.

Greek particle indices $\alpha, \beta, \gamma,$ {\it et cetera} represent
states of definite flavor.  Latin particle indices $i,j,k,$ {\it et cetera}
represent states of definite mass.

We will explicitly write out all {\it flavor} sums, since repeated flavor 
indices are not necessarily summed over.  

Repeated {\it Lorentz} indices $\mu$, $\nu$, and $a$ are always summed over.

$\sum_{i\neq j}$ represents a single sum over the index
$i$, omitting the term for which $i=j$.  Sums over both indices will be
represented by $\sum_j \sum_{i\neq j}$.

In this paper, we use the ``west-coast metric'' in which $g_{00} = 1$,
$g_{aa}=-1$, and the other elements are zero.

$e=\sqrt{4\pi\alpha}$ in this paper, where $\alpha$ here is the fine-structure
constant.  In standard particle physics units ($c=\hbar=1$), our $e$ is the
electric charge.  In other units it is proportional to the electric charge.

$\gupo, \gupa, \gupb, \gupc, \mbox{ and } \gdnf$ are Dirac {\it gamma
matrices}, discussed in detail in Appendix~\ref{gammapp}

$\partial_{\mu} \equiv \frac{\partial}{\partial x_{\mu}}$ is the 
Lorentz-covariant partial derivative.

$\theta_W$ is the Weinberg angle, defined by equation (\ref{thetaW}).

$P_L = \frac{1-\gdnf}{2}$ is the left-handed projection operator, and

$P_R = \frac{1+\gdnf}{2}$ is the right-handed projection operator.  They are
discussed in detail in Appen\-dix~\ref{handsec}.

$\vec{p}\equiv {\bf p}$ is the 3-vector portion of the four-vector p.

{\boldmath $p \sigma \equiv p \cdot \sigma$} 
is the dot product of the vectors ${\bf p}$ and {\boldmath $\sigma$}.

$\psi$ is used to denote a generic spinor, or the spinor containing all
neutrino states, described below.

$\psi^*$ is a complex-conjugated spinor.

$\psi^{\dagger}=\psi^{*T}$ is a Hermitian-conjugated spinor.

The symbols $^*$ and $ ^{\dagger}$ have the 
same effects on other objects as they do on spinors.

$\ol{\psi}=\psi^{\dagger} \gupo$ is a Dirac-conjugated spinor.

$\smc{\psi}=C\ol{\psi}^T$ is a charge-conjugated spinor.

$C=i \gupb \gupo$ is the charge conjugation operator, discussed in detail in
Appendix~\ref{Csec}.

$\nua_L$ is the left-handed active neutrino of flavor $\alpha=e,\mu,\tau$.

$\nuaRc$ is its CP-conjugate antineutrino of flavor $\alpha$; it is
right-handed.

$N_{\alpha R}$ is the right-handed sterile partner to $\nua_L$.  It has not yet
been observed.

$\NaLc$ is the left-handed sterile partner to $\nuaRc$.  It is also the
CP-conjugate of $N_{\alpha R}$ and has not yet been observed.

$\alpha_L$ is the left-handed charged lepton of flavor $\alpha$.

$\alpha_R$ is its right-handed partner.

$\smc{\alpha}_R$ is the CP-conjugate of $\alpha_L$.

$\smc{\alpha}_L$ is the CP-conjugate of $\alpha_R$.

$\psi_{\alpha L} = \left( \ba{c} \nua_L \\ \alpha_L \ea \right)$ is left-handed
lepton doublet of flavor $\alpha$.

$\nuL$ is an $n_L$-dimensional vector containing all $n_L$ active
left-handed neutrinos.  $\nuRc$ similarly contains all $n_L$ of the active
right-handed antineutrinos.

$N_R$ is an $n_R$-dimensional vector containing all $n_R$ sterile
right-handed neutrinos.  $\NLc$ similarly contains all $n_R$ sterile
left-handed antineutrinos.

$\psi$, in addition to being a generic spinor, represents the $n$-dimensional
vector containing $\nuL$, $N_R$, and any additional sterile fields not related
to active fields.

$l_{L,R}$ is the $3$-dimensional vector containing all three flavors of charged
lepton spinors $\alpha_{L,R}$.

$\phi_D$ is the Higgs doublet which couples to down-type quarks and charged
leptons.

$\tilde{\phi}_D$ is the Higgs doublet which couples to up-type quarks and
 neutrinos.


\chapter{The Dirac Equation, Wave Functions, and Transformations}
\label{notn}


In this Appendix we review many aspects of field theory as they relate to
neutrinos.  As in the Introduction, the presentation here is synthesized from
many sources.  For a more in-depth treatment of field theory and
weak-interaction physics, the reader is encouraged to consult any or all of the
references \cite{Grein3}, \cite{Nact}, \cite{MS}, \cite{Itzykson},
\cite{fieldtheory}, \cite{Ramond}.


\section{Gamma Matrices}
\label{gammapp}


The Dirac gamma matrices $\gupu$ are defined by their appearance in the Dirac
equation:
\beq
\left[i\gupu \partial_{\mu} - m \right] \psi (x) = 0.
\label{Diraceqn}
\eeq
The spinor $\psi (x)$ is a four-component spinor, and the gamma matrices are $4
\times 4$.
For Dirac's equation to give the Klein-Gordon equation when its operation is
performed twice, 
the Dirac matrices must obey the anticommutation relations \cite{Nact}
\beq
\gupu \gupv + \gupv \gupu = 2 g_{\mu \nu},
\label{anticom}
\eeq
where $g_{\mu \nu}$ is the metric, defined in Appendix~\ref{key}.  
Dirac noted that the matrices
\beq
\gupo = \left( \ba{cc} \openone & 0 \\ 0 & -\openone \ea \right) 
\mbox{\ \ and \ \ } 
\gamma^a = \left( \ba{cc} 0 & \sigma_a \\ - \sigma_a & 0 \ea \right)
\label{gammadef}
\eeq
satisfy the anticommutation requirement.  Each element of the matrices in
equation (\ref{gammadef}) represents a $2 \times 2$ submatrix; 
$\sigma_a$ are the Pauli
matrices,
\beq
\sigma_1 = \left( \ba{cc} 0 & 1 \\ 1 & 0 \ea \right), \mbox{\ \ \ \ \ }
\sigma_2 = \left( \ba{cc} 0 & -i \\ i & 0 \ea \right), \mbox{\ \ and \ \ }
\sigma_3 = \left( \ba{cc} 1 & 0 \\ 0 & -1 \ea \right),
\label{Paulimat}
\eeq
and $\openone$ is the $2 \times 2$ unit matrix.  The product of the four gamma
matrices, $\gdnf \equiv i \gupo \gupa \gupb \gupc$, is useful also.  In the
above {\it standard representation} it has the form
\beq
\gdnf = \left( \ba{cc} 0 & \openone \\ \openone & 0 \ea \right).
\eeq

The representation of the gamma matrices chosen by Dirac is not unique.  If a
set of $\gupu$ satisfy equation (\ref{anticom}), then so will any
transformation of that set by an invertible matrix $A \gupu A^{-1}$.  Two
other popular representations include the {\it Majorana representation} 
\cite{MP},
$$
\gupo = \left( \ba{cc} 0 & \sigma_2 \\ \sigma_2 & 0 \ea \right), 
\mbox{\ \ \ \ \ }
\gupa = \left( \ba{cc} i \sigma_3 & 0 \\ 0 & i \sigma_3 \ea \right),
\mbox{\ \ \ \ \ }
\gupb = \left( \ba{cc} 0 & -\sigma_2 \\ \sigma_2 & 0 \ea \right),
$$
\beq
\gupc = \left( \ba{cc} -i \sigma_1 & 0 \\ 0 & -i \sigma_1 \ea \right),
\mbox{\ \ and \ \ } \gdnf = \left( \ba{cc} \sigma_2 & 0 \\ 0 & - \sigma_2 \ea
\right),
\label{Majgammadef}
\eeq
in which Majorana particles have pure real spinors, and the {\it chiral
representation} \cite{Rolnick},
\beq
\gupo = \left( \ba{cc} 0 & \openone \\ \openone & 0 \ea \right), 
\mbox{\ \ \ \ \ }
\gamma^a = \left( \ba{cc} 0 & -\sigma_a \\ \sigma_a & 0 \ea \right), 
\mbox{\ \ and \ \ }
\gdnf = \left( \ba{cc} \openone & 0 \\ 0 & -\openone \ea \right).
\label{chirgammadef}
\eeq


\section{Neutrino Wave Functions}
\label{wvfcnsec}


Neutrino solutions to the Dirac equation (\ref{Diraceqn})
are described by {\it field operators} of the form \cite{MP}
\beq
\psi(x)=\int \frac{d^3p}{(2 \pi)^{\frac{3}{2}}} \sum_{s=\pm 1/2} \left[ e^{ipx}
v_s(p) {\bf b_s^{\dagger}(p)} + e^{-ipx}u_s(p) {\bf a_s(p)} \right].
\eeq
$u_s(p)$ and $v_s(p)$ are single-particle {\it spinors}, and $p x$ is shorthand
for the Lorentz invariant $p_{\mu} x^{\mu}$.  Using the standard
representation for the gamma matrices, the spinors have the forms
\beq \ba{ccc}
u_s(p)=\sqrt{\frac{E+m}{2E}} \left( \ba{c} \chi_s \\ 
\frac{\mbox{\boldmath $\sigma \cdot p$}}{E+m} \chi_s \ea \right), & \mbox{ and } &
v_s(p)=\sqrt{\frac{E+m}{2E}} \left( \ba{c} 
\frac{\mbox{\boldmath $\sigma \cdot p$}}{E+m} \chi_s  \\ \chi_s \ea \right),
\ea \eeq
with
\beq \ba{cc}
\chi_{\half} = \left( \ba{c} 1\\0  \ea \right), \mbox{ and}&
\chi_{-\half} = \left( \ba{c} 0\\1 \ea \right).
\ea \eeq
${\bf b_s^{\dagger}(p)}$ and ${\bf a_s(p)}$ are the appropriate creation operator
for antiparticle (negative momentum) fields and the destruction operator of
particle(positive momentum) fields.  {\boldmath $\sigma$} are the Pauli matrices,
given in equation (\ref{Paulimat}).

For simplicity, we will take the momentum along the $+\hat{z}$
direction,\footnote{
This choice reflects the convention to quantize spin along the z-axis.  When
discussing kinematics, we follow the convention to choose {\bf p} along
the x-axis, which may at first seem to contradict our first assumption.
To be consistent throughout, we could either choose p always along the
z-axis so {\bf p} $\djdot$ {x} $= pz$ in our kinematic discussion, or we could
choose to quantize spin along the x-axis and redefine the Pauli matrices
so $\sigma_1$ is diagonal with $\pm 1$ eigenvalues.  Either consistent
treatment would yield the same results as ours, so we will stick with our
convention-motivated choices, inconsistent though they seem.
}
 and
the mass to be much smaller than the energy of the neutrino.  The quantity 
$\frac{\mbox{\boldmath $\sigma \cdot p$}}{E+m}$ 
appearing in the spinor definitions then becomes
\beq
\frac{\mbox{\boldmath $\sigma \cdot p$}}{E+m} = \frac{|p|}{E+m} \sigma_3 \approx (1-\epsilon)
\sigma_3,
\label{sigdotp}
\eeq
where $\epsilon=\frac{m}{E}$.  
Using the standard representation for the gamma matrices, the neutrino wave
function now has the form
\beq
\psi(t,{\bf x}) = \int \frac{d^3p}{(2 \pi)^{\frac{3}{2}}} \sqrt{\frac{E+m}{2E}}
\left( \ba{c} 
(1-\epsilon) e^{ipx}{\bf b_+^{\dagger}(p)} + e^{-ipx} {\bf a_+(p)} \\
-(1-\epsilon) e^{ipx}{\bf b_-^{\dagger}(p)} + e^{-ipx} {\bf a_-(p)} \\
e^{ipx}{\bf b_+^{\dagger}(p)} + (1-\epsilon) e^{-ipx} {\bf a_+(p)} \\
e^{ipx}{\bf b_-^{\dagger}(p)} - (1-\epsilon) e^{-ipx} {\bf a_-(p)} 
\ea \right),
\label{psiposp}
\eeq
where the subscripts $+$ and $-$ on the creation and annihilation operators
represent the spin states $+ \half$ and $- \half$, respectively.

If we reflect the coordinate system,
the momentum is in the $-\hat{z}$ direction and
the quantity in equation (\ref{sigdotp}) changes sign.  The wave function for the reflected system is
\beq
\psi(t,-{\bf x}) = \int \frac{d^3p}{(2 \pi)^{\frac{3}{2}}} \sqrt{\frac{E+m}{2E}}
\left( \ba{c} 
-(1-\epsilon) e^{ipx}{\bf b_+^{\dagger}(-p)} + e^{-ipx} {\bf a_+(-p)} \\
(1-\epsilon) e^{ipx}{\bf b_-^{\dagger}(-p)} + e^{-ipx} {\bf a_-(-p)} \\
e^{ipx}{\bf b_+^{\dagger}(-p)} - (1-\epsilon) e^{-ipx} {\bf a_+(-p)} \\
e^{ipx}{\bf b_-^{\dagger}(-p)} + (1-\epsilon) e^{-ipx} {\bf a_-(-p)} 
\ea \right).
\label{psinegp}
\eeq
The exponentials do not change under the transformation because $x' p'=
tE-(-{\bf x})(-{\bf p}) = x p$.  The differential $d^3 p$ changes sign, but the
limits of the integration change sign too, and those two effects cancel each
other.


\section{Helicity, Chirality, Parity, and Parity Violation}


\subsection{A Review of Handedness}
\label{handsec}


The wave function in equation (\ref{psiposp}) contains all spin states.  We may,
however separate the states with right-handed, or positive, spin from those
with left-handed, or negative, spin through the helicity projectors $P'_R$
and $P'_L$.  The handedness of states is determined by considering the component
of spin along the momentum axis.  When one puts the right thumb along the
direction of momentum, if the fingers of the right hand curl in the direction
of the spin, the state has positive helicity.  If the spin is instead in the
direction of the fingers of the left hand, the state has negative helicity.

The helicity projection operators are given by \cite{Grein5}
\beq
P'_R = \half \left( \openone + \frac{{\bf \Sigma \djdot p}}{|{\bf p}|} \right), \mbox{\ \ and \ \ } 
P'_L = \half \left(\openone - \frac{{\bf \Sigma \djdot p}}{|{\bf p}|} \right),
\eeq
with
\beq
{\bf \Sigma} \equiv \left( \ba{cc} \mbox{\boldmath $\sigma$} & 0 \\
0 & \mbox{\boldmath $\sigma$} \ea \right)
\eeq
in the standard representation, and they 
are functions of the momentum and energy \cite{Grein5}:
\beq
P'_{R,L} = \half \left(\openone \pm \gdnf \frac{E-\beta m_0}{p}\right),
\eeq
where $\beta$ is the particles velocity in terms of $c$.
These operators are not manifestly Lorentz-invariant, but in the
relativistic limit they become
\beq
\half (\openone \pm \gdnf).
\eeq
The relativistic
limits of the helicity projection operators are Lorentz-invariant and are called
the {\it chirality projection operators}, $P_{R,L}$:
\beq
P_R = \half (\openone+\gdnf), \mbox{\ \ and \ \ } P_L = \half (\openone-\gdnf).
\eeq
The chirality operators have the properties
\beq
P_R P_R = P_R, \mbox{\ \ \ \ } P_L P_L = P_L, 
\mbox{\ \ and \ \ } P_L P_R = P_R P_L = 0,
\eeq
which are consistent with their roles of projecting out particular states.
The helicity is only the same as the chirality if the particle mass is zero,
which means the $\epsilon$ in the wave function equations becomes zero.
We will restrict ourselves to that case here for simplicity.  Due to the
smallness of the neutrino masses, we also use the words helicity and chirality
interchangeably in the text.  
In the standard basis, the chirality projection operators have the forms
\beq
P_R = \half \left( \ba{cc} \openone & \openone \\ 
                             \openone & \openone \ea \right), \mbox{\ \ and \ \ }
P_L = \half \left( \ba{cc} \openone & -\openone \\ 
                             -\openone & \openone \ea \right).
\eeq
Applying these projections to the wave function of equation (\ref{psiposp}) 
and taking $\epsilon$ to zero, we find
\bea
P_R \psi (x) & = & \int \frac{d^3p}{(2 \pi)^{\frac{3}{2}}} \sqrt{\frac{E+m}{2E}}
\left( \ba{c} 
e^{ipx}{\bf b_+^{\dagger}(p)} + e^{-ipx} {\bf a_+(p)} \\
0\\
e^{ipx}{\bf b_+^{\dagger}(p)} + e^{-ipx} {\bf a_+(p)} \\
0 
\ea \right), \mbox{\ \ and \ \ } 
\label{PRpsiposp} \\
P_L \psi(x) & = & \int \frac{d^3p}{(2 \pi)^{\frac{3}{2}}} \sqrt{\frac{E+m}{2E}}
\left( \ba{c} 
0 \\
- e^{ipx}{\bf b_-^{\dagger}(p)} + e^{-ipx} {\bf a_-(p)} \\
0 \\
e^{ipx}{\bf b_-^{\dagger}(p)} -  e^{-ipx} {\bf a_-(p)} 
\ea \right).
\label{PLpsi}
\eea
The only states surviving the right-handed projection are those with spin
pointing in the same direction as the momentum, and those which survive the 
left-handed projection have spin pointing in the opposite direction.  If we
were to apply the projection operators to the reflected wave function of
equation (\ref{psinegp}), the right-handed projection would again choose the
spin states aligned with the momentum, but it those would have negative spin
since the momentum is negative:
\beq
P_R \psi(t,-{\bf x}) = 
\int \frac{d^3p}{(2 \pi)^{\frac{3}{2}}} \sqrt{\frac{E+m}{2E}} 
\left( \ba{c} 
0 \\
e^{ipx}{\bf b_-^{\dagger}(-p)} + e^{-ipx} {\bf a_-(-p)} \\
0 \\
e^{ipx}{\bf b_-^{\dagger}(-p)} +  e^{-ipx} {\bf a_-(-p)} 
\ea \right).
\label{PRpsinegp}
\eeq
Operating on the reflected wave function with $P_L$ pulls out the positive spin
states in a similar manner.  One must be especially careful to change the
argument of the {\it original} wave function (\ref{psiposp}) and form the new
wave function (\ref{psinegp}) {\it before} using the projection operators. 
The expression in equation (\ref{PRpsinegp}) will never fall out of any
operation on the wave function in equation (\ref{PRpsiposp}), since the
projection operation has already thrown out the negative spin states in
equation (\ref{PRpsiposp}).


\subsection{Parity Transformations}


The operation of space reflection is called a {\it parity transformation}.  Two
parity operations must return a system to its original state, so the reflected
annihilation and creation operators
${\bf a'_s(- p)}$ and ${\bf b'^{\dagger}_s(- p)}$ may differ from the original
fields only by a phase:
\beq
{\bf a'_s(- p)} = {\bf a_s(- p)} = \eta_a {\bf a_s(p)}, \mbox{\ \ and \ \ } 
{\bf b'^{\dagger}_s(-p)} = {\bf b^{\dagger}_s(-p)} = \eta_b {\bf b^{\dagger}_s(p)}.
\eeq
The parity-transformed wave function in the new coordinates $x'= (t,-{\bf x})$
and $p'= (E,-{\bf p})$ is then
\bea
\smp{\psi}(x') & = & \int \frac{d^3p}{(2 \pi)^{\frac{3}{2}}} \sqrt{\frac{E+m}{2E}}
\left( \ba{c} 
-(1-\epsilon) e^{ipx} {\bf b_+^{'\dagger}(-p)} + e^{-ipx} {\bf a'_+(-p)} \\
(1-\epsilon) e^{ipx}{\bf b_-^{'\dagger}(-p)} + e^{-ipx} {\bf a'_-(-p)} \\
e^{ipx}{\bf b_+^{'\dagger}(-p)} - (1-\epsilon) e^{-ipx} {\bf a'_+(-p)} \\
e^{ipx}{\bf b_-^{'\dagger}(-p)} + (1-\epsilon) e^{-ipx} {\bf a'_-(-p)} 
\ea \right) \nonumber \\
& = & \int \frac{d^3p}{(2 \pi)^{\frac{3}{2}}} \sqrt{\frac{E+m}{2E}}
\left( \ba{c} 
-(1-\epsilon) e^{ipx}\eta_b {\bf b_+^{\dagger}(p)} + e^{-ipx} \eta_a {\bf a_+(p)} \\
(1-\epsilon) e^{ipx}\eta_b {\bf b_-^{\dagger}(p)} + e^{-ipx} \eta_a {\bf a_-(p)} \\
e^{ipx}\eta_b {\bf b_+^{\dagger}(p)} - (1-\epsilon) e^{-ipx} \eta_a {\bf a_+(p)} \\
e^{ipx}\eta_b {\bf b_-^{\dagger}(p)} + (1-\epsilon) e^{-ipx} \eta_a {\bf a_-(p)} 
\ea \right)
\label{psinew}
\eea
As mentioned above, $e^{ip'x'} = e^{ipx}$.

The transformed wave function should be related to the original wave function
of equation (\ref{psiposp}) by simple matrix operations, so each element of the
new wave function should be proportional to the corresponding element of the
old wave function.  Comparison of equations (\ref{psinew}) and (\ref{psiposp})
reveals that $\eta_a$ must equal $-\eta_b$.  Choosing $\eta_a=1$, we have
\bea
\smp{\psi}(x') & = & \int \frac{d^3p}{(2 \pi)^{\frac{3}{2}}} \sqrt{\frac{E+m}{2E}}
\left( \ba{c} 
(1-\epsilon) e^{ipx} {\bf b_+^{\dagger}(p)} + e^{-ipx} {\bf a_+(p)} \\
-(1-\epsilon) e^{ipx} {\bf b_-^{\dagger}(p)} + e^{-ipx} {\bf a_-(p)} \\
-e^{ipx} {\bf b_+^{\dagger}(p)} - (1-\epsilon) e^{-ipx} {\bf a_+(p)} \\
-e^{ipx} {\bf b_-^{\dagger}(p)} + (1-\epsilon) e^{-ipx} {\bf a_-(p)} 
\ea \right) \nonumber \\
& = & \gupo \psi(x).
\eea
So a parity transformation of a wave function multiplies the wave function by 
$\gupo$, and changes the sign of the
three-vector position and momentum which also changes the
sign on the antiparticle creation operators.

Operating with parity 
on a state of definite chirality is equivalent to flipping the
chirality of the state and using the reflected coordinates.  
For example, the parity conjugate $\smp{\psi}_L (x')$ 
of a left-handed state $\psi_L (x) = P_L \psi (x)$ is
\bea
\smp{\psi}_L (x') & = & \gupo \psi_L (x) = \gupo P_L \psi (x) = \nonumber \\
& = &\int \frac{d^3p}{(2 \pi)^{\frac{3}{2}}} \sqrt{\frac{E+m}{2E}}
\left( \ba{c} 
0 \\
- e^{ipx}{\bf b_-^{\dagger}(p)} + e^{-ipx} {\bf a_-(p)} \\
0 \\
-e^{ipx}{\bf b_-^{\dagger}(p)} +  e^{-ipx} {\bf a_-(p)} 
\ea \right),
\eea
where we have used the result of equation (\ref{PLpsi}).  This parity conjugate
wave function is identical to the one found in equation (\ref{PRpsinegp}):
\bea
\psi_R (x') = P_R \psi(x') & = & 
\int \frac{d^3p}{(2 \pi)^{\frac{3}{2}}} \sqrt{\frac{E+m}{2E}}  
\left( \ba{c} 
0 \\
e^{ipx}{\bf b_-^{\dagger}(-p)} + e^{-ipx} {\bf a_-(-p)} \\
0 \\
e^{ipx}{\bf b_-^{\dagger}(-p)} +  e^{-ipx} {\bf a_-(-p)} 
\ea \right) \nonumber \\
& = &
\int \frac{d^3p}{(2 \pi)^{\frac{3}{2}}} \sqrt{\frac{E+m}{2E}} 
\left( \ba{c} 
0 \\
e^{ipx}(-{\bf b_-^{\dagger}(p)}) + e^{-ipx} {\bf a_-(p)} \\
0 \\
e^{ipx}(-{\bf b_-^{\dagger}(p)}) +  e^{-ipx} {\bf a_-(p)} 
\ea \right).
\eea
Thus we may represent the parity conjugate field $\smp{\psi}_L (x')$ as
$\psi_R (x')$.

This chirality flip due to parity may also be seen by considering the properties
of the gamma matrices.  $\gupo$ commutes with $\openone$ but anticommutes
with $\gdnf$, so
multiplying a projected state by $\gupo$ changes its chirality:
\beq
\gupo P_L \psi = P_R \gupo \psi,
\eeq
which is a right-handed state.  Because of the presence of the matrix $\gupo$
in the Dirac conjugation operation,
taking the Dirac conjugate of a particle swaps its chirality:
\beq
\ol{\psi_{L,R}} = \psi^{\dagger} P_{L,R}^{\dagger} \gupo = \psi^{\dagger}
\frac{1 \mp \gdnf}{2} \gupo = \psi^{\dagger} \gupo \frac{ 1 \pm \gdnf}{2} = 
\ol{\psi} P_{R,L}.
\label{olhand}
\eeq
{\it Ergo} a left-handed outgoing particle is equivalent to a right-handed projection
on an incoming particle.


\subsection{Parity Violation in Weak Interactions}


The weak charged current involves only left-handed particles, so it is
explicitly parity-violating.  Consider a charged-current 
two-vertex lepton interaction represented by an effective Lagrangian,
${\cal J}^{\mu}_{CC}{\cal J}^{\dagger}_{CC \mu}$, 
as discussed in Section \ref{SMsec}.
As discussed in Section \ref{mattersec}, a typical charged-current interaction
can be Fierzed to produce an interaction of the form
$$
\lag_{eff}^{CC} = 
-\frac{G}{\sqrt{2}} {\cal J}^{\mu}_{CC}{\cal J}^{\dagger}_{CC \mu} = 
-\frac{G}{\sqrt{2}} \left[ \ol{\nue} \gupu (1-\gdnf) \nue
\right] \left[ \ol{e} \gdnu (1-\gdnf) e \right].
\eqno{(\ref{mLeff})}
$$
This coupling explicitly violates parity through the presence of the
left-handed projection operator $P_L$.  
If we were to not {\it a priori} assume the Standard Model form of the
Lagrangian, we would need
a more general interaction
\cite{Grein5}
\beq
\lag_{eff}^{CC} = -\frac{G}{\sqrt{2}} \left[ \ol{\nua} \gupu O^i 
\left(C_i - C'_i \gdnf \right) \nub \right] \left[\ol{\beta} O_i \alpha
\right],
\label{GreinLeff}
\eeq
where $O_i$ can be any of the complete set of couplings:  
$1$(Scalar), $\gdnu$(Vector), 
$\sigma^{\mu \nu}$(Tensor), $\gdnu \gdnf$(Axial vector), or 
$\gdnf$(Pseudoscalar).  The expression 
$C_i - C'_i \gdnf$ appears only in the first factor in equation 
(\ref{GreinLeff}).  Such an expression could be placed in the second factor as
well, but it would not introduce any new freedom in the equation. 
The coefficients of the S, T, and P couplings must be zero at tree-level to preserve
gauge-invariance.  Deviations in the A and V couplings from common strength
would be evidence of some right-handed current.
Experiments measure observables, such as interaction rates and energy
distributions, not effective Lagrangians.  Michel and
Bouchiat \cite{MB}, \cite{Nact} showed that the observable results of an
interaction of the form (\ref{GreinLeff}) 
depend only on four so-called {\it Michel parameters}: $\rho$, $\eta$, $\xi$,
and $\delta$.  These parameters are functions of the coefficients
$C_i$ and $C'_i$, and they are predicted to be equal to $\frac{3}{4}$, $0$,
$1$, and $\frac{3}{4}$, respectively, for the Standard Model's completely
parity-violating weak current.  Experimental measurements yield
values consistent with the Standard Model predictions \cite{Grein5}:
\bea
\rho &=& 0.7517 \pm 0.0026, \nonumber \\
\eta &=& -0.12 \pm 0.21, \\
\xi &=& 0.972 \pm 0.013, \mbox{\ \ and \ \ } \nonumber \\
\delta &=& 0.7551 \pm 0.0085. \nonumber
\eea
Such a marked agreement makes the introduction of right-handed currents in
extensions of the Standard Model quite difficult.


\section{Charge Conjugation}
\label{Csec}


In this paper, we will distinguish between a charge-conjugate particle
$\smc{\psi}$ and an antiparticle.  Dirac originally postulated the existence of
antiparticles as the negative-energy solutions to the Dirac equation having the
same mass and opposite charge as their associated particles.  The original
Dirac equation was charge-conjugation-invariant, so the term antiparticle became associated
with the charge-conjugate 
state,\footnote{
Our notation here differs from that of reference \cite{BP} since they
apparently use $\smc{\psi}_L$ to represent the charge-conjugate field of the
{\it right-handed} field, while we use it to represent the conjugate of the
left-handed field.}
\beq
\smc{\psi}_{L,R} = C \ol{\psi_{R,L}}^T,
\label{psic}
\eeq
where $C$ is the charge-conjugation operator, discussed below.
  With the
discovery of charged current interactions in the late 1950s, 
however, came the discovery that
charge-conjugation and parity are not good symmetries independently.  
The combination of
charge-conjugation with parity, CP, seemed to be a good symmetry,
so the new negative-energy solution to the Dirac equation was thought to have 
the form
\beq
\smcp{(\psi_{L,R} (x'))} = C \ol{\gupo \psi_{L,R}(x)}^T = C
\ol{\psi_{R,L} (x')}^T = \smc{\psi}_{R,L} (x')
\label{CPfield}
\eeq
until the early 1960s.  With the study of kaon systems came the discovery that
even CP is not conserved exactly.  We now believe that 
CPT is a good symmetry, where the T stands for time reversal.  
The physical antiparticles
therefore correspond to the CPT-conjugate states $\smcpt{\psi} (-x)$.
Because the violation of CP is so small, and could possibly be zero in the
lepton sector, we will loosely use the term antineutrino to refer to the
CP-conjugate
\beq
\nuRc \equiv \smcp{(\nu_L)} = C \ol{\nuL}^T.
\label{nuRcdef}
\eeq

Having clarified the terminology, let us return to the definition of the
charge-conjugate field in equation (\ref{psic}).
$C$ has the following properties \cite{BP}:
\beq
C^T = C^{\dagger} = C^{-1} = -C, 
\mbox{\ \ \ \ \ } C C^{\dagger} = \openone, \mbox{\ \ and \ \ }
C \gdnu^T C^{-1} = -\gdnu.
\label{Cprop}
\eeq
Using Dirac's standard representation of the gamma matrices, $C$ has the form
\beq
C = i \gupb \gupo.
\eeq
This identification of the matrix $C$ is, however, representation-dependent
\cite{MP}, so care should be used when applying it.

The charge-conjugate $\smc{\psi}_{R,L}$ of a field of definite chirality 
$\psi_{R,L}$ will be composed of the opposite-chirality field but transform 
like the
original field \cite{Ramond}.  For example,
\beq
\smc{\psi}_L = C \ol{\psi_R}^T
\eeq
is formed from $\psi_R$ but transforms as a left-handed field since 
$C$ commutes with the projection operators.  In the standard representation,
\beq
P_{R,L} C = \frac{1 \pm \gdnf}{2} i \gupb \gupo = 
i \gupb \frac{1 \mp \gdnf}{2} \gupo = i \gupb \gupo \frac{1 \pm \gdnf}{2} =
C P_{R,L},
\eeq
so the charge-conjugate field $\smc{\psi}_L$ has the chirality of the
Dirac-conjugate field $\ol{\psi_R}$.  The latter field is shown to be
left-handed in equation (\ref{olhand}), so the
field $\smc{\psi}_L$ transforms as a left-handed field, as claimed
above.

Using the properties of $C$ and the definition (\ref{CPfield}), 
we can find an expression for 
$\ol{\smc{\psi}_{L,R}}$
\bea
\ol{\smc{\psi}_{L,R}} = {\smc{\psi}}^{\dagger}_{L,R} \gupo 
& = & \left( C \ol{\psi_{R,L}}^T \right)^{\dagger} \gupo
= \psi_{R,L}^T \gupo^{T \dagger} C^{\dagger} \gupo
= \psi_{R,L}^T C^{\dagger} C \gupo C^{\dagger} \gupo \nonumber \\
& = & \psi_{R,L}^T C^{\dagger} (-\gupo^T) \gupo
= - \ \psi_{R,L}^T C^{\dagger}.
\label{olpsic}
\eea
Applied to the neutrino, we find
\beq
\ol{\nuRc} = - \nuL^T C^{\dagger}.
\label{olnuc}
\eeq


\chapter{A More Realistic Treatment of Neutrino Oscillations}
\label{appc}


The common treatment of neutrino oscillation
makes many sometimes contradictory assumptions, yet manages to arrive at an
equation for the transition probability, (\ref{oscillation}), that is
consistent with more realistic treatments 
to leading order in neutrino masses.  These assumptions include (a) {\em the
treatment of neutrinos as plane waves rather than wave packets},  (b) {\em all
neutrino mass states have a common momentum but different energy} (energy is
conserved for the interaction, but momentum conservation is ignored), (c)
neutrinos are relativistic, so {\boldmath $t \approx |{\bf x}|$}, and 
(d) again the neutrino is
relativistic, so {\boldmath $E_i-p \approx \frac{m_i^2}{2 p}$}.  In addition,
we define the x-axis to be in the direction of the neutrino
three-momentum ${\bf p}_i$, so ${\bf p}_i \cdot {\bf x}_i = p_i x_i.$  This final
relationship is merely a definition of the coordinate system, so we do not lose
any information making it.  Under these
assumptions, the phase difference of Chapter~\ref{osc},
\beq
\Phi_{ij} = \half \left[ \left( E_i t_i - E_j t_j \right) 
- \left( {\bf p}_i \cdot {\bf x}_i - {\bf p}_j \cdot {\bf x}_j \right) \right]
\eeq
becomes
\beq
\Phi_{ij}^{historic}
 =\half \left[ \left(E_i-p\right) x - \left(E_j-p\right) x \right]
 = \frac{m_i^2-m_j^2}{4p}x.
\label{phihist}
\eeq
Many authors have addressed the nature of
the assumptions (a)-(d).  Most notably, the plane-wave approximation 
(a) is addressed in references \cite{Kays2}, 
\cite{Nussinov}, \cite{Kim} and \cite{GKL}.  
Reference \cite{GL} argues that a replacement of
assumption (b) with (b') {\em  all neutrino mass states have a common energy but
different momentum} is far more logical for most experiments which measure the
distance traveled by a neutrino, not the time taken to travel.  
When either $p$ or $E$ is common to all mass states, 
the wave-packet treatment yields the
traditional $\phi_{ij}^{historic}$ of equation (\ref{phihist}), as will be 
shown in Section~\ref{wvpktsec}.

Neither of the assumptions (b) or (b') are correct or even preferable,
according to Goldman, the author of reference 
\cite{Goldman}.  Goldman's work 
provides a relativistic treatment of oscillations, conserving
both energy and momentum, and points out
that (c) and (d) are inconsistent assumptions, since the difference between $t$
and $x$ is, to first order, 
the same as the difference between $p$ and $E$.  Goldman allows 
different mass states to have both unique
energies and unique momenta, but arrives at the common expression 
(\ref{oscillation})
to leading order in neutrino masses.  We examine his work and extend it to an
arbitrary number of neutrino flavors in Section~\ref{relativsec}.


\section{Neutrinos as Wave-Packets}
\label{wvpktsec}


Nussinov \cite{Nussinov} was perhaps the first to examine neutrino oscillations
using wave packets rather than plane waves to describe the neutrino states. 
Kayser extended Nussinov's work in reference~\cite{Kays2}.  Kayser's work laid
the groundwork for many more detailed treatments, such as \cite{GKL} and
\cite{GL}.  Reference \cite{Kim} provides a useful summary of the wave-packet
approach.  The probability for the transition $\nua \rightarrow \nub$ for wave
packets is given by reference \cite{GKL} to be
\beq
\Pab = A \sum_i \sum_j \Baibj e^{i 2 \Phi_{ij}^{hist} B} C e^{-D-F}.
\eeq
$A$ is a normalization constant,
\beq
A = \left( \sum_k \frac{|V_{\alpha k}|^2}{|v_k|} \right)^{-1},
\eeq
where $v_k$ represents the velocity of the $k$th mass state.
\beq
B = \frac{v_i+v_j}{v_i^2+v_j^2} - 
\frac{\langle p_i \rangle-\langle p_j \rangle}
  {\langle E_i \rangle-\langle E_j \rangle}
\eeq
represents the deviation of the phase from
the traditional phase. $C$ and $F$ both result from the time-averaging 
performed in reference \cite{GKL}.  
\beq
C=\sqrt{\frac{2}{v_i^2+v_j^2}}
\eeq
arises because the probability to find a mass state at a detector is inversely
proportional to the speed of the mass state.
\beq
F=\frac{\left( \langle E_i\rangle - \langle E_j \rangle \right)^2}
{4 \sigma_p^2 \left(v_i^2+v_j^2\right)},
\eeq
where $\sigma_p$ is the width of the Gaussian wave packet in momentum space,
insures that energy is conserved within the uncertainty of the wave packet
\cite{GKL}.
Finally, 
\beq
D=\frac{x^2}{4 \sigma_x^2} \frac{\left(v_i-v_j\right)^2}{v_i^2+v_j^2},
\eeq
where $\sigma_x$ is the width of the Gaussian wave packet in coordinate space,
is a damping factor which accounts for the separation of the mass
states as they propagate.  As they separate, the overlap decreases and the
oscillatory behavior disappears.  This behavior gives rise to the definition
of a coherence length,
\beq
L_{ij}^{coh} \equiv 4 \sigma_x 
\sqrt{ \frac{v_i^2+v_j^2}{\left(v_i-v_j\right)^2} }.
\eeq
As long as the coherence length is much greater than the size of the wave
packet, the different mass states will interfere, and oscillations will occur
\cite{GKL}.

In the traditional approach, $AC = B \approx 1$, and $D+F \approx 0$.


\section{Neutrinos in a Relativistically Correct Light}
\label{relativsec}


Everything in Section~\ref{eomsec} is relativistically correct, so we may start
with equations (\ref{nuinit}) and (\ref{xfinal}):
$$
\nu_L (t=0,x=0) = \nua_L = \sum_i V_{\alpha i} \nui_L,
\eqno{\ref{nuinit}}
$$
and
$$
\nu_L (t, x) = \sum_{i=1}^{n} \Vai \nui e^{-i\left(E_i t_i - p_i x_i\right)} 
\equiv \sum_{i=1}^{n} \Vai \nui e^{-i \phi_i},
\eqno{\ref{xfinal}}
$$
The probability for transition $\Pab$, as shown in Section~\ref{vacoscsec},
depends on the phase difference
\beq
\Phi_{ij} = \half \left[ \left( E_i t_i - E_j t_j \right) 
  - \left( p_i x_i - p_j x_j \right) \right].
\eeq
To find this phase in terms of the neutrino masses we must consider
relativistic kinematics.

The $p_i$ and $E_i$ are of course related by $E_i^2 =
p_i^2+m_i^2$, and they depend on the other energies and momentum of the process
which created the neutrino.  For simplicity, we will work in the rest frame of
the parent particle.  The invariant mass squared of the parent particle is
$M^2$, and the decay products other than the neutrino will have total mass
squared $M_f^2$ and momenta summing to $-{\bf p}_i$.  For oscillations to occur, we
must not be able to measure the momenta or energies 
of the other particles closely enough
to distinguish between the different possibilities of $p_i$.  We also must not
be able to resolve $\Delta t_{ij} \equiv t_i-t_j$ 
between the arrival times of the different mass states.

Energy conservation for the production of the neutrino requires that
\beq
M = \left(E_{\nu} + E_{other}\right) 
= \sqrt{p_i^2 + m_i^2} + \sqrt{p_i^2 + M_f^2},
\eeq
so
\beq
\left(M-\sqrt{p_i^2 + m_i^2}\right)^2 
= M^2 -2 M \sqrt{p_i^2 + m_i^2} + p_i^2 + m_i^2 = p_i^2 + M_f^2,
\eeq
and
\beq
E_i = \sqrt{p_i^2 + m_i^2} = \frac{M^2 - M_f^2 + m_i^2}{2 M}.
\eeq
We may extract $p_i$ from this equation:
\bea
p_i^2 & = & \frac{\left(M^2-M_f^2\right)^2+2m_i^2\left(M^2-M_f^2\right)+m_i^4}
{4 M^2} - \frac{4 M^2 m_i^2}{4 M^2}, \mbox{\ \ so \ \ } \nonumber \\
p_i & = & \sqrt{\frac{
\left(M^2-M_f^2\right)^2 - 2 m_i^2 M^2 - 2 m_i^2 M_f^2 + m_i^4}{4M^2}}
\nonumber \\
& = & 
\frac{\left(M^2-M_f^2\right)}{2M}\sqrt
{1-\frac{2m_i^2\left(M^2+M_f^2\right)}{\left(M^2-M_f^2\right)^2}
+\frac{m_i^4}{\left(M^2-M_f^2\right)^2}}.
\eea
Using the Taylor expansion $\sqrt{1+x} = 1+\frac{x}{2}-\frac{x^2}{8}+ \cdots$
and keeping terms to fourth order in the neutrino mass $m_i$, we find
\bea
p_i & = & \frac{ \left( M^2-M_f^2 \right) }{2M}\left[
1-\half \frac{2m_i^2 \left( M^2+M_f^2\right)+m_i^4 }{\left( M^2-M_f^2 \right)^2}
-\frac{m_i^4}{8} \frac{4 \left( M^2+M_f^2 \right)^2 }
  {\left( M^2-M_f^2 \right)^4} + {\cal O}(m_i^6)\right]  \nonumber \\
& = & \frac{\left(M^2-M_f^2\right)}{2M}\left[
1-\frac{m_i^2\left(M^2+M_f^2\right)}{\left(M^2-M_f^2\right)^2}
+\frac{m_i^4}{2}\frac{\left(M^2-M_f^2\right)^2-\left(M^2+M_f^2\right)^2}
  {\left(M^2-M_f^2\right)^4}
+ {\cal O}(m_i^6)\right] \nonumber \\
& = & \frac{\left(M^2-M_f^2\right)}{2M}\left[
1-m_i^2\frac{\left(M^2+M_f^2\right)}{\left(M^2-M_f^2\right)^2}
-m_i^4\frac{2 M^2 M_f^2}{\left(M^2-M_f^2\right)^4}
+ {\cal O}(m_i^6)\right]
\eea
We will later want the differences between two energies and between two
momenta:
\bea
E_i-E_j & = & \frac{M^2 - M_f^2 + m_i^2}{2 M} - \frac{M^2 - M_f^2 + m_j^2}{2 M}
= \frac{m_i^2 - m_j^2}{2M}, \mbox{\ \ and \ \ } 
\label{Eij} \\
p_i-p_j & = & \frac{\left(M^2-M_f^2\right)}{2M}\left[
1-\frac{m_i^2\left(M^2+M_f^2\right)}{\left(M^2-M_f^2\right)^2}
-\frac{m_i^4}{2}\frac{4 M^2 M_f^2}{\left(M^2-M_f^2\right)^4} \right.\nonumber \\
&&\left. -\  1+\frac{m_j^2\left(M^2+M_f^2\right)}{\left(M^2-M_f^2\right)^2}
+\frac{m_j^4}{2}\frac{4 M^2 M_f^2}{\left(M^2-M_f^2\right)^4} \right]\label{pij}\\
& = & \frac{\left(M^2-M_f^2\right)}{2M}
\left[ -\left(m_i^2-m_j^2\right) \frac{M^2+M_f^2}{\left(M^2-M_f^2\right)^2}
-\left(m_i^2-m_j^2\right)\left(m_i^2+m_j^2\right) 
\frac{2 M^2 M_f^2}{\left(M^2-M_f^2\right)^4} \right].
\nonumber 
\eea

$\Phi_{ij}$ does not, however, depend only on $E$ and $p$; it also depends on
$x$ and $t$, and deciding what values to use for these variables is rather
tricky.  As pointed out in reference \cite{GL}, experiments measure the
distance $x$ between the neutrino source and the detector, so all mass states
must have traveled the same distance $x=x_i=x_j$.  One could then find the
individual times from the relativistic relation $t_i = \frac{E_i}{p_i} x$ and
arrive at the phase
\beq
\hat{\Phi}_{ij} = 
\half 
\left[\left(E_i\frac{E_i}{p_i}-p_i\right)-\left(E_j\frac{E_j}{p_j}-p_j\right)
\right] x = 
\left[\frac{m_i^2}{p_i}-\frac{m_j^2}{2 p_j}\right]x.
\eeq
To leading order in mass-squared, this phase is twice the phase obtained by
the traditional approach:
\beq
\hat{\Phi}_{ij} \approx \frac{m_i^2-m_j^2}{2p_{ave}} = 2 \Phi^{historic}_{ij},
\eeq

Goldman reproduces the historic phase to leading order in reference
\cite{Goldman} by using the average time of travel, 
\bea
t_{ave} &=& \half \left( \frac{E_i}{p_i} - \frac{E_j}{p_j} \right) x
= \half \left[\sqrt{1+\frac{m_i^2}{p_i^2}} 
        + \sqrt{1+\frac{m_j^2}{p_j^2}}\right] x \nonumber \\
&=& \half \left[ 2+\frac{m_i^2}{2 p_i^2}+\frac{m_j^2}{2 p_j^2} 
+ {\cal O}(m_i^4)\right] x \nonumber \\
& = & \left[ 1 + 
\frac{m_i^2}{4} \frac{4 M^2}{\left(M^2-M_f^2\right)^2} 
  + \frac{m_j^2}{4} \frac{4 M^2}{\left(M^2-M_f^2\right)^2}
+ {\cal O}(m_i^4) \right] x,
\label{tave}
\eea
for all mass states rather than the individual travel times $t_i$.  This choice
may be justified on the grounds of uncertainty 
since we cannot resolve between the different travel
times if oscillations are to occur.  A combined treatment of wave packets and
correct relativistic kinematics would of course illustrate the correct phase,
but that is a topic for a different paper.  In this paper, Gentle Reader, we
will stick with Goldman's $t_{ave}$.

Now we may calculate $\Phi_{ij}$ using equations (\ref{Eij}), (\ref{pij}), and
(\ref{tave}):
\bea
\Phi_{ij} & = & \half \left[ \left(E_i-E_j\right) t_{ave} 
- \left(p_i-p_j\right) x \right]  
= \half \left[\left(E_i-E_j\right) \left( 1 
+ \frac{\left(m_i^2+m_j^2\right) M^2}{\left(M^2-M_f^2\right)^2} \right)
-\left(p_i-p_j\right)\right] x \nonumber \\
&=&
\frac{m_i^2-m_j^2}{4M} \left[1  + \frac{M^2+M_f^2}{M^2-M_f^2}
  + \frac{\left(m_i^2+m_j^2\right)M^2 }{\left(M^2-M_f^2\right)^2}
  + \frac{2 M^2 M_f^2\left(m_i^2+m_j^2\right)}{\left(M^2-M_f^2\right)^3}
\right] x
\nonumber \\
& = &
\frac{m_i^2-m_j^2}{4M} \left[\frac{2 M^2}{M^2-M_f^2} 
+ \frac{m_i^2+m_j^2}{\left(M^2-M_f^2\right)^3}
   \left[M^2\left(M^2-M_f^2\right) + 2M^2M_f^2\right]
\right] x \nonumber \\
& = &
\frac{m_i^2-m_j^2}{4} \left[\frac{2 M}{M^2-M_f^2} 
+ \frac{\left(m_i^2+m_j^2\right) M\left(M^2+M_f^2\right)}
  {\left(M^2-M_f^2\right)^3}
\right] x
\eea
The first term in the brackets, $\frac{2 M}{M^2-M_f^2}$ is just the reciprocal
of the leading
term in all $p_i$, so we recover
\beq
\Phi_{ij} \approx \frac{m_i^2-m_j^2}{4p} x = \Phi_{ij}^{historic}
\label{approxphi}
\eeq
for small $m_i$, as promised above.


\chapter{Neutrinoless Double-Beta Decay}
\label{appd}


We demonstrated in Section~\ref{relphasesec} that 
oscillation experiments will not distinguish Majorana neutrinos from Dirac
neutrinos.  One experiment which can distinguish between the two
types is neutrinoless double-beta decay.  As mentioned in the Introduction,
neutrinos were first discovered by their role in beta decay:
\beq
n \rightarrow p + e + \smc{\nu}_e.
\eeq
Double-beta decay is a higher-order process and therefore occurs less often than
single beta decay. Figure~\ref{BBfig}a illustrates a typical double-beta decay
interaction.
This process is identical to one with only one of the $\smc{\nu}_e$ going out,
and the other outgoing $\smc{\nu}_e$ being replaced by an incoming $\nue$, as
shown in Figure~\ref{BBfig}b.  If neutrinos are Dirac particles, no transition
between $\smc{\nu}_e$ and $\nue$ may occur, so two distinct neutrinos must
participate in double-beta decay.  But if neutrinos are Majorana particles, 
$\smc{\nu}_e$ and $\nue$ are two spin states of the same particle, so the
$\smc{\nu}_e$ emitted by the top neutron could flip into a $\nu_e$ and be
absorbed by the bottom neutron.  This possibility is illustrated in
Figure~\ref{BBnonufig}.

\begin{figure}[htb]
\centerline{\hbox{
\psfig{figure=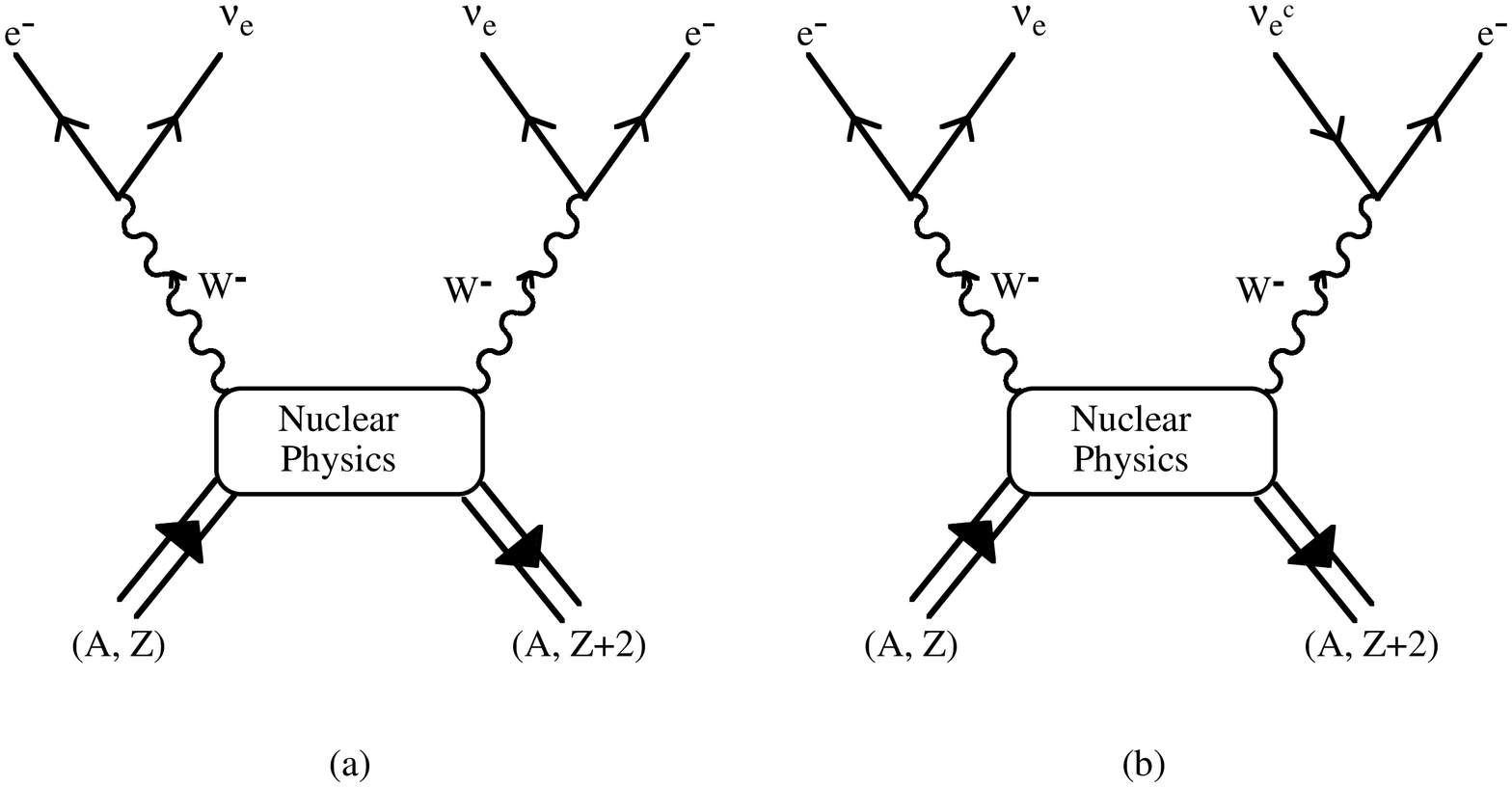,height=3 in}
}}
\caption[A typical double-beta decay process.]
{A typical double-beta decay process. (a) and (b) are equivalent diagrams by
the Feynman rules of particle-antiparticle exchange. \label{BBfig}}
\end{figure}

\begin{figure}[htb]
\centerline{\hbox{
\psfig{figure=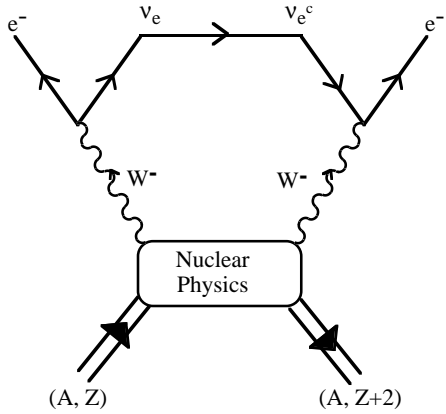,height=3 in}
}}
\caption[Neutrinoless double-beta decay.]
{Neutrinoless double-beta decay. The outgong antineutrino flips helicity states
to become a neutrino, and is absorbed.  \label{BBnonufig}}
\end{figure}

Neutrinoless double beta decay is overwhelmed by single beta decay when the
latter is possible, so searches for the neutrinoless decay use
nuclei in which the regular beta decay channel is energetically forbidden
\cite{Grein5}.  The probability that a Majorana neutrino will flip helicity
states so it can be absorbed in the neutrinoless process goes as
$\left(\frac{m}{E}\right)^2$, which suppresses the neutrinoless channel with
respect to the two-neutrino channel.  
But the neutrinoless process contains only two
final-state leptons rather than the four involved in the two-neutrino process,
so the neutrinoless channel gets relatively enhanced due to the extra
phase-space available.  If the neutrino mass is on the order of a few eV, these
two effects cancel each other \cite{Grein5}.  

Other explanations for neutrinoless double-beta decay exist, but those decays
caused by neutrino mixing would have a unique final state.  If the
neutrinoless decay
were caused by the exchange of supersymmetric particles with broken R-parity, 
the final state would contain extra particles in addition to the 
electrons; if it were due to the production of a Majoran by the
two neutrinos, the final state would include a Majoran which would share energy
with the electrons \cite{BBpaper}.  Only the neutrino mixing solution would
produce two electrons with equal but opposite momenta in the rest frame of the
decay.

The calculation of neutrino masses from neutrinoless double-beta decay involves
nuclear matrix elements and is outside the focus of this work.  The interested
reader may turn to references \cite{Grein5}, \cite{MP}, and \cite{KimPev} for a
more detailed study of neutrinoless double-beta decay.


\chapter{A Graphical Representation of Boxes}
\label{graphapp}


Many of the relationships between boxes developed in Chapter~\ref{boxes} were
originally derived using a graphical representation.  In this method,
boxes in the numerator of a product are represented by two vertical lines; boxes in
the denominator are represented by two horizontal lines.  Lines exit the locations
of the matrix elements which are not complex conjugated 
and enter the locations of the complex-conjugated
matrix elements.  For example, the box $\B{11}{22}=V_{11} V^*_{12} V_{22}
V^*_{21}$ is represented by a
vertical line pointing from $V_{11}$ to $V_{21}$ and a vertical line pointing
from $V_{22}$ to $V_{12}$, as shown in Figure~\ref{graphfig}a.  The inverse box
$\frac{1}{\B{11}{22}} = \left(\B{11}{22}\right)^{-1}$ is represented by a
horizontal line pointing from $V_{11}$ to $V_{12}$ and one pointing from 
$V_{22}$ to $V_{21}$, as shown in Figure~\ref{graphfig}b.  The
complex-conjugated box $\Bs{11}{22}$ is equal to $\B{12}{21}$, so one just
reverses the arrows to complex conjugate a box, 
as shown in Figure~\ref{graphfig}c.

\begin{figure}[htb]
\vsp
\begin{centering}
\begin{picture}(310,90)(-5,5)
\thicklines
\multiput(0,90)(30,0){3}{\circle*{4}}
\multiput(0,60)(30,0){3}{\circle*{4}}
\multiput(0,30)(30,0){3}{\circle*{4}}
\put(0,87){\vector(0,-1){24}}
\put(30,63){\vector(0,1){24}}
\put(24,10){(a)}
\multiput(120,90)(30,0){3}{\circle*{4}}
\multiput(120,60)(30,0){3}{\circle*{4}}
\multiput(120,30)(30,0){3}{\circle*{4}}
\put(123,90){\vector(1,0){24}}
\put(147,60){\vector(-1,0){24}}
\put(144,10){(b)}
\multiput(240,90)(30,0){3}{\circle*{4}}
\multiput(240,60)(30,0){3}{\circle*{4}}
\multiput(240,30)(30,0){3}{\circle*{4}}
\put(240,63){\vector(0,1){24}}
\put(270,87){\vector(0,-1){24}}
\put(264,10){(c)}
\end{picture}
\caption{The graphical representation for (a) $\B{11}{22}$, \ (b)
$\left(\B{11}{22}\right)^{-1}$, \ and (c) $\Bs{11}{22}$.
\label{graphfig}}
\end{centering}
\end{figure}
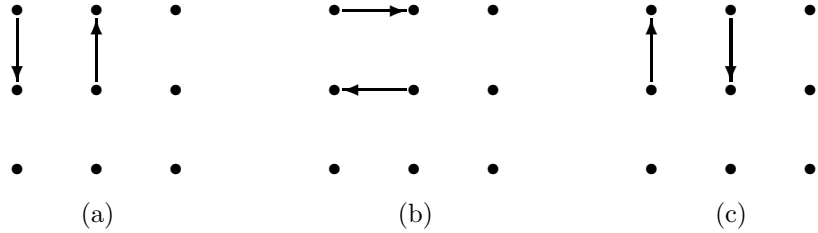

To multiply boxes together graphically, one merely draws the lines corresponding
to each factor on the same grid.  Horizontal arrows entering or leaving a point
cancel out vertical arrows entering or leaving, respectively, that point. 
Uncanceled arrows entering a point signify the survival of the
complex-conjugated matrix element associated with that point.  Those leaving a
point signify the survival of the ordinary matrix element of that point. 
Figure~\ref{cancelfig}a represents the product 
$\B{11}{22} \B{12}{33} \left(\B{23}{32}\right)^{-1}$.  The horizontal arrows do
not cancel vertical arrows at any point, so no simplification may occur.  The
upward arrow at $V_{22}$ represents that element in the numerator.  The
horizontal arrow represents the element $V^*_{22}$ in the denominator, which
does not cancel.  Counting off the arrows at each vertex, we find the expression
\beq
\B{11}{22} \B{12}{33} \left(\B{23}{32}\right)^{-1} = 
\frac{V_{11} V_{12} V^*_{12} V^*_{13} V^*_{21} V_{22} V^*_{32} V_{33}}
{V^*_{22} V_{23} V_{32} V^*_{33}},
\eeq
which agrees with the definitions of boxes.  

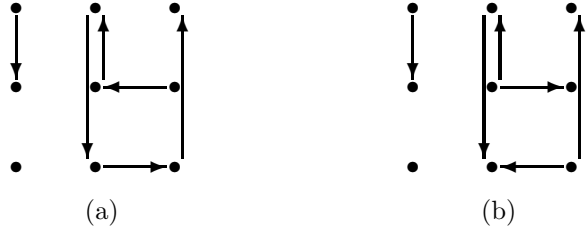
\begin{figure}[thb]
\begin{centering}
\begin{picture}(220,100)(-5,5)
\thicklines
\multiput(0,90)(30,0){3}{\circle*{4}}
\multiput(0,60)(30,0){3}{\circle*{4}}
\multiput(0,30)(30,0){3}{\circle*{4}}
\put(0,87){\vector(0,-1){24}}
\put(27,87){\vector(0,-1){54}}
\put(33,63){\vector(0,1){24}}
\put(63,33){\vector(0,1){54}}
\put(57,60){\vector(-1,0){24}}
\put(33,30){\vector(1,0){24}}
\put(26,10){(a)}
\multiput(150,90)(30,0){3}{\circle*{4}}
\multiput(150,60)(30,0){3}{\circle*{4}}
\multiput(150,30)(30,0){3}{\circle*{4}}
\put(150,87){\vector(0,-1){24}}
\put(177,87){\vector(0,-1){54}}
\put(183,63){\vector(0,1){24}}
\put(213,33){\vector(0,1){54}}
\put(207,30){\vector(-1,0){24}}
\put(183,60){\vector(1,0){24}}
\put(176,10){(b)}
\end{picture}
\caption{The graphical representation for the products 
a) $\B{11}{22} \B{12}{33} \left(\B{23}{32}\right)^{-1}$, and 
b)$\B{11}{22} \B{12}{33} \left(\B{22}{33}\right)^{-1}$.
\label{cancelfig}}
\end{centering}
\end{figure}

Figure~\ref{cancelfig}b represents
$\B{11}{22} \B{12}{33} \left(\B{22}{33}\right)^{-1}$, a product in which
some canceling does occur.  Picking out the uncanceled arrows at each vertex, we
are left with
\beq
\B{11}{22} \B{12}{33} \left(\B{22}{33}\right)^{-1} = 
\frac{V_{11} V_{12} V^*_{12} V^*_{13} V^*_{21}}{V^*_{23}}.
\eeq

The graphical method is a powerful tool for finding relationships between boxes.
Consider the degenerate boxes $\B{\alpha i}{\alpha i} = |V_{\alpha i}|^4$. 
Trying to obtain the equations~(\ref{vfour}) and (\ref{vfour2}) presented in
Chapter~\ref{boxes} without graphs required quite a bit of running down dead
ends.  Using the graphical method, we need only find a series of arrows which
cancel for every point except $(\alpha, i)$ and leave two incoming and two
outgoing vertical arrows at that point.  Consider $|V_{21}|^4$ as an example. 
We choose to draw all of the arrows involving $V_{21}$ pointing downward, as
shown in Figure~\ref{V21fig}a.  These arrows must be part of boxes, so in
Figure~\ref{V21fig}b we add the arrows to finish those boxes.  
Next we draw two horizontal boxes in Figure~\ref{V21fig}c to
cancel the extra arrows in the first column of the matrix.  This still leaves
$V_{22}$ and $V_{23}$ with two sets of uncanceled arrows apiece.  In
Figure~\ref{V21fig}d we draw two more horizontal boxes to compensate.  This adds
arrows to our previously clean $V_{12}$, $V_{13}$, $V_{32}$, and $V_{33}$;
drawing the final vertical box in Figure~\ref{V21fig}e cancels those.  Recapping
what we have done, we see that step (b) completes $\B{11}{22}$, $\B{11}{23}$,
$\B{21}{32}$, and $\B{21}{33}$ in the numerator.  Step (c) divides by
$\B{11}{32}$ and $\B{11}{33}$, and step (d) divides by $\B{12}{23}$ and
$\B{22}{33}$.  Step (e) multiplies by $\B{12}{33}$, leaving only the point
$V_{21}$ with uncanceled arrows.  It has two vertical arrows coming in and two
leaving, so our graph represents the equation
$$
|V_{21}|^4 = \frac{\B{21}{32}\B{11}{22}\B{21}{33}\B{11}{23}\B{12}{33}}
{\B{11}{32}\B{11}{33}\B{22}{33}\B{12}{23}}
\eqno{(\ref{V21four})}
$$
of Chapter~\ref{boxes}.
Other examples of this representation are included in that Chapter.

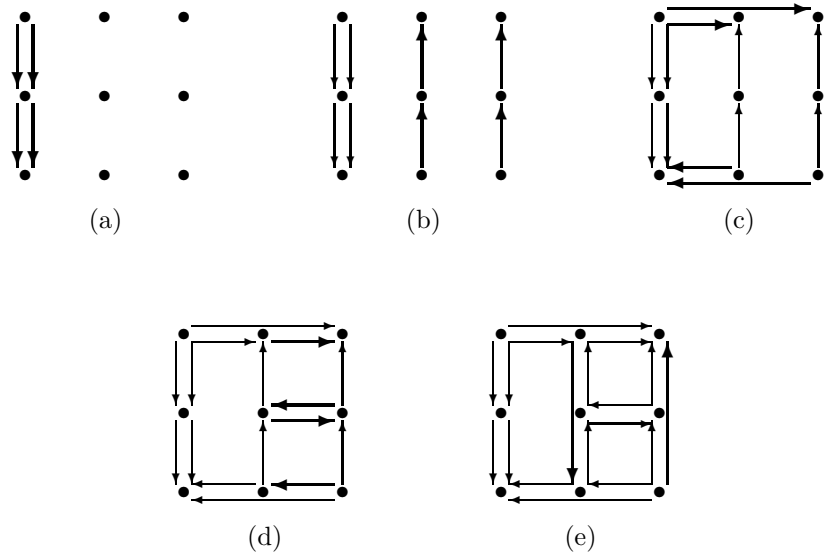
\begin{figure}[t]
\begin{centering}
\begin{picture}(310,220)(-5,5)
\thicklines
\multiput(0,210)(30,0){3}{\circle*{4}}
\multiput(0,180)(30,0){3}{\circle*{4}}
\multiput(0,150)(30,0){3}{\circle*{4}}
\put(-3,207){\vector(0,-1){24}}
\put(3,207){\vector(0,-1){24}}
\put(-3,177){\vector(0,-1){24}}
\put(3,177){\vector(0,-1){24}}
\put(24,130){(a)}
\multiput(120,210)(30,0){3}{\circle*{4}}
\multiput(120,180)(30,0){3}{\circle*{4}}
\multiput(120,150)(30,0){3}{\circle*{4}}
\thinlines
\put(117,207){\vector(0,-1){24}}
\put(123,207){\vector(0,-1){24}}
\put(117,177){\vector(0,-1){24}}
\put(123,177){\vector(0,-1){24}}
\thicklines
\put(150,183){\vector(0,1){24}}
\put(180,183){\vector(0,1){24}}
\put(150,153){\vector(0,1){24}}
\put(180,153){\vector(0,1){24}}
\put(144,130){(b)}
\multiput(240,210)(30,0){3}{\circle*{4}}
\multiput(240,180)(30,0){3}{\circle*{4}}
\multiput(240,150)(30,0){3}{\circle*{4}}
\thinlines
\put(237,207){\vector(0,-1){24}}
\put(243,207){\vector(0,-1){24}}
\put(237,177){\vector(0,-1){24}}
\put(243,177){\vector(0,-1){24}}
\put(270,183){\vector(0,1){24}}
\put(300,183){\vector(0,1){24}}
\put(270,153){\vector(0,1){24}}
\put(300,153){\vector(0,1){24}}
\thicklines
\put(243,213){\vector(1,0){54}}
\put(243,207){\vector(1,0){24}}
\put(297,147){\vector(-1,0){54}}
\put(267,153){\vector(-1,0){24}}
\put(264,130){(c)}
\multiput(60,90)(30,0){3}{\circle*{4}}
\multiput(60,60)(30,0){3}{\circle*{4}}
\multiput(60,30)(30,0){3}{\circle*{4}}
\thinlines
\put(57,87){\vector(0,-1){24}}
\put(63,87){\vector(0,-1){24}}
\put(57,57){\vector(0,-1){24}}
\put(63,57){\vector(0,-1){24}}
\put(90,63){\vector(0,1){24}}
\put(120,63){\vector(0,1){24}}
\put(90,33){\vector(0,1){24}}
\put(120,33){\vector(0,1){24}}
\put(63,93){\vector(1,0){54}}
\put(63,87){\vector(1,0){24}}
\put(117,27){\vector(-1,0){54}}
\put(87,33){\vector(-1,0){24}}
\thicklines
\put(93,87){\vector(1,0){24}}
\put(117,63){\vector(-1,0){24}}
\put(93,57){\vector(1,0){24}}
\put(117,33){\vector(-1,0){24}}
\put(84,10){(d)}
\multiput(180,90)(30,0){3}{\circle*{4}}
\multiput(180,60)(30,0){3}{\circle*{4}}
\multiput(180,30)(30,0){3}{\circle*{4}}
\thinlines
\put(177,87){\vector(0,-1){24}}
\put(183,87){\vector(0,-1){24}}
\put(177,57){\vector(0,-1){24}}
\put(183,57){\vector(0,-1){24}}
\put(213,63){\vector(0,1){24}}
\put(237,63){\vector(0,1){24}}
\put(213,33){\vector(0,1){24}}
\put(237,33){\vector(0,1){24}}
\put(183,93){\vector(1,0){54}}
\put(183,87){\vector(1,0){24}}
\put(237,27){\vector(-1,0){54}}
\put(207,33){\vector(-1,0){24}}
\put(213,87){\vector(1,0){24}}
\put(237,63){\vector(-1,0){24}}
\put(213,56){\vector(1,0){24}}
\put(237,33){\vector(-1,0){24}}
\thicklines
\put(207,87){\vector(0,-1){54}}
\put(243,33){\vector(0,1){54}}
\put(204,10){(e)}
\end{picture}
\caption{The steps to obtaining $|V_{21}|^4$ as a function of ordered,
non-degenerate boxes.  The additions in each step are designated by the thicker
arrows.
\label{V21fig}}
\end{centering}
\end{figure}


\end{document}